\documentclass[a4paper,10pt]{article}

\usepackage{graphicx}

\newtheorem{thm}{Theorem}

\newtheorem{tab}{Table}
\newtheorem{fig}{Figure}

\def\leurre{\noindent\leftskip0pt\small\baselineskip 10pt}
\def\grostrait{\ligne{\vrule height 1pt depth 1pt width \hsize}}
\def\demitrait{\ligne{\vrule height 0.5pt depth 0.5pt width \hsize}}

\def\encadre#1#2{%
\setbox100=\hbox{\kern#1{#2}\kern#1}
\dimen100=\ht100 \advance \dimen100 by #1
\dimen101=\dp100 \advance \dimen101 by #1
\setbox100=\hbox{\vrule height \dimen100 depth \dimen101\box100\vrule}
\setbox100=\vbox{\hrule\box100\hrule}
\advance \dimen100 by .4pt \ht100=\dimen100
\advance \dimen101 by .4pt \dp100=\dimen101
\box100
\relax
}

\def\ligne#1{\hbox to \hsize{#1}}
\def\PlacerEn#1 #2 #3 {\rlap{\kern#1\raise#2\hbox{#3}}}

\font\itix=cmti9
\font\rmx=cmr10
\font\rmxii=cmr12

\font\itx=cmti10
\font\ttx=cmtt10

\font\ttix=cmtt9

\font\rmix=cmr9 \font\mmix=cmmi9 \font\symix=cmsy9
\def\mathix{\textfont0=\rmix \textfont1=\mmix \textfont2=\symix}

\title{A family of weakly universal cellular automata in the hyperbolic plane 
with two states
\vskip 15pt
\rmxii
\ligne{\hfill Maurice Margenstern\hfill} 
\vskip 15pt
\rmx\baselineskip=12pt
\ligne{\hfill
Laboratoire d'Informatique Th\'eorique et Appliqu\'ee, EA 3097,\hfill}
\ligne{\hfill Universit\'e de Lorraine,\hfill}
\ligne{\hfill Campus du Saulcy,\hfill}
\ligne{\hfill 57045 Metz Cedex, France,\hfill}
\ligne{\hfill {\itx email:} {\ttx maurice.margenstern@univ-lorraine.fr}\hfill}
}

\begin{document}
\maketitle

\vskip 10pt
\begin{abstract}
In this paper, we construct a family of weakly universal cellular automaton
for all grids $\{p,3\}$ of the hyperbolic plane for $p\geq 13$.
The scheme is general for $p\geq 17$ and for $13\leq p<17$, we give 
such a cellular automaton for $p=13$, which is enough. Also, an important
property of this family is that the set of cells of the cellular automaton
which are subject to changes is actually a planar set.
The problem for $p<13$ for a truly planar construction is still open.
The best result, for $p=7$, is four states and was obtained by the same author. 
\end{abstract}
\vskip 10pt

\def\cqfd{\hbox{\kern 2pt\vrule height 6pt depth 2pt width 8pt\kern 1pt}}
\def\Hii{$I\!\!H^2$}
\def\Hiii{$I\!\!H^3$}
\def\Hiv{$I\!\!H^4$}
\def\norm{\hbox{$\vert\vert$}}
\section{Introduction}

   The first result about universality for cellular automata in a grid $\{p,3\}$
of the hyperbolic plane was obtained by the author and Y. Song, see~\cite{mmsyENTCS}.
The result was just constructed for $p=7$, {\it i.e.} the grid~$\{7,3\}$ called
the heptagrid. As there is no grid of this family 
when $p<7$, the construction in the heptagrid can easily be extended to the 
whole family. A bit later, the author obtained a significant improvement the heptagrid,
with four states only. It is still the best result if we consider a truly
planar cellular automata, which means that the cells which change during the
execution of the cellular automaton constitute a planar graph with many cycles.
In~\cite{mmRP2010}, the author has proved that there is a weakly universal
cellular automaton with two states only in all tilings $\{p,q\}$ of the hyperbolic 
plane. Now, the result was obtained by embedding the elementary cellular automaton 
rule~110 into the tiling. The problem is that this embedding produces a 
one-dimensional structure. 

   This paper is a first answer to this latter question about a planar weakly cellular
automaton with two states actually living in the hyperbolic plane. In this paper,
I construct such an automaton in infinitely many tilings of the hyperbolic plane.
More precisely, I do this for all tilings $\{p,3\}$ when $p\geq13$. The lower
bound $p=13$ leaves the question open for the cases when $7\leq p<13$.

   In this paper, as in many papers of the author alone or with
coauthors, see~\cite{fhmmTCS,mmsyBristol,mmsyPPL,mmbook2,mmsyENTCS}, 
I use the railway model to
define a weakly universal cellular automaton. As in~\cite{fhmmTCS} it is proved that
using such a model, one can simulate a two-register machine, it is enough to implement
such a simulation by a cellular automaton in the considered grid. This model
was never used in the Euclidean plane for cellular automata, but is used
for the same purpose in the hyperbolic plane.

    In planar simulations of the railway model, especially when we try to reduce the 
number of states, the difficult point is to model the crossings. This point will be
clear in the figures of Section~\ref{railway} for readers who are not familiar with 
the model. The main features of this model are explained in Section~\ref{railway}.
Note that in the $3D$-space, this problem vanishes as crossings can be easily
replaced by bridges. This is why simulations of the same model in the
hyperbolic $3D$-space have always be performed with less states than in the
planar cases. This is also why the threshold of two states with a true spatial
cellular automaton was reached by the author in a regular tiling of the 
hyperbolic $3D$-space, see~\cite{mmRP2011} before this paper. Now, it is interesting
to notice that if a previous result in the hyperbolic $3D$-space has taken benefit
from an idea used in a result in the hyperbolic plane, this paper takes benefit
from an idea used in~\cite{mmRP2011}. However, as there is no crossing in the
$3D$-simulation, something new had to be found for the plane.  
Subsection~\ref{scenario} explains this new idea and Section~\ref{implement} 
thoroughly describes the implementation.

    As most often, a computer program was used here too in order to check the
coherence of the set of rules, the cellular automaton obtained in the paper 
being rotation invariant. The rules and their construction are explained in
Section~\ref{rules}. Traces of executions of various pieces of the simulation
are given in Section~\ref{traces}. 
Section~\ref{rulesgene} gives a uniform rules for $\{p,3\}$ when
\hbox{$p\geq17$}. With this section, we shall reach the end of the proof
of the following result:

\begin{thm}\label{universal} {\rm(Margenstern)} $-$ 
There is a rotation invariant cellular automaton on all grids $\{p,3\}$ of 
the hyperbolic plane, with $p\geq13$, which is weakly universal and which 
has two states. There is a uniform set of rules for the automaton when $p\geq17$.
The 
initial configuration of the automaton is 
infinite: it is ultimately periodic along two 
different rays of mid-points~$r_1$ and~$r_2$ 
of the tiling~$\{p,3\}$
and finite in the complement of the parts attached to~$r_1$ and~$r_2$. The
set of cells which change their state at least once during the computation
is a planar graph with infinitely many cycles.
\end{thm}

   Section~\ref{hypgeom} remembers the reader with the main features of hyperbolic
geometry which are needed in order to understand the implementation.

\section{The railway circuit}
\label{railway}

   As initially devised in~\cite{stewart} and then mentioned
in~\cite{mmCSJMtrain,fhmmTCS,mmsyBristol,mmsyENTCS,mmbook2},
the circuit uses tracks represented by lines and quarters of circles and switches.
There are three kinds of switches: the {\bf fixed}, the {\bf memory} and the
{\bf flip-flop} switches. They are represented by the schemes given in
Fig.~\ref{aiguillages}.

\vtop{
\vspace{10pt}
\ligne{\hfill\includegraphics[scale=0.84]{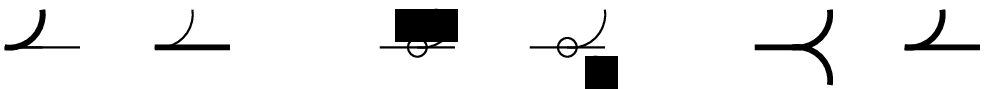}
\hfill}
\vspace{-5pt}
\begin{fig}
\label{aiguillages}
\leurre
The three kinds of switches. From left to right: fixed, flip-flop and memory switches.
\end{fig}
}

\vskip 10pt

   Note that a switch is an oriented structure: on one side, it has a single
track~$u$ and, on the the other side, it has two tracks~$a$ and~$b$. This 
defines two ways of crossing a switch. Call the way from~$u$ to~$a$ or~$b$
{\bf active}. Call the other way, from~$a$ or~$b$ to~$u$ {\bf passive}. The 
names comes from the fact that in a passive way, the switch plays no role on 
the trajectory of the locomotive. On the contrary, in an active
crossing, the switch indicates which track between~$a$ and~$b$ will be followed by
the locomotive after running on~$u$: the new track is called the {\bf selected}
track. 

   As indicated by its name, the {\bf fixed switch} is left unchanged by the 
passage of the locomotive. It always remains in the same position: when
actively crossed by the locomotive, the switch always sends it onto the same 
track. The flip-flop switch is assumed to be crossed actively only. Now,
after each crossing by the locomotive, it changes the selected track. 
The memory switch can be crossed by the locomotive actively and passively.
In an active passage, the locomotive is sent onto the selected track. Now, the
selected track is defined by the track of the last passive crossing by the 
locomotive. Of course, at initial time, the selected track is fixed.

\vtop{
\vspace{10pt}
\ligne{\hfill\includegraphics[scale=1.2]{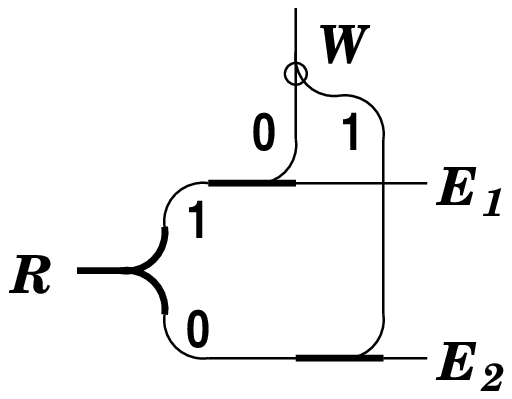}
\hfill}
\vspace{-5pt}
\begin{fig}
\label{element}
\leurre
The elementary circuit.
\end{fig}
}
\vskip 10pt

   With the help of these three kind of switches, we define an 
{\bf elementary circuit} as in~\cite{stewart}, which exactly contains one bit of 
information. The circuit is illustrated by Fig.~\ref{element}, above.
It can be remarked that the working of the circuit strongly depends on how
the locomotive enters it. If the locomotive enters the circuit through~$R$,
it leaves the circuit through~$E_1$ or~$E_2$, depending on the selected track
of the memory switch which stands near~$R$. If the locomotive enters through~$W$,
the application of the given definitions shows that the selected track at the
switches near~$R$ and~$W$ are both changed: the switch at~$W$ is a flip-flop which
is changed by the very active passage of the locomotive and the switch at~$R$
is a memory one which is changed because it is passively crossed by the 
locomotive and through the non-selected track. The just described actions of
the locomotive correspond to a {\bf read} and a {\bf write} operation on the 
bit contained by the circuit which consists of the configurations of the 
switches at~$R$ and at~$L$. It is assumed that the write operation is triggered
when we know that we have to change the bit which we wish to rewrite.

\vtop{
\vspace{5pt}
\ligne{\hfill
\includegraphics[scale=1.0]{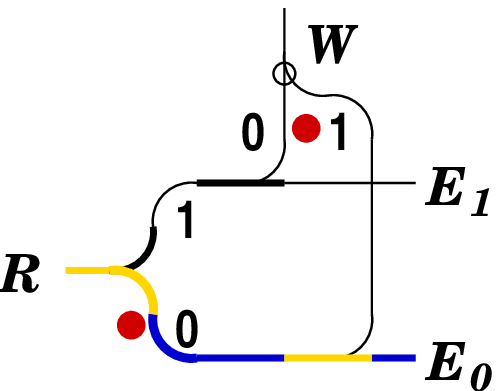}
\hskip 15pt
\includegraphics[scale=1.0]{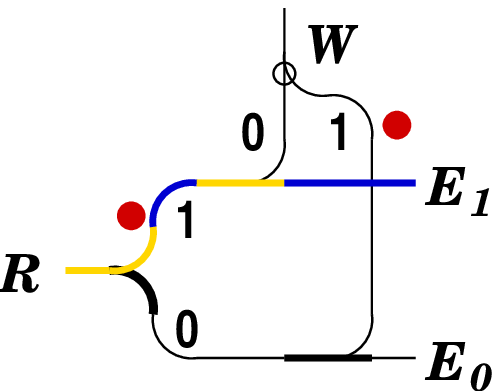}
\hfill}
\vspace{10pt}
\ligne{\hfill
\includegraphics[scale=0.5]{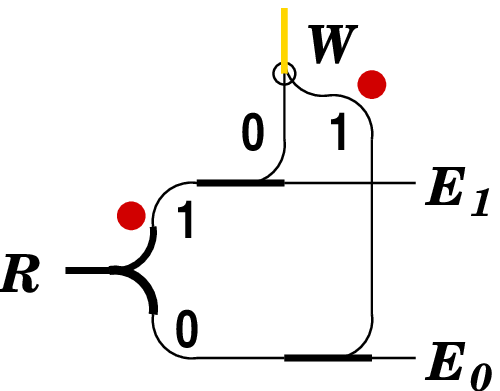}
\includegraphics[scale=0.5]{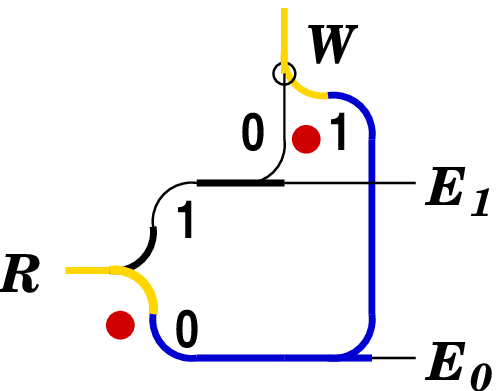}
\hskip 15pt
\includegraphics[scale=0.5]{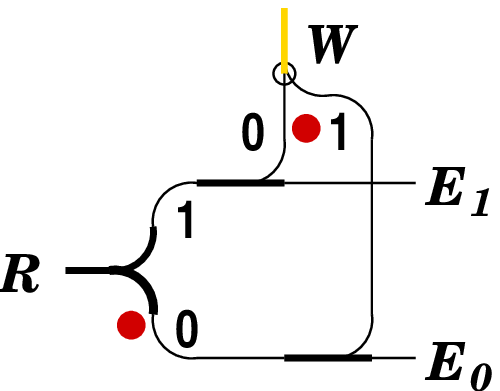}
\includegraphics[scale=0.5]{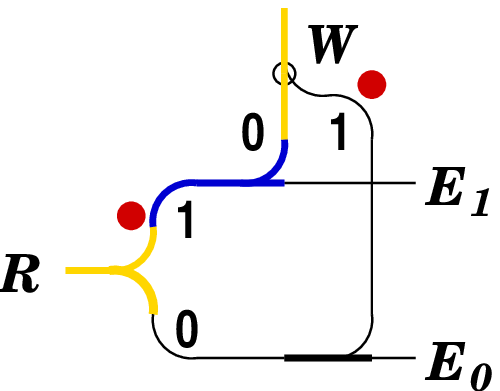}
\hfill}
\begin{fig}\label{elemexplic}
\leurre
Working of the elementary circuit.
Above: reading. Below: writing. To left, changing~$0$ into~$1$. To right,
changing~$1$ into~$0$.
\end{fig}
}

\vtop{
\vspace{10pt}
\ligne{\hfill
\raise-2pt\hbox{\includegraphics[scale=0.6]{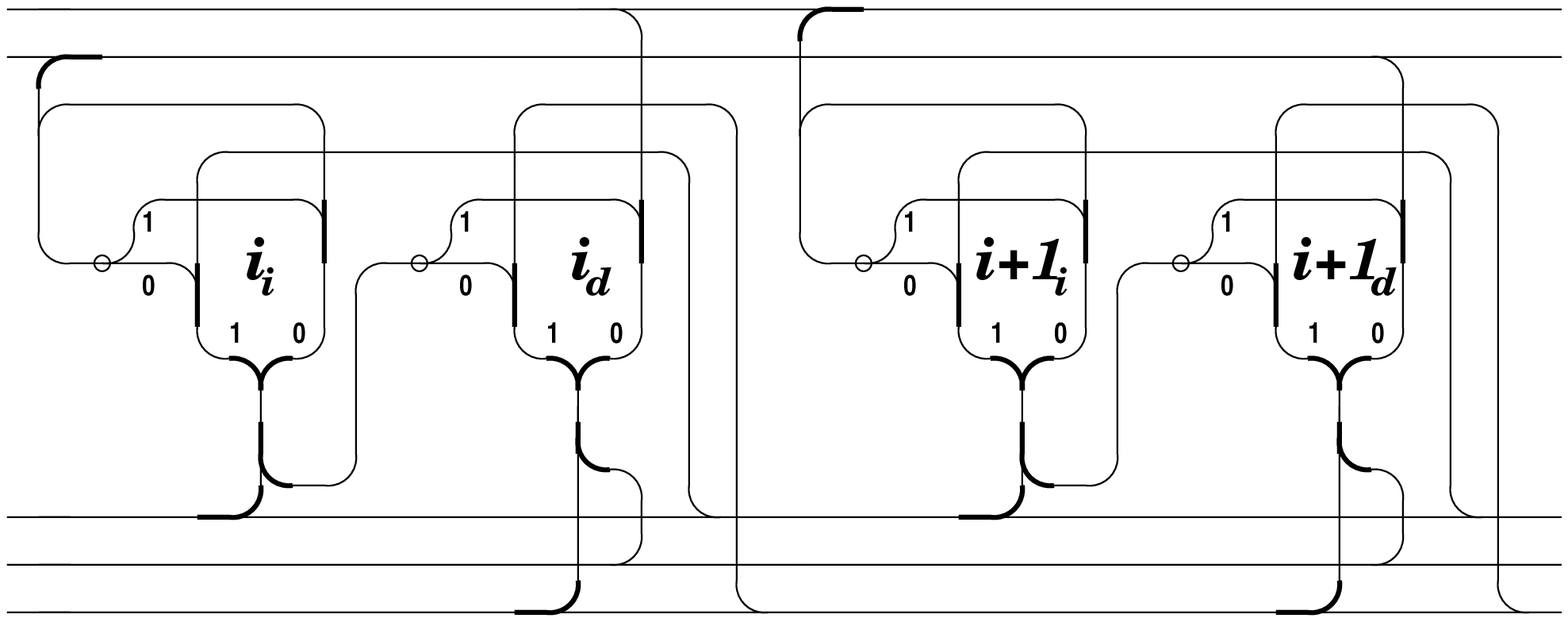}}
\PlacerEn {-335pt} {0pt} \box110
\PlacerEn {-328pt} {24pt} {\mathix$i$}
\PlacerEn {-328pt} {14.5pt} {\mathix$d$}
\PlacerEn {-328pt} {5pt} {\mathix$r$}
\PlacerEn {-328pt} {137pt} {\mathix$j_1$}
\PlacerEn {-328pt} {127pt} {\mathix$j_2$}
}
\vspace{-5pt}
\begin{fig}
\label{unit}
\leurre
Here, we have two consecutive units of a register. A register contains 
infinitely many copies of units. Note the tracks $i$, $d$, $r$, $j_1$ and~$j_2$.
For incrementing, the locomotive arrives at a unit through~$i$ and it leaves the
unit through~$r$. For decrementing, it arrives though~$d$ and it leaves 
also through~$r$ if decrementing the register was possible, otherwise, it leaves
through~$j_1$ or~$j_2$.
\end{fig}
}
\vskip 10pt
   Figure~\ref{elemexplic} illustrates this working.
As mentioned in the caption, the unit can be used in two ways. The reading way 
enters the circuit through the switch~$R$ and exits through the track~$E_1$
or~$E_2$ depending on what is read at~$R$. The writing way enters the circuit
through~$W$ and exits through~$R$. This is a special writing: it changes
the content of what is found into the opposite value. This means that
before writing we have to test what is in the unit. If we have to write the same
thing, nothing has to be done. If we have to write the other value, then
the enter through~$W$ is appropriate.
 
   The combination of such elementary circuits allows us to define more complex
structures which are useful to control the motion of the locomotive through the
circuit. As an example, Fig.~\ref{unit} illustrates
an implementation of a unit of a register.

   Other parts of the needed circuitry are described 
in~\cite{mmCSJMtrain,fhmmTCS}. The main idea in these different parts is
to organize the circuit in possibly visiting several elementary circuits
which represent the bits of a configuration which allow the whole system
to remember the last visit  of the locomotive. The use of this technique is
needed for the following two operations.

\vtop{
\vspace{10pt}
\ligne{\hfill
\includegraphics[scale=1.2]{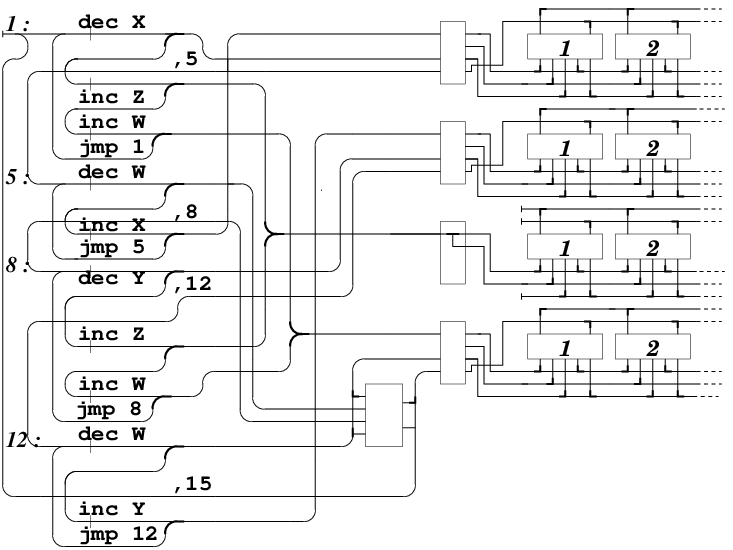}
\hfill}
\begin{fig}
\label{example}
\leurre
An example of the implementation of a small program of a register machine.
On the left-hand side of the figure, the part of the sequencer. It can be noticed
how the tracks are attached to each instruction of the program. Note that there 
are four decrementing instructions for~$W$: this is why a selector gathers 
the arriving tracks before sending the locomotive to the
control of the register. On the way back, the locomotive is sent on the right
track. 
\end{fig}
}
\vskip 10pt

   When the locomotive arrives to a register~$R$, it arrives either to 
increment~$R$ or to decrement it. As can be seen on Fig.~\ref{unit}, when the
instruction is performed, the locomotive goes back from the register by the
same track. Accordingly, we need somewhere to keep track of the fact whether
the locomotive incremented~$R$ or it decremented~$R$. This is one type of control.
The other control comes from the fact that several instructions usually apply
to the same register. Again, when the locomotive goes back from~$R$,
in general it goes back to perform a new instruction which depends on the one
it has just performed on~$R$. Again this can be controlled by what we called
the {\bf selector} in~\cite{mmCSJMtrain,fhmmTCS}.

   At last, the dispatching of the locomotive on the right track for the next 
instruction is performed by the {\bf sequencer}, a circuit whose main structure
looks like its implementation in the classical models of cellular automata such 
as the game of life or the billiard ball model. The reader is referred to the 
already quoted papers for full details on the circuit. Remember that this 
implementation is performed in the Euclidean plane, as clear from 
Fig.~\ref{example} which illustrates the case of a few lines of a program of 
a register machine.

   Now, before turning to the implementation in the hyperbolic plane, we provides the
reader with the minimal properties of hyperbolic geometry for a better understanding
of the paper.

\section{Short introduction to hyperbolic geometry}
\label{hypgeom}

   Hyperbolic geometry appeared in the first half of the 19$^{\rm th}$ century,
in the last attempts to prove the famous parallel axiom of Euclid's {\it Elements} 
from the remaining ones. Independently, Lobachevsky and Bolyai discovered a 
new geometry by assuming that in the plane, from a point out of a given line, 
there are at least two lines which are parallel to the given line. Later, 
models of the new
geometry were found, in particular Poincar\'e's model, which is the frame of
all this study.

\subsection{Poincar\'es disc model}
\label{Poincaredisc}

   In this model, the hyperbolic plane is the set of points
which lie in a fixed open disc~$\cal D$ of the Euclidean plane.
Let $\cal C$~be the border of~$\cal D$.
The lines of the hyperbolic plane in Poincar\'e's disc
model are either the trace of diametral lines or the trace of circles
which are orthogonal to~$\cal C$, see Fig.~\ref{model}.
We say that the considered lines or circles {\bf support} the hyperbolic
line, simply {\bf line} for short, 
\ligne{\hfill}

\vtop{
\ligne{\hfill
\includegraphics[scale=1]{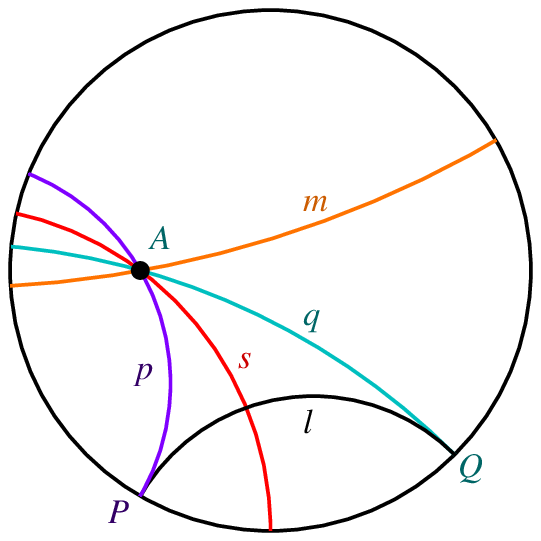}
\hfill}
\vskip 0pt
\begin{fig}\label{model}
\leurre
The lines $p$ and $q$ are {\bf parallel} to the line~$\ell$, with points at
infinity~$P$ and~$Q$, on the border of the unit disc. The $h$-line $m$ is
{\bf non-secant} with $\ell$: it can be seen that there are infinitely 
many such lines.
\end{fig}
}

\noindent
when there is no ambiguity, $h$-{\bf line}
when it is needed to avoid it. Fig.~\ref{model}
illustrates the notion of {\bf parallel} and {\bf non-secant} lines in
this setting. The points of~$\cal C$ which do not belong to the hyperbolic plane
are called {\bf points at infinity}. If $P$~is such a point, and if a the circle
which supports a line~$\ell$ of the model passes through~$P$, we say that $\ell$
passes through~$P$ and that $P$ is a point at infinity of~$\ell$. Points at infinity
play an important role in the hyperbolic plane. They define a kind of direction.

   The angle between two $h$-lines are defined as the Euclidean angle between
the tangents to their support. The reason for choosing the Poincar\'e's model
is that hyperbolic angles between $h$-lines are, in a natural way, the 
Euclidean angle between the corresponding supports. In particular, orthogonal circles 
support perpendicular $h$-lines.

   An important difference between Euclidean and hyperbolic geometries is the notion
of similarity. In Euclidean spaces, a figure can exist in different shapes. This is so
familiar that most probably, the reader does need a precise definition to understand
about what it is speaking. As an example, a square exists in infinitely many sizes: just
change the length of the side. This is not at all the case in hyperbolic spaces.
As an example, there is no square, there. But instead, there is regular convex 
pentagon with right angles between consecutive sides. Now, for this pentagon,
there is a unique length for the side. This leads us to say that in the hyperbolic
plane, a shape has a definite size. This does not mean, however, that there is not
at all similarity in the hyperbolic plane. Something can be said about that but this
is outside the scope of this paper.

\subsection{The family of tilings $\{p,3\}$}

   Remember that in the Euclidean plane and up to similarities,
there are only three kinds of tilings based on the recursive replication of 
a regular polygon by reflection in its sides and of the images in their sides. 
In the hyperbolic plane, where the notion of similarity is not very meaningful,
there are infinitely many such tilings. In this paper, we consider the
family of tilings $\{p,3\}$, $p\geq7$. They are defined on the basis of a regular 
convex polygon with an angle of $\displaystyle{{2\pi}\over 3}$ between consecutive
angles. It is known that such a polygon exists in the hyperbolic plane starting
from $p=7$. It is also known that for all these values of~$p$, it is possible to
tile the plane by replicating the polygon in its sides and, recursively, 
by replicating the images in their sides. This provides a tiling: there is no overlap
and no hole. As a consequence of the angle of the polygon, there are exactly three
of them around any vertex. The smallest polygon of the family is the heptagon, 
defined with $p=3$. It gives rise to the {\bf heptagrid} $\{7,3\}$.
Fig.~\ref{hepta} and~\ref{eclate_73} give an illustrative representation of this
tiling. We refer the reader to\cite{mmbook1} and to~\cite{mmDMTCS} for more details
and references.

   The left-hand side of Fig.~\ref{hepta} illustrates the heptagrid.
But, besides the occurrence of a lot of symmetries, nothing can be grasped on the
structure of the tiling from this mere picture. The right-hand side picture of
Fig.~\ref{hepta} illustrates the main tool to make the structure visible.
There, we can see two lines which we call {\bf mid-point lines} as they
join mid-points of edges of consecutive heptagons belonging to the tiling. On 
the figure,
a half of each line is drawn with a thicker stroke. It is a {\bf ray} issued
from the common point of these lines: here, a mid-point of an edge of the central
heptagon of the figure. We shall say a {\bf ray of mid-points}. These two rays 
define an angle, and the set of tiles
whose all mid-points of the edges fall inside the angle is called a {\bf sector}.

   Fig.~\ref{hepta} and~\ref{eclate_73} sketchily remember that the tiling 
is spanned by a generating tree. In fact, as can be noticed on both the right-hand
side of Fig.~\ref{hepta} and the left-hand side of Fig.~\ref{eclate_73}, the
set of tiles constituting a sector is spanned by a Fibonacci tree, 
see~\cite{mmbook1,mmDMTCS} for references. The name of the tree comes from
the fact that the number of nodes on a given level~$n$ is $f_{2n+1}$, where
$\{f_n\}$ denotes the Fibonacci sequence with $f_1=1$, $f_2=2$.

Now, as indicated in Fig.~\ref{eclate_73},
seven sectors around a central tile allow us to exactly cover the
hyperbolic plane with the heptagrid which is the tessellation
obtained from the regular heptagon described above and easily seen on the figures.

\vskip 14pt
\vtop{
\ligne{\hfill
\includegraphics[scale=0.45]{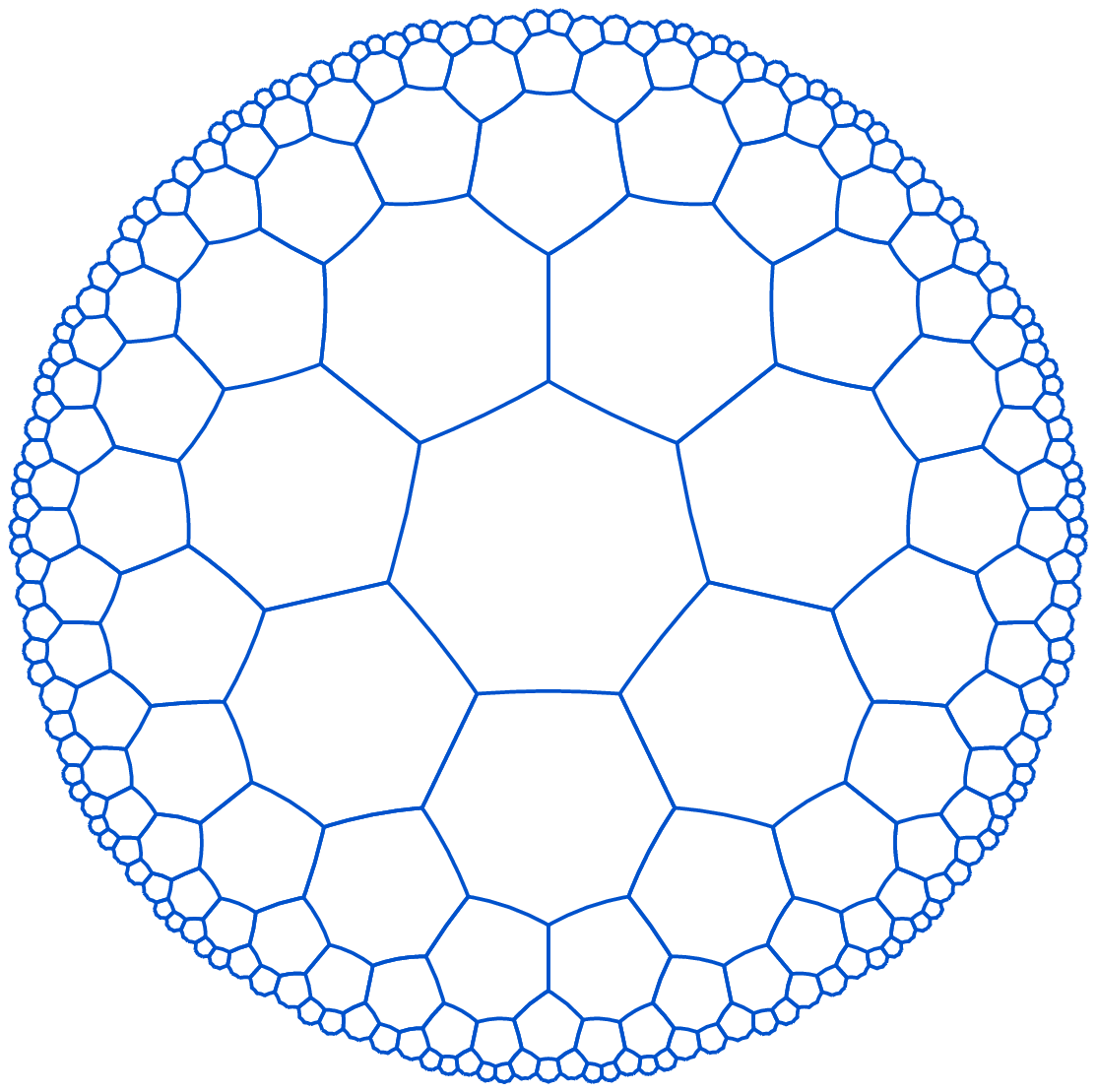}
\includegraphics[scale=0.6]{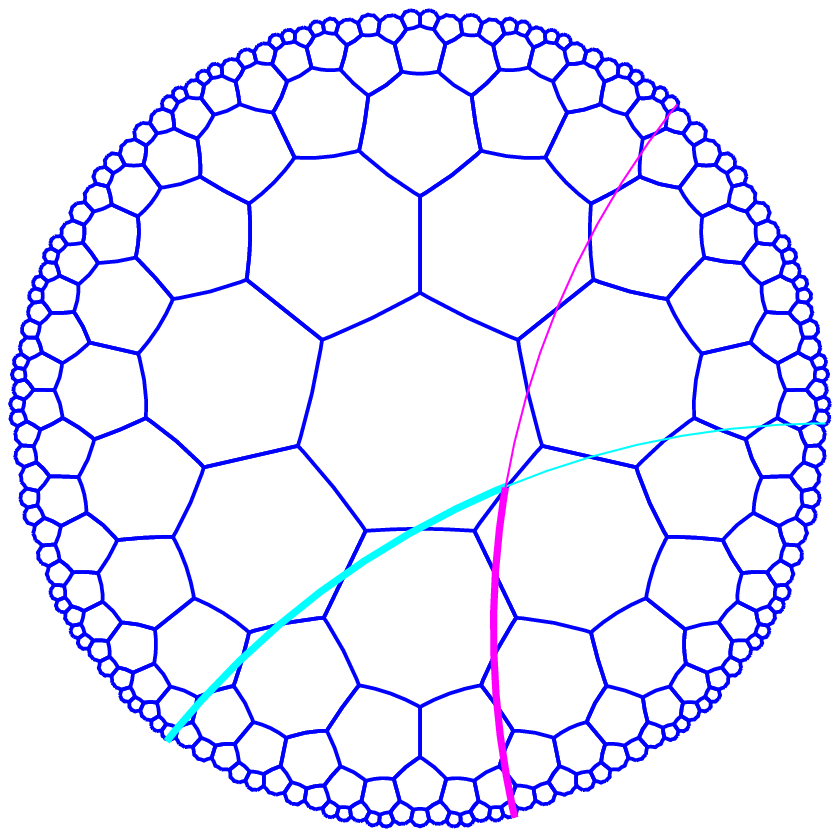}
\hfill}
\begin{fig}\label{hepta}
\leurre
On the left: the tiling; on the right: the delimitation of the sectors
which are spanned by a tree. Note the rays of mid-points. They are issued
from the same point: a mid-point of an edge of the central cell of the figure.
\end{fig}
}
\vskip 10pt

   In the left-hand side picture of Fig.~\ref{eclate_73},
we represent the sectors in terms of tiles. The tiles are in bijection 
with the tree which is represented on the right-hand side part of the figure.
This allows us to define the coordinates in a sector of the heptagrid, 
see \cite{mmbook1}. We number
the nodes of the tree, starting from the root and going on, level
by level and, on each level, from the left to the right. Then, we
represent each number in the basis defined by the quoted Fibonacci sequence,
taking the maximal representation, see\cite{mmJUCSii,mmbook1}.

\vskip 10pt
\vtop{
\ligne{\hfill
\includegraphics[scale=0.4]{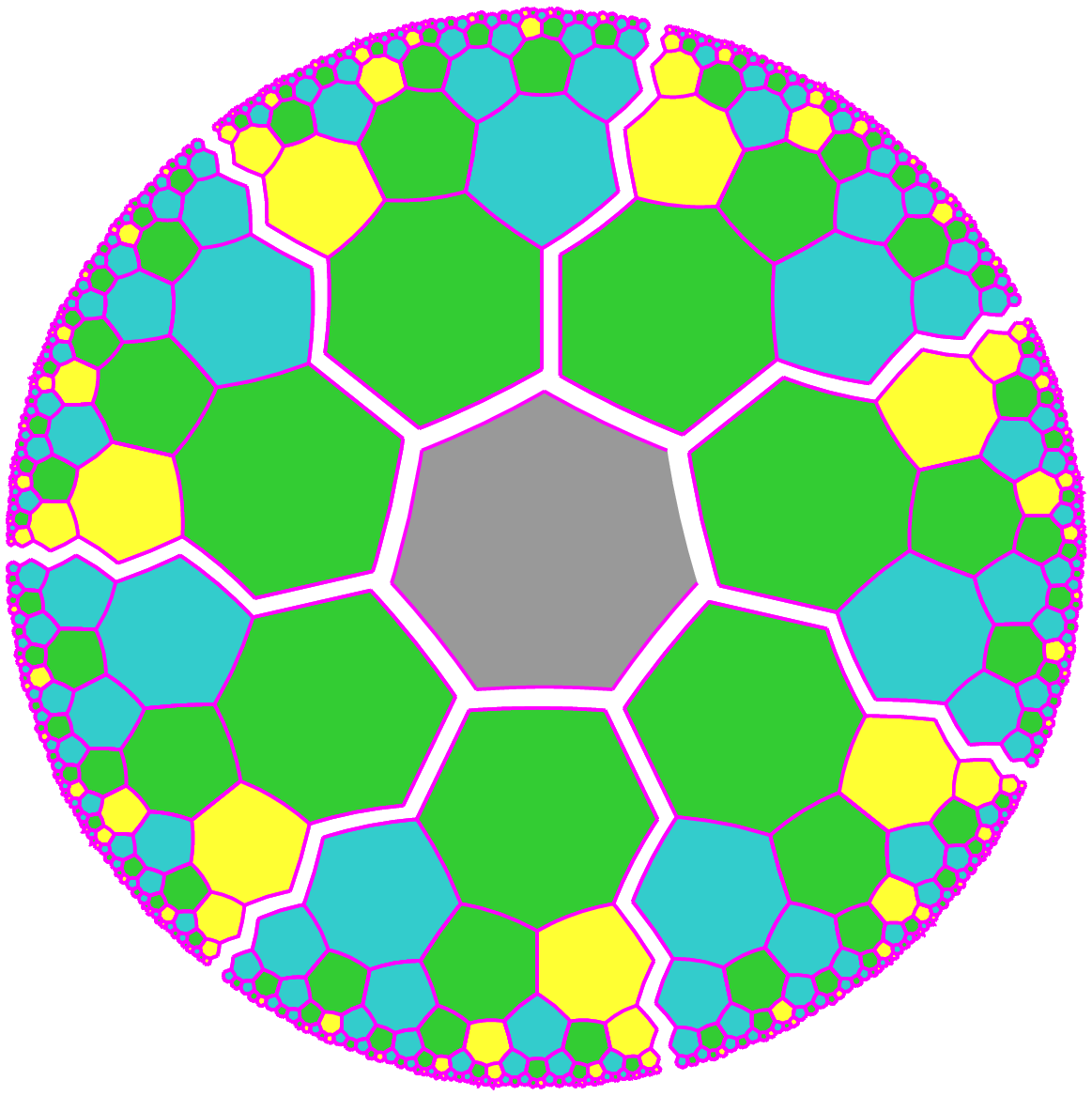}
\includegraphics[scale=0.95]{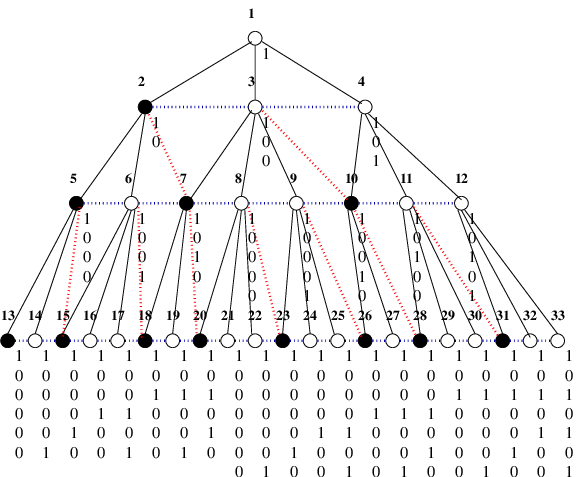}
\hfill}
\begin{fig}\label{eclate_73}
\leurre
On the left: seven sector around a central tile;
on the right: the representations of the numbers attached to the
nodes of the Fibonacci tree.
\end{fig}
}
\vskip 10pt

  One of the reasons to use this system of coordinates
is that from any cell, we can find out the coordinates of its
neighbours in linear time with respect to the coordinate of the cell.
Also in linear time from the coordinate of the cell, we can compute
the path which goes from the central cell to the cell. These properties
are established in \cite{mmASTC03,mmbook1} and they rely on a particular property
of the coordinates in the tree which allow to compute the coordinate of the 
father of a node in constant time from the coordinate of the node. In the paper,
the coordinate of a cell is of the form $\nu(\sigma)$ where $\sigma$
is the number of the sector where the cell is and $\nu$ is its number
in the Fibonacci tree which spans the sector.

   What was said for the heptagrid can be extended to any tiling $\{p,3\}$ with
$p\geq 7$, see~\cite{mmbook1}. We give here the main properties which can be used in 
Section~\ref{implement}. 

   Consider $p$ as fixed with $p\geq 7$. 
   Around a central cell, we can manage $p$~sectors defined by mid-point rays exactly
as this was the case in the heptagrid, illustrated by Fig.~\ref{eclate_73}. Here too,
there is a tree~$\cal T$ 
in bijection with the tiles which have all their mid-points in the angle defined 
by the rays. This tree generalizes the one given in the right-hand side of 
Fig.~\ref{eclate_73}. Let $u_n$ be the number of tiles which are on the level~$n$ of
the tree, with $u_0=1$ as level~0 is that of the root of the tree. We
have $p$$-$4 sons for the root which defines \hbox{$u_1=p$$-$$4$} and further,
\hbox{$u_{n+2}=(p$$-$$2)u_{n+1}-u_n$}. From this sequence, we can define a 
representation of the integers, see~\cite{fraenkel,hollander,mmbook1} as
follows: each positive number~$n$ can be written as 
$n=\displaystyle{\sum\limits_{i=1}^ka_iu_i}$ with $a_i\in[0..p$$-$$3]$.
In general, this representation is not unique, but it can be unique by requiring
it to be the maximal one in size. We can make this restriction more clear
as follows. Say that $\cal T$~has two kind of nodes: {\bf white} sons have
$p$$-$4 sons and {\bf black} sons have $p$$-$5 of them. The root is a white node.
Now, there is a precise rule for the position of black and white nodes. 
Each node has exactly one black node and, when running through the sons of a node
from left to right, the black node is the penultimate node. Number the nodes of
the tree level by level, 1 being given to the root and then, increasingly on
each level from left to right. Say that the maximal representation of~$n$ as
above defined is the {\bf coordinate} of the node numbered with~$n$. Then, we have
that the black nodes are exactly those whose coordinate end in~0. Similarly, 
the coordinate of the black son of a node is obtained by appending one~0 to the
coordinate of the node. This property allows us to define the path from a node 
to the root in linear time from the size of the coordinate of the node. 

   As we said, the nodes of~$\cal T$ are in bijection with the tiles of a sector.
The bijection allows us to consider as {\bf coordinate of a tile} the coordinate
of the node associated to the tile by the bijection. Say that two tiles are
{\bf neighbour} of each other if and only if they share a common side.
The main problem we have to solve is to compute the coordinates of the neighbour
of a tile~$\tau$ from the coordinate of~$\tau$. We refer the reader to~\cite{mmbook1}
for the result and the corresponding explanation. We display the results in 
Table~\ref{neighbours} in Section~\ref{implement}.

   Now, as the system of coordinates is fixed, we can turn to
the application to the implementation of cellular automata on the
heptagrid.

\section{Cellular automata on the grids $\{p,3\}$}
\label{hypCA}

   A cellular automaton  on a grid $\{p,3\}$ is defined by a {\bf local
transition function} which can be put in form of a table. Each
row of the table defines a {\bf rule} and the table has $p$+2 columns
numbered from~0 to~$p$+1, each entry of the table containing a state of the
automaton. On each row, column~0 contains the state of the cell
to which the rule applies. The rule applies because columns~1 to~$p$
contain the states of the neighbours of the cell defined in the following
way. For the central cell, its neighbour~1 is fixed once and for all. 
For another cell, its neighbour~1 is its father. In all cases, the other 
neighbours are increasingly numbered from~2 to~$p$ while counter-clockwise turning 
around the cell starting from side~1. The representation mentioned in 
Section~\ref{hypgeom}  allows to find the coordinates of
the neighbours from that of the coordinate of the cell in linear time.
As promised in that section, Table~\ref{neighbours} indicates how to compute 
the coordinates of the neighbours of a cell~$\nu$ from the coordinate of~$\nu$.
The list of states on a row, from column~0 to~$p$ is called the {\bf context}
of a rule. It is required that two different rules have different contexts.
We say that the cellular automaton is {\bf deterministic}. As there is a
single row to which a rule can be applied to a given cell, the state of
column~$p$+1 defines the {\bf new state} of the cell. The local transition function
is the function which transforms the state of a cell into its new one, also
depending on the states of the neighbours as just mentioned.
In Table~\ref{neighbours}, we introduce a function $\chi(\nu)$ which indicates
the lowest digit of the coordinate of~$\nu$. It also defines the notion of 
{\bf common ancestor} of two cells: 

   An important case in the study of cellular automata is what are called
{\bf rotation invariant} cellular automata. To define this notion,
we consider the following transformation on the rules. Say that
the context of a rule is the {\bf rotated image} of another one if and only
if both contexts have the same state in column~0 and if one context is obtained 
from the other by a {\bf circular} permutation on the contents of columns~1 to~$p$. 
Now, a cellular automaton is {\bf rotation invariant} if and only if its table of 
transition~$T$ possesses the following properties:

{\leftskip 20pt\parindent 0pt
- for each row~$\rho$ of~$T$, $T$ also contains $p$$-$1 rules exactly
whose contexts are the rotated image of that of~$\rho$ and whose new state
is that of~$\rho$;

- if $\rho_1$ and~$\rho_2$ are two rules of~$T$ whose contexts are 
the rotated image of each other, then their column~$p$+1 contains the same state.
\par}

   In the rest of the paper, sometimes we shall have to write the rules
of the automaton for a precise situation. The rules can be written according to the 
following format:
\vskip 3pt
\ligne{\hfill 
$\eta_0$, $\eta_1$, $\ldots$, $\eta_{p} \rightarrow \eta_0^1$,\hfill}
\vskip 2pt
\noindent 
where $\eta_0$ is the state of the cell,
$\eta_i$ the state of its neighbour~$i$ and~$\eta_0^1$ is its new state.

   However, in tables and also in order to have a more compact notation, a
rule will be written as a word. The above is rewritten as the following
word: 
$\underline{\eta_0}\eta_1\ldots\eta_{p}\underline{\eta_0^1}$, using the same 
notations. 

   The name of rotation invariance comes from the fact that a rotation around
a tile~$T$ leaving the tiling globally invariant is characterized by a circular
permutation on the neighbours of~$T$ defined as above.

   Note that the universal cellular automata devised 
in~\cite{fhmmTCS,mmsyBristol,mmsyENTCS} are rotation invariant while the one 
of~\cite{mmkmTCS} is not. 
For the question of rotation invariance for cellular
automata on the heptagrid, we refer the reader to~\cite{mmACMC07}.

\def\ttV{\vrule depth 6pt height 12pt width 0pt}
\def\ttH{\hrule depth 0pt height 0pt width \hsize}
\def\cette_rangee #1 #2 #3 {\ligne{\ttV\hbox to 50pt{\hfill#1\hfill}
                                \ttV\hbox to 60pt{\hfill#2\hfill}
                                \ttV\hbox to 210pt{\hskip 10pt#3\hfill}
                                \ttV}
                        }
\vtop{
\begin{tab}\label{neighbours}
\leurre
Computation of the coordinates of the neighbours of a node.
\end{tab}
\vspace{-12pt}
\grostrait
\vspace{-8pt}
\vtop{\offinterlineskip\leftskip0pt\parindent0pt\hsize=330pt
             \ttH
             \cette_rangee {$\chi(\nu)$} {$\chi(f$+1)} {neighbours}
}
\vspace{0pt}
\demitrait
\vspace{-6pt}
\vtop{\offinterlineskip\leftskip0pt\parindent0pt\hsize=330pt
             \ttH
             \cette_rangee 0 {} {$f$, $c_1$+1, $c_1$+2, $\ldots$, $c_1$+$p$$-$4,
                              $c$, $c$+1, $c$+2}
             \ttH
             \cette_rangee 1 1 {$f$, $c_1$+2, $\ldots$, $c_1$+$p$$-$3,
                              $c$, $c$+1, $c$+2}
             \cette_rangee {} {3,...,$p$$-$3,0} {$f$, $c_1$+2, $\ldots$,
                              $c_1$+$p$$-$3, $c$, $c$+1, $f$+1}
             \cette_rangee {} 2 {$st(g)=0$: $f$, $c_1$+2, $\ldots$,
                              $c_1$+$p$$-$3, $c$, $c$+1, $c$+2}
             \cette_rangee {} {} {$st(g)=1$: $f$, $c_1$+2, $\ldots$,
                              $c_1$+$p$$-$3, $c$, $c$+1, $f$+1$^\circ$}
             \ttH
             \cette_rangee 2 2 {$f$, $f$$-$1, $c_1$+2, $\ldots$, $c_1$+$p$$-$3,
                              $c$, $c$+1}
             \cette_rangee {} {4,...,$p$$-$3,0,1} {$f$, $c_1$+1, $c_1$+2,
                              $\ldots$, $c_1$+$p$$-$3, $c$, $c$+1}
             \cette_rangee {} 3 {$st(g_1)=0$: $f$, $f$$-$1, $c_1$+2,
                                $\ldots$, $c_1$+$p$$-$3, $c$, $c$+1}
             \cette_rangee {} {} {$st(g_1)=1$: $f$, $c_1$+1, $c_1$+2,
                                $\ldots$, $c_1$+$p$$-$3, $c$, $c$+1$^\circ$}
             \ttH
             \cette_rangee {3,...,$p$$-$$3^\ast$} {} {$f$, $c_1$+1, $c_1$+2,
                                $\ldots$, $c_1$+$p$$-$3, $c$, $c$+1}
             \ttH
    }
\vskip 2pt
\demitrait
\vspace{0pt}
\noindent
{\leurre {\itix Note.} Here,
$\nu$ is the node, $f$ its father, $c$ its black son;
$c_1$ is the black son of $\nu$$-$$1$.
when $\chi(\nu)=1$ or $\chi(nu)=2$, define $h$ as the common ancestor
of $\nu$ and $\nu$$+$$1$, when $\chi(\nu)=1$,
of $\nu$$-$$1$ and $\nu$, when $\chi(nu)=2$; we denote by
$g$ and $g_1$ the sons of~$h$ on respectively the left-hand side, right-hand
side branch issued from~$h$.\vskip0pt
\noindent
$^\ast$ When $f$ is black, $\chi(\nu)$ ranges only in $[3..p$$-$$4]$.
\vskip0pt
\noindent
$\circ$ When $\nu$ is on the border, $c_1$$+$$1$ is to be considered on the
left-hand side tree and $f$$+$$1$ is to be considered on the right-hand side tree.
\par}
}
  
   Now, we can turn to the simulation of the railway circuit by a cellular
automaton.

\section{Implementation in the hyperbolic plane}
\label{implement}

   In this section, Subsection~\ref{hyprailway} will first present
the general features of the implementation
of a railway circuit which are shared by all the simulations of the previous
papers, see~\cite{fhmmTCS,mmsyBristol,mmsyPPL,mmbook2,mmsyENTCS,mmarXiv3D}, as well
as the implementation of this paper. Then, in Subsection~\ref{scenario}, we describe 
the main lines of the present implementation. Then, we develop the presentation
thoroughly in Subsections~\ref{tracks}, \ref{crossings} and~\ref{switches}.

\subsection{The implementation of the railway circuit}
\label{hyprailway}

   In~\cite{fhmmTCS}, the various elements of the circuit mentioned 
in~\cite{mmCSJMtrain} are implemented. In fact, the paper does not give an exact
description of the implementation: it only gives the guidelines, but with
enough details, so that an exact implementation is useless. In this paper,
we take the same model, and we repeat the main lines of implementation mentioned
in~\cite{fhmmTCS,mmsyENTCS}. So that we refer the reader to these papers for more
precise details. Just to help him/her to have a better of view of the overall
configuration, we refer the reader to Fig.~\ref{hypexample}. The figure provides
a simplified illustration of the implementation of the example given by 
Fig.~\ref{example} which takes place in the heptagrid. 
  
   If the reader carefully looks at the figure, he/she will notice that
the tracks mostly follow branches of a Fibonacci tree and sets of nodes which
are on the same level of the Fibonacci tree. In this implementation, we have
to pay a very precise attention to this situation. We shall tune it a bit
with the help of an intermediary structure. As it will be used for the initial
configuration only, there is no need to translate this structure into the states
of the automaton.

\vtop{
\ligne{\hfill
\includegraphics[scale=0.35]{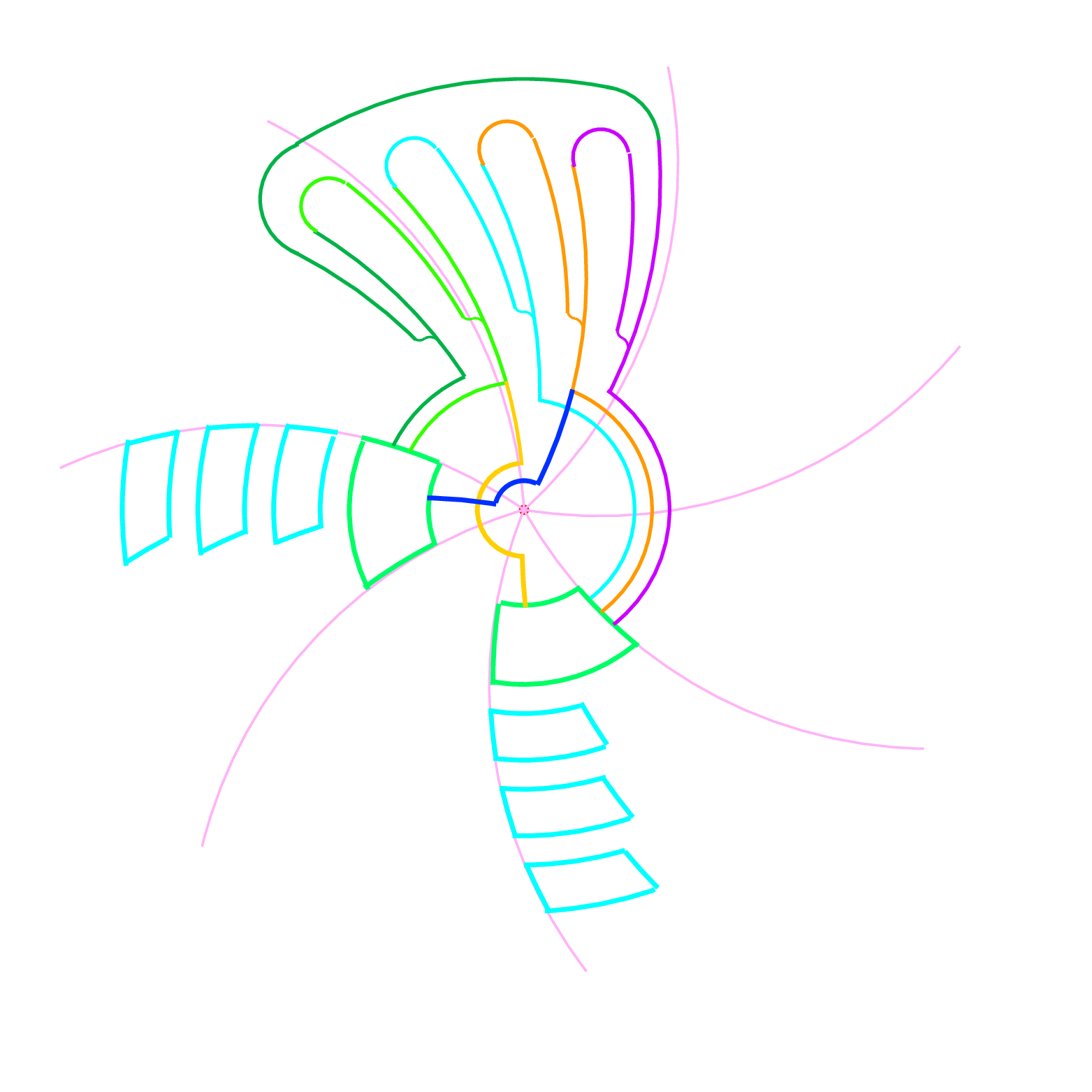}
\hfill}
\vspace{-15pt}
\begin{fig}\label{hypexample}
\leurre
The implementation, on the heptagrid, of the example of Fig.~{\rm\ref{example}}.
In sector~$1$, also overlapping onto sector~$2$, the sequencing of the 
instructions of the program of the register machine. In sector~$3$, we can see
the first register and,  in sector~$5$, the second one. For simplicity, the 
figure represents two registers only.
\vskip 0pt
Note the instructions which arrive to the control of the register
through tracks in the shape of an arc of circle. Also note the return from the 
controller of the register when decrementing a register fails, because its content
was zero.
\end{fig}
}

   The intermediary structure which we shall use is illustrated in 
Fig~\ref{hortandvert} in the case of the heptagrid. It consists in defining 
{\bf horizontals} and {\bf verticals}. Horizontals and verticals are a familiar 
way to define a rectangular grid. However, we can easily understand that we can 
distort the structure provided that the cycles defined by the rectangles are 
preserved. We can also alter the structure provided that we keep enough room for 
infinitely many zones of arbitrary sizes. In the hyperbolic plane, we have no 
rectangle but, in the tilings $\{p,3\}$ of the hyperbolic plane, we have a generic
way to define a kind of horizontal: it is a set of tile whose complement in the plane
has two components exactly and two of our horizontals either coincide or do not meet.
Also, we can guarantee the existence of only half verticals: they are 
sets of tiles which follow a half-line and they cross infinitely many horizontals. 
Fig.~\ref{hortandvert} illustrates this notion. This is enough for our purpose: 
the intersections between verticals and horizontals allow us to define quadrangle 
as large as needed to enclose the structures we have to define to implement the
circuits described in Section~\ref{railway}.

\vskip 15pt
\vtop{
\ligne{\hfill
\includegraphics[scale=0.5]{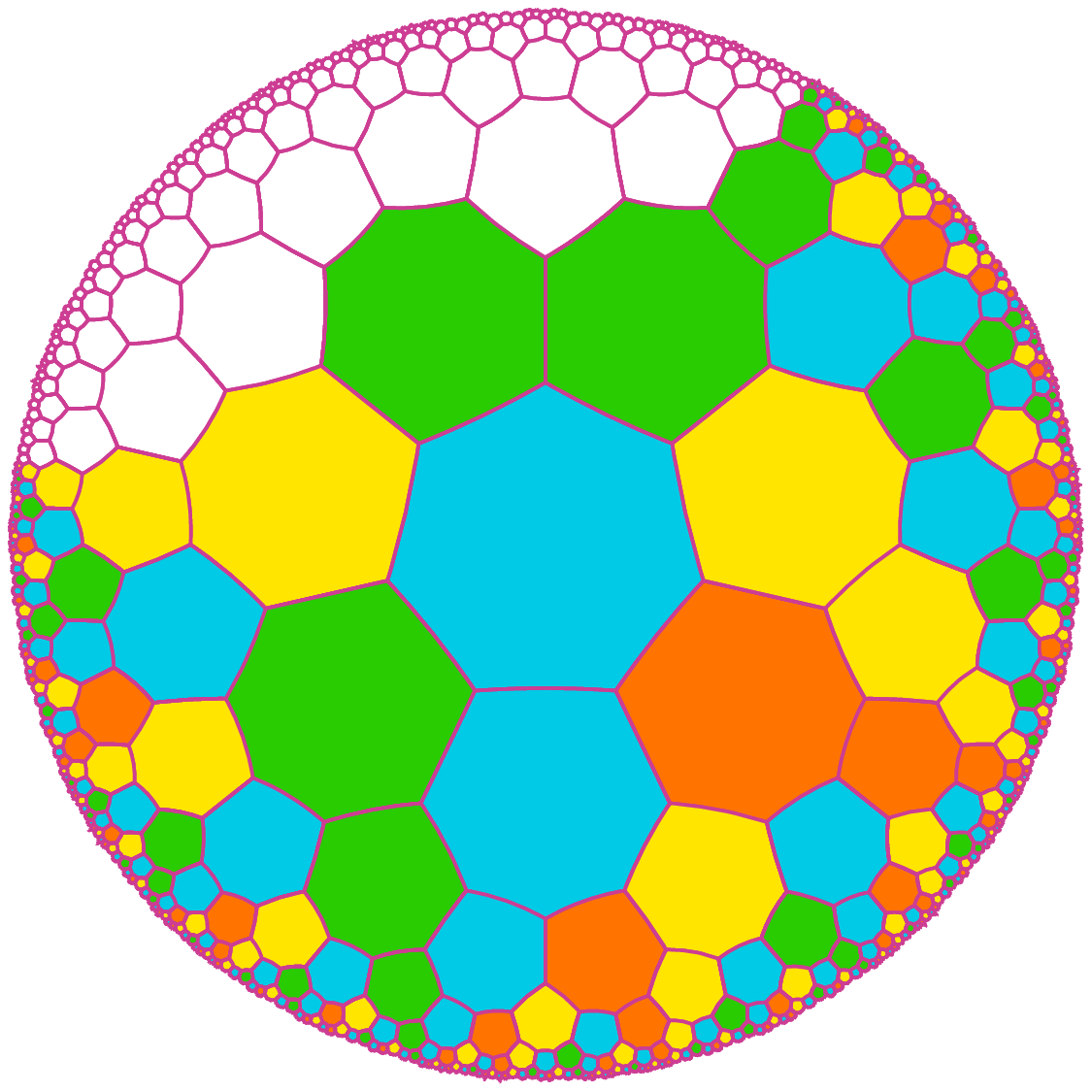}
\includegraphics[scale=0.5]{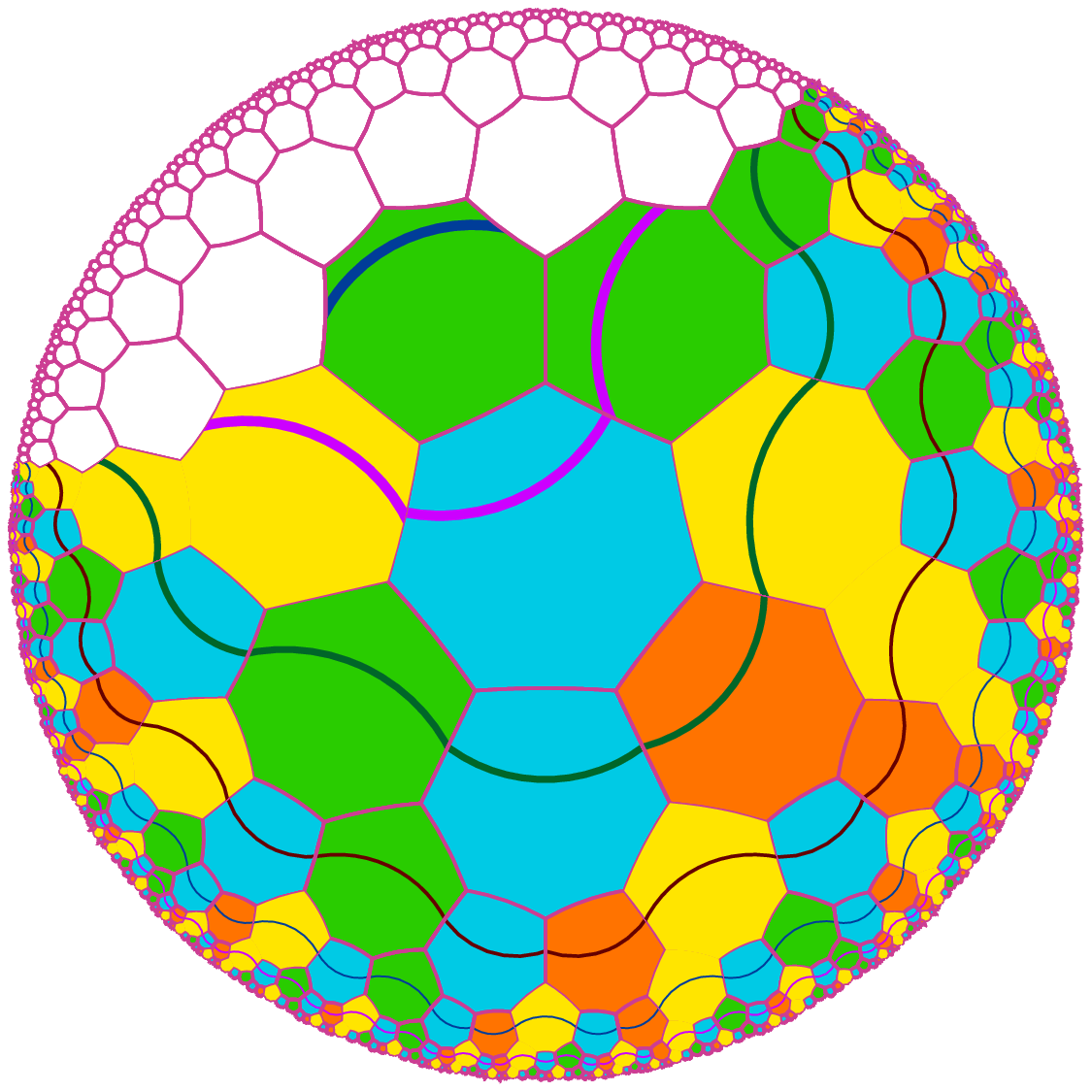}
\hfill}
\begin{fig}\label{hortandvert}
\leurre
The definition of horizontals and verticals:
\vskip 0pt
On the left-hand side, the coloration which allows to define the horizontals,
which are drawn on the right-hand side.
The verticals are represented by rays of yellow tiles. On the right-hand 
side picture, note that the common side of adjacent yellow tiles is not drawn: 
rays of yellow tiles appear as solid blocks of tiles. 
\end{fig}
}
\vskip 10pt
   Fig.~\ref{hortandvert} is also based on the tree represented in 
Fig.~\ref{eclate_73}. Now, the colours of the picture have a meaning.
Each tile defines the colours of its sons according to the following rules:

\vskip 5pt
\ligne{\hskip 40pt
$G \rightarrow YBG$,\hfill $Y \rightarrow YBG$,\hfill $O \rightarrow YBO$,
\hfill $B\rightarrow BO$,\hskip 40pt}
\noindent
where $G$, $Y$, $O$ and~$B$ have obvious meaning.

   The horizontals are defined by the levels of the tree. Now, a level is delimited
by the leftmost and the rightmost branches of the tree. In order to constitute an 
infinite horizontal, the levels are glued by considering an infinite sequence
of growing trees whose roots are set along an infinite vertical. The union of these
trees is the whole hyperbolic plane. Then the level~$n$ of a tree is the level~$n$+1 
of the smallest tree of the sequence which contains it, see~\cite{mmbook1}. Now, 
each branch of any tree can be a vertical. We restrict the choice to the branches
issued from a yellow tile and which consists of yellow tiles only, see 
Fig.~\ref{hortandvert}.

   It is important to notice that the fact that our verticals are rays only does not
prevent them from constituting a {\bf grid} with the horizontals we defined. We have
the fact that in between two verticals starting from a horizontal~$\iota$, new 
verticals appear as we go down from a horizontal to the next one, starting 
from~$\iota$. We may ignore these new verticals if we do not need them. And so, 
these verticals with the piece of~$\iota$ and a piece of another deeper level at 
which we decide to stop constitute a figure which we may call a {\bf quadrangle}. 
These quadrangles allow us to implement the pieces of circuitry described in 
Section~\ref{railway}. For the registers, it is enough to display such quadrangles 
in such a way that the quadrangles have a side along the same yellow branch of a 
tree. This is enough to see that we can consider the setting of 
Fig.~\ref{hypexample} as enough for our purpose.

   This situation of the heptagrid can be transported to any grid $\{p,3\}$
with $p\geq 7$. The same definition of the mid-point lines allows us to transport
the constructions. The difference with the heptagrid is that in $\{p,3\}$, the 
larger~$p$, the larger the number of sons for a node and so, the higher the exponential
rate of the growth of a disc around a tile.

Later on, we shall illustrate our construction in the
tiling $\{13,3\}$, but we shall not represent it in Poincar\'e's model as we can see 
almost nothing in such a representation. Instead, we shall use a symbolic 
representation with circles: a circle will represent a 13-gon and a common side 
of two 13-gons will be defined by a tangential contact between the circles 
representing the 13-gons. A common vertex to three 13-gons will appear as a
kind of triangle between the tangency points of three circles which are pairwise 
tangent. As an example, Fig.~\ref{treize} represents a 13-gons with its neighbours.

\vtop{
\ligne{\hfill
\includegraphics[scale=0.80]{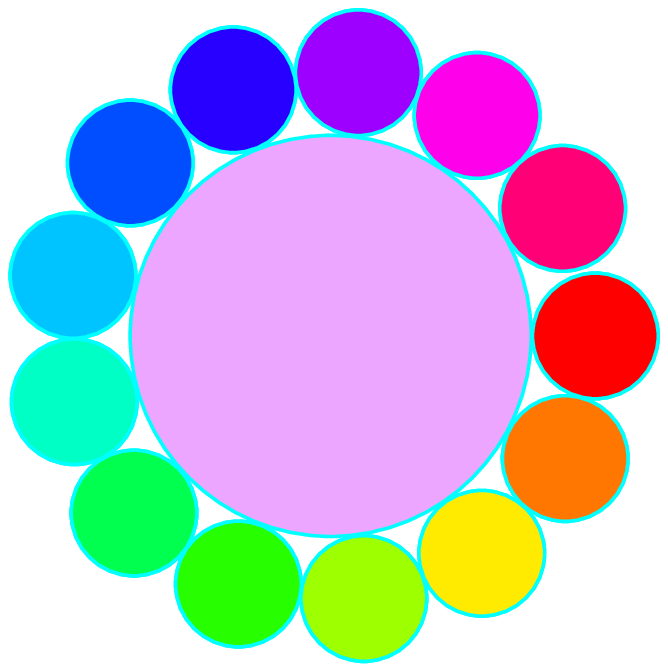}
\hfill}
\vspace{-35pt}
\begin{fig}\label{treize}
\leurre
Representation of a cell in the tiling $\{13,3\}$ with its neighbours using
circles. Note the role of the tangency points.
\end{fig}
}

\subsection{The scenario of the present implementation}
\label{scenario}

   The present scenario is based on three ideas. One appeared in~\cite{mmsyENTCS}.
The previous implementations represented the tracks as a linear structure on which
the locomotive appeared, the structure being defined by a colour, different from that
of the background. In~\cite{mmsyENTCS}, we introduced a new implementation of the 
tracks. This time, the track does not differ from the background: it is simply 
delimited by regularly dispatched milestones on both sides of the track. In this
new representation, the locomotive was represented by a block of two contiguous cells,
with two colours, different from that of the background. In~\cite{mmarXiv3D}, a new 
idea allowed us to establish the existence of a spatial two-state universal cellular 
automaton. With two states, it is no more possible to use two colours for the 
locomotive. Moreover, with a single cell, it is impossible to distinguish the front
from the rear on the locomotive itself. This is why the locomotive was there reduced to
a single cell, and it was now called a particle. The counterpart was to change 
the implementation of the tracks. It was defined by the fact that a track is now
one-way. Accordingly, in the cases when the tracks have to be crossed by the locomotive
in both directions, the corresponding tracks are implemented as two one-way track: one
in one direction, the other, in the opposite direction. This raises a change in the
switches which are illustrated by Fig.~\ref{newswitches}. 

   Indeed, as the flip-flop switch must be crossed actively only, it is implemented
by one-way tracks only, see the red pattern of the memory switch in 
Fig.~\ref{newswitches}. Due to its definition, the fixed switch has a passive one-way
switch, see Fig.~\ref{newswitches} and an active one-way track with no switch at all.

\vtop{
\ligne{\hfill
\includegraphics[scale=0.80]{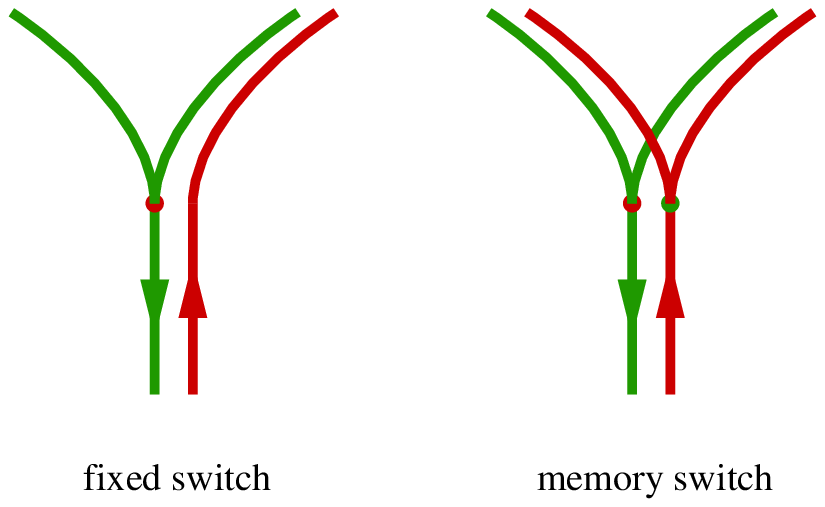}
\hfill}
\vspace{-35pt}
\begin{fig}\label{newswitches}
\leurre
The new switches. Note that the red part of the memory switch also implements
the flip-flop switch.
\end{fig}
}

   On the contrary, the memory switch is the superposition of two switches:
an active and a passive one. On one hand, the active switch is passive: it is 
in fact a {\bf programmable} fixed switch. The selected track depends from the 
last passive passage of the particle. When it is fixed, the active switch is
also fixed. On the other hand, the passive switch is in fact active: it may
react to the passage of the particle. If the particle goes through the selected
track, nothing happens. Now, if it arrives through the non-selected track, then
the switch changes the selected track to a non-selected one and conversely.
But the switch also acts actively: it triggers a signal sent to the passive part
in order to change the selection also there. 

   Thanks to the bridges in the hyperbolic $3D$ space, this new idea was enough to
go down to two states for a universal cellular automaton.

   Now, for the plane, it is needed to find a new idea as crossings cannot be avoided.
And so, for this implementation, we introduce a new pattern which we call a
{\bf round-about}: it incorporates a motorway pattern to the railway circuits. 
The pattern deals with the crossing of two one-way tracks, see 
Fig.~\ref{rondpointsimple}.

\vtop{
\vspace{-40pt}
\ligne{\hfill
\includegraphics[scale=0.80]{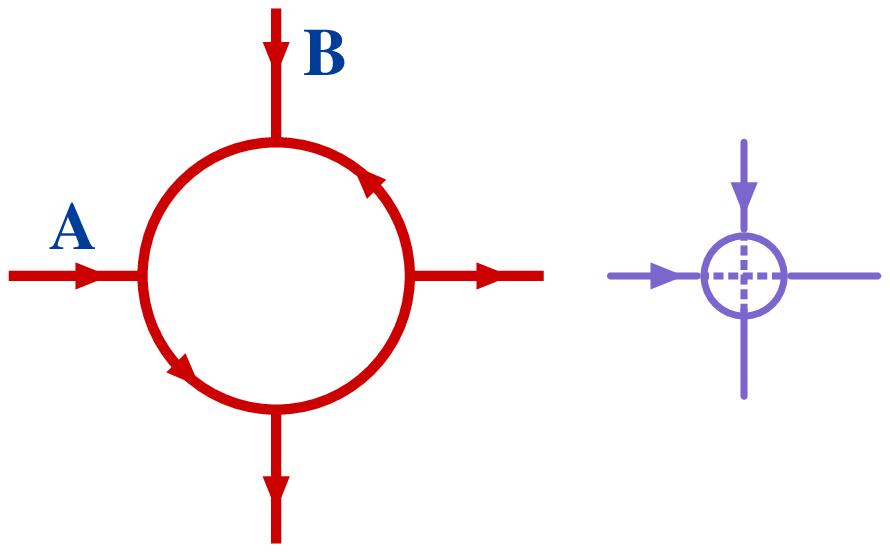}
\hfill}
\vspace{-35pt}
\begin{fig}\label{rondpointsimple}
\leurre
Left-hand side: scheme of the simple round-about. Right-hand side: its symbolization
for later use. 
\end{fig}
}

Note that whether the particle arrives from~$A$ or from~$B$ to the round-about, 
it has to exit at the second track meeting the round-about. And so we have to count
up to~2. But we have two states only. The solution is the following: when the 
particle arrives at the round about, an additional particle is appended to it. This
pair of consecutive particles goes on along the round-about. When it meets the
first way going away or arriving to the round-about, the first particle vanishes and
the second one follows the round about. This time, when a single particle arrives
to a track which goes out from the round-about, the particle follows this track
and so, it leaves the round-about on the right way.

   It is now not very difficult to define the crossing of two two-way tracks, which
is illustrated in Fig.~\ref{rondpoint}.

\vtop{
\vspace{-40pt}
\ligne{\hfill
\includegraphics[scale=0.70]{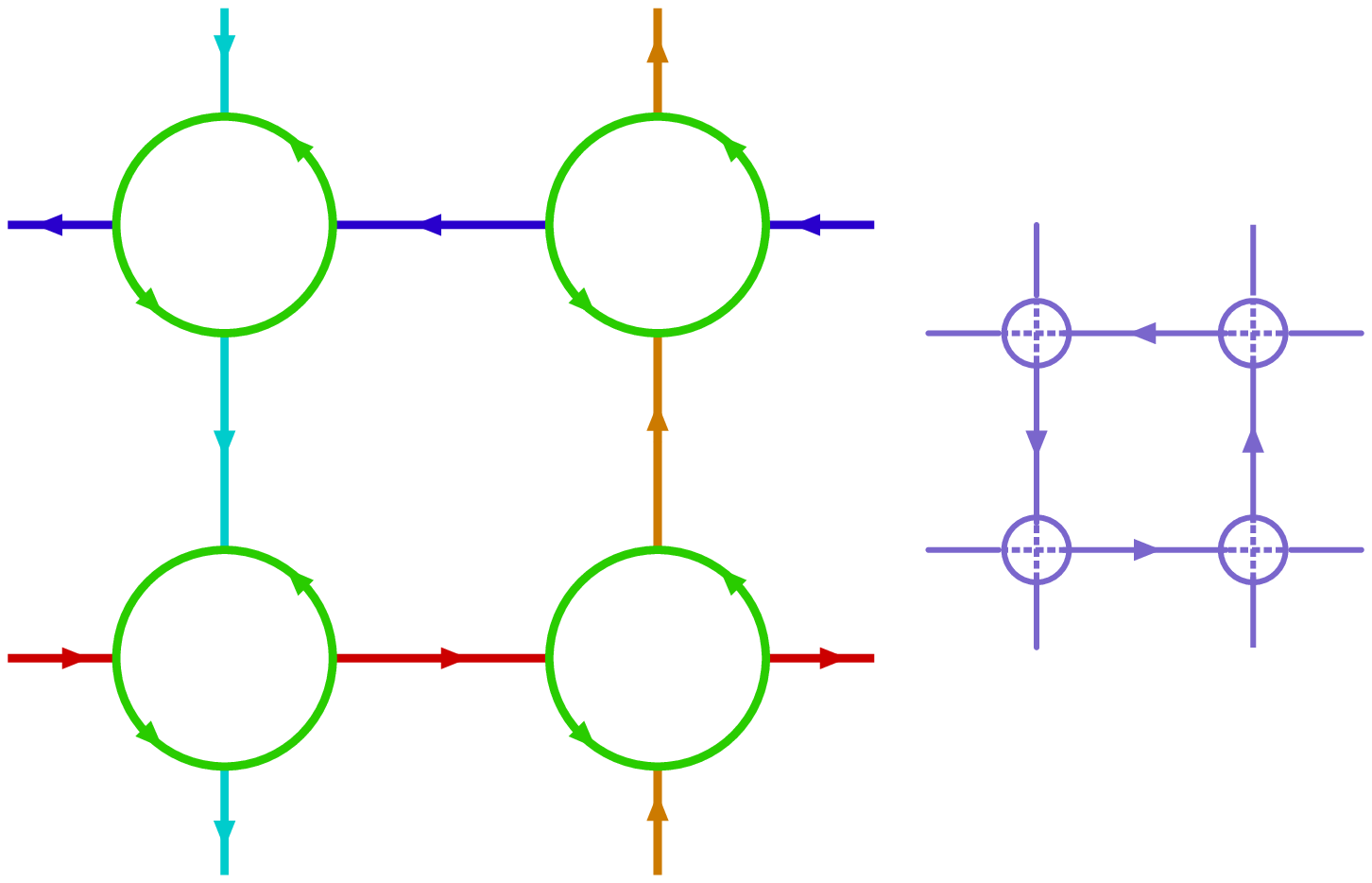}
\hskip 15pt}
\vspace{-25pt}
\begin{fig}\label{rondpoint}
\leurre
Left-hand side:
the round-about for the crossing of two two-way tracks. Four one-way track 
round-abouts allow us to solve the problem. \vskip 0pt
Right-hand side: using the symbolic representation of the one-way round-about.
\end{fig}
}

   Note the symbolic representation used for the crossing of one-way tracks. We shall
use it in our further representations of the switches as the memory switch will
need them.

   In the next subsection we thoroughly look at the configurations needed to
obtain the expected behaviour of the particle. We first deal with the tracks in
Subsection~\ref{tracks}, then with the crossings in Subsection~\ref{crossings}
and then with the switches in Subsection~\ref{switches}. This study will allow us
to construct the rules, a process of which an account is given in Section~\ref{rules}.

\subsection{The implementation of the tracks}
\label{tracks}

   From now on, we call tiles {\bf cells}. Most cells are in a quiescent state
which we also call the {\bf blank}. In the following figures of the paper,
it is represented by a light colour, not necessarily the same in order to facilitate
the understanding of the configurations. In our setting we have another
colour which we call black but which will have several dark or bright colours in 
the figures. Remember that cells are said to be neighbours if and only
if they share a side.

   As already mentioned, the track is here implemented with milestones as
in~\cite{mmsyENTCS,mmarXiv3D}. However, as the tracks must be one-way, and
as the locomotive is a particle, represented by a single cell, the direction
must be defined by the context visited by the particle: it cannot be deduced 
from the particle itself. We have met this problem in~\cite{mmarXiv3D}. It was solved
in a simple way which cannot be projected on the plane as, in most cases, the 
projection would map both ways on the single track.

   In Fig.~\ref{trackcells}, we illustrate the solution in the case of
the tiling $\{13,3\}$. For most of the tiles, we have the configurations represented
by the upper row of the figure. On one side we have the pattern for one direction
while the other side illustrates the pattern for the other direction. A track consists
in a sequence of such cells which in some sense follow a path of the tiling from
one cell to another, see Fig.~\ref{voie1}.

\vtop{
\vspace{-30pt}
\ligne{\hfill
\includegraphics[scale=0.70]{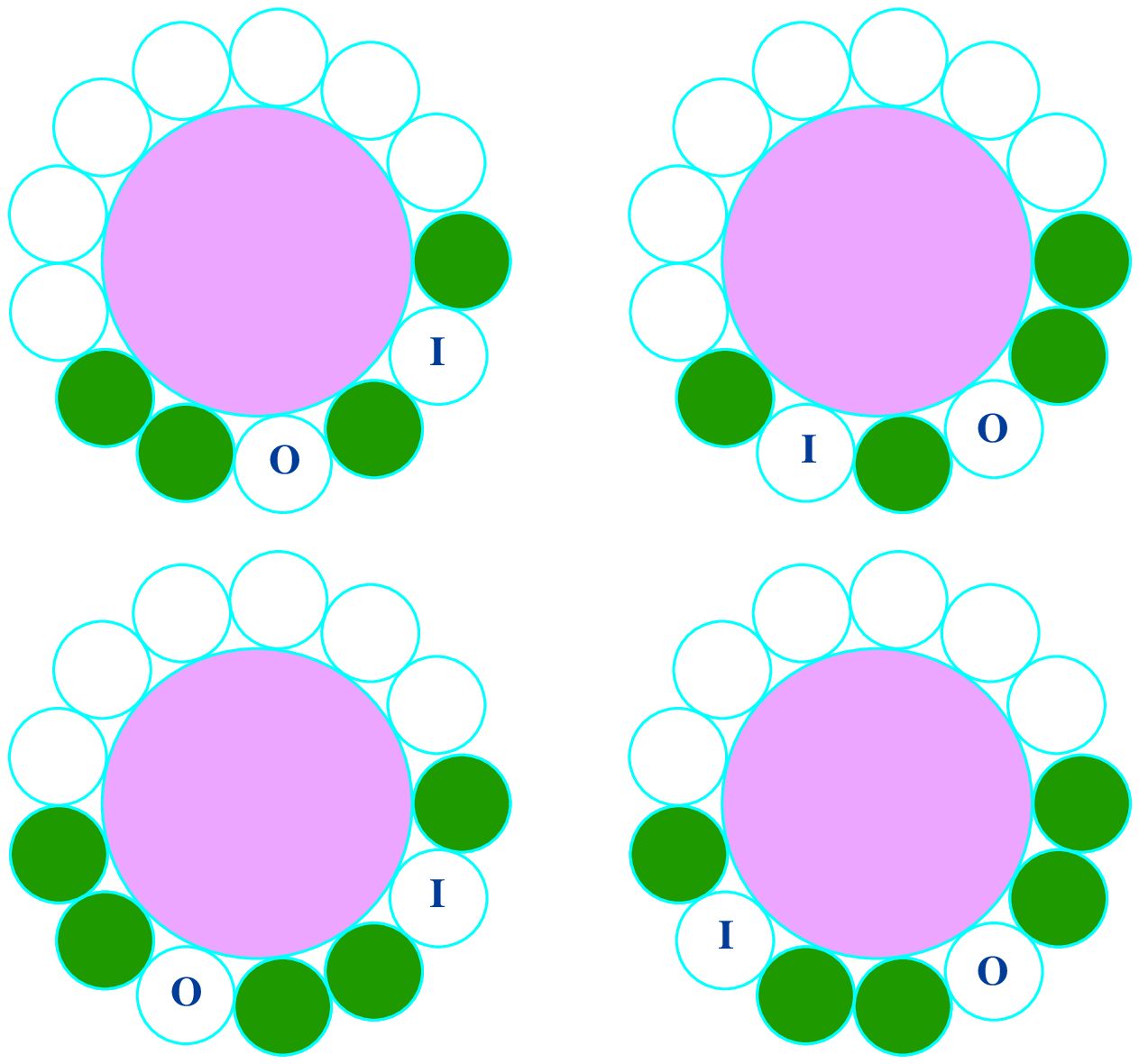}
\hfill}
\vspace{-5pt}
\begin{fig}\label{trackcells}
\leurre
Above: the most common cells of the tracks. To left: in one direction;
to right: in the opposite direction.\vskip 0pt
Below: a cell which has necessarily two 'supporting' milestones, see
Fig.~{\rm\ref{voie1}}. To left: in one direction; to right: in the opposite direction.
\vskip 0pt
In all these patterns, $I$~indicates the neighbour of the cell through which the
particle enters the cell: this neighbour plays the role of {\bf input}.
Similarly, $O$~indicates the neighbour through which the particle leaves the cell:
this other neighbour plays the role of {\bf output}.
\end{fig}
}

   To go from a tile~$P$ to the tile~$Q$, we first take a shortest path~$\pi$
in the tiling going from~$P$ to~$Q$. If we consider the neighbours of the cells
of the path which do not belong to the path, we can share these cells into two cells:
one of them, say $\tau_1$, is on one side of the path from~$P$ to~$Q$ while the 
cells of the other set~$\tau_2$ are on the other side of the path, 
see Fig.~\ref{voie1}. Now, all cells of~$\pi$ are milestones. They correspond to
the black cell which is in between the cells~$I$ and~$O$ in the first row
of Fig.~\ref{trackcells}.

   In Fig.~\ref{voie1}, say that the cells of~$\tau_1$ go from~$P$ to~$Q$. We can see
that the patterns of cells~1, 2 and~3 which belong to~$\tau_1$ correspond to the
pattern defined by the left-hand side picture in the first row of 
Fig.~\ref{trackcells}. We can also see that if we want that the particle goes
from~$Q$ to~$P$ along the track~$\tau_2$, we have to use the pattern which is
on the right-hand side of the first row of Fig.~\ref{trackcells}. This can be seen
by the small cells below cell~4{} in Fig.~\ref{voie1}. 

\vtop{
\vspace{-40pt}
\ligne{\hfill
\includegraphics[scale=0.70]{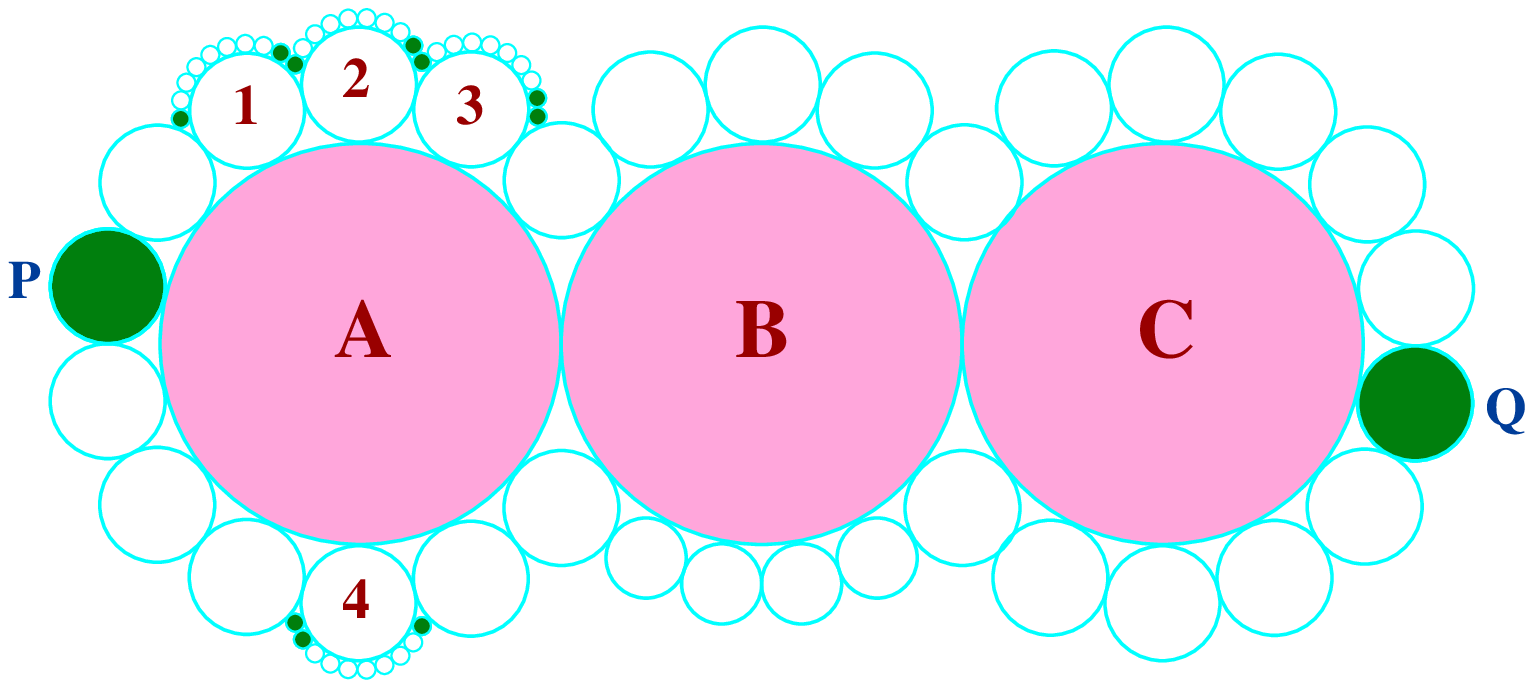}
\hfill}
\vspace{-30pt}
\begin{fig}\label{voie1}
\leurre
Above~$A$, $B$ and~$C$, the cells of the track~$\tau_1$ going from~$P$ to~$Q$.
The additional patterns are indicated for cells~$1$, $2$ and~$3$ of the track.
\vskip 0pt
Below~$A$, $B$ and~$C$, we have the track~$\tau_2$ from~$Q$ to~$P$.
The additional patterns are indicate for cell~$4$ of~$\tau_2$.
\end{fig}
}

   Now, in Fig.~\ref{voie1}, we can see that a cell of~$\tau_1$ has two consecutive
milestones which are not compatible with the already considered patterns. For these
particular cells, we use the other patterns defined by the lower row of 
Fig~\ref{trackcells}.

   Accordingly, Fig.~\ref{voie1} illustrates how to define a general algorithm
to define a two-way track going from a given cell of the tiling $\{p,3\}$
to another one. However, when a one-way track only is needed, we have some
flexibility to tune the path using the same method as is illustrated by
Fig.~\ref{voie2}.

\vtop{
\vspace{-40pt}
\ligne{\hfill
\includegraphics[scale=0.70]{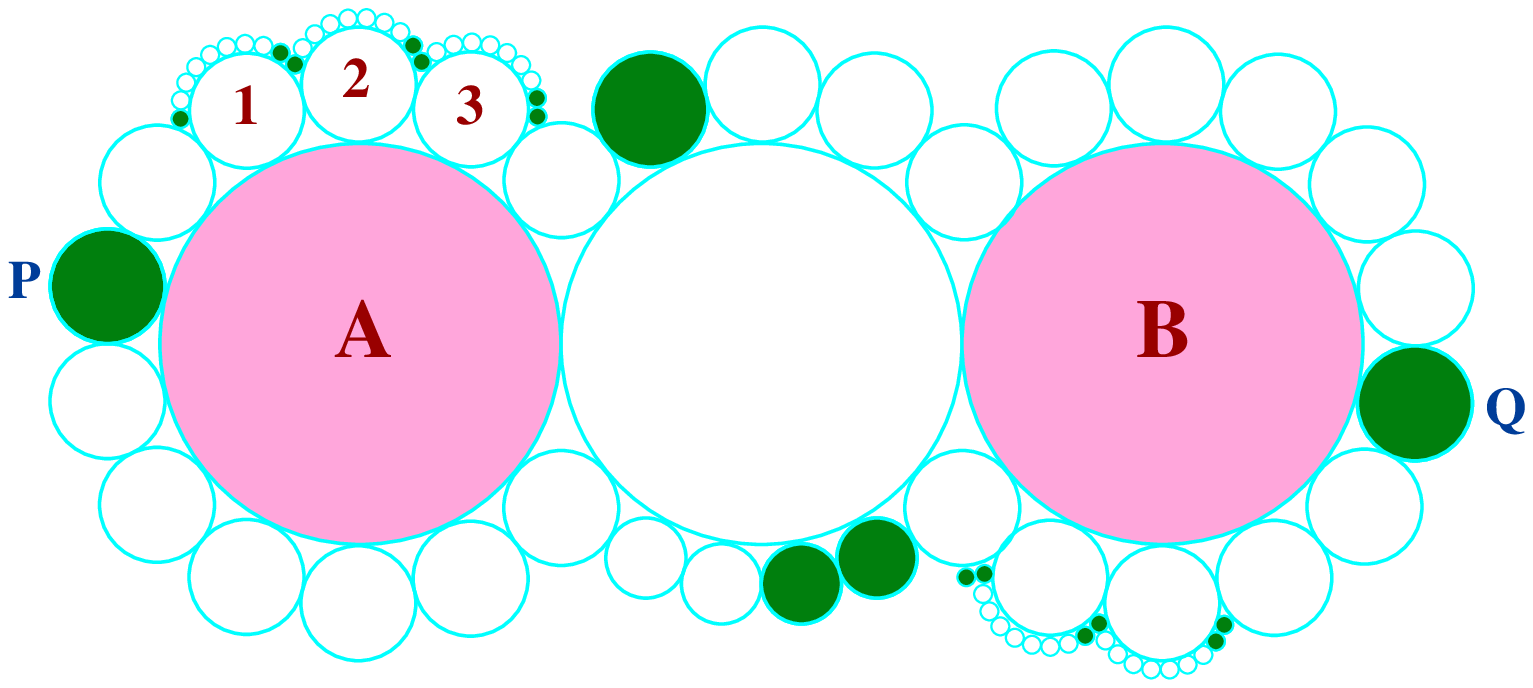}
\hfill}
\vspace{-30pt}
\begin{fig}\label{voie2}
\leurre
Here, a one way track going from~$P$ to~$Q$. Note that it crosses the initial
shortest path~$\pi$ joining~$P$ to~$Q$. One cell of~$\pi$ is used by the track
and its neighbours on~$\pi$ are milestones.
\end{fig}
}

As Fig.~\ref{voie2} suggests, for a one way track, it is possible to use one or more
cells of the shortest path to go from one given tile to another given one. 

The rules attached to the corresponding motions of the particle are given in 
Section~\ref{rules}.

\subsection{The crossings}   
\label{crossings}  

   Now, we are in the position to study the crossing of two tracks. Thanks to 
what we have seen in Subsection~\ref{tracks}, especially with Fig.~\ref{rondpoint},
it is enough to look at the crossing of two one-way tracks. Following 
Fig.~\ref{rondpointsimple} and the principle defined for the tracks in
Subsection~\ref{tracks}, we may consider that the ring which materializes the
round-about consists of cells which are around a single cell which call the 
{\bf core}. 

\vtop{
\vspace{-10pt}
\ligne{\hfill
\includegraphics[scale=0.90]{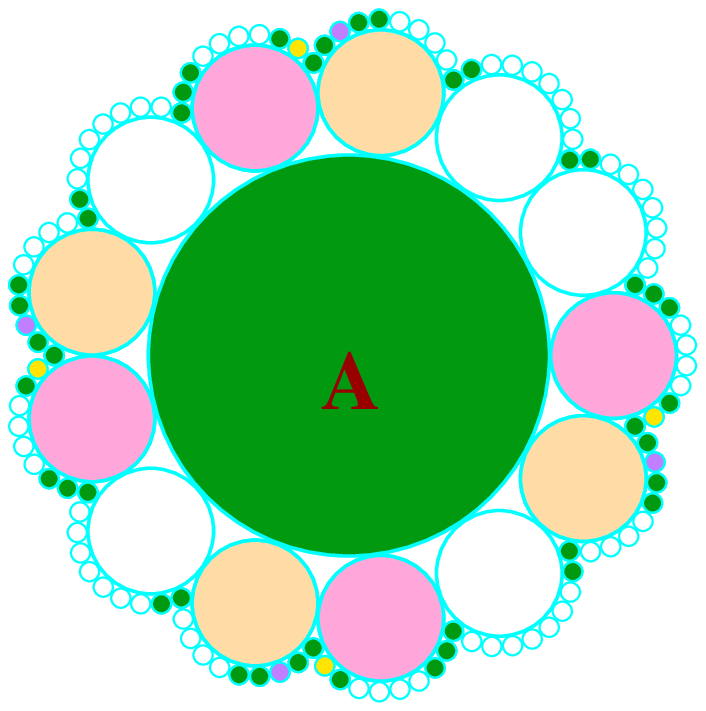}
\hfill}
\vspace{-10pt}
\begin{fig}\label{crossconfig}
\leurre
The general view of a round-about. We have four pairs of pink and pale orange
cells, each one marking the branching of a way with the round-about.
\end{fig}
}

\vtop{
\vspace{5pt}
\ligne{\hfill
\includegraphics[scale=0.40]{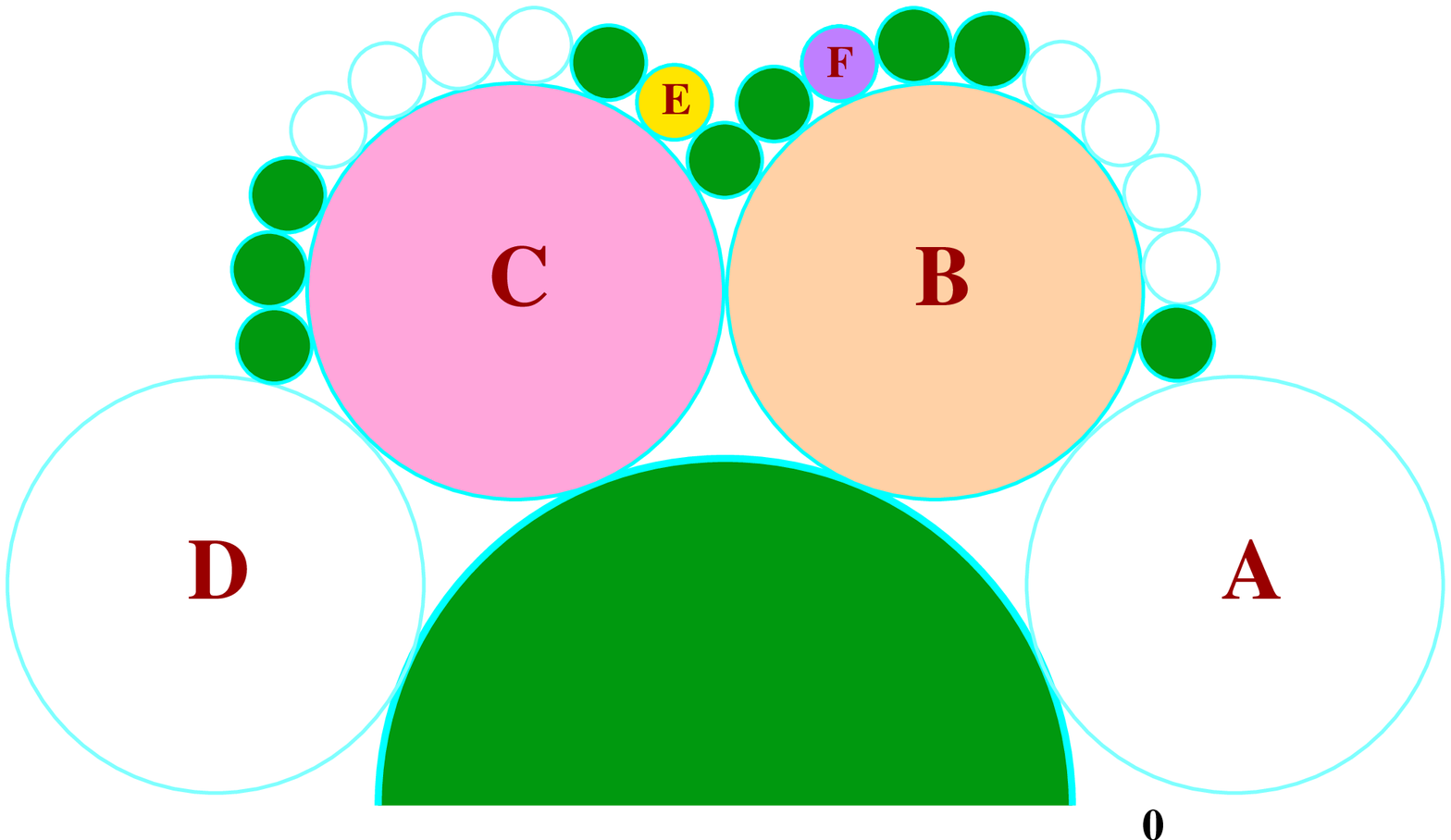}
\hfill}
\begin{fig}\label{crossing0}
\leurre
Zoom on a branching of the round-about. The cell~$C$ is concerned with the
arrival of a particle from a way abutting the round-about at the branching.
The cell~$D$ is concerned by the exit or continuation of the particle arriving
at a branching from the round-about.
\end{fig}
}

   We can see an illustration of the configuration on Fig.~\ref{crossconfig}.
The cells where the way meet with the round-about is called a {\bf branching}.
The cells in pink and pale orange colours mark the branchings.
The figure illustrates the configuration as it is when the particle is away from it.
In Fig.~\ref{crossing0} we can see a zoom which focuses our attention
on a branching and its particular surrounding. On the zoomed image, the pink cell
is called~$C$ and the pale orange one is called~$B$. 

   We shall look on three situations regarding the zoomed image:

{\leftskip 20pt\rightskip 20pt\parindent 0pt
- a particle arrives from a way which abuts the round-about at~$C$ through
the cell~$E$, the yellow neighbour of~$C$;

- two contiguous particles arrive at~$A$;

- a single particle arrive at~$A$.
\par}
\vskip 5pt
   The second situation is a consequence of the first one: when the particle arrives
at~$C$, $C$~appends a new particle to the visiting one. Consequently, at the next
branching, we have two consecutive particles: this is the second situation. At this 
stage, the second particle is cancelled so that at the next branching, a single 
particle arrives at~$A$: this is the third situation. We can notice that while 
turning around the round-about, starting from any arriving, the second branching
is the expected exit.

   Let us look at what happens at each branching in these situations.

The arrival of the particle to the round-about is illustrated by Fig.~\ref{crossing1}
where we have a zoom at the concerned branching.

\vtop{
\vspace{5pt}
\ligne{\hskip-10pt
\includegraphics[scale=0.425]{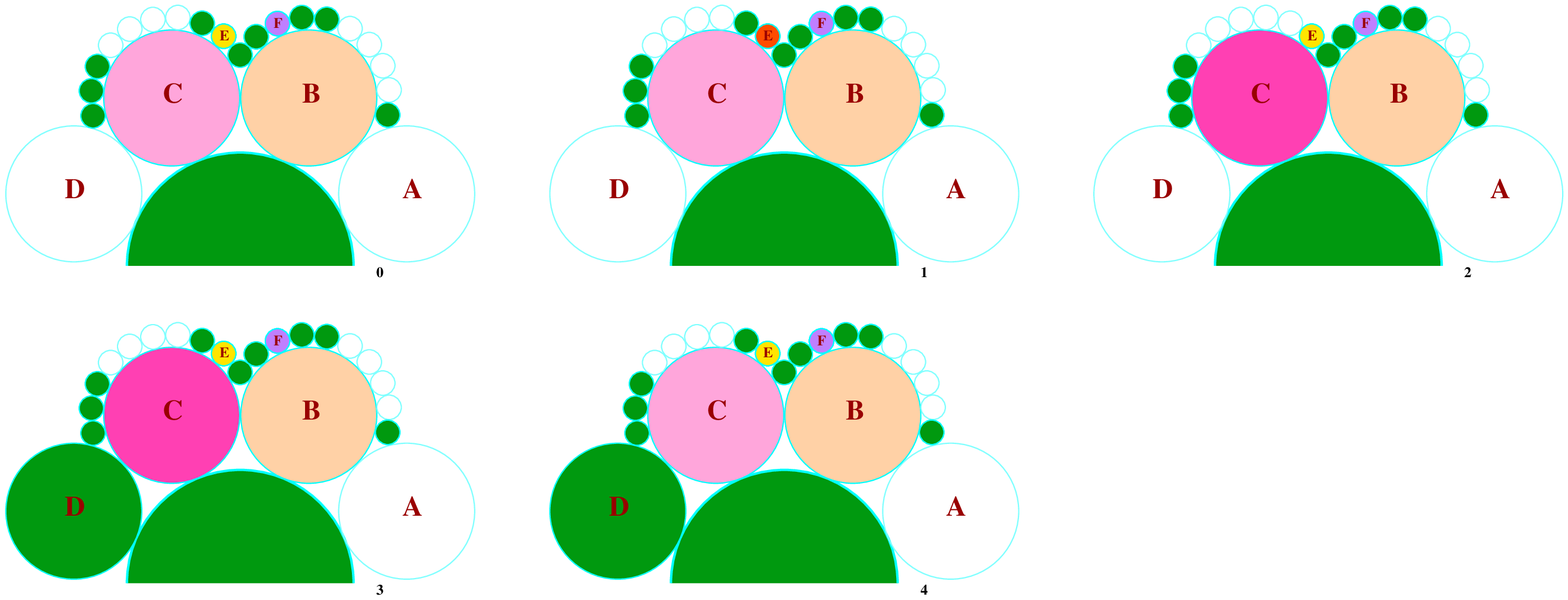}
\hfill}
\begin{fig}\label{crossing1}
\leurre
Zoom on a branching when the particle arrives at the round-about. Note the creation
of a second particle which is contiguous to the initial one on the tracks.
\end{fig}
}

   As can be seen in the figure, the particle crosses~$E$ through which it arrives
at~$C$ on the round-about. Number the neighbours of~$C$ from~1 to~13, neighbour~1
being the milestone which is common to all the cells of the round-about and
the numbers being increasing while counter-clockwise turning around~$C$. 
We shall do the same fro~$B$, so that we shall denote the neighbours of these
cells by $C.i$ and~$B.i$ respectively. The neighbour~$C.2$
is~$B$ and $E$~is $C.4$. Now, the occurrence of the particle at~$C$ triggers the
change of~$C.5$ which becomes white. When the particle
is at~$C$, the cell can see that $C.4$ and $C.5$ are both white: it is the
signal for the creation of the additional particle. This means that while the first
particle is in~$D$, $C$~is still black: $C$~returns to white the next time only.
Later on, both particles travel together along the round-about in the 
counter-clockwise direction.

   Accordingly, both particles arrive at the next branching of the round-about. What
now happens is illustrated by Fig.~\ref{crossing2}. Due to the number of surroundings
we have to define for the cells which play a particular role, we have to decide that
$F$~is an ordinary cell of the tracks. Note that $F$~is $B.10$ and that $B.11$, 
which is a milestone for~$F$, can see both~$F$ and~$B$. So that $B.11$ can detect
whether there are one or two particles arriving at the branching from the round-about.
We decide that $B.11$ does not change if there is a single particle: as $F$ 
is an ordinary cell of the tracks, the motion will go on and the particle will
leave the round-about. If $B.11$ detects two particles, then it changes from
black to white. We say that $B.11$ flashes. This flash destroys the two particles
at the next time. But at the next time, $B.11$ returns to black
and $B.12$ flashes as it has witnessed the flash of~$B.11$. The flash of~$B.12$ 
triggers a new particle in~$C$.

\vtop{
\vspace{5pt}
\ligne{\hskip-10pt
\includegraphics[scale=0.425]{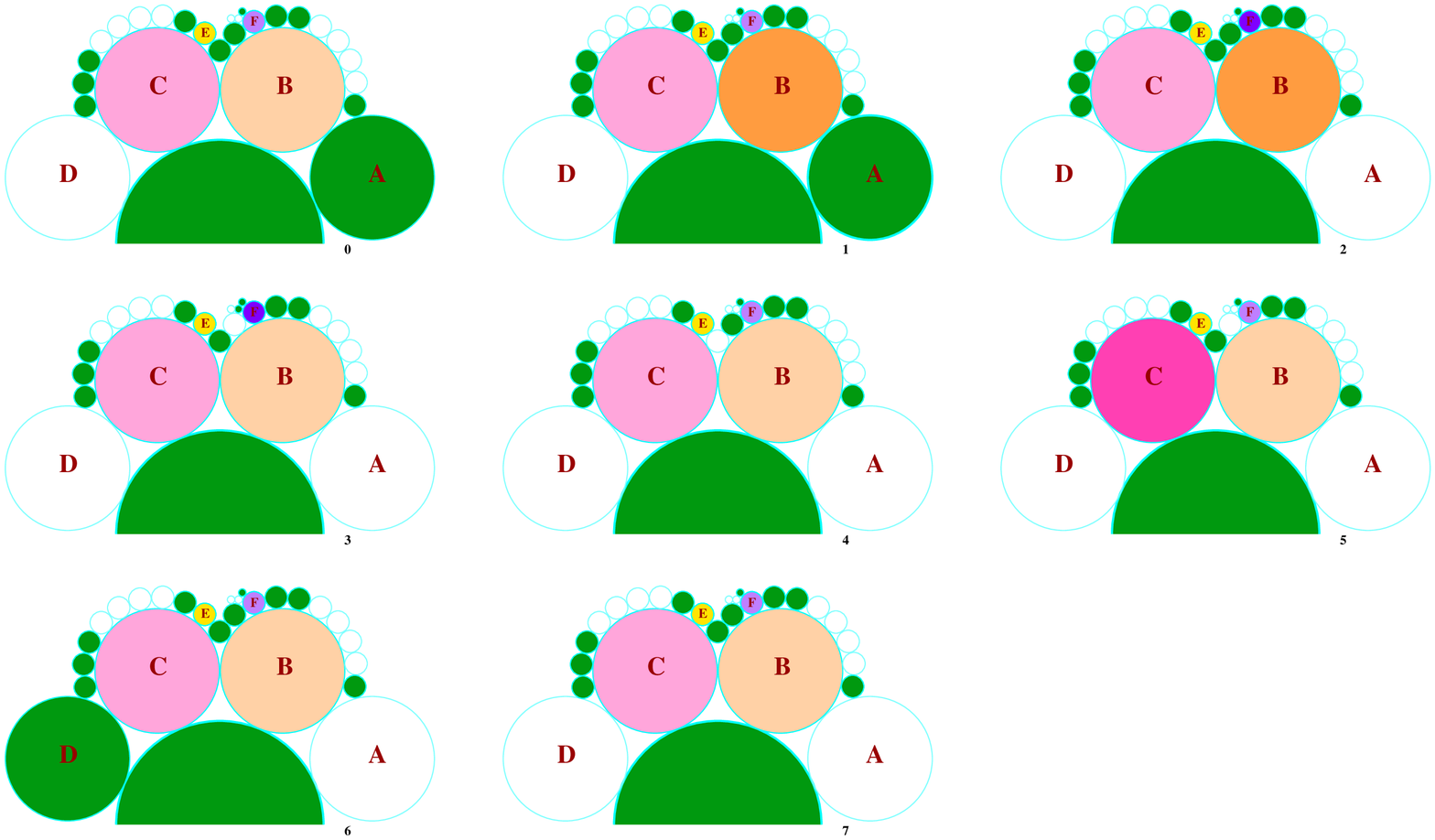}
\hfill}
\begin{fig}\label{crossing2}
\leurre
Zoom on a branching when two particle arrive at a branching from the round-about. 
Note that both particles are cancelled and that a new particle is created
in~$C$ which goes on along the round-about.
\end{fig}
}

\vtop{
\vspace{-25pt}
\ligne{\hskip-10pt
\includegraphics[scale=0.425]{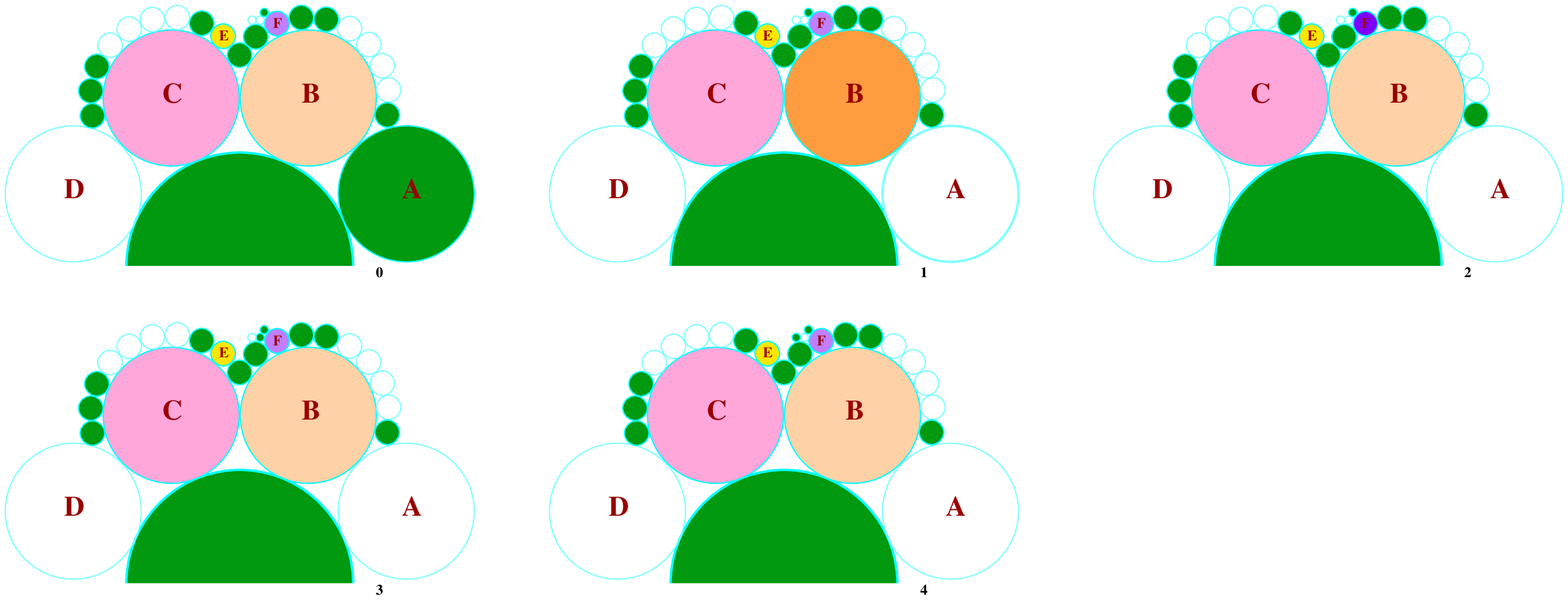}
\hfill}
\begin{fig}\label{crossing3}
\leurre
Zoom on a branching when the particle arrives alone at the round-about. Note 
that it continues its motion on the tracks leaving the round-about.
\end{fig}
}

   In Fig.~\ref{crossing2}, we can see that $B.11$ flashes again after the flash
of~$B.12$: this is required by the rules which were devised globally, so that
the flash is induced by a rule created for another context which also applies to
this one by the rotation effect. Once a particle is present in~$C$, it goes on
its way along the round-about, until it meets the next branching. As already mentioned,in this case, when the single particle arrives at~$F$, $B.11$ detects that the
particle is alone and so, it remains black. Accordingly, the particle goes
on its way on the tracks leaving the round-about: this is exactly what was expected.
Such a motion is illustrated by Fig.~\ref{crossing3}.

\subsection{The switches}
\label{switches}

   We can now turn to the description of the configurations needed by the switches.
As in the case of the crossing, we first present the configuration when the particle
is at large which we call the {\bf idle configuration}, see Figure~\ref{idle_configs}.
Later, we present several pictures illustrating the motion of the particle through 
the configuration with all the changes it induces with respect to the idle 
configuration. We shall successively look at the fixed switch, the flip-flop and 
the memory switch.

\vtop{
\vspace{5pt}
\ligne{\hfill
\includegraphics[scale=0.425]{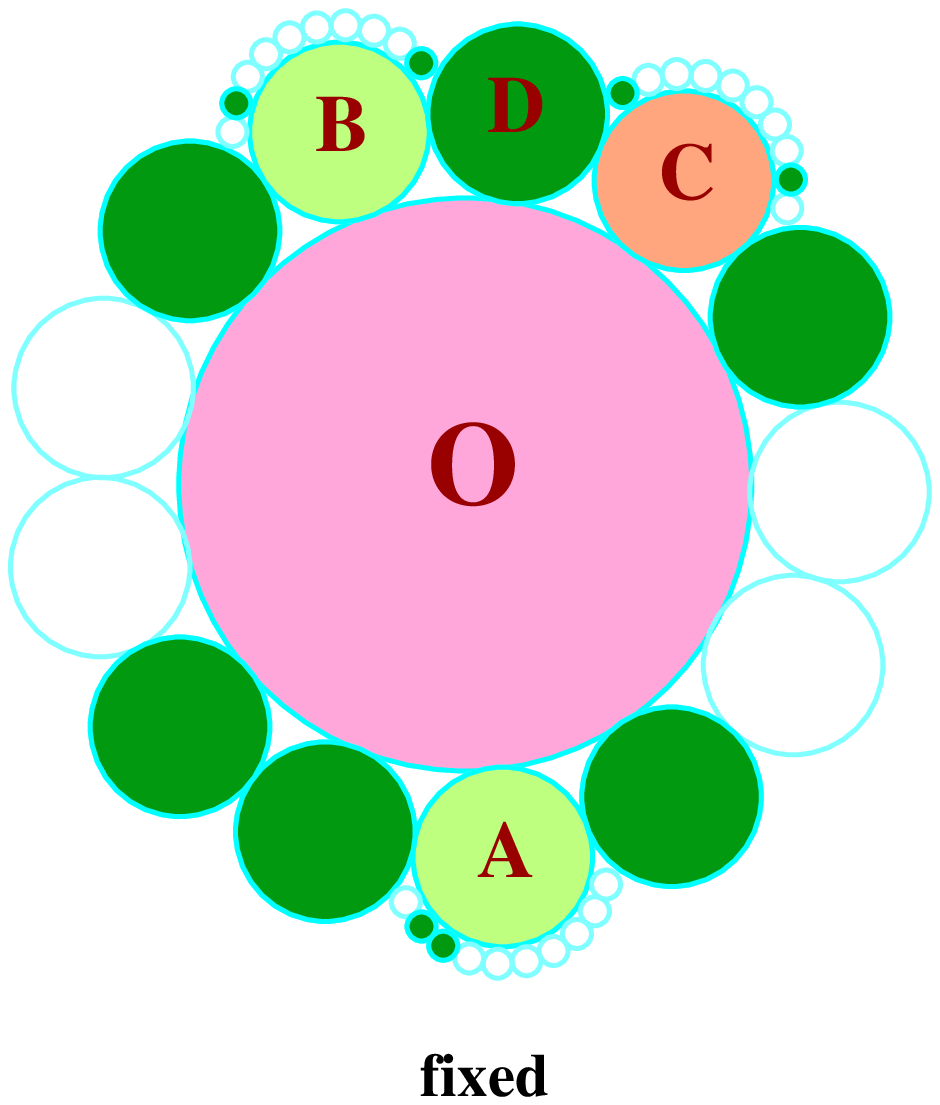}
\includegraphics[scale=0.425]{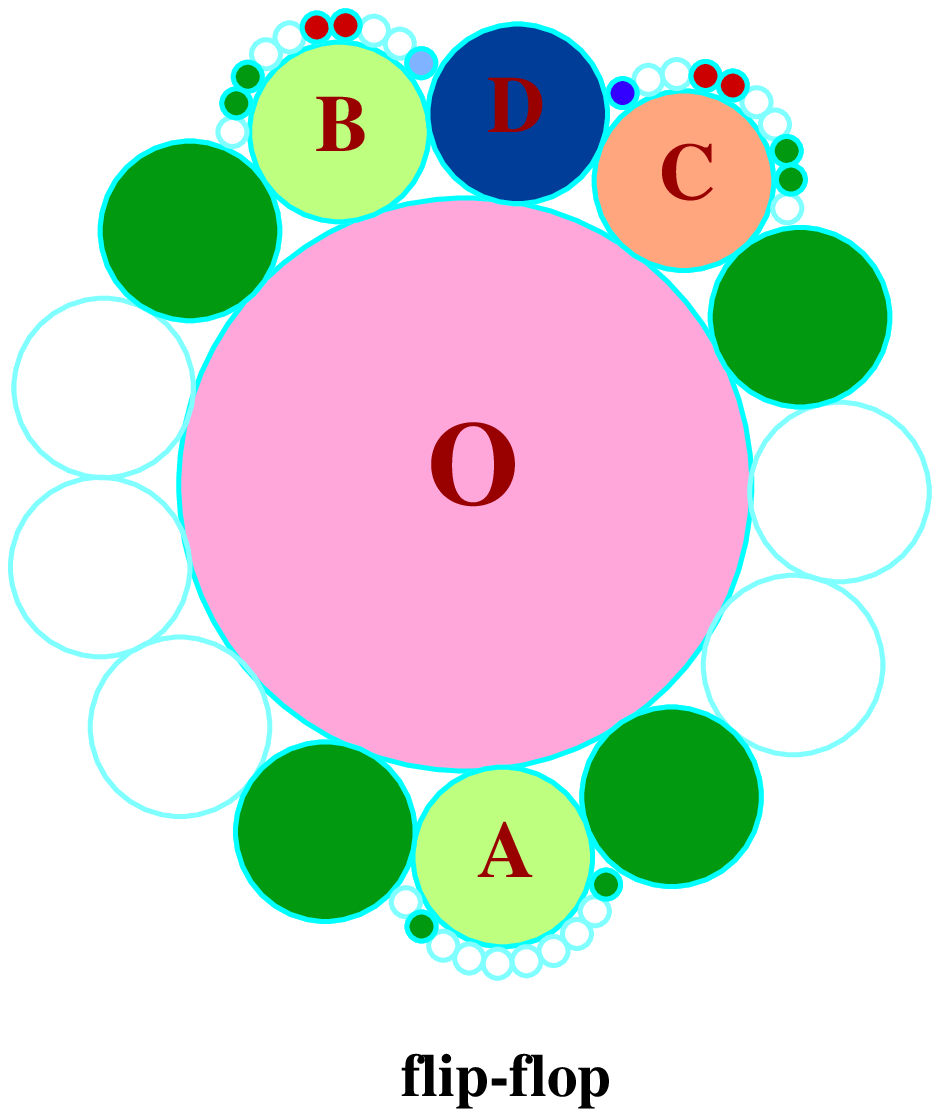}
\hfill}
\vspace{-15pt}
\ligne{\hfill
\includegraphics[scale=0.325]{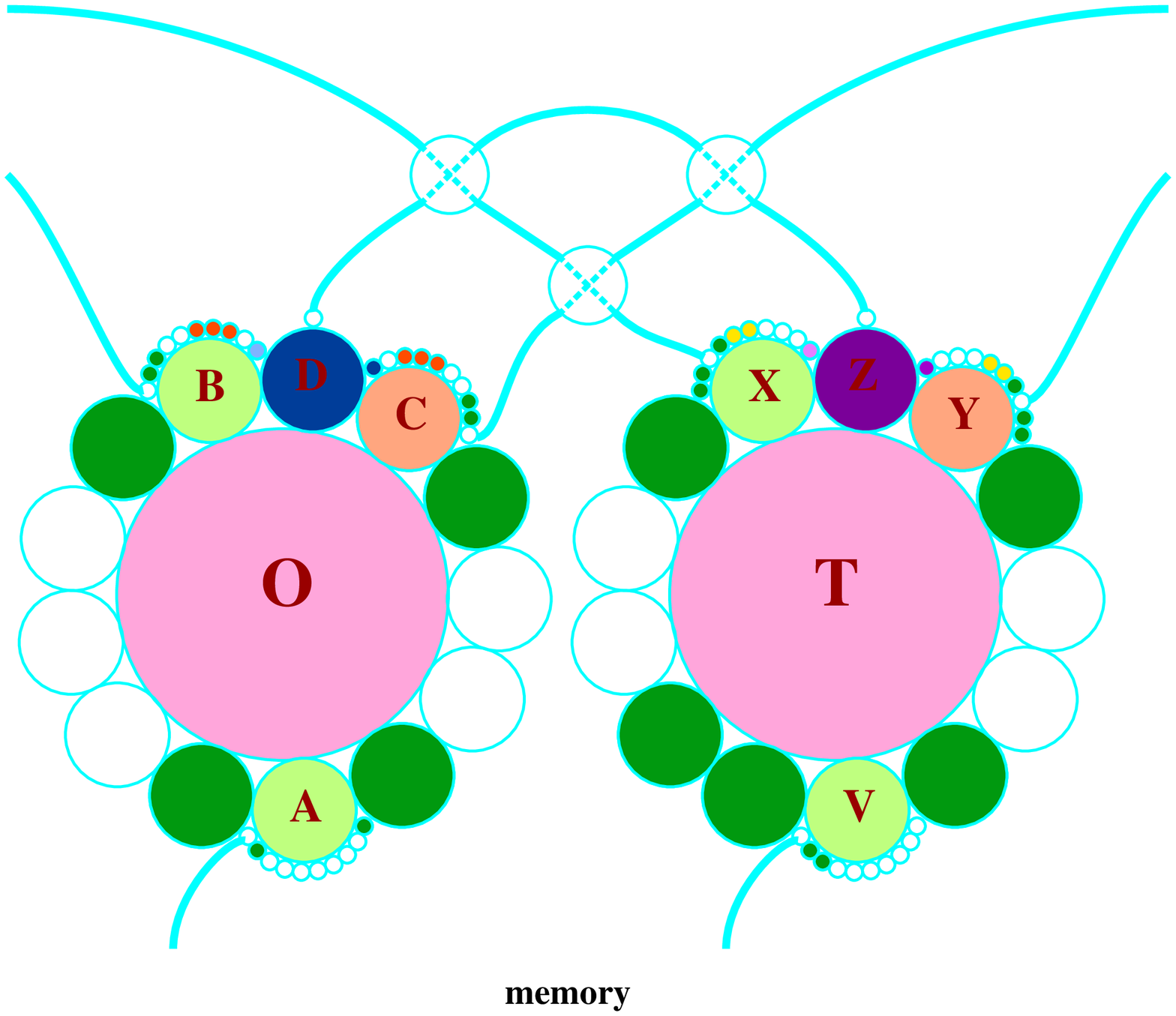}
\hfill}
\vspace{-5pt}
\begin{fig}\label{idle_configs}
\leurre
The idle configurations of the switches: above, the fixed one and the flip-flop;
below, the memory switch. For the flip-flop and for the memory switch, only
the switch where the left-hand side is selected is illustrated.
\end{fig}
}

   The fixed switch and the flip-flop are simple switches: they both involve a single
one-way switch. It is a passive switch in the case of the fixed switch, it is an 
active switch in the case of the flip-flop. In the memory switch we have a more 
complex structure: we have two one-way switches and one of them has an influence on 
the other.

   The one-way switches have the same structure: three paths arrive to a centre.
In the case of the fixed switch, there is no consequence after the passive crossing.
In the case of the flip-flop, the active crossing only is implemented but the
crossing induces a change.

    As in~\cite{mmsyENTCS}, the centre of the switch is a blank cell: it belongs to
the tracks. But its milestones define a pattern which is very different from that of
the cells of the tracks. Figure~\ref{idle_configs} show the patterns used
to distinguish each kind of switch. In the fixed switch, the black cells around
the centre of the switch are all milestones and the cells named $A$, $B$ and~$C$
which represent the abutting tracks are ordinary cells of the tracks. In the flip-flop
and in the memory switch we have new cells which we call {\bf sensors} and
{\bf controllers}. We shall study them in Subsubsections~\ref{flip_flop}
and~\ref{memory}.

   In the figures illustrating the switches, we represent only one case: the
case when the selected track is the left-hand side one.

   Now, we shall consider that the particle arrives at a switch. Each case is
examined in an appropriate Sub-subsection.

\subsubsection{Fixed switches}
\label{fixed}

   The crossing of a fixed switch is illustrated by Fig.~\ref{fixed_motion}.
It is a passive crossing where the selected track is not mentioned as in such
a crossing, the side from where the particle comes does not matter. The matter
is only for the selected track in the active crossing of the switch. Now,
Fig.~\ref{newswitches} suggests that we might choose the active track in such a 
way that the active tracks do not cross passive ones. This is not true. If we fix an
order for the directions in two-way tracks, it may happen that the active
tracks has to cross passive ones. This is not difficult to implement as in
this case, the crossing involves a one-way crossing as illustrated by 
Fig.~\ref{fixed2ways}.

\vtop{
\ligne{\hfill
\includegraphics[scale=0.8]{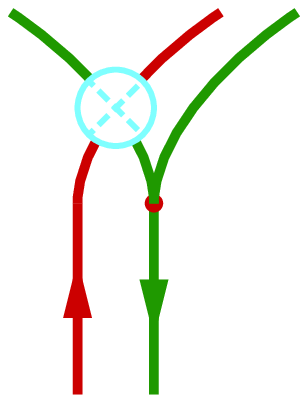}
\hfill}
\vspace{-10pt}
\begin{fig}\label{fixed2ways}
\leurre
The implementation of a fixed switch for which the active way has to cross a
passive one.
\end{fig}
}

   We note that in Fig.~\ref{fixed2ways}, the centre of the switch must stand
at a large enough distance from the crossing: 6~cells from the centre
of the switch to the centre of the round-about is a secure enough distance.

   From Fig.~\ref{fixed_motion}, we note that all black cells around the centre
of the fixed switch can be simple milestones. As we shall see in Section\ref{rules},
the motion rules of the tracks will be enough to ensure that the crossing can be
performed without additional marking: that of the central cell of the switch is
enough. This will spare us a few patterns which will be useful to distinguish
the sensors and controllers of the flip-flop from those of the memory switch,
remembering that in the latter switch we have two one-way ones which differ 
between each other.

\vtop{
\vspace{5pt}
\ligne{\hfill
\includegraphics[scale=0.37]{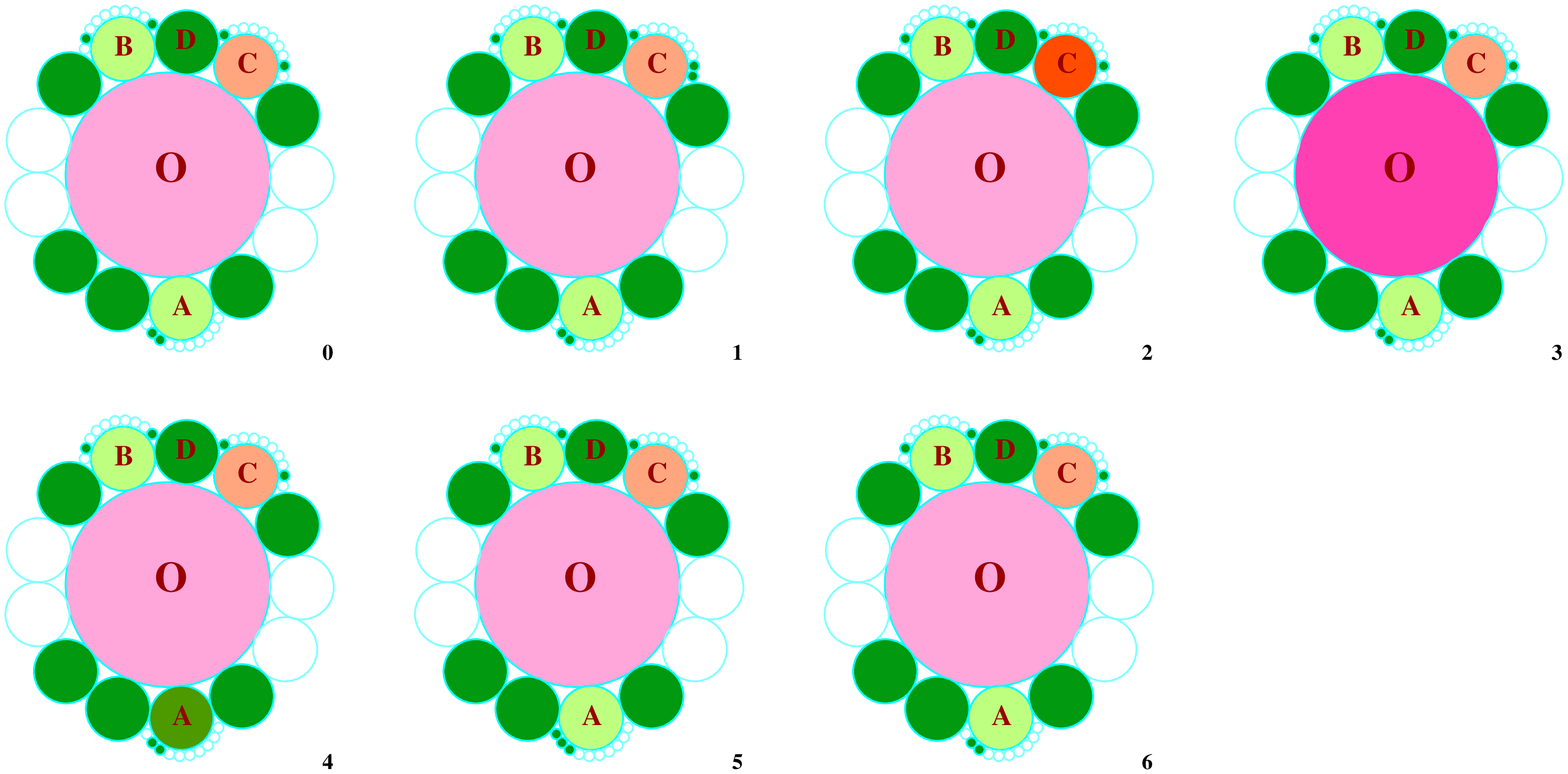}
\hfill}
\vspace{-15pt}
\begin{fig}\label{fixed_motion}
\leurre
The passive crossing of a fixed switch by the particle.
\end{fig}
}

\subsubsection{Flip-flop switches}
\label{flip_flop}

   We have seen the idle configuration of a flip-flop in Fig.~\ref{idle_configs}.
It also appear as the first picture in Fig.~\ref{flip_flop_motion} which
illustrates the motion of the particle through the flip-flop. Here two,
we denote by~$A$, $B$ and~$C$ the cells of the tracks in contact with the
centre of the switch. Here, it is important to specify the role of each cell.
The cell~$A$ is the cell from which the particle arrives at the switch. It leaves
the switch either through the cell~$B$ or through the cell~$C$, depending on which
is the selected track.

\vtop{
\vspace{-55pt}
\ligne{\hfill
\includegraphics[scale=0.37]{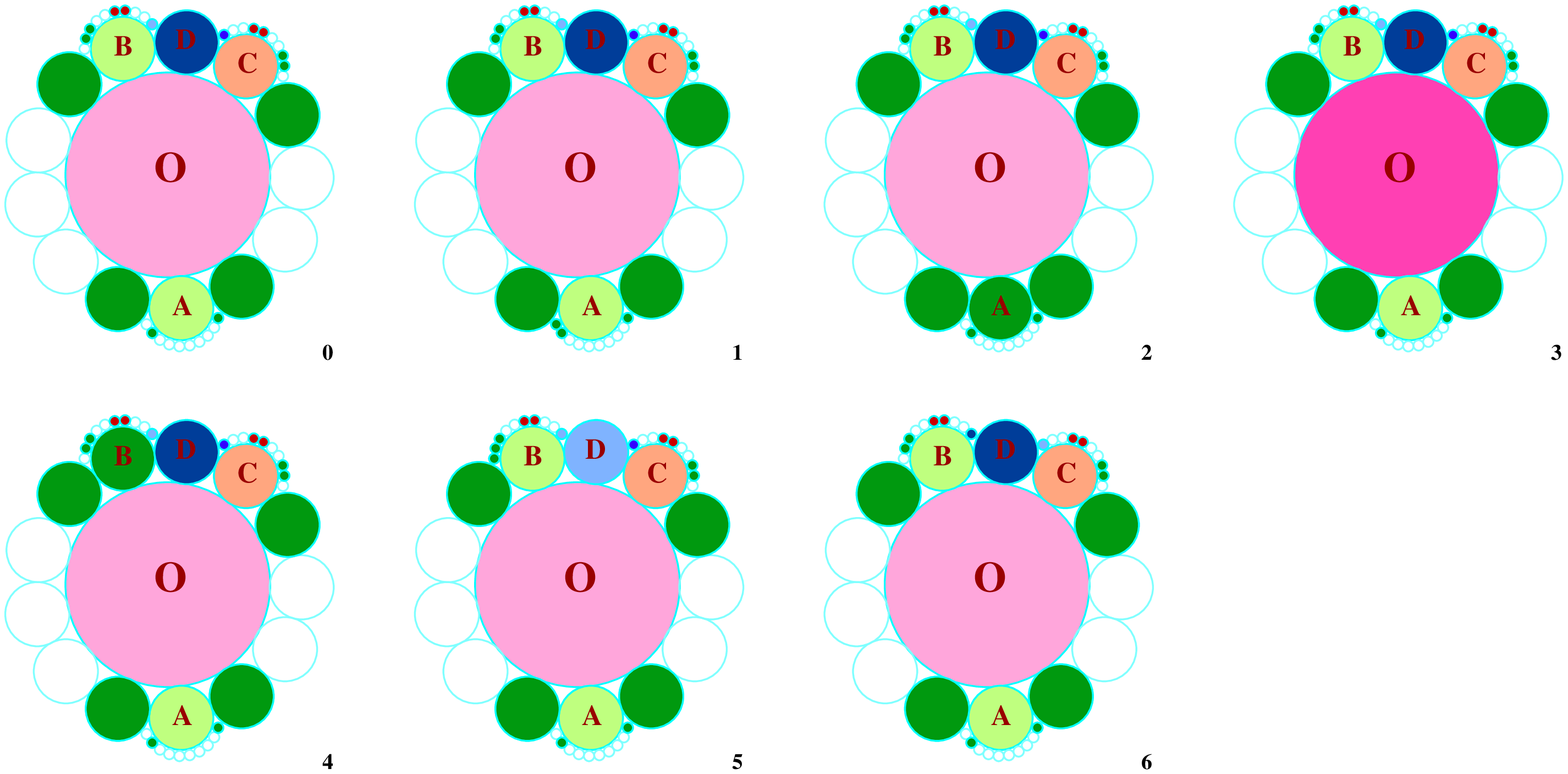}
\hfill}
\vspace{-15pt}
\begin{fig}\label{flip_flop_motion}
\leurre
The active crossing of a flip-flop by the particle. Here, the selected track
is the left-hand side one.
\end{fig}
}

   In the configurations of Fig.~\ref{idle_configs} and Fig~\ref{flip_flop_motion}, 
we mark the non-selected track instead of the selected one. The difference 
between~$B$ and~$C$ is that there is an additional black neighbour in the cell
of the non-selected track: the lack of this mark indicates that the particle must 
enter this cell; the presence to this mark forbids the particle to enter the cell.
This can be seen in the mentioned figures: the mark is a small cell neighbouring
the cell~$D$. If $D.1$~is the central cell, the cells possibly supporting the
marks are $D.3$ and~$D.12$. This means that one of these two cells is black while
the other is white. As $C$~is $D.2$ and $B$~is $D.13$, $D$ can see both
$D.3$ and~$C$ as well as both $D.12$ and~$B$. The particle enters the
side where the mark is missing. This means that $D$~is in position to detect the
presence of the mark in the appropriate side. When this is the case, $D$~flashes
and, at the next time, $D.3$ and~$D.12$ both change their state. This automatically
changes the side of the selected track as required. The last configuration of
Fig.~\ref{flip_flop_motion} is the idle one where the selected track is the 
righ-hand side one.

\subsubsection{Memory switches}
\label{memory}

   The idle configuration of a memory switch is very different from the already studied
configurations of the fixed and the flip-flop switches as can already be noticed
from Fig.~\ref{idle_configs}. As this switch involves both a passive one-way switch
and an active one-way switch, it might seem possible to simply take copies of the
previous switches to perform the needed action. This is not possible for several
reasons. First, the passive switch has to send a signal to the active one, which is
a difference with the fixed switch. And so, the cell~$Z$ of the passive memory switch
cannot be the cell~$D$ of the fixed switch. Moreover, the active switch of the memory
switch is in fact {\it passive}: the selected track is fixed as long as there
is no passive crossing on the other one-way switch. This means that the active
switch should be viewed as a {\bf programmable} one-way fixed switch. And so,
in this case, the cell~$D$ in the active one-way switch of the memory switch 
acts as a controller while $D.3$ and~$D.11$ are flexible markers. 

\vtop{
\vspace{-20pt}
\ligne{\hskip 20pt
\includegraphics[scale=0.37]{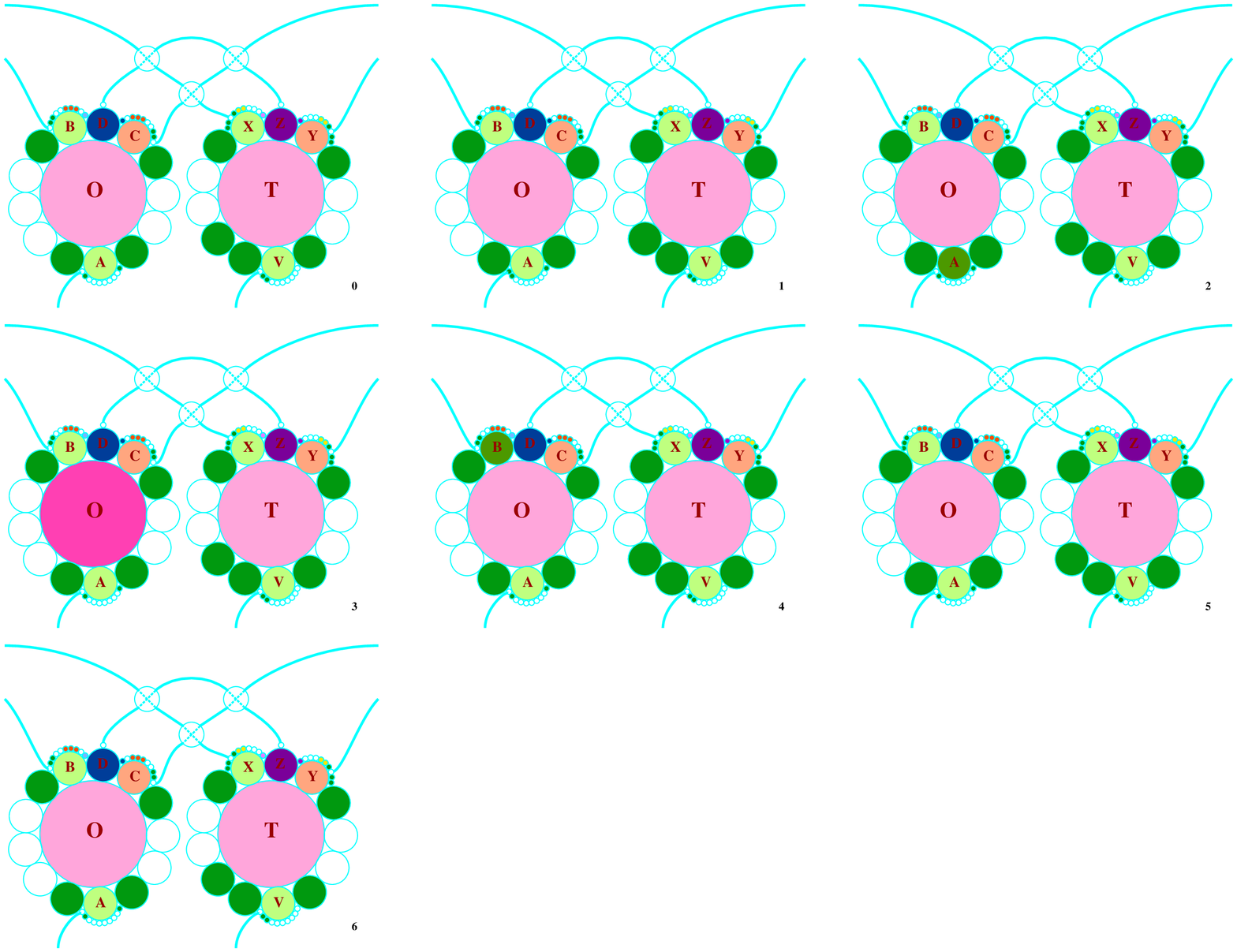}
\hfill}
\vspace{-25pt}
\begin{fig}\label{memory1}
\leurre
The active crossing of a memory switch by the particle. Here, the selected track
is the left-hand side one.
\end{fig}
}

   An active passage of the particle is illustrated by Fig.~\ref{memory1}. We can see
that nothing is changed in the markers $D.3$ and~$D.11$.

\vtop{
\vspace{-20pt}
\ligne{\hskip 20pt
\includegraphics[scale=0.35]{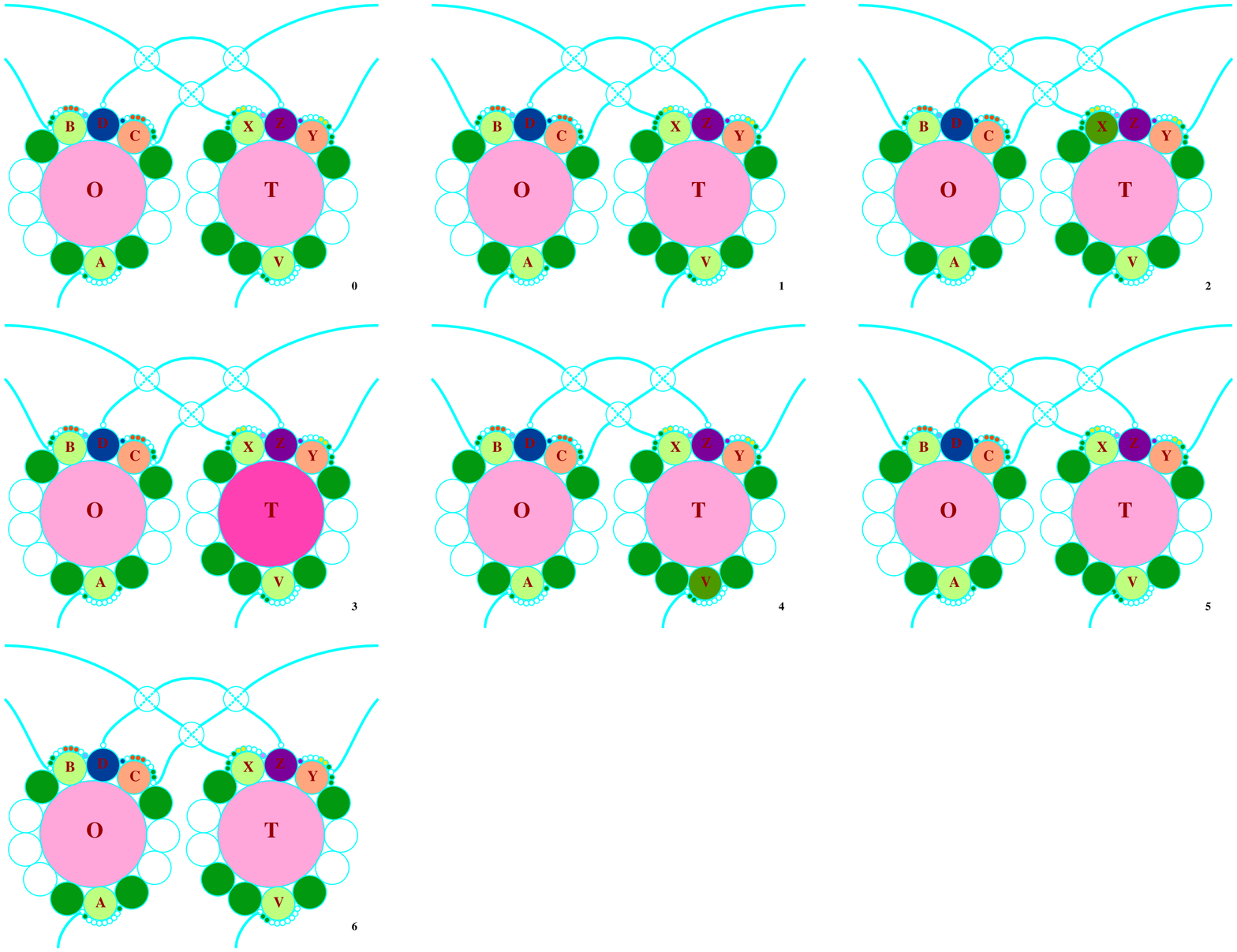}
\hfill}
\vspace{-25pt}
\begin{fig}\label{memory2}
\leurre
The passive crossing of a memory switch by the particle through the selected track.
Here, the selected track is the left-hand side one.
\end{fig}
}

\vtop{
\vspace{-10pt}
\ligne{\hskip 20pt
\includegraphics[scale=0.35]{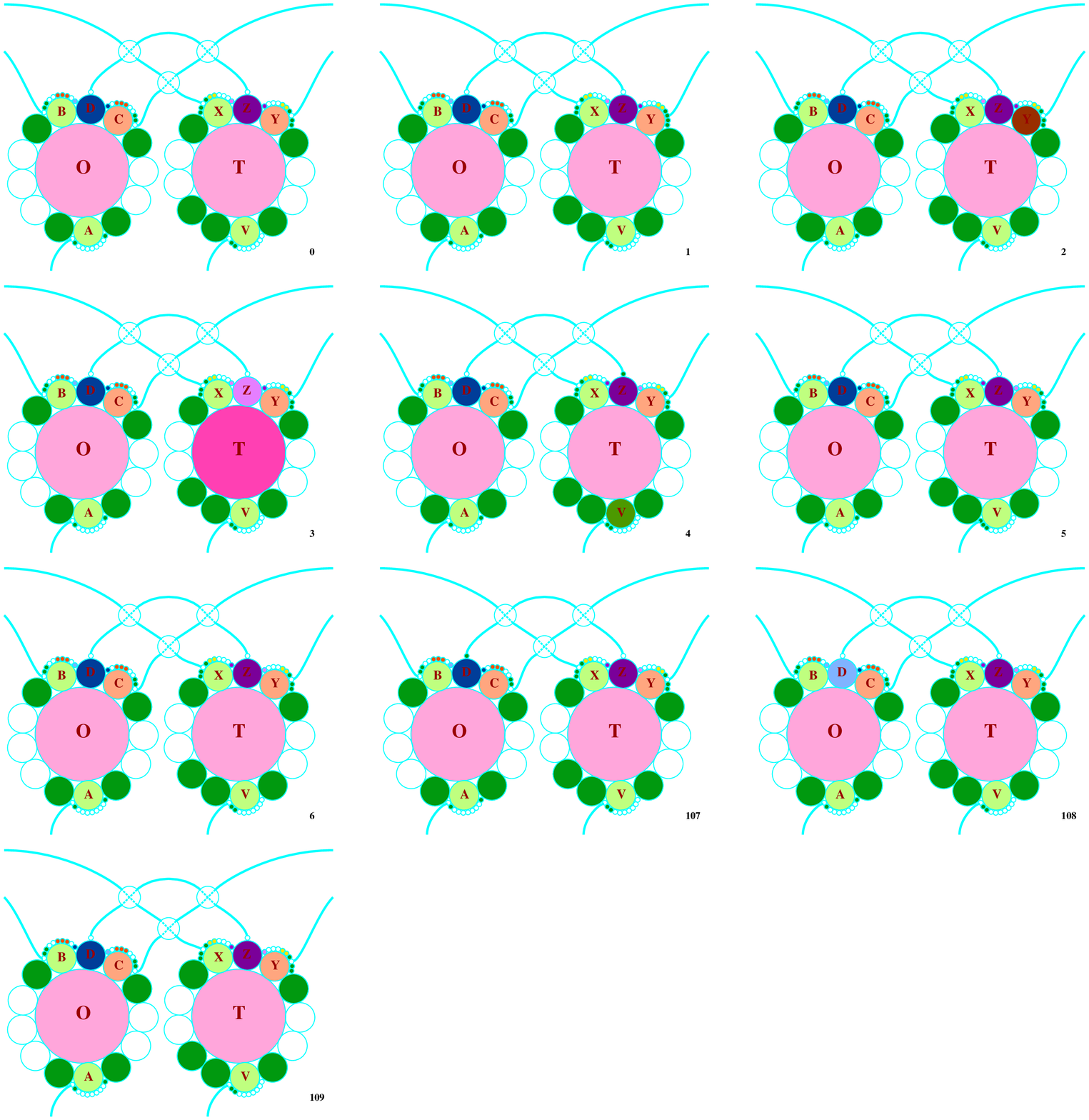}
\hfill}
\vspace{-15pt}
\begin{fig}\label{memory3}
\leurre
The passive crossing of a memory switch by the particle through the non-selected track.
Here, the selected track is the left-hand side one. 
Note the change of selection
when the particle leaves the cell. Note the delay of the change in the active
one-way switch.
\end{fig}
}

   For what is the passive one-way switch, it has to detect whether a passage
occurs through the selected or through the non-selected tracks. This can be
detected by the cell~$Z$ which can see at the same time the cells~$X$, $Y$, $Z.3$ 
and~$Z.11$. If $X$~is black and $Z.11$ is white, or if $Y$ is black and $Z.3$
is white, then it is a passage through the selected track. No change of state occurs
in $Z.3$ or $Z.11$ and the particle leaves the switch through~$V$. This situation
is illustrated by Fig.~\ref{memory2}.

If $V$ and $Z.11$ are both black or if $U$ and $Z.3$ are both black,
then $Z$ must flash: this triggers the change of state in both $Z.3$ and~$Z.11$
and it also sends a particle to~$D$ through $Z.8$. This second particle travels
through two crossings to~$D.7$. When $D.7$ is black, $D$~flashes which
triggers the change of state in both $D.3$ and~$D.11$. This is illustrated
by Fig.~\ref{memory3}. Note that the change of selection occurs in the figure
and also, note the delay of the transmission of the signal from~$Z.8$ to~$D.7$.
The actual delay can be made less than the one given in the figure. The last 
configuration in Fig.~\ref{memory3} is the idle configuration of memory switch
where the selected track is the right-hand side one.

   We can now turn to the rules: in Section~\ref{rules} we look at the
case of the tiling $\{13,3\}$; in Section~\ref{rulesgene}, we uniformly define the
rules for the tilings $\{p,3\}$ when $p\geq17$.

\section{The rules for a cellular automaton on $\{13,3\}$}
\label{rules}

It is clear that the
solution for $=13$ can be extended to the case $\{p,3\}$ for any $p\geq13$. 
However, the solution
for $p\geq17$ can be given a more simple expression which allows us to check
its correctness easily: this will be ween in Section\ref{rulesgene}.

\def\displayrule #1 #2 #3 #4 #5 #6 #7 {%
\hbox{\ttix\hbox to 15pt{#1\hfill}\hbox to 15pt{#2\hfill}\hbox to 15pt{#3\hfill}%
\hbox to 15pt{#4\hfill}\hbox to 15pt{#5\hfill}\hbox to 15pt{#6\hfill}%
\hbox to 15pt{#7\hfill}
}
}

The set of rules of the automaton are displayed as tables in 
Subsections~\ref{trackrules}, \ref{crossrules} and~\ref{switchrules}.
Here, we indicate how the rules where computed by a computer program and 
how the rules are represented and dispatched in the table. After a precision to
Section~\ref{hypCA} about the format of the rules, we follow the discussion 
of Subsections~\ref{tracks}, \ref{crossings} and~\ref{switches} in order
to establish the rules for all possible situations.

\subsection{The format of the rules and rotation invariance}

   We remind that the format of the rules is the one introduced in
Section~\ref{hypCA}. First, we look at the rules in $\{13,3\}$ and
then in $\{p,3\}$.

   This means that a rule is presented as a word of the form:

\ligne{\hskip 20pt
\hbox{$\underline{\eta_0}\eta_1\eta_2\eta_3\eta_4\eta_5\eta_6\eta_7\eta_8%
\eta_9\eta_{10}\eta_{11}\eta_{12}\eta_{13}\underline{\eta^1_0}$}.
\hfill\hskip 20pt}

We remind that $\eta_0$ is the current state of the cell, that 
$\eta_i$ is the state of the neighbour~$i$ and that $\eta^1_0$ is the new state
of the cell after the rule has been applied to it.

   In all this section and in the tables of
Subsections~\ref{trackrules}, \ref{crossrules} and~\ref{switchrules},
the states of the automaton
are {\tt W} and {\tt B}. We decide that {\tt W} is the quiescent state, which means
that the rule
\hbox{\tt$\underline{\hbox{\tt W}}$WWWWWWWWWWWWW$\underline{\hbox{\tt W}}$}
is in the table. We shall also write
\hbox{\tt$\underline{\hbox{\tt W}}$W$^{13}$$\underline{\hbox{\tt W}}$} which will be
called a {\bf condensed format}. Similarly we decide that any
neighbourhood with less than three rules does not change the current state.
This induces the following rules:

\ligne{\hfill\tt
\hbox{\tt$\underline{\hbox{\tt X}}$W$^{13}$$\underline{\hbox{\tt X}}$},
\hbox{\tt$\underline{\hbox{\tt X}}$BW$^{12}$$\underline{\hbox{\tt X}}$},
\hbox{\tt$\underline{\hbox{\tt X}}$BBW$^{11}$$\underline{\hbox{\tt X}}$},
\hfill}

\ligne{\hfill\tt
\hbox{\tt$\underline{\hbox{\tt X}}$BWBW$^{10}$$\underline{\hbox{\tt X}}$},
\hbox{\tt$\underline{\hbox{\tt X}}$BW$^{2}$BW$^{9}$$\underline{\hbox{\tt X}}$},
\hbox{\tt$\underline{\hbox{\tt X}}$BW$^{3}$BW$^{8}$$\underline{\hbox{\tt X}}$},
\hbox{\tt$\underline{\hbox{\tt X}}$BW$^{4}$BW$^{7}$$\underline{\hbox{\tt X}}$},
\hbox{\tt$\underline{\hbox{\tt X}}$BW$^{5}$BW$^{6}$$\underline{\hbox{\tt X}}$},
\hfill}

\noindent 
where {\tt X $\in$ $\{$B,W$\}$}. Also, a rule 
\hbox{\tt$\underline{\hbox{\tt X}}$Y$\underline{\hbox{\tt X}}$}, 
with {\tt Y $\in$ $\{$B,W$\}^{13}$} means that both rules
\hbox{\tt$\underline{\hbox{\tt W}}$Y$\underline{\hbox{\tt W}}$} and
\hbox{\tt$\underline{\hbox{\tt B}}$Y$\underline{\hbox{\tt B}}$} are in the table.
Note that the rules 
\hbox{\tt$\underline{\hbox{\tt X}}$BW$^{5}$BW$^{6}$$\underline{\hbox{\tt X}}$}
and
\hbox{\tt$\underline{\hbox{\tt X}}$BW$^{6}$BW$^{5}$$\underline{\hbox{\tt X}}$}
are the same under the rotation invariance condition.

   The just described rules are not listed in the tables
of Subsections~\ref{trackrules}, \ref{crossrules} and~\ref{switchrules}
which contain all the other rules for $\{13,3\}$ in a condensed notation.

   For the general case, we shall denote the rules 
as \hbox{\tt$\underline{\hbox{\tt X}}$Z$\underline{\hbox{\tt X}}$}
where {\tt X},{\tt Y}~$\in$~$\{${\tt W},{\tt B}$\}$ and 
{\tt Z}~$\in$~$\{${\tt W},{\tt B}$\}^p$. As previously,
the rules \hbox{\tt$\underline{\hbox{\tt X}}$BW$^{p-1}$Y$\underline{\hbox{\tt X}}$}
and \hbox{\tt$\underline{\hbox{\tt X}}$Z$\underline{\hbox{\tt X}}$},
with {\tt X} as before and {\tt Z}~$\in$~$\{${\tt W},{\tt B}$\}^p$
with $\vert${\tt Z}$\vert_{\tt B}=2$ are in the table, $\vert{\tt Z}\vert_{\tt B}$
being the number of occurrences of~{\tt B} in~{\tt Z}.
We say that \hbox{\tt$\underline{\hbox{\tt X}}$Z} is the {\bf context} of the
rule.

\vskip 5pt
   The rules for the case $\{13,3\}$ were defined and checked with the help of a
computer program. In particular, the program checked the rotation invariance of
the rules. This was performed as described in~\cite{mmsyENTCS,mmTCSHCA4st}.
The principle is as follows: in a rotation invariant cellular automaton,
if a rule \hbox{$\underline{\eta_0}\eta_1\eta_2...
\eta_{13}\underline{\eta^1_0}$} belongs
to the table of the automaton, 
the rule
\hbox{$\underline{\eta_0}\eta_{\pi(1)}\eta_{\pi(2)}...
\eta_{\pi(13)}\underline{\eta^1_0}$} also belongs to the table for any
circular permutation~$\pi$ on the numbers from~1 up to~13. in this
case, we say that \hbox{$\underline{\eta_0}\eta_{\pi(1)}\eta_{\pi(2)}...
\eta_{\pi(13)}\underline{\eta^1_0}$} is a {\bf rotated image} of the rule
\hbox{$\underline{\eta_0}\eta_1\eta_2...
\eta_{13}\underline{\eta^1_0}$}. Now, if we consider the contexts 
\hbox{$\underline{\eta_0}\eta_{\pi(1)}\eta_{\pi(2)}...
\eta_{\pi(13)}$}, for all circular permutations~$\pi$, there is a
{\bf minimal} one with respect to the lexicographic order. 
The rule \hbox{$\underline{\eta_0}\eta_{\pi(1)}\eta_{\pi(2)}...
\eta_{\pi(13)}\underline{\eta^1_0}$} is called the {\bf minimal rotated form}
of \hbox{$\underline{\eta_0}\eta_{\pi(1)}\eta_{\pi(2)}...
\eta_{\pi(13)}\underline{\eta^1_0}$} for the $\pi$ which realizes the minimal
rotated context.
Now, it is clear
that two rules are rotated images of each other if and only if
they have the same minimal rotated form. From this, we easily deduce an algorithm
for checking the coherence of any new rule introduced in an already coherent
set of rules. The programming of the algorithm raises no difficulty.
   The computer program was written in~$ADA$.

   The program uploads the initial configuration of the crossings and of
the switches from a file and puts the corresponding information into a 
table~0. In this table, each row represents a cell. The 
entries of the row indicate the coordinates of the neighbours of the cell as 
well as the states of the cell and of its neighbours. The program also contains 
a copy of table~0 with no state in the cells which we call 
table~1. The set of rules is in a file under an appropriate format, close to 
the one which was just depicted.

    During the construction of the set of rules, the program works as follows.

First, the program reads the file of the rules which, initially
contains the rule $\underline{\tt W}${\tt W$^{13}$}$\underline{\tt W}$ which
says that a cell in the quiescent state whose neighbours are all quiescent
remains quiescent. The program scans the cells of the list one after the other. 
It takes the context~$\kappa$ of the considered cell~$c$ in table~0. Then, 
it compares~$\kappa$ with the contexts of the rules of the file. 
If it finds a match, it copies the new state of the rule at the address 
of~$c$ in table~1, under column~0. If it does not find a match, it asks for a 
new rule which the user writes into the file. To help the user, the program 
indicates the context of the cell. The user enters the new state only. Then the 
program resumes its computation: it reads again table~0 from the initial 
configuration and performs the computation as far as possible. If it can 
compute
the new state of all cells of table~0, it completes table~1 by computing the
new states of the neighbours of each cell. When this task is over, the program
copies table~1 onto table~0: a new step of the computation of the cellular 
automaton can be processed. This cycle is repeated until no new rule is 
required and until the fixed in advance number of steps is reached. 

   Now, when a new rule is entered by the user on a cell~$c$, it may happen 
that the new rule is in conflict with the previously entered rules. 
This happens when there is a rule~$\eta$ whose context is a rotated form 
of the context of~$c$, but the state suggested by the user is not the new 
state of the rule. In this case, the program stops with an error message 
which also displays the rule with which the program have found a mismatch. 
If the rule constructed on the context of the cell and the state indicated 
by the user is a rotated form of an already existing rule, it is appended to 
the set of rules.

   When the program can be run without asking a new rule nor indicating
any error, we know that the set of rules is computed.

In the display of the tables of
Subsections~\ref{trackrules}, \ref{crossrules} and~\ref{switchrules},
we assume that the minimal rotated 
forms of the rules are pairwise distinct. 
This means that on the thirteen rotated forms which should be present for each 
rule, we keep only those needed by the tested configurations. We find 207~rules
and most of them are rotation independent from the others.

\subsection{Implementation of the configurations}   
   
    For each configuration of the crossings and the switches,
   the number of cells to explore would be rather important and only a very few 
number of them is supposed to change during the computation. And so, the idea
is to consider only the set of cells which are possibly changing. 
We have a two-dimensional table. The first column defines the set of cells
under inspection by the program. This defines these sets as an ordered list~$L$
from~1 to~$\vert L\vert$, the number of elements of~$L$. Each row corresponds to 
the same cell of this set. It contains the cell itself and the thirteen 
neighbours of the cell. For the cell and each neighbour, the corresponding entry 
contains two fields: {\it num} and {\it state}. For a cell, considered as 
neighbour~0  of its row, {\it num} is the number of the cell in $[1..\vert L\vert]$.
For a neighbour, {\it num}~$=$~0 if the neighbour never changes its state.
Otherwise, the considered neighbour is a cell of the list, say whose number
is~$m$, so that in this case, {\it num}~$=$~$m$. This table allows us to easily
apply the rules to any cell~$c$. It is enough to compute the minimal form of the rule
and also to compute the minimal context defined by the state of~$c$ and those
of its neighbours: it is enough to take them in the row associated to~$c$.
We look in the table after a rule such that the context of its minimal form
matches the minimal context of the neighbours of~$c$. When the rule is found,
the new state of~$c$ is found.

    Before turning to the various parts of a configuration, we shall divide the
rules which we shall define into two classes. First, we consider the 
{\bf conservative} rules which are applied to a cell without changing its state.
This is the case, in particular, for the cells of an idle configuration. The rules
applied to each cell must not change the state. The other rules, which we call
{\bf active}, do change the state of the cells to which they are applied.
This is in particular the case of the rules which apply to cells directly connected 
to the motion of the particle: we call them the {\bf motion} rules.

   We can now turn to the various configurations we have to study.

\subsection{Rules for the tracks}
\label{trackrules}

    Looking at Fig.~\ref{trackcells}, we first define its conservative rules.
The milestones are applied the rule $\underline{\tt B}{\tt W}^{13}\underline{\tt B}$
as well as the rule $\underline{\tt B}{\tt BW}^{12}\underline{\tt B}$ and even the
rule $\underline{\tt B}{\tt BWBW}^{10}\underline{\tt B}$ for the case when the     
cell has two 'supporting' milestones. There are four conservative rules for a cell 
of the tracks itself, as the number of pictures in Fig.~\ref{trackcells}: 

\ligne{\hfill
$\underline{\tt W}{\tt BBWBWBW}^{7}\underline{\tt W}$,\hskip 20pt
$\underline{\tt W}{\tt BBW}^7{\tt BWBW}\underline{\tt W}$,
\hfill}
\ligne{\hfill
$\underline{\tt W}{\tt BBWBBWBW}^{6}\underline{\tt W}$,\hskip 20pt
$\underline{\tt W}{\tt BBWBBW}^6{\tt BW}\underline{\tt W}$,
\hfill}

   Indeed, the four patterns of Fig.~\ref{trackcells} are rotationally different
as already noticed, so that the minimal forms of the rules are also different as
we can now see.

For the white neighbours
of the cell and for the white neighbours of the milestones, the rules
are: $\underline{\tt W}{\tt W}^{13}\underline{\tt W}$,
$\underline{\tt W}{\tt BW}^{12}\underline{\tt W}$ and
$\underline{\tt W}{\tt BBW}^{11}\underline{\tt W}$.

    Now, taking into account the definitions of the cells, we have the following
motion rules, first for one direction and then, the two last lines, for the
opposite direction:

\ligne{\hfill
$\underline{\tt W}{\tt B}^3{\tt W}^{7}{\tt BBW}\underline{\tt B}$,\hskip 15pt
$\underline{\tt B}{\tt BBWBWBW}^{7}\underline{\tt W}$,\hskip 15pt
$\underline{\tt W}{\tt B}^4{\tt WBW}^{7}\underline{\tt W}$
\hfill}

\ligne{\hfill
$\underline{\tt W}{\tt B}^4{\tt W}^{6}{\tt BBW}\underline{\tt B}$,\hskip 15pt
$\underline{\tt B}{\tt BBWBBWBW}^{6}\underline{\tt W}$,\hskip 15pt
$\underline{\tt W}{\tt B}^5{\tt WBW}^{6}\underline{\tt W}$
\hfill}

\ligne{\hfill
$\underline{\tt W}{\tt B}^3{\tt WBBW}^{7}\underline{\tt B}$,\hskip 15pt
$\underline{\tt B}{\tt BBW}^{7}{\tt BWBW}\underline{\tt W}$,\hskip 15pt
$\underline{\tt W}{\tt B}^4{\tt W}^{7}{\tt BW}\underline{\tt W}$
\hfill}

\ligne{\hfill
$\underline{\tt W}{\tt B}^4{\tt WBBW}^{6}\underline{\tt B}$,\hskip 15pt
$\underline{\tt B}{\tt BBWBBW}^{6}{\tt BW}\underline{\tt W}$,\hskip 15pt
$\underline{\tt W}{\tt B}^5{\tt W}^{6}{\tt BW}\underline{\tt W}$
\hfill}

Taking the same order for the non idle configurations, we can display them
as follows:

\vtop{
\vspace{10pt}
\ligne{\hskip 20pt
\includegraphics[scale=0.5]{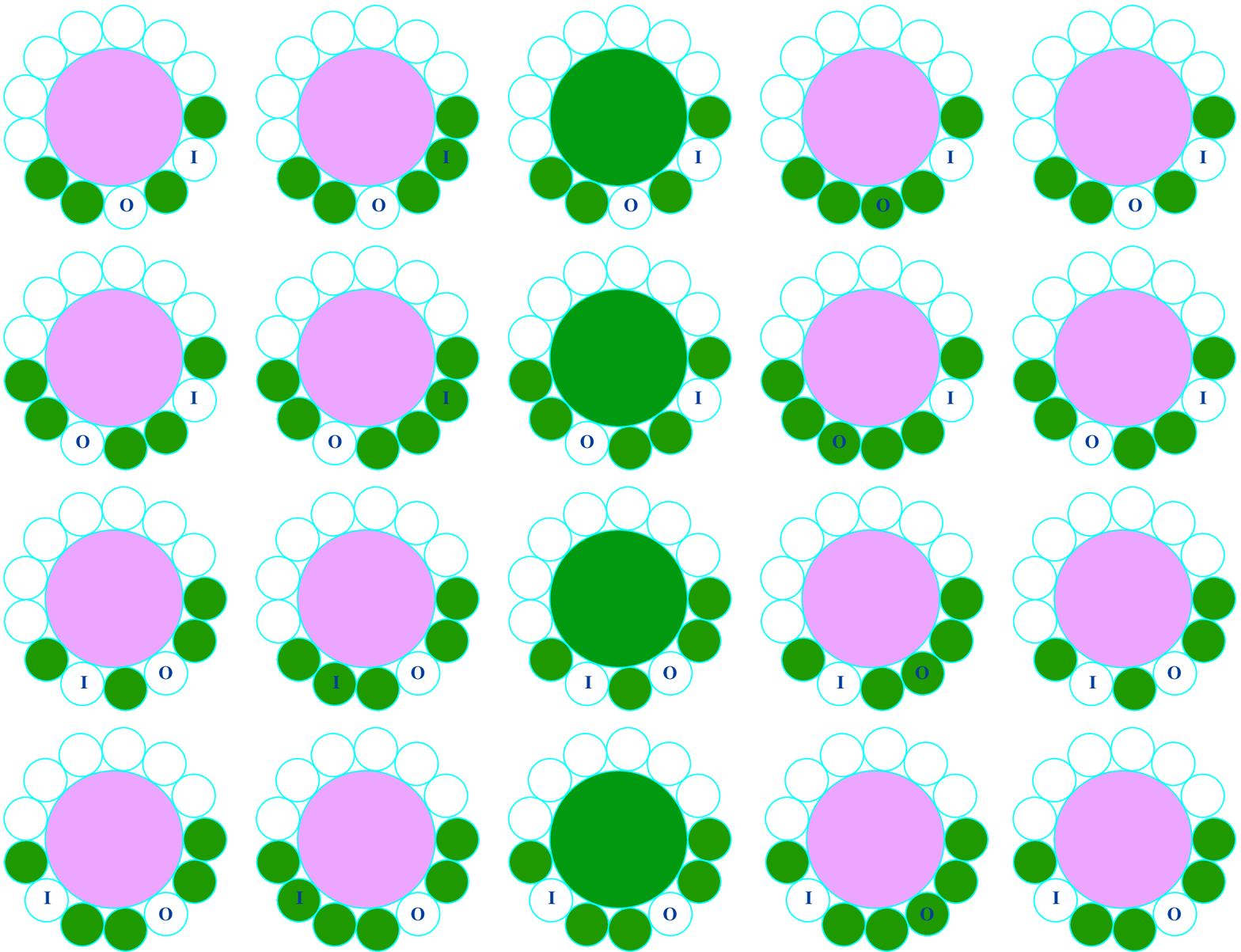}
\hfill}
\begin{fig}\label{trackmove}
\leurre
Illustration of the motion rules for the tracks. In the first and the last pictures 
of each row, we have the idle configuration. It is assumed that~$I$ and~$O$ are
both cells of the tracks. Number the neighbours from the first block of two 
consecutive black cells while turning around the cell~$c$. Let $c.1$~be the leftmost 
cell in this first block. The input
of~$c$ is~$c.5$ or~$c.6$ in the first two rows, $c.11$ or~$c.13$ in the last
two rows. Similarly, the output of~$c$ is~$c.3$ in the first two rows,
while it is~$c.13$ or~$c.3$ in the last two rows. 
\end{fig}
}

Now, we have to append a few rules for the case when two contiguous particles
travel on the tracks, which happens for instance in the tracks which runs along
a round-about.

\ligne{\hfill
$\underline{\tt B}{\tt B}^3{\tt W}^{7}{\tt BBW}\underline{\tt B}$,\hskip 15pt
$\underline{\tt B}{\tt B}^4{\tt W}^{6}{\tt BBW}\underline{\tt B}$,\hskip 15pt
$\underline{\tt B}{\tt B}^3{\tt WBBW}^{7}\underline{\tt B}$,\hskip 15pt
$\underline{\tt B}{\tt B}^4{\tt WBBW}^{6}\underline{\tt B}$,
\hfill}

\ligne{\hfill
$\underline{\tt B}{\tt B}^4{\tt WBW}^{7}\underline{\tt W}$,\hskip 15pt
$\underline{\tt B}{\tt B}^5{\tt WBW}^{6}\underline{\tt W}$,\hskip 15pt
$\underline{\tt B}{\tt B}^4{\tt W}^{7}{\tt BW}\underline{\tt W}$,\hskip 15pt
$\underline{\tt B}{\tt B}^5{\tt W}^{6}{\tt BW}\underline{\tt W}$.
\hfill}

   In these situations, these cells are black and there is another black cell,
either at the entrance of the particle. As mentioned in the caption of 
Fig.~\ref{trackmove}, the entrance of a cell~$c$ of the tracks is $c.5$ or~$c.6$ in 
one direction, while it the output is~$c.3$. In the other direction, the entrance 
is $c.11$ or~$c.13$ while the output is $c.13$ or~$c.3$ respectively. In the first 
case, the cell remains black, in the second one, it turns black to white.

It is not difficult to see that the above rules are in a minimal form, assuming
that {\tt B}~$<$~{\tt W}. As an easy corollary, they are rotation independent. 

   The flexibility of the way with which we can define tracks shows us that there
is no need of further simulations in order to check the correctness of these
rules.

\subsection{Rules for the crossings}
\label{crossrules}

   This time, we deduce the rules from the illustrations given in
Fig.\ref{idle_configs} and Fig.\ref{crossing0}. The figures already show that
most of the conservative rules are induced by the rules of the form {\tt XZX} where 
$\vert{\tt Z}\vert=13$ and $\vert{\tt Z}\vert_{\tt B}\leq2$.

   Here, the active rules are those of the tracks, in particular $A$, $B$, $C$, $D$,
$E$, $F$, $B.11$, $B.12$ and $C.5$, see Fig.~\ref{crossing0} and 
Subsection~\ref{crossings}. As $A$ and~$D$ are cells of the tracks, the rules
applying to them have already be seen in Subsection~\ref{trackrules}.
Now, $B$ and~$C$ have particular surroundings as they have a particular behaviour
with respect to cells of the tracks. Also, $E$ and~$F$ are cells of the 
tracks but at least one of their milestones is not permanently black: it may flash
at a definite event. This means that the milestone becomes white at a certain time,
turning back to black at the next time. Of course, $B.11$, $B.12$ and~$C.5$
have to be studied separately.

   In order to find out the rules, we indicate the neighbours of the cell
according to what was said above and then, we explain the construction of the
rules with the help of an appropriate table. The rows of the table follow 
the trajectory of the particle: the row below the row associated to time~$t$
is associated to time~$t$+1. We decide that the first row is always the 
conservative rule associated to the considered cell. It is considered that the
last rule is followed by the conservative rule so that the conservative rule
appears only once in the representation of a motion. 

\def\lignetitre{%
{\hsize=335pt
\ligne{\hbox to 20pt{\hfill 0\hfill}
\hbox to 20pt{\hfill 1\hfill}
\hbox to 20pt{\hfill 2\hfill}
\hbox to 20pt{\hfill 3\hfill}
\hbox to 20pt{\hfill 4\hfill}
\hbox to 20pt{\hfill 5\hfill}
\hbox to 20pt{\hfill 6\hfill}
\hbox to 20pt{\hfill 7\hfill}
\hbox to 20pt{\hfill 8\hfill}
\hbox to 20pt{\hfill 9\hfill}
\hbox to 20pt{\hfill 10\hfill}
\hbox to 20pt{\hfill 11\hfill}
\hbox to 20pt{\hfill 12\hfill}
\hbox to 20pt{\hfill 13\hfill}
\hbox to 20pt{\hfill \hfill}
\hfill}
}
}

\def\legendetrack{%
{\hsize=335pt
\ligne{\hbox to 20pt{\hfill 0\hfill}
\hbox to 20pt{\hfill \tt B\hfill}
\hbox to 20pt{\hfill \tt B\hfill}
\hbox to 20pt{\hfill $O$\hfill}
\hbox to 20pt{\hfill \tt B\hfill}
\hbox to 20pt{\hfill $I$\hfill}
\hbox to 20pt{\hfill \tt B\hfill}
\hbox to 20pt{\hfill \tt W\hfill}
\hbox to 20pt{\hfill \tt W\hfill}
\hbox to 20pt{\hfill \tt W\hfill}
\hbox to 20pt{\hfill \tt W\hfill}
\hbox to 20pt{\hfill \tt W\hfill}
\hbox to 20pt{\hfill \tt W\hfill}
\hbox to 20pt{\hfill \tt W\hfill}
\hbox to 20pt{\hfill $n$\hfill}
\hfill}
}
}
\def\tempstrack act #1 enI #2 enO #3 nov #4 {%
{\hsize=335pt
\ligne{\hbox to 20pt{\hfill\tt  #1\hfill}
\hbox to 20pt{\hfill\tt B\hfill}
\hbox to 20pt{\hfill\tt B\hfill}
\hbox to 20pt{\hfill\tt #3\hfill}
\hbox to 20pt{\hfill\tt B\hfill}
\hbox to 20pt{\hfill\tt #2\hfill}
\hbox to 20pt{\hfill\tt B\hfill}
\hbox to 20pt{\hfill\tt W\hfill}
\hbox to 20pt{\hfill\tt W\hfill}
\hbox to 20pt{\hfill\tt W\hfill}
\hbox to 20pt{\hfill\tt W\hfill}
\hbox to 20pt{\hfill\tt W\hfill}
\hbox to 20pt{\hfill\tt W\hfill}
\hbox to 20pt{\hfill\tt W\hfill}
\hbox to 20pt{\hfill\tt #4\hfill}
\hfill}
}
}
\vtop{
\begin{tab}\label{track_tab}
\leurre
Rules for a simple cell of the tracks
\end{tab}
\vspace{-10pt}
\grostrait
\ligne{\hfill
\lignetitre
\hfill}
\ligne{\hfill
\legendetrack
\hfill}
\demitrait
\tempstrack act W enI W enO W nov W 
\tempstrack act W enI B enO W nov B 
\tempstrack act B enI W enO W nov W 
\tempstrack act W enI W enO B nov W 
\demitrait
\vspace{10pt}
}

As an example, 
Table~\ref{track_tab} represents such a table for a simple cell of the tracks.
Accordingly, the table has four rows and we can see that the particle arriving
near~$c$ visits~$c.5$ at time~$t$+1, visits~$c$ at time~$t$+2 and then
visits~$c.3$ at time~$t$+3. It is clear that at times~$t$ and~$t$+3 the conservative
rule applies. We shall follow this convention in all tables.
Moreover, in all tables, we indicate in bold characters the particle as well as the
flashes of the sensors and the controllers. Also, in order to better handle the
rules, we give each of them the number it has in the program.

Our next table is the table devoted to the cell~$B$, see Table~\ref{cross_tabB}.
The crossing of a cell requires in fact three times as previously. Here, however,
the possible presence of a particle adds one more step and the flashes of two
cells brings in two additional steps.

\def\demirangeea #1 #2 #3 #4 #5 #6 #7 {%
\hbox to 15pt{\hfill\tt #1\hfill}
\hbox to 15pt{\hfill\tt #2\hfill}
\hbox to 15pt{\hfill\tt #3\hfill}
\hbox to 15pt{\hfill\tt #4\hfill}
\hbox to 15pt{\hfill\tt #5\hfill}
\hbox to 15pt{\hfill\tt #6\hfill}
\hbox to 15pt{\hfill\tt #7\hfill}
}
\def\demirangeeb #1 #2 #3 #4 #5 #6 #7 #8 {%
\hbox to 15pt{\hfill\tt #1\hfill}
\hbox to 15pt{\hfill\tt #2\hfill}
\hbox to 15pt{\hfill\tt #3\hfill}
\hbox to 15pt{\hfill\tt #4\hfill}
\hbox to 15pt{\hfill\tt #5\hfill}
\hbox to 15pt{\hfill\tt #6\hfill}
\hbox to 15pt{\hfill\tt #7\hfill}
\hbox to 15pt{\hfill\tt #8\hfill}
}

\vtop{
\begin{tab}\label{cross_tabB}
\leurre
Rules for $B$. The cell is 
concerned in three different
situations which are clearly
separated.
\end{tab}
\vspace{-10pt}
\grostrait
\ligne{\hfill
\hbox to 288pt{%
\demirangeea 0 1 2 3 4 5 6  \demirangeeb {7} {8} {9} {10} {11} {12} {13}  {}
\hbox to 50pt{\hfill $n$\hskip 20pt}
\hfill}\hfill}
\ligne{\hfill
\hbox to 288pt{%
\demirangeea {} {} {\bf A} {} {} {} {}  \demirangeeb {} {} {} {\bf F} 
{\small \bf B$^{11}$} {\small\bf B$^{12}$} {\bf C}  {}
\hbox to 50pt{\hfill}
\hfill}\hfill}
\vspace{-4pt}
\demitrait
\vspace{4pt}
\ligne{\hfill\tt
-I- a particle arrives to the round-about from outside
\hfill}
\demitrait
\vspace{4pt}
\ligne{\hfill
\hbox to 288pt{%
\demirangeea W B W B W W W   \demirangeeb W B B W B B W W 
\hbox to 50pt{\hfill 9\hskip 20pt}
\hfill}\hfill}
\ligne{\hfill
\hbox to 288pt{%
\demirangeea W B W B W W W   \demirangeeb W B B W B B {\small\bf B} W 
\hbox to 50pt{\hfill 13\hskip 20pt}
\hfill}\hfill}
\ligne{\hfill
\hbox to 288pt{%
\demirangeea W B W B W W W   \demirangeeb W B B W B B {\small\bf B} W 
\hbox to 50pt{\hfill 13\hskip 20pt}
\hfill}\hfill}
\demitrait
\vspace{4pt}
\ligne{\hfill\tt
-II- two particles arrive from the round-about
\hfill}
\demitrait
\vspace{4pt}
\ligne{\hfill
\hbox to 288pt{%
\demirangeea W B W B W W W   \demirangeeb W B B W B B W W 
\hbox to 50pt{\hfill 9\hskip 20pt}
\hfill}\hfill}
\ligne{\hfill
\hbox to 288pt{%
\demirangeea W B {\small\bf B} B W W W   \demirangeeb W B B W B B W B 
\hbox to 50pt{\hfill 10\hskip 20pt}
\hfill}\hfill}
\ligne{\hfill
\hbox to 288pt{%
\demirangeea {\small\bf B} B {\small\bf B} B W W W   \demirangeeb W B B W B B W B 
\hbox to 50pt{\hfill 11\hskip 20pt}
\hfill}\hfill}
\ligne{\hfill
\hbox to 288pt{%
\demirangeea {\small\bf B} B W B W W W   \demirangeeb W B B {\small\bf B} B B W W 
\hbox to 50pt{\hfill 12\hskip 20pt}
\hfill}\hfill}
\ligne{\hfill
\hbox to 288pt{%
\demirangeea W B W B W W W   \demirangeeb W B B {\small\bf B} {\small\bf W} B W W 
\hbox to 50pt{\hfill 19\hskip 20pt}
\hfill}\hfill}
\ligne{\hfill
\hbox to 288pt{%
\demirangeea W B W B W W W   \demirangeeb W B B W B {\small\bf W} W W 
\hbox to 50pt{\hfill 8\hskip 20pt}
\hfill}\hfill}
\ligne{\hfill
\hbox to 288pt{%
\demirangeea W B W B W W W   \demirangeeb W B B W {\small\bf W} B B W 
\hbox to 50pt{\hfill 14\hskip 20pt}
\hfill}\hfill}
\demitrait
\vspace{4pt}
\ligne{\hfill\tt
-III- a single particle arrives from the round-about
\hfill}
\demitrait
\vspace{4pt}
\ligne{\hfill
\hbox to 288pt{%
\demirangeea W B W B W W W   \demirangeeb W B B W B B W W 
\hbox to 50pt{\hfill 9\hskip 20pt}
\hfill}\hfill}
\ligne{\hfill
\hbox to 288pt{%
\demirangeea W B {\small\bf B} B W W W   \demirangeeb W B B W B B W B 
\hbox to 50pt{\hfill 10\hskip 20pt}
\hfill}\hfill}
\ligne{\hfill
\hbox to 288pt{%
\demirangeea {\small\bf B} B W B W W W   \demirangeeb W B B W B B W W 
\hbox to 50pt{\hfill 15\hskip 20pt}
\hfill}\hfill}
\ligne{\hfill
\hbox to 288pt{%
\demirangeea W B W B W W W   \demirangeeb W B B {\small\bf B} B B W W 
\hbox to 50pt{\hfill 21\hskip 20pt}
\hfill}\hfill}
\demitrait\vspace{15pt}
}

   Note that in the part -II- of the table, the last time is triggered by the
new particle created in~$C$ after the destruction of the two particles which
passed through~$F$: as $B$ and~$C$ are in contact, this rule is also needed.

   We can see in Table~\ref{cross_tabB} that the cell~$B$ is mainly concerned 
when one or two particles arrive to the branching from the round-about. It 
witnesses~$F$ and also the two sensors $B.11$ and~$B.12$ which discriminate
the distinction between the two cases and rule the situation when a new particle
is generated for~$C$. Rule~12 can see that the conditions for the flashing of~$B.11$
are present and rule~19 witnesses the flash.
As~$C$ can be seen from~$B$, $B$~is also a witness of the creation of the single
particle which goes on on the round-about. This is witnessed by rule~14.
The flashes of~$B.11$ and $B.12$ are clearly visible: rule~19 and~14 for~$B.11$,
rule~8 for~$B.12$.

\vtop{
\begin{tab}\label{cross_tabC}
\leurre
Rules for $C$. The cell is 
concerned in three different
situations which are clearly
separated.
\end{tab}
\vspace{-10pt}
\grostrait
\ligne{\hfill
\hbox to 288pt{%
\demirangeea 0 1 2 3 4 5 6  \demirangeeb {7} {8} {9} {10} {11} {12} {13}  {}
\hbox to 50pt{\hfill $n$\hskip 20pt}
\hfill}\hfill}
\ligne{\hfill
\hbox to 288pt{%
\demirangeea {} {} {\bf B} {\small\bf B$^{12}$} {\bf E} {\small\bf C$^5$} {}  
\demirangeeb {} {} {} {} {} {} {\bf D}  {}
\hbox to 50pt{\hfill}
\hfill}\hfill}
\vspace{-4pt}
\demitrait
\vspace{4pt}
\ligne{\hfill\tt
-I- a particle arrives to the round-about from outside
\hfill}
\demitrait
\vspace{4pt}
\ligne{\hfill
\hbox to 288pt{%
\demirangeea W B W B W B W   \demirangeeb W W W B B B W W 
\hbox to 50pt{\hfill 19\hskip 20pt}
\hfill}\hfill}
\ligne{\hfill
\hbox to 288pt{%
\demirangeea W B W B {\small\bf B} B W   \demirangeeb W W W B B B W B 
\hbox to 50pt{\hfill 16\hskip 20pt}
\hfill}\hfill}
\ligne{\hfill
\hbox to 288pt{%
\demirangeea {\small\bf B} B W B W {\small\bf W} W   \demirangeeb W W W B B B W B 
\hbox to 50pt{\hfill 17\hskip 20pt}
\hfill}\hfill}
\ligne{\hfill
\hbox to 288pt{%
\demirangeea {\small\bf B} B W B W B W   \demirangeeb W W W B B B {\small\bf B} W 
\hbox to 50pt{\hfill 12\hskip 20pt}
\hfill}\hfill}
\ligne{\hfill
\hbox to 288pt{%
\demirangeea W B W B W B W   \demirangeeb W W W B B B {\small\bf B} W 
\hbox to 50pt{\hfill 21\hskip 20pt}
\hfill}\hfill}
\demitrait
\vspace{4pt}
\ligne{\hfill\tt
-II- two particles arrive to the round-about 
\hfill}
\demitrait
\vspace{4pt}
\ligne{\hfill
\hbox to 288pt{%
\demirangeea W B W B W B W   \demirangeeb W W W B B B W W 
\hbox to 50pt{\hfill 19\hskip 20pt}
\hfill}\hfill}
\ligne{\hfill
\hbox to 288pt{%
\demirangeea W B {\small\bf B} B W B W   \demirangeeb W W W B B B W W 
\hbox to 50pt{\hfill 22\hskip 20pt}
\hfill}\hfill}
\ligne{\hfill
\hbox to 288pt{%
\demirangeea W B {\small\bf B} B W B W   \demirangeeb W W W B B B W W 
\hbox to 50pt{\hfill 22\hskip 20pt}
\hfill}\hfill}
\ligne{\hfill
\hbox to 288pt{%
\demirangeea W B W B W B W   \demirangeeb W W W B B B W W 
\hbox to 50pt{\hfill 19\hskip 20pt}
\hfill}\hfill}
\ligne{\hfill
\hbox to 288pt{%
\demirangeea W B W {\small\bf W} W B W   \demirangeeb W W W B B B W {\small\bf B} 
\hbox to 50pt{\hfill 18\hskip 20pt}
\hfill}\hfill}
\ligne{\hfill
\hbox to 288pt{%
\demirangeea {\small\bf B} B W B W B W   \demirangeeb W W W B B B W W 
\hbox to 50pt{\hfill 20\hskip 20pt}
\hfill}\hfill}
\ligne{\hfill
\hbox to 288pt{%
\demirangeea W B W B W B W   \demirangeeb W W W B B B {\small\bf B} W 
\hbox to 50pt{\hfill 21\hskip 20pt}
\hfill}\hfill}
\demitrait
\vspace{4pt}
\ligne{\hfill\tt
-III- a single particle arrives to the round-about 
\hfill}
\demitrait
\vspace{4pt}
\ligne{\hfill
\hbox to 288pt{%
\demirangeea W B W B W B W   \demirangeeb W W W B B B W W 
\hbox to 50pt{\hfill 19\hskip 20pt}
\hfill}\hfill}
\ligne{\hfill
\hbox to 288pt{%
\demirangeea W B {\small\bf B} B W B W   \demirangeeb W W W B B B W W 
\hbox to 50pt{\hfill 22\hskip 20pt}
\hfill}\hfill}
\demitrait\vspace{15pt}
}

   We can see that the cell~$C$ is not very much concerned by the case when a single 
particle arrives to the branching from the round about. The cell is mainly concerned
when a particle arrives at the round-about from outside and when two particles
arrive to the branching from the round-about.

   In the first case, we can see that two particles are created, rules~16 and~17.
Rule~16 transfers the particle from~$E$ to~$C$. Rule~17 creates a new particle
as it can see the flash of~$C.5$. Next, rules~12 and~21 control the motion of
two particles together.

   When two particles arrive from he round-about, $C$ is concerned by the creation
of the single particle. We note that rule~22 is applied twice: it simply witnesses
that two particles pass through~$B$. The flash of~$B.12$ allows rule~18 to create
the single particle. Rules~20 and~21 act as simple motion rules adapted to the context
of~$C$.

   Note that rule~19 appeared already in Table~\ref{cross_tabB}. Here, in
Table~\ref{cross_tabB} it is a conservative rule while in Table~\ref{cross_tabB}
it allows us to witness the flash of~$B.11$. In fact, the two different forms
of the rule are compatible because they are rotated forms of each other.

\vtop{
\begin{tab}\label{cross_tabE}
\leurre
Rules for $E$. The cell is 
concerned in two different
situations which are clearly
separated.
\end{tab}
\vspace{-10pt}
\grostrait
\ligne{\hfill
\hbox to 288pt{%
\demirangeea 0 1 2 3 4 5 6  \demirangeeb {7} {8} {9} {10} {11} {12} {13}  {}
\hbox to 50pt{\hfill $n$\hskip 20pt}
\hfill}\hfill}
\ligne{\hfill
\hbox to 288pt{%
\demirangeea {} {\small\bf B$^{12}$} {} {} {} {} {}  
\demirangeeb {} {} {} {} {$tr$} {\small\bf C$^5$} {\bf C}  {}
\hbox to 50pt{\hfill}
\hfill}\hfill}
\vspace{-4pt}
\demitrait
\vspace{4pt}
\ligne{\hfill\tt
-I- a particle arrives to the round-about from outside
\hfill}
\demitrait
\vspace{4pt}
\ligne{\hfill
\hbox to 288pt{%
\demirangeea W B B W W W W   \demirangeeb W W W B W B W W
\hbox to 50pt{\hfill 94\hskip 20pt}
\hfill}\hfill}
\ligne{\hfill
\hbox to 288pt{%
\demirangeea W B B W W W W   \demirangeeb W W W B {\small\bf B} B W B
\hbox to 50pt{\hfill 95\hskip 20pt}
\hfill}\hfill}
\ligne{\hfill
\hbox to 288pt{%
\demirangeea {\small\bf B} B B W W W W   \demirangeeb W W W B W B W W
\hbox to 50pt{\hfill 171\hskip 20pt}
\hfill}\hfill}
\ligne{\hfill
\hbox to 288pt{%
\demirangeea W B B W W W W   \demirangeeb W W W B W {\small\bf W} {\small\bf B} W
\hbox to 50pt{\hfill 189\hskip 20pt}
\hfill}\hfill}
\ligne{\hfill
\hbox to 288pt{%
\demirangeea W B B W W W W   \demirangeeb W W W B W {\small\bf B} B W
\hbox to 50pt{\hfill 96\hskip 20pt}
\hfill}\hfill}
\demitrait
\vspace{4pt}
\ligne{\hfill\tt
-II- two particles arrive from the round-about
\hfill}
\demitrait
\vspace{4pt}
\ligne{\hfill
\hbox to 288pt{%
\demirangeea W B B W W W W   \demirangeeb W W W B W B W W
\hbox to 50pt{\hfill 94\hskip 20pt}
\hfill}\hfill}
\ligne{\hfill
\hbox to 288pt{%
\demirangeea W {\small\bf W} B W W W W   \demirangeeb W W W B W B W W
\hbox to 50pt{\hfill 49\hskip 20pt}
\hfill}\hfill}
\ligne{\hfill
\hbox to 288pt{%
\demirangeea W B B W W W W   \demirangeeb W W W B W {\small\bf B} B W
\hbox to 50pt{\hfill 96\hskip 20pt}
\hfill}\hfill}
\demitrait\vspace{15pt}
}

   In Table~\ref{cross_tabE}, we can see that rule~91 is the conservative rule
attached to the configuration of the neighbours around~$E$. 

   When a particle arrives to the round-about from the tracks, it runs 
over~$E$. In the caption of the table below the numbers of the neighbours, we 
can see that $E.11$~is a cell of the track arriving to~$E$. In fact, $E$ has 
the configuration of an ordinary cell of the track and rule~94 is simply a 
rotated form of the first rule in Table~\ref{track_tab}. Now, rules~95 and~181
convey the particle to~$E$ and then to~$C$, see also rule~16{} in
Table~\ref{cross_tabC}. This makes~$C.5$ flash which is witnessed by
rule~199 which triggers the creation of the second particle in~$C$. 
Rule~96 also witnesses the second particle in~$C$.

   In Table~\ref{cross_tabF}, the conservative rule is rule~2. This explains why we 
said that cell~$F$ is almost a cell of the tracks. We said almost because one of
its milestones is in fact a sensor: it is the cell~$B.11$ which can see both~$F$
and~$B$ and thus, is able to distinguish the case when two particles arrive at~$B$
from the case when a single one arrives at~$B$. This is witnessed by rule~4 
for the presence of two particles on the round-about, while rule~44 witnesses the
flash of~$B.11$. The second flash of~$B.11$ is witnessed by the rule~47.

   The second part of the table is devoted to the case when a single particle
arrives at~$B$. In this case, the rule~58 witnesses that there is a single particle.
We can notice that in this case, as $B.11$~remains black, the cell~$F$ behaves
as an ordinary cell of the tracks. Note that the rules are different from those
related to rule~91. Indeed, in both cases we have rules for cells of the tracks
but one cell is in one direction while the other cell is in the opposite one.

\vtop{
\begin{tab}\label{cross_tabF}
\leurre
Rules for $F$. The cell is 
concerned in two different
situations which are clearly
separated.
\end{tab}
\vspace{-10pt}
\grostrait
\ligne{\hfill
\hbox to 288pt{%
\demirangeea 0 1 2 3 4 5 6  \demirangeeb {7} {8} {9} {10} {11} {12} {13}  {}
\hbox to 50pt{\hfill $n$\hskip 20pt}
\hfill}\hfill}
\ligne{\hfill
\hbox to 288pt{%
\demirangeea {} {} {} {$tr$} {\small\bf B$^{11}$} {\bf B} {}  
\demirangeeb {} {} {} {} {} {} {}  {}
\hbox to 50pt{\hfill}
\hfill}\hfill}
\vspace{-4pt}
\demitrait
\vspace{4pt}
\ligne{\hfill\tt
-II- two particle arrive from the round-about
\hfill}
\demitrait
\vspace{4pt}
\ligne{\hfill
\hbox to 288pt{%
\demirangeea W B B W B W B   \demirangeeb W W W W W W W W 
\hbox to 50pt{\hfill 2\hskip 20pt}
\hfill}\hfill}
\ligne{\hfill
\hbox to 288pt{%
\demirangeea W B B W B {\small\bf B} B   \demirangeeb W W W W W W W B 
\hbox to 50pt{\hfill 3\hskip 20pt}
\hfill}\hfill}
\ligne{\hfill
\hbox to 288pt{%
\demirangeea {\small\bf B} B B W B {\small\bf B} B   \demirangeeb W W W W W W W B 
\hbox to 50pt{\hfill 4\hskip 20pt}
\hfill}\hfill}
\ligne{\hfill
\hbox to 288pt{%
\demirangeea {\small\bf B} B B {\small\bf B} {\small\bf W} W B   
\demirangeeb W W W W W W W W 
\hbox to 50pt{\hfill 44\hskip 20pt}
\hfill}\hfill}
\ligne{\hfill
\hbox to 288pt{%
\demirangeea W B B W B W B   \demirangeeb W W W W W W W W 
\hbox to 50pt{\hfill 2\hskip 20pt}
\hfill}\hfill}
\ligne{\hfill
\hbox to 288pt{%
\demirangeea W B B W {\small\bf W} W B   \demirangeeb W W W W W W W W 
\hbox to 50pt{\hfill 47\hskip 20pt}
\hfill}\hfill}
\demitrait
\vspace{4pt}
\ligne{\hfill\tt
-III- a single particle arrives from the round-about
\hfill}
\demitrait
\vspace{4pt}
\ligne{\hfill
\hbox to 288pt{%
\demirangeea W B B W B W B   \demirangeeb W W W W W W W W 
\hbox to 50pt{\hfill 2\hskip 20pt}
\hfill}\hfill}
\ligne{\hfill
\hbox to 288pt{%
\demirangeea W B B W B {\small\bf B} B   \demirangeeb W W W W W W W B 
\hbox to 50pt{\hfill 3\hskip 20pt}
\hfill}\hfill}
\ligne{\hfill
\hbox to 288pt{%
\demirangeea {\small\bf B} B B W B W B   \demirangeeb W W W W W W W W 
\hbox to 50pt{\hfill 58\hskip 20pt}
\hfill}\hfill}
\ligne{\hfill
\hbox to 288pt{%
\demirangeea W B B {\small\bf B} B W B   \demirangeeb W W W W W W W W 
\hbox to 50pt{\hfill 5\hskip 20pt}
\hfill}\hfill}
\demitrait\vspace{15pt}
}

   Table~\ref{cross_tabB11} is devoted to the rules needed by the cell~$B.11$.
This cell shares with~$F$ the fact that it is not concerned by the arrival
of a particle to the round-about from outside. In fact, as explained in
Section~\ref{implement}, the role of this cell is to detect the occurrence
of one or two particles to~$F$. Rule~24 witnesses the arrival of the particle at~$B$
and rule~26 can see that there is a particle at~$F$ and a particle at~$B$. This is
why the rule makes~$B.11$ flash. Rule~27 witnesses this flash and
rule~167 witnesses the consequence of this flash: both the cancellation of
the particles on the tracks which go around~$B.11$ and the flash of~$B.12$
caused by the previous flash of~$B.11$. And the rule makes~$B.11$ flash again.
The reason is that rule~167 is also used in another context where it makes
a black cell flash in the memory switch: see further Table~\ref{mm_tabI}. 

   When a single particle arrives at~$B$, $B.11$ can see the particle travel on four
cells of the tracks which are its neighbours, in consecutive positions: $B$, $F$
and the cells called~$tr_1$ and~$tr_2$ in the caption of the table under the
numbers of the neighbours. We can see that the particle visits the four cells
successively, the one after the other.

   It is the point to note that the branching is different, depending on whether 
the branching concerns tracks which arrive at the round-about or tracks
which leave the round-about. If the tracks arrive at the round-about, the arrival
at the round-about is~$C$ and the tracks pass through~$E$. In this case, there
is no need to define cells of the tracks after~$tr_1$ and~$tr_2$ which are
needed by the test for discriminating the number of arriving particles. On the
contrary, when the branching concerns tracks which leave the round-about,
the neighbour of~$E$ need not be a cell of the tracks and the tracks which
leave the round-about goes through~$B$, $F$, $tr_1$ and~$tr_2$.  

\vtop{
\begin{tab}\label{cross_tabB11}
\leurre
Rules for $B11$. The cell is 
concerned in two different
situations which are clearly
separated.
\end{tab}
\vspace{-10pt}
\grostrait
\ligne{\hfill
\hbox to 288pt{%
\demirangeea 0 1 2 3 4 5 6  \demirangeeb {7} {8} {9} {10} {11} {12} {13}  {}
\hbox to 50pt{\hfill $n$\hskip 20pt}
\hfill}\hfill}
\ligne{\hfill
\hbox to 288pt{%
\demirangeea {} {\bf B} {\bf F} {$tr_1$} {$tr_2$} {} {}  
\demirangeeb {} {} {} {} {} {} {\small\bf B$^{12}$}  {}
\hbox to 50pt{\hfill}
\hfill}\hfill}
\vspace{-4pt}
\demitrait
\vspace{4pt}
\ligne{\hfill\tt
-II- two particle arrive from the round-about
\hfill}
\demitrait
\vspace{4pt}
\ligne{\hfill
\hbox to 288pt{%
\demirangeea B W W W W B B   \demirangeeb B B W B B W B B 
\hbox to 50pt{\hfill 23\hskip 20pt}
\hfill}\hfill}
\ligne{\hfill
\hbox to 288pt{%
\demirangeea B {\small\bf B} W W W B B   \demirangeeb B B W B B W B B 
\hbox to 50pt{\hfill 24\hskip 20pt}
\hfill}\hfill}
\ligne{\hfill
\hbox to 288pt{%
\demirangeea B {\small\bf B} {\small\bf B} W W B B   \demirangeeb B B W B B W B W 
\hbox to 50pt{\hfill 26\hskip 20pt}
\hfill}\hfill}
\ligne{\hfill
\hbox to 288pt{%
\demirangeea {\small\bf W} W {\small\bf B} {\small\bf B} W B B   
\demirangeeb B B W B B W B B 
\hbox to 50pt{\hfill 27\hskip 20pt}
\hfill}\hfill}
\ligne{\hfill
\hbox to 288pt{%
\demirangeea B W W W W B B   \demirangeeb B B W B B W {\small\bf W} W
\hbox to 50pt{\hfill 157\hskip 20pt}
\hfill}\hfill}
\ligne{\hfill
\hbox to 288pt{%
\demirangeea {\small\bf W} W W W W B B   \demirangeeb B B W B B W B B 
\hbox to 50pt{\hfill 29\hskip 20pt}
\hfill}\hfill}
\demitrait
\vspace{4pt}
\ligne{\hfill\tt
-III- a single particle arrives from the round-about
\hfill}
\demitrait
\vspace{4pt}
\ligne{\hfill
\hbox to 288pt{%
\demirangeea B W W W W B B   \demirangeeb B B W B B W B B 
\hbox to 50pt{\hfill 23\hskip 20pt}
\hfill}\hfill}
\ligne{\hfill
\hbox to 288pt{%
\demirangeea B {\small\bf B} W W W B B   \demirangeeb B B W B B W B B 
\hbox to 50pt{\hfill 24\hskip 20pt}
\hfill}\hfill}
\ligne{\hfill
\hbox to 288pt{%
\demirangeea B W {\small\bf B} W W B B   \demirangeeb B B W B B W B B 
\hbox to 50pt{\hfill 25\hskip 20pt}
\hfill}\hfill}
\ligne{\hfill
\hbox to 288pt{%
\demirangeea B W W {\small\bf B} W B B   \demirangeeb B B W B B W B B 
\hbox to 50pt{\hfill 30\hskip 20pt}
\hfill}\hfill}
\ligne{\hfill
\hbox to 288pt{%
\demirangeea B W W W {\small\bf B} B B   \demirangeeb B B W B B W B B 
\hbox to 50pt{\hfill 31\hskip 20pt}
\hfill}\hfill}
\demitrait\vspace{15pt}
}

   Table~\ref{cross_tabB12} deals with the cell~$B.12$ and as this cell is
a common neighbour of~$B$ and~$C$, it is concerned by all the cases we have to
consider for a crossing. And so, we have the splitting into three cases already
seen in Tables~\ref{cross_tabB} and~\ref{cross_tabC}. The table is a bit shorter
than Table~\ref{cross_tabC} for instance. 

    The conservative rule for the cell~$B.12$ is rule~32. It is of course
different from the that of~$B.11$. We can notice that the pattern attached to
the neighbourhood of~$B.12$ consists of almost black cells only for the cells
which are invariant.
  
    In the first part of the table, the particle arrives
at the round-about through the cell~$E$. This is witnessed by rule~34 and rule~35
witnesses the presence of the particle at~$C$. The rule is applied twice as
a second particle is created in~$C$.

    In the second part, the role of~$B.12$ is important: it conveys the signal
sent by~$B.11$ that two particles arrived to~$C$, in order to trigger a new particle
which, from~$C$, will go to the next branching. Rule~36 applies twice as two 
particles arrive at~$B$. Rule~37 recognizes the flash of~$B.11$ and makes $B.12$
flash also. When $B.11$~flashes again, the new particle is in~$C$: this
allows rule~38 to cancel the flash of~$B.12$ and to ignore the second flash 
of~$B.11$.

    In the third part, as $B.11$ always remains black, rule~36 simply witnesses
that the particle passed through~$B$.

\vtop{
\begin{tab}\label{cross_tabB12}
\leurre
Rules for $B12$. The cell is 
concerned in three different
situations which are clearly
separated.
\end{tab}
\vspace{-10pt}
\grostrait
\ligne{\hfill
\hbox to 288pt{%
\demirangeea 0 1 2 3 4 5 6  \demirangeeb {7} {8} {9} {10} {11} {12} {13}  {}
\hbox to 50pt{\hfill $n$\hskip 20pt}
\hfill}\hfill}
\ligne{\hfill
\hbox to 288pt{%
\demirangeea {} {\bf B} {\small\bf B$^{11}$} {} {} {} {} 
\demirangeeb {} {} {} {} {} {\bf E} {\bf C}  {}
\hbox to 50pt{\hfill}
\hfill}\hfill}
\vspace{-4pt}
\demitrait
\vspace{4pt}
\ligne{\hfill\tt
-I- a particle arrives to the round-about from outside
\hfill}
\demitrait
\vspace{4pt}
\ligne{\hfill
\hbox to 288pt{%
\demirangeea B W B B B W B   \demirangeeb B B B B W W W B 
\hbox to 50pt{\hfill 32\hskip 20pt}
\hfill}\hfill}
\ligne{\hfill
\hbox to 288pt{%
\demirangeea B W B B B W B   \demirangeeb B B B B W {\small\bf B} W B 
\hbox to 50pt{\hfill 34\hskip 20pt}
\hfill}\hfill}
\ligne{\hfill
\hbox to 288pt{%
\demirangeea B W B B B W B   \demirangeeb B B B B W W {\small\bf B} B 
\hbox to 50pt{\hfill 35\hskip 20pt}
\hfill}\hfill}
\ligne{\hfill
\hbox to 288pt{%
\demirangeea B W B B B W B   \demirangeeb B B B B W W {\small\bf B} B 
\hbox to 50pt{\hfill 35\hskip 20pt}
\hfill}\hfill}
\demitrait
\vspace{4pt}
\ligne{\hfill\tt
-II- two particles arrive from the round-about
\hfill}
\demitrait
\vspace{4pt}
\ligne{\hfill
\hbox to 288pt{%
\demirangeea B W B B B W B   \demirangeeb B B B B W W W B 
\hbox to 50pt{\hfill 32\hskip 20pt}
\hfill}\hfill}
\ligne{\hfill
\hbox to 288pt{%
\demirangeea B {\small\bf B} B B B W B   \demirangeeb B B B B W W W B 
\hbox to 50pt{\hfill 36\hskip 20pt}
\hfill}\hfill}
\ligne{\hfill
\hbox to 288pt{%
\demirangeea B {\small\bf B} B B B W B   \demirangeeb B B B B W W W B 
\hbox to 50pt{\hfill 36\hskip 20pt}
\hfill}\hfill}
\ligne{\hfill
\hbox to 288pt{%
\demirangeea B W {\small\bf W} B B W B   \demirangeeb B B B B W W W W 
\hbox to 50pt{\hfill 37\hskip 20pt}
\hfill}\hfill}
\ligne{\hfill
\hbox to 288pt{%
\demirangeea {\small\bf W} W B B B W B   \demirangeeb B B B B W W W B 
\hbox to 50pt{\hfill 33\hskip 20pt}
\hfill}\hfill}
\ligne{\hfill
\hbox to 288pt{%
\demirangeea B W {\small\bf W} B B W B   \demirangeeb B B B B W W B B 
\hbox to 50pt{\hfill 38\hskip 20pt}
\hfill}\hfill}
\demitrait
\vspace{4pt}
\ligne{\hfill\tt
-III- a single particle arrives from the round-about
\hfill}
\demitrait
\vspace{4pt}
\ligne{\hfill
\hbox to 288pt{%
\demirangeea B W B B B W B   \demirangeeb B B B B W W W B 
\hbox to 50pt{\hfill 32\hskip 20pt}
\hfill}\hfill}
\ligne{\hfill
\hbox to 288pt{%
\demirangeea B {\small\bf B} B B B W B   \demirangeeb B B B B W W W B 
\hbox to 50pt{\hfill 36\hskip 20pt}
\hfill}\hfill}
\demitrait\vspace{15pt}
}

Table~\ref{cross_tabC5} gives the rules for~$C.5$ during a crossing. This time,
the cell is a neighbour of~$C$ but it cannot see~$B$ neither~$B.12$. And so,
it is concerned by the arrival of the particle at the round-about from the tracks
only and by the arrival of two particles at~$B$. In the latter case, this triggers
an ultimate creation of a particle in~$C$, a feature that~$C.5$ can detect.
Consequently, the table has two parts.

   In the first part of the table, we can see that the conservative rule is
rule~39. It indicates a new pattern for the black neighbours which remain invariant.
Rule~40 detects the arrival of the particle at the cell of the tracks which is just
before~$E$. Rule~41 detects the presence of the particle in~$E$ so that the rule
makes~$C.5$ flash when the particle is in~$C$. This is witnessed by rule~42
which can see that the particle is in~$C$ and which makes~$C.5$ turn back to 
black. Now, this flash of~$C.5$ triggers the presence of a new particle in~$C$.
The particle which was seen by rule~42 is now further on the round-about so that the
particle which is seen in~$C$ by rule~43 is the second one. The rule simply
witnesses this fact and the state of~$C.5$ is unchanged. The second particle follows
the first one on their way to the next branching.

   In the second part, the particle witnesses the creation of the particle in~$C$:
again rule~43 applies. The particle goes on its way to the next branching.

\vtop{
\begin{tab}\label{cross_tabC5}
\leurre
Rules for $C5$. The cell is 
concerned in two different
situations which are clearly
separated.
\end{tab}
\vspace{-10pt}
\grostrait
\ligne{\hfill
\hbox to 288pt{%
\demirangeea 0 1 2 3 4 5 6  \demirangeeb {7} {8} {9} {10} {11} {12} {13}  {}
\hbox to 50pt{\hfill $n$\hskip 20pt}
\hfill}\hfill}
\ligne{\hfill
\hbox to 288pt{%
\demirangeea {} {\bf C} {\bf E} {$tr$} {} {} {} 
\demirangeeb {} {} {} {} {} {} {}  {}
\hbox to 50pt{\hfill}
\hfill}\hfill}
\vspace{-4pt}
\demitrait
\vspace{4pt}
\ligne{\hfill\tt
-I- a particle arrives to the round-about from outside
\hfill}
\demitrait
\vspace{4pt}
\ligne{\hfill
\hbox to 288pt{%
\demirangeea B W W W B B B   \demirangeeb W B B B W B W B 
\hbox to 50pt{\hfill 39\hskip 20pt}
\hfill}\hfill}
\ligne{\hfill
\hbox to 288pt{%
\demirangeea B W W {\small\bf B} B B B   \demirangeeb W B B B W B W B 
\hbox to 50pt{\hfill 40\hskip 20pt}
\hfill}\hfill}
\ligne{\hfill
\hbox to 288pt{%
\demirangeea B W {\small\bf B} W B B B   \demirangeeb W B B B W B W W 
\hbox to 50pt{\hfill 41\hskip 20pt}
\hfill}\hfill}
\ligne{\hfill
\hbox to 288pt{%
\demirangeea {\small\bf W} {\small\bf B} W W B B B   \demirangeeb W B B B W B W B 
\hbox to 50pt{\hfill 42\hskip 20pt}
\hfill}\hfill}
\ligne{\hfill
\hbox to 288pt{%
\demirangeea B {\small\bf B} W W B B B   \demirangeeb W B B B W B W B 
\hbox to 50pt{\hfill 43\hskip 20pt}
\hfill}\hfill}
\demitrait
\vspace{4pt}
\ligne{\hfill\tt
-II- two particles arrive from the round-about
\hfill}
\demitrait
\vspace{4pt}
\ligne{\hfill
\hbox to 288pt{%
\demirangeea B W W W B B B   \demirangeeb W B B B W B W B 
\hbox to 50pt{\hfill 39\hskip 20pt}
\hfill}\hfill}
\ligne{\hfill
\hbox to 288pt{%
\demirangeea B {\small\bf B} W W B B B   \demirangeeb W B B B W B W B 
\hbox to 50pt{\hfill 43\hskip 20pt}
\hfill}\hfill}
\demitrait\vspace{15pt}
}

   Now we have seen that all rules of Tables~\ref{cross_tabB},
\ref{cross_tabC}, \ref{cross_tabE}, \ref{cross_tabF}, \ref{cross_tabB11},
\ref{cross_tabB12} and~\ref{cross_tabC5} realize the implementation of the
scenario described in~Subsection~\ref{scenario}.

\subsection{The rules for the switches}
\label{switchrules}

   This subsection deals with the fixed switch and with the flip-flop. We first
look at the first one, a very simple situation, and then at the second one, a
bit more complex situation, involving many cells, as in the case of the crossings.

\subsubsection{Rules for the fixed switch}

   Table~\ref{fixB_tabB} gives the rule for the cell~$B$ when the particle
goes through~$B$, see Fig.\ref{idle_configs} and Fig.\ref{fixed_motion}.

\vtop{
\begin{tab}\label{fixB_tabB}
\leurre
Passive fixed switch. Rules for $B$. The particle comes from~$B$.
\end{tab}
\vspace{-10pt}
\grostrait
\ligne{\hfill
\hbox to 288pt{%
\demirangeea 0 1 2 3 4 5 6  \demirangeeb {7} {8} {9} {10} {11} {12} {13}  {}
\hbox to 50pt{\hfill $n$\hskip 20pt}
\hfill}\hfill}
\ligne{\hfill
\hbox to 288pt{%
\demirangeea {} {} {} {} {} {} {} \demirangeeb {} {} {} {} {$tr$} {} {\bf O}  {}
\hbox to 50pt{\hfill}
\hfill}\hfill}
\vspace{-4pt}
\demitrait
\vspace{4pt}
\ligne{\hfill
\hbox to 288pt{%
\demirangeea W B B W W W W   \demirangeeb W W W B W B W W
\hbox to 50pt{\hfill 94\hskip 20pt}
\hfill}\hfill}
\ligne{\hfill
\hbox to 288pt{%
\demirangeea W B B W W W W   \demirangeeb W W W B {\small\bf B} B W B
\hbox to 50pt{\hfill 95\hskip 20pt}
\hfill}\hfill}
\ligne{\hfill
\hbox to 288pt{%
\demirangeea {\small\bf B} B B W W W W   \demirangeeb W W W B W B W W
\hbox to 50pt{\hfill 171\hskip 20pt}
\hfill}\hfill}
\ligne{\hfill
\hbox to 288pt{%
\demirangeea W B B W W W W   \demirangeeb W W W B W B {\small\bf B} W
\hbox to 50pt{\hfill 96\hskip 20pt}
\hfill}\hfill}
\demitrait\vspace{15pt}
}

   We note that the conservative rule is rule~91: this rule is also the conservative
rule of~$E$ in the crossings. The reason is that it is a cell of the tracks and that
it is oriented like~$E$. The difference with~$E$ is that here, the black cells
of the idle configuration are milestones. They always stay black. The table
simply involves ordinary motion rules. The rules for~$C$, when the particle goes
through~$C$, are given by Table~\ref{fixC_tabC}. This time the conservative rule
is rule~2. Here too, we can see purely motion rules.

\vtop{
\begin{tab}\label{fixC_tabC}
\leurre
Passive fixed switch. Rules for $C$. The particle comes from~$C$.
\end{tab}
\vspace{-10pt}
\grostrait
\ligne{\hfill
\hbox to 288pt{%
\demirangeea 0 1 2 3 4 5 6  \demirangeeb {7} {8} {9} {10} {11} {12} {13}  {}
\hbox to 50pt{\hfill $n$\hskip 20pt}
\hfill}\hfill}
\ligne{\hfill
\hbox to 288pt{%
\demirangeea {} {} {} {\bf O} {} {$tr$} {} \demirangeeb {} {} {} {} {} {} {}  {}
\hbox to 50pt{\hfill}
\hfill}\hfill}
\vspace{-4pt}
\demitrait
\vspace{4pt}
\ligne{\hfill
\hbox to 288pt{%
\demirangeea W B B W B W B   \demirangeeb W W W W W W W W 
\hbox to 50pt{\hfill 2\hskip 20pt}
\hfill}\hfill}
\ligne{\hfill
\hbox to 288pt{%
\demirangeea W B B W B {\small\bf B} B   \demirangeeb W W W W W W W B 
\hbox to 50pt{\hfill 3\hskip 20pt}
\hfill}\hfill}
\ligne{\hfill
\hbox to 288pt{%
\demirangeea {\small\bf B} B B W B W B   \demirangeeb W W W W W W W W 
\hbox to 50pt{\hfill 58\hskip 20pt}
\hfill}\hfill}
\ligne{\hfill
\hbox to 288pt{%
\demirangeea W B B {\small\bf B} B W B   \demirangeeb W W W W W W W W 
\hbox to 50pt{\hfill 5\hskip 20pt}
\hfill}\hfill}
\demitrait\vspace{15pt}
}

\vtop{
\begin{tab}\label{fix_tabO}
\leurre
Passive fixed switch. Rules for $O$. 
\end{tab}
\vspace{-10pt}
\grostrait
\ligne{\hfill
\hbox to 288pt{%
\demirangeea 0 1 2 3 4 5 6  \demirangeeb {7} {8} {9} {10} {11} {12} {13}  {}
\hbox to 50pt{\hfill $n$\hskip 20pt}
\hfill}\hfill}
\ligne{\hfill
\hbox to 288pt{%
\demirangeea {} {} {} {\bf A} {} {} {} \demirangeeb {} {\bf C} {} {\bf B} {} {} {}  {}
\hbox to 50pt{\hfill}
\hfill}\hfill}
\vspace{-4pt}
\demitrait
\vspace{4pt}
\ligne{\hfill\tt
-I- the particle comes from $B$
\hfill}
\demitrait
\vspace{4pt}
\ligne{\hfill
\hbox to 288pt{%
\demirangeea W B B W B W W   \demirangeeb B W B W B W W W
\hbox to 50pt{\hfill 97\hskip 20pt}
\hfill}\hfill}
\ligne{\hfill
\hbox to 288pt{%
\demirangeea W B B W B W W   \demirangeeb B W B {\small\bf B} B W W B
\hbox to 50pt{\hfill 101\hskip 20pt}
\hfill}\hfill}
\ligne{\hfill
\hbox to 288pt{%
\demirangeea {\small\bf B} B B W B W W   \demirangeeb B W B W B W W W
\hbox to 50pt{\hfill 98\hskip 20pt}
\hfill}\hfill}
\ligne{\hfill
\hbox to 288pt{%
\demirangeea W B B {\small\bf B} B W W   \demirangeeb B W B W B W W W
\hbox to 50pt{\hfill 99\hskip 20pt}
\hfill}\hfill}
\demitrait
\vspace{4pt}
\ligne{\hfill\tt
-II- the particle comes from $C$
\hfill}
\demitrait
\vspace{4pt}
\ligne{\hfill
\hbox to 288pt{%
\demirangeea W B B W B W W   \demirangeeb B W B W B W W W
\hbox to 50pt{\hfill 97\hskip 20pt}
\hfill}\hfill}
\ligne{\hfill
\hbox to 288pt{%
\demirangeea W B B W B W W   \demirangeeb B {\small\bf B} B W B W W B
\hbox to 50pt{\hfill 100\hskip 20pt}
\hfill}\hfill}
\ligne{\hfill
\hbox to 288pt{%
\demirangeea {\small\bf B} B B W B W W   \demirangeeb B W B W B W W W
\hbox to 50pt{\hfill 98\hskip 20pt}
\hfill}\hfill}
\ligne{\hfill
\hbox to 288pt{%
\demirangeea W B B {\small\bf B} B W W   \demirangeeb B W B W B W W W
\hbox to 50pt{\hfill 99\hskip 20pt}
\hfill}\hfill}
\demitrait\vspace{15pt}
}

Table~\ref{fix_tabO} gives the rules for the cell~$O$ which is the central
cell of the switch: the three one-way tracks meet there. It is better to consider 
that the tracks coming from~$B$ and~$C$ merge at~$O$ from where they constitute
a unique track leaving the switch through~$A$. When the particle comes through~$B$,
rule~101 attracts the particle to~$O$. Rule~98 witnesses that the particle
leaves the cell and rule~99 confirms that the leaving happens through~$A$.
When the particle comes through~$C$, this time rule~100 attracts the particle to~$O$.
Once it is in~$O$, the same rules, 98 and~99 apply which witness the leaving of the
particle from the switch through~$A$.

   Note that here, the conservative rule is rule~97 and that the pattern of the 
neighbours is very different from the one which we have met up to now.

   We do not give the rules for~$A$: it is an ordinary cell of the tracks and
the rules for it have already be given. In order to complete the set of rules,
we have to check the rules for~$B$ and~$C$ when the particle crosses the switch
from the other side. This is done by Table~\ref{fix_passBC}: we can see that
the involved rules have already been used in Table~\ref{fixB_tabB} for rules~94
and~96 and in Table~\ref{fixC_tabC} for rules~2 and~5.
   
\vtop{
\begin{tab}\label{fix_passBC}
\leurre
Passive fixed switch. Rules for $B$ and~$C$ when they are silent. 
\end{tab}
\vspace{-10pt}
\grostrait
\ligne{\hfill
\hbox to 288pt{%
\demirangeea 0 1 2 3 4 5 6  \demirangeeb {7} {8} {9} {10} {11} {12} {13}  {}
\hbox to 50pt{\hfill $n$\hskip 20pt}
\hfill}\hfill}
\demitrait
\vspace{4pt}
\ligne{\hfill\tt
-I- at $B$ when the particle comes from $C$
\hfill}
\demitrait
\vspace{4pt}
\ligne{\hfill
\hbox to 288pt{%
\demirangeea W B B W W W W   \demirangeeb W W W B W B W W
\hbox to 50pt{\hfill 94\hskip 20pt}
\hfill}\hfill}
\ligne{\hfill
\hbox to 288pt{%
\demirangeea W B B W W W W   \demirangeeb W W W B W B {\small\bf B} W
\hbox to 50pt{\hfill 96\hskip 20pt}
\hfill}\hfill}
\demitrait
\vspace{4pt}
\ligne{\hfill\tt
-II- at $C$ when the particle comes from $B$
\hfill}
\demitrait
\vspace{4pt}
\ligne{\hfill
\hbox to 288pt{%
\demirangeea W B B W B W B   \demirangeeb W W W W W W W W 
\hbox to 50pt{\hfill 2\hskip 20pt}
\hfill}\hfill}
\ligne{\hfill
\hbox to 288pt{%
\demirangeea W B B {\small\bf B} B W B   \demirangeeb W W W W W W W W 
\hbox to 50pt{\hfill 5\hskip 20pt}
\hfill}\hfill}
\demitrait\vspace{10pt}
}

\subsubsection{Rules for the flip-flop}
\label{flip_flop_rules}

   We now turn to the rules for the flip-flop, a more complex structure than that
of the fixed switch. It relies on the figures we have seen in the study of this
switch, see Fig.~\ref{idle_configs} and Fig.~\ref{flip_flop_motion}.

   Remember that in this switch, outside the cells~$A$, $B$, $C$ and~$O$ already
considered in the fixed switch, which are cells on the tracks, we have
two sensors, $H$ and~$K$ and a controller~$D$. Here, $A$~is an ordinary cell
of the track, so there is no need to give the concerned cells: we have already
seen them. Now, for~$B$ and~$C$, they are not ordinary cells of the track as
the sensors are necessary one of their neighbours. As there must be an exchange
between the states of the sensor, the simplest way to implement this is the situation
indicated in Fig.~\ref{idle_configs} and Fig.~\ref{flip_flop_motion}. This
entails that here, $B$ and~$C$ have a specific pattern. This is a fortiori the
case for~$D$, $H$ and~$K$.

   This time, we start with the rules for~$O$ given by Table~\ref{ff_tabO}.

   We can see that the conservative rule for~$O$ is rule~51 which also
differs from the configuration induced by rule~94 which is the idle
configuration of the fixed switch.

    The particle arrives by~$A$ as witnessed by rule~53. The rule attracts the
particle to~$O$ and rule~52 witnesses that it leaves~$O$. When appropriate,
rule~54 shows that the particle leaves~$O$ through~$C$ and rule~55 that it leaves
through~$B$. The difference cannot be directly detected by~$O$: the selection is 
performed by~$B$ and~$C$ themselves as we have seen in Subsubsection~\ref{flip_flop}.

\vtop{
\begin{tab}\label{ff_tabO}
\leurre
Flip-flop. Rules for $O$. 
\end{tab}
\vspace{-10pt}
\grostrait
\ligne{\hfill
\hbox to 288pt{%
\demirangeea 0 1 2 3 4 5 6  \demirangeeb {7} {8} {9} {10} {11} {12} {13}  {}
\hbox to 50pt{\hfill $n$\hskip 20pt}
\hfill}\hfill}
\ligne{\hfill
\hbox to 288pt{%
\demirangeea {} {} {\bf A} {} {} {} {} \demirangeeb {\bf C} {} {\bf B} {} {} {} {}  {}
\hbox to 50pt{\hfill}
\hfill}\hfill}
\vspace{-4pt}
\demitrait
\vspace{4pt}
\ligne{\hfill
\hbox to 288pt{%
\demirangeea W B W B W W B   \demirangeeb W B W B W W W W 
\hbox to 50pt{\hfill 51\hskip 20pt}
\hfill}\hfill}
\ligne{\hfill
\hbox to 288pt{%
\demirangeea W B {\small\bf B} B W W B   \demirangeeb W B W B W W W B 
\hbox to 50pt{\hfill 53\hskip 20pt}
\hfill}\hfill}
\ligne{\hfill
\hbox to 288pt{%
\demirangeea {\small\bf B} B W B W W B   \demirangeeb W B W B W W W W 
\hbox to 50pt{\hfill 52\hskip 20pt}
\hfill}\hfill}
\demitrait
\vspace{4pt}
\ligne{\hfill\tt
-I- the particle goes to $C$
\hfill}
\demitrait
\vspace{4pt}
\ligne{\hfill
\hbox to 288pt{%
\demirangeea W B W B W W B   \demirangeeb {\small\bf B} B W B W W W W 
\hbox to 50pt{\hfill 54\hskip 20pt}
\hfill}\hfill}
\demitrait
\vspace{4pt}
\ligne{\hfill\tt
-II- the particle goes to $B$
\hfill}
\demitrait
\vspace{4pt}
\ligne{\hfill
\hbox to 288pt{%
\demirangeea W B W B W W B   \demirangeeb W B {\small\bf B} B W W W W 
\hbox to 50pt{\hfill 55\hskip 20pt}
\hfill}\hfill}
\demitrait\vspace{15pt}
}

\vtop{
\begin{tab}\label{ffB_tabB}
\leurre
Flip-flop. Rules for $B$, the particle goes to $C$
\end{tab}
\vspace{-10pt}
\grostrait
\ligne{\hfill
\hbox to 288pt{%
\demirangeea 0 1 2 3 4 5 6  \demirangeeb {7} {8} {9} {10} {11} {12} {13}  {}
\hbox to 50pt{\hfill $n$\hskip 20pt}
\hfill}\hfill}
\ligne{\hfill
\hbox to 288pt{%
\demirangeea {} {} {} {$tr$} {} {\bf O} {\bf D} 
\demirangeeb {\bf H} {} {} {} {} {} {}  {}
\hbox to 50pt{\hfill}
\hfill}\hfill}
\vspace{-4pt}
\demitrait
\vspace{4pt}
\ligne{\hfill
\hbox to 288pt{%
\demirangeea W B B W B W B   \demirangeeb B W W B B W W W 
\hbox to 50pt{\hfill 66\hskip 20pt}
\hfill}\hfill}
\ligne{\hfill
\hbox to 288pt{%
\demirangeea W B B W B {\small\bf B} B   \demirangeeb B W W B B W W W 
\hbox to 50pt{\hfill 67\hskip 20pt}
\hfill}\hfill}
\ligne{\hfill
\hbox to 288pt{%
\demirangeea W B B W B W B   \demirangeeb B W W B B W W W 
\hbox to 50pt{\hfill 66\hskip 20pt}
\hfill}\hfill}
\ligne{\hfill
\hbox to 288pt{%
\demirangeea W B B W B W {\small\bf W}   \demirangeeb B W W B B W W W 
\hbox to 50pt{\hfill 63\hskip 20pt}
\hfill}\hfill}
\ligne{\hfill
\hbox to 288pt{%
\demirangeea W B B W B W B   \demirangeeb {\small\bf W} W W B B W W W 
\hbox to 50pt{\hfill 61\hskip 20pt}
\hfill}\hfill}
\demitrait\vspace{15pt}
}

\vtop{
\begin{tab}\label{ffC_tabC}
Flip-flop. Rules for $C$. The particle goes to $B$
\end{tab}
\vspace{-10pt}
\grostrait
\ligne{\hfill
\hbox to 288pt{%
\demirangeea 0 1 2 3 4 5 6  \demirangeeb {7} {8} {9} {10} {11} {12} {13}  {}
\hbox to 50pt{\hfill $n$\hskip 20pt}
\hfill}\hfill}
\ligne{\hfill
\hbox to 288pt{%
\demirangeea {} {} {} {} {} {} {} 
\demirangeeb {} {} {\bf K} {\bf D} {\bf O} {} {$tr$}  {}
\hbox to 50pt{\hfill}
\hfill}\hfill}
\vspace{-4pt}
\demitrait
\vspace{4pt}
\ligne{\hfill
\hbox to 288pt{%
\demirangeea W B B W W B B   \demirangeeb W W B B W B W W
\hbox to 50pt{\hfill 66\hskip 20pt}
\hfill}\hfill}
\ligne{\hfill
\hbox to 288pt{%
\demirangeea W B B W W B B   \demirangeeb W W B B {\small\bf B} B W W
\hbox to 50pt{\hfill 76\hskip 20pt}
\hfill}\hfill}
\ligne{\hfill
\hbox to 288pt{%
\demirangeea W B B W W B B   \demirangeeb W W B B W B W W
\hbox to 50pt{\hfill 66\hskip 20pt}
\hfill}\hfill}
\ligne{\hfill
\hbox to 288pt{%
\demirangeea W B B W W B B   \demirangeeb W W B {\small\bf W} W B W W
\hbox to 50pt{\hfill 72\hskip 20pt}
\hfill}\hfill}
\ligne{\hfill
\hbox to 288pt{%
\demirangeea W B B W W B B   \demirangeeb W W {\small\bf W} B W B W W
\hbox to 50pt{\hfill 70\hskip 20pt}
\hfill}\hfill}
\demitrait\vspace{15pt}
}

   Tables~\ref{ffB_tabB} and~\ref{ffC_tabB} show the rules for the cell~$B$
while Tables~\ref{ffC_tabC} and~\ref{ffB_tabC} show the rules for the cell~$C$.
We shall first study what happens for a cell which is not on the selected
path for the particle. This means that for this cell, the sensor is black:
$H$~is black if $B$~is not selected while $K$~is black if $C$~is not selected.
Consequently, we first look at Tables~\ref{ffB_tabB} and~\ref{ffC_tabB}.

   We can see that the conservative rule is rule~66 for both~$B$ and~$C$.
However, the place of the common neighbours, here~$O$ and~$D$ is not the same
as clear from the tables. Rule~67 for~$B$ and rule~76 for~$C$ witness the occurrence
of the particle at~$O$. Now, at the next time, the particle disappears: it went
to~$C$ in Table~\ref{ffB_tabB} and to~$B$ in Table~\ref{ffC_tabC}. This is why
the conservative rule~66 applies in both cases. Now, as $D$~has seen the
particle in its appropriate place, it flashes, which is witnessed by rule~63 for~$B$
and rule~72 for~$C$. The flash of~$D$ make both~$H$ and~$K$ change their states.
This is witnessed by rule~61 for~$B$ and rule~70 for~$C$. Note that these rules
are the conservative rules for~$B$ and~$C$ respectively when the sensor is white.

\vtop{
\begin{tab}\label{ffC_tabB}
Flip-flop. Rules for $B$. The particle goes to $B$
\end{tab}
\vspace{-10pt}
\grostrait
\ligne{\hfill
\hbox to 288pt{%
\demirangeea 0 1 2 3 4 5 6  \demirangeeb {7} {8} {9} {10} {11} {12} {13}  {}
\hbox to 50pt{\hfill $n$\hskip 20pt}
\hfill}\hfill}
\ligne{\hfill
\hbox to 288pt{%
\demirangeea {} {} {} {$tr$} {} {\bf O} {\bf D} 
\demirangeeb {\bf H} {} {} {} {} {} {}  {}
\hbox to 50pt{\hfill}
\hfill}\hfill}
\vspace{-4pt}
\demitrait
\vspace{4pt}
\ligne{\hfill
\hbox to 288pt{%
\demirangeea W B B W B W B   \demirangeeb W W W B B W W W 
\hbox to 50pt{\hfill 61\hskip 20pt}
\hfill}\hfill}
\ligne{\hfill
\hbox to 288pt{%
\demirangeea W B B W B {\small\bf B} B   \demirangeeb W W W B B W W B 
\hbox to 50pt{\hfill 65\hskip 20pt}
\hfill}\hfill}
\ligne{\hfill
\hbox to 288pt{%
\demirangeea {\small\bf B} B B W B W B   \demirangeeb W W W B B W W W 
\hbox to 50pt{\hfill 62\hskip 20pt}
\hfill}\hfill}
\ligne{\hfill
\hbox to 288pt{%
\demirangeea W B B {\small\bf B} B W {\small\bf W}   \demirangeeb W W W B B W W W 
\hbox to 50pt{\hfill 64\hskip 20pt}
\hfill}\hfill}
\ligne{\hfill
\hbox to 288pt{%
\demirangeea W B B W B W B   \demirangeeb {\small\bf B} W W B B W W W 
\hbox to 50pt{\hfill 66\hskip 20pt}
\hfill}\hfill}
\demitrait\vspace{15pt}
}

\vtop{
\begin{tab}\label{ffB_tabC}
\leurre
Flip-flop. Rules for $C$. The particle goes to $C$
\end{tab}
\vspace{-10pt}
\grostrait
\ligne{\hfill
\hbox to 288pt{%
\demirangeea 0 1 2 3 4 5 6  \demirangeeb {7} {8} {9} {10} {11} {12} {13}  {}
\hbox to 50pt{\hfill $n$\hskip 20pt}
\hfill}\hfill}
\ligne{\hfill
\hbox to 288pt{%
\demirangeea {} {} {} {} {} {} {} 
\demirangeeb {} {} {\bf K} {\bf D} {\bf O} {} {$tr$}  {}
\hbox to 50pt{\hfill}
\hfill}\hfill}
\vspace{-4pt}
\demitrait
\vspace{4pt}
\ligne{\hfill
\hbox to 288pt{%
\demirangeea W B B W W B B   \demirangeeb W W W B W B W W
\hbox to 50pt{\hfill 70\hskip 20pt}
\hfill}\hfill}
\ligne{\hfill
\hbox to 288pt{%
\demirangeea W B B W W B B   \demirangeeb W W W B {\small\bf B} B W B
\hbox to 50pt{\hfill 74\hskip 20pt}
\hfill}\hfill}
\ligne{\hfill
\hbox to 288pt{%
\demirangeea {\small\bf B} B B W W B B   \demirangeeb W W W B W B W W
\hbox to 50pt{\hfill 71\hskip 20pt}
\hfill}\hfill}
\ligne{\hfill
\hbox to 288pt{%
\demirangeea W B B W W B B   \demirangeeb W W W {\small\bf W} W B B W
\hbox to 50pt{\hfill 73\hskip 20pt}
\hfill}\hfill}
\ligne{\hfill
\hbox to 288pt{%
\demirangeea W B B W W B B   \demirangeeb W W {\small\bf B} B W B W W 
\hbox to 50pt{\hfill 66\hskip 20pt}
\hfill}\hfill}
\demitrait\vspace{15pt}
}

   This is what we can see in the first row of Tables~\ref{ffC_tabB}
and Table~\ref{ffB_tabC}. We are know in the case when $B$ or~$C$ is on the
selected tracks. As previously, the particle occurs in~$O$, which is witnessed
by new rules, 65 and~74 respectively which make the particle enter the cell.
Rules~62 and~71 make the particle leave~$B$ or~$C$, respectively, and
to go on on the selected tracks. This is witnessed by rules~64 for~$B$ and~73
for~$C$ which also witness the flash of~$D$. Now, as $H$ and~$K$  both change 
their states, which is witnessed by rule~66{} in both cases, we find again the
idle configuration of the previous tables with, of course, the same rule.

   Now, we can turn to~$D$, the controller of the flip-flop, making the sensors
change their states when the particle has taken the selected tracks. The table
for the rules of this cell is Table~\ref{ff_tabD}.

   When the selected track is~$C$, $B$ respectively, see first, second part of 
Table~\ref{ff_tabD}, respectively,
the conservative rule is rule~80, rule~78 respectively. Then rule~82, rule~83,
respectively, witnesses the occurrence of the particle in~$O$.
Then rule~84, rule~85, respectively, witness the entrance of the particle
on the right tracks, $C$ or~$B$, respectively. Both rules make~$D$ flash.
Then, rule~81, rule~79, respectively, make the cell return to black.
Then, rule~78, rule~80, witness the change of states in both~$H$ and~$K$.
Accordingly, we have the conservative rules which we noted at the initial step.

\vtop{
\begin{tab}\label{ff_tabD}
\leurre
Flip-flop. Rules for $D$. 
\end{tab}
\vspace{-10pt}
\grostrait
\ligne{\hfill
\hbox to 288pt{%
\demirangeea 0 1 2 3 4 5 6  \demirangeeb {7} {8} {9} {10} {11} {12} {13}  {}
\hbox to 50pt{\hfill $n$\hskip 20pt}
\hfill}\hfill}
\ligne{\hfill
\hbox to 288pt{%
\demirangeea {} {\bf O} {\bf C} {\bf K} {} {} {} 
\demirangeeb {} {} {} {} {} {\bf H} {\bf B}  {}
\hbox to 50pt{\hfill}
\hfill}\hfill}
\vspace{-4pt}
\demitrait
\vspace{4pt}
\ligne{\hfill\tt
-I- the particle goes to $C$
\hfill}
\demitrait
\vspace{4pt}
\ligne{\hfill
\hbox to 288pt{%
\demirangeea B W W W B W W   \demirangeeb B W W B W B W B
\hbox to 50pt{\hfill 80\hskip 20pt}
\hfill}\hfill}
\ligne{\hfill
\hbox to 288pt{%
\demirangeea B {\small\bf B} W W B W W   \demirangeeb B W W B W B W B
\hbox to 50pt{\hfill 82\hskip 20pt}
\hfill}\hfill}
\ligne{\hfill
\hbox to 288pt{%
\demirangeea B W {\small\bf B} W B W W   \demirangeeb B W W B W B W W
\hbox to 50pt{\hfill 84\hskip 20pt}
\hfill}\hfill}
\ligne{\hfill
\hbox to 288pt{%
\demirangeea {\small\bf W} W W W B W W   \demirangeeb B W W B W B W B
\hbox to 50pt{\hfill 81\hskip 20pt}
\hfill}\hfill}
\ligne{\hfill
\hbox to 288pt{%
\demirangeea B W W {\small\bf B} B W W   \demirangeeb B W W B W {\small\bf W} W B
\hbox to 50pt{\hfill 78\hskip 20pt}
\hfill}\hfill}
\demitrait
\vspace{4pt}
\ligne{\hfill\tt
-II- the particle goes to $B$
\hfill}
\demitrait
\vspace{4pt}
\ligne{\hfill
\hbox to 288pt{%
\demirangeea B W W B B W W   \demirangeeb B W W B W W W B
\hbox to 50pt{\hfill 78\hskip 20pt}
\hfill}\hfill}
\ligne{\hfill
\hbox to 288pt{%
\demirangeea B {\small\bf B} W B B W W   \demirangeeb B W W B W W W B
\hbox to 50pt{\hfill 83\hskip 20pt}
\hfill}\hfill}
\ligne{\hfill
\hbox to 288pt{%
\demirangeea B W W B B W W   \demirangeeb B W W B W W {\small\bf B} W
\hbox to 50pt{\hfill 85\hskip 20pt}
\hfill}\hfill}
\ligne{\hfill
\hbox to 288pt{%
\demirangeea {\small\bf W} W W B B W W   \demirangeeb B W W B W W W B
\hbox to 50pt{\hfill 79\hskip 20pt}
\hfill}\hfill}
\ligne{\hfill
\hbox to 288pt{%
\demirangeea B W W {\small\bf W} B W W   \demirangeeb B W W B W {\small\bf B} W B
\hbox to 50pt{\hfill 80\hskip 20pt}
\hfill}\hfill}
\demitrait\vspace{15pt}
}

   We conclude this subsection with the rules for~$H$ and~$K$, see
Tables~\ref{ff_tabH} and~\ref{ff_tabK} respectively.

   As we know from Subsections~\ref{scenario} and~\ref{flip_flop}, these cells
behave as markers of the non-selected tracks. If the marker is black, the access
to its neighbour belonging to the path is forbidden for the particle. This what we
have seen on Tables~\ref{ffB_tabB}, \ref{ffC_tabC}. When the marker is white,
the access to its neighbour belonging to the path is performed when the particle
is in~$O$. We have seen the appropriate rules in Tables~\ref{ffB_tabC} 
and~\ref{ffC_tabB}. We also have seen in Table~\ref{ff_tabD} that the flash of~$D$
triggers the change of state in both~$H$ and~$K$. And so, we have to complete the
set of rules by the rules especially dedicated to these cells.

   In the first part of Table~\ref{ff_tabH}, the conservative rule is rule~87. As 
the state of~$H$ is black, the particle cannot go to~$B$, so that it goes to~$BC$.
Accordingly, the column
devoted to~$B$ remains with {\tt W} in this part of the table. Rule~90 witnesses
the flash of~$D$ so that it makes the state of~$H$ change to white: rule~86 is the 
conservative rule associated to this state, as confirmed by the first row in
the second part of the rules.

   In the second part of Table~\ref{ff_tabH}, where the conservative rule is
rule~86, as the state of~$H$ is white, the particle cannot go to~$C$ as
it goes to~$B$ when it is present at~$O$. The entrance to~$B$ is witnessed
by rule~88. Rule~89 detects the flash of~$D$ making the sate of~$H$ change to
black. Rule~87 keeps this situation as it is the conservative rule for~$H$ when
its state is black.

\vtop{
\begin{tab}\label{ff_tabH}
\leurre
Flip-flop. Rules for $H$. 
\end{tab}
\vspace{-10pt}
\grostrait
\ligne{\hfill
\hbox to 288pt{%
\demirangeea 0 1 2 3 4 5 6  \demirangeeb {7} {8} {9} {10} {11} {12} {13}  {}
\hbox to 50pt{\hfill $n$\hskip 20pt}
\hfill}\hfill}
\ligne{\hfill
\hbox to 288pt{%
\demirangeea {} {\bf B} {\bf D} {} {} {} {} 
\demirangeeb {} {} {} {} {} {} {}  {}
\hbox to 50pt{\hfill}
\hfill}\hfill}
\vspace{-4pt}
\demitrait
\vspace{4pt}
\ligne{\hfill\tt
-I- the particle goes to $C$
\hfill}
\demitrait
\vspace{4pt}
\ligne{\hfill
\hbox to 288pt{%
\demirangeea B W B B B B B   \demirangeeb B W W W W W W B
\hbox to 50pt{\hfill 87\hskip 20pt}
\hfill}\hfill}
\ligne{\hfill
\hbox to 288pt{%
\demirangeea B W {\small\bf W} B B B B   \demirangeeb B W W W W W W W
\hbox to 50pt{\hfill 90\hskip 20pt}
\hfill}\hfill}
\ligne{\hfill
\hbox to 288pt{%
\demirangeea {\small\bf W} W B B B B B   \demirangeeb B W W W W W W W
\hbox to 50pt{\hfill 86\hskip 20pt}
\hfill}\hfill}
\demitrait
\vspace{4pt}
\ligne{\hfill\tt
-II- the particle goes to $B$
\hfill}
\demitrait
\vspace{4pt}
\ligne{\hfill
\hbox to 288pt{%
\demirangeea W W B B B B B   \demirangeeb B W W W W W W W
\hbox to 50pt{\hfill 86\hskip 20pt}
\hfill}\hfill}
\ligne{\hfill
\hbox to 288pt{%
\demirangeea W {\small\bf B} B B B B B   \demirangeeb B W W W W W W W
\hbox to 50pt{\hfill 88\hskip 20pt}
\hfill}\hfill}
\ligne{\hfill
\hbox to 288pt{%
\demirangeea W W {\small\bf W} B B B B   \demirangeeb B W W W W W W B
\hbox to 50pt{\hfill 89\hskip 20pt}
\hfill}\hfill}
\ligne{\hfill
\hbox to 288pt{%
\demirangeea {\small\bf B} W B B B B B   \demirangeeb B W W W W W W B
\hbox to 50pt{\hfill 87\hskip 20pt}
\hfill}\hfill}
\demitrait\vspace{15pt}
}

\vtop{
\begin{tab}\label{ff_tabK}
\leurre
Flip-flop. Rules for $K$. 
\end{tab}
\vspace{-10pt}
\grostrait
\ligne{\hfill
\hbox to 288pt{%
\demirangeea 0 1 2 3 4 5 6  \demirangeeb {7} {8} {9} {10} {11} {12} {13}  {}
\hbox to 50pt{\hfill $n$\hskip 20pt}
\hfill}\hfill}
\ligne{\hfill
\hbox to 288pt{%
\demirangeea {} {\bf C} {} {} {} {} {} 
\demirangeeb {} {} {} {} {} {} {\bf D}  {}
\hbox to 50pt{\hfill}
\hfill}\hfill}
\vspace{-4pt}
\demitrait
\vspace{4pt}
\ligne{\hfill\tt
-I- the particle goes to $C$
\hfill}
\demitrait
\vspace{4pt}
\ligne{\hfill
\hbox to 288pt{%
\demirangeea W W W B B B B   \demirangeeb B W W W W W B W
\hbox to 50pt{\hfill 91\hskip 20pt}
\hfill}\hfill}
\ligne{\hfill
\hbox to 288pt{%
\demirangeea W {\small\bf B} W B B B B   \demirangeeb B W W W W W B W
\hbox to 50pt{\hfill 93\hskip 20pt}
\hfill}\hfill}
\ligne{\hfill
\hbox to 288pt{%
\demirangeea W W W B B B B   \demirangeeb B W W W W W {\small\bf W} B
\hbox to 50pt{\hfill 89\hskip 20pt}
\hfill}\hfill}
\ligne{\hfill
\hbox to 288pt{%
\demirangeea {\small\bf B} W W B B B B   \demirangeeb B W W W W W B B
\hbox to 50pt{\hfill 92\hskip 20pt}
\hfill}\hfill}
\demitrait
\vspace{4pt}
\ligne{\hfill\tt
-II- the particle goes to $B$
\hfill}
\demitrait
\vspace{4pt}
\ligne{\hfill
\hbox to 288pt{%
\demirangeea B W W B B B B   \demirangeeb B W W W W W B B
\hbox to 50pt{\hfill 92\hskip 20pt}
\hfill}\hfill}
\ligne{\hfill
\hbox to 288pt{%
\demirangeea B W W B B B B   \demirangeeb B W W W W W {\small\bf W} W
\hbox to 50pt{\hfill 90\hskip 20pt}
\hfill}\hfill}
\ligne{\hfill
\hbox to 288pt{%
\demirangeea {\small\bf W} W W B B B B   \demirangeeb B W W W W W B W
\hbox to 50pt{\hfill 91\hskip 20pt}
\hfill}\hfill}
\demitrait\vspace{15pt}
}

   Similarly, Table~\ref{ff_tabK} gives the rule for~$K$ which is the marker
of the selection/non-selection of the tracks whose initial cell is~$C$.

   We can see that the idle configuration is different from that of~$H$.
They both have a block of five consecutive black cells which remain black.
However, this block is not at the same place with respect with~$D$ while turning
around the cell in the same direction and, in this direction, starting from the first
black cell of the block.

    When $K$~is black, second part of Table~\ref{ff_tabK}, the conservative rule is
rule~92. Then rule~90 detects the flash of~$D$ when it occurs. Note that
as $K$ cannot see~$O$ nor~$B$, it knows that the particle entered the switch by
the flash of~$D$ only. Rule~90 makes $K$~turn to white and rule~91 keeps this
new state as it is the corresponding conservative rule.

   When $K$ is white, first part of Table~\ref{ff_tabK}, we check that the
conservative rule is rule~91. Now, rule~93 detects that the particle entered
cell~$C$. At the next time, rule~89 detects the flash of~$D$, making the
state of~$K$ change from white to black. Rule~92 keeps this state as it is the
conservative rule for~$K$ when its state is black. This has previously been noted.

\subsubsection{Rules for the memory switch}
\label{memory_rules}

   We arrive to the final set of rules which deal with the memory switch. The rules
dedicated to this switch fall into two subsets: those needed for the active switch
and those needed to the passive one. 

\vspace{7pt}
\noindent
$\underline{\hbox{\bf The active memory switch}}$
\vspace{7pt}
\label{active_memo_rules}

   This subsection is parallel to Subsection~\ref{flip_flop_rules}. The main
reason is that the configuration of the active memory switch looks like that
of the flip-flop, see Fig.~\ref{idle_configs}. This is also why we use the same names
for the cells for which we show the rules. In fact, we use the same cells~$A$ and~$B$
and~$C$ for the tracks as in the flip-flop, which means with the same rules
and a few additional ones.

   First, we look at the rules for~$O$, as those for~$A$ are simple motion rules
for an ordinary cell of the tracks. 

\vtop{
\begin{tab}\label{mm_tabO}
\leurre
Rules for $O$. The particle goes to $C$
\end{tab}
\vspace{-10pt}
\grostrait
\ligne{\hfill
\hbox to 288pt{%
\demirangeea 0 1 2 3 4 5 6  \demirangeeb {7} {8} {9} {10} {11} {12} {13}  {}
\hbox to 50pt{\hfill $n$\hskip 20pt}
\hfill}\hfill}
\ligne{\hfill
\hbox to 288pt{%
\demirangeea {} {} {\bf A} {} {} {} {} 
\demirangeeb {\bf C} {} {\bf B} {} {} {} {}  {}
\hbox to 50pt{\hfill}
\hfill}\hfill}
\vspace{-4pt}
\demitrait
\vspace{4pt}
\ligne{\hfill
\hbox to 288pt{%
\demirangeea W B W B W W B   \demirangeeb W B W B W W W W 
\hbox to 50pt{\hfill 51\hskip 20pt}
\hfill}\hfill}
\ligne{\hfill
\hbox to 288pt{%
\demirangeea W B {\small\bf B} B W W B   \demirangeeb W B W B W W W B 
\hbox to 50pt{\hfill 53\hskip 20pt}
\hfill}\hfill}
\ligne{\hfill
\hbox to 288pt{%
\demirangeea {\small\bf B} B W B W W B   \demirangeeb W B W B W W W W 
\hbox to 50pt{\hfill 52\hskip 20pt}
\hfill}\hfill}
\demitrait
\vspace{4pt}
\ligne{\hfill\tt
-I- the particle goes to $C$
\hfill}
\demitrait
\vspace{4pt}
\ligne{\hfill
\hbox to 288pt{%
\demirangeea W B W B W W B   \demirangeeb {\small\bf B} B W B W W W W 
\hbox to 50pt{\hfill 54\hskip 20pt}
\hfill}\hfill}
\demitrait
\vspace{4pt}
\ligne{\hfill\tt
-II- the particle goes to $B$
\hfill}
\demitrait
\vspace{4pt}
\ligne{\hfill
\hbox to 288pt{%
\demirangeea W B W B W W B   \demirangeeb W B {\small\bf B} B W W W W 
\hbox to 50pt{\hfill 55\hskip 20pt}
\hfill}\hfill}
\demitrait\vspace{15pt}
}

Table~\ref{mm_tabO} gives us the rules
and the reader may notice that this table contains exactly the same rules
as those of Table~\ref{ff_tabO} for the rules for~$O$ in a flip-flop.
   The reason is very simple: at this stage, as the configuration of~$O$ is here
the same as its configuration in the flip-flop, the particle does not even know
that it is in a memory switch. This will be come clear later, when the particle
will go either to~$B$ or to~$C$. Here, we take the same cells~$B$ and~$C$ 
as in the flip-flop switch, although there is a slight difference in the
working in the active memory switch as we shall see a bit later.
The reason is that $D$, which is also different, does not behave in the same way as 
in the flip-flop: $D$~does not flash when the particle is in the exit of~$B$ or~$C$. 
It flashes in a situation when $B$ and~$C$ are both idle. However, as $H$ 
can see only~$B$ and~$D$ and as~$K$
can see only~$D$ with $C$, we keep $H$ and~$K$ the same as in the flip-flop. 

\vtop{
\begin{tab}\label{mmB_tabB}
\leurre
Active memory switch. Rules for $B$. The particle goes to $C$
\end{tab}
\vspace{-10pt}
\grostrait
\ligne{\hfill
\hbox to 288pt{%
\demirangeea 0 1 2 3 4 5 6  \demirangeeb {7} {8} {9} {10} {11} {12} {13}  {}
\hbox to 50pt{\hfill $n$\hskip 20pt}
\hfill}\hfill}
\ligne{\hfill
\hbox to 288pt{%
\demirangeea {} {} {} {$tr$} {} {\bf O} {\bf D} 
\demirangeeb {\bf H} {} {} {} {} {} {}  {}
\hbox to 50pt{\hfill}
\hfill}\hfill}
\vspace{-4pt}
\demitrait
\vspace{4pt}
\ligne{\hfill
\hbox to 288pt{%
\demirangeea W B B W B W B   \demirangeeb B W W B B W W W
\hbox to 50pt{\hfill 66\hskip 20pt}
\hfill}\hfill}
\ligne{\hfill
\hbox to 288pt{%
\demirangeea W B B W B {\small\bf B} B   \demirangeeb B W W B B W W W
\hbox to 50pt{\hfill 67\hskip 20pt}
\hfill}\hfill}
\demitrait
}

\vtop{
\begin{tab}\label{mmC_tabC}
Active memory switch. Rules for $C$. The particle goes to $B$
\end{tab}
\vspace{-10pt}
\grostrait
\ligne{\hfill
\hbox to 288pt{%
\demirangeea 0 1 2 3 4 5 6  \demirangeeb {7} {8} {9} {10} {11} {12} {13}  {}
\hbox to 50pt{\hfill $n$\hskip 20pt}
\hfill}\hfill}
\ligne{\hfill
\hbox to 288pt{%
\demirangeea {} {} {} {} {} {} {} 
\demirangeeb {} {} {\bf K} {\bf D} {\bf O} {} {$tr$}  {}
\hbox to 50pt{\hfill}
\hfill}\hfill}
\vspace{-4pt}
\demitrait
\vspace{4pt}
\ligne{\hfill
\hbox to 288pt{%
\demirangeea W B B W W B B   \demirangeeb W W B B W B W W
\hbox to 50pt{\hfill 66\hskip 20pt}
\hfill}\hfill}
\ligne{\hfill
\hbox to 288pt{%
\demirangeea W B B W W B B   \demirangeeb W W B B {\small\bf B} B W W
\hbox to 50pt{\hfill 76\hskip 20pt}
\hfill}\hfill}
\demitrait\vspace{15pt}
}

  Compared with the rules of Tables~\ref{ffB_tabB} and~\ref{ffC_tabC}, we can see
that the behaviour of~$B$ and~$C$ for an active passage when they are not on the
selected track is much simpler than in the case of the flip-flop where the flash
of~$D$ is always triggered. Here, rule~66 is the conservative rule. Rules~67 
and~76 simply witness the passage of the particle through~$O$. 

   Now, as shown by Tables~\ref{mmC_tabB} and~\ref{mmB_tabC}, the rules for~$B$
and~$C$ are a bit different from those of the cells with the same names in the 
flip-flop: here, $B$ and~$C$ behave passively as $D$~does not react to the passage
of the particle through them. Consequently, here we have simple motion rules
adapted to the pattern of the cells.

   The conservation rules are now rules~61 and~70
for~$B$ and~$C$ respectively, as in the flip-flop as $H$ and~$K$ are now white in 
both tables, of course applied at different times.
Rules~65 and~74 attract the particle from~$O$ to~$B$ and~$C$ respectively, again as
in the flip-flop. Also as in the flip-flop, rules~62 and~71 return the cell to white.
Now, as there is no flash of~$D$, the similarity with the flip-flop stops here.
Rule~68 and~134 witness that the particle went to the next cell of the tracks
and then, in both cases, we have the conservation rules, rules~61 and~70 respectively,
which are not mentioned according to our convention. Let us remark that rules~68
and~134 were not used in the working of the flip-flop.

\vtop{
\begin{tab}\label{mmC_tabB}
Active memory switch. Rules for $B$. The particle goes to $B$
\end{tab}
\vspace{-10pt}
\grostrait
\ligne{\hfill
\hbox to 288pt{%
\demirangeea 0 1 2 3 4 5 6  \demirangeeb {7} {8} {9} {10} {11} {12} {13}  {}
\hbox to 50pt{\hfill $n$\hskip 20pt}
\hfill}\hfill}
\ligne{\hfill
\hbox to 288pt{%
\demirangeea {} {} {} {$tr$} {} {\bf O} {\bf D} 
\demirangeeb {\bf H} {} {} {} {} {} {}  {}
\hbox to 50pt{\hfill}
\hfill}\hfill}
\vspace{-4pt}
\demitrait
\vspace{4pt}
\ligne{\hfill
\hbox to 288pt{%
\demirangeea W B B W B W B   \demirangeeb W W W B B W W W
\hbox to 50pt{\hfill 61\hskip 20pt}
\hfill}\hfill}
\ligne{\hfill
\hbox to 288pt{%
\demirangeea W B B W B {\small\bf B} B   \demirangeeb W W W B B W W B
\hbox to 50pt{\hfill 65\hskip 20pt}
\hfill}\hfill}
\ligne{\hfill
\hbox to 288pt{%
\demirangeea {\small\bf B} B B W B W B   \demirangeeb W W W B B W W W
\hbox to 50pt{\hfill 62\hskip 20pt}
\hfill}\hfill}
\ligne{\hfill
\hbox to 288pt{%
\demirangeea W B B {\small\bf B} B W B   \demirangeeb W W W B B W W W
\hbox to 50pt{\hfill 68\hskip 20pt}
\hfill}\hfill}
\demitrait
}

\vtop{
\begin{tab}\label{mmB_tabC}
\leurre
Active memory switch. Rules for $C$. The particle goes to $C$
\end{tab}
\vspace{-10pt}
\grostrait
\ligne{\hfill
\hbox to 288pt{%
\demirangeea 0 1 2 3 4 5 6  \demirangeeb {7} {8} {9} {10} {11} {12} {13}  {}
\hbox to 50pt{\hfill $n$\hskip 20pt}
\hfill}\hfill}
\ligne{\hfill
\hbox to 288pt{%
\demirangeea {} {} {} {} {} {} {} 
\demirangeeb {} {} {\bf K} {\bf D} {\bf O} {} {$tr$}  {}
\hbox to 50pt{\hfill}
\hfill}\hfill}
\vspace{-4pt}
\demitrait
\vspace{4pt}
\ligne{\hfill
\hbox to 288pt{%
\demirangeea W B B W W B B   \demirangeeb W W W B W B W W
\hbox to 50pt{\hfill 70\hskip 20pt}
\hfill}\hfill}
\ligne{\hfill
\hbox to 288pt{%
\demirangeea W B B W W B B   \demirangeeb W W W B {\small\bf B} B W B
\hbox to 50pt{\hfill 74\hskip 20pt}
\hfill}\hfill}
\ligne{\hfill
\hbox to 288pt{%
\demirangeea {\small\bf B} B B W W B B   \demirangeeb W W W B W B W W
\hbox to 50pt{\hfill 71\hskip 20pt}
\hfill}\hfill}
\ligne{\hfill
\hbox to 288pt{%
\demirangeea W B B W W B B   \demirangeeb W W W B W B {\small\bf B} W
\hbox to 50pt{\hfill 134\hskip 20pt}
\hfill}\hfill}
\demitrait\vspace{15pt}
}

   We can check that the idle configuration of the neighbours of~$B$ and~$C$
are the same as those of their homonyms in the flip-flop.

   Table~\ref{mm_tabD} gives the rules for~$D$ and we can see that it strongly
differs from the rules of Table~\ref{ff_tabD} for the similar situation in the
flip-flop. Also note that the pattern of the neighbours of~$D$ is different 
from what it is in the flip-flop. The reason is the difference of behaviour 
of the cell which is passive here. 

\vtop{
\begin{tab}\label{mm_tabD}
\leurre
Active memory switch. Rules for $D$. 
\end{tab}
\vspace{-10pt}
\grostrait
\ligne{\hfill
\hbox to 288pt{%
\demirangeea 0 1 2 3 4 5 6  \demirangeeb {7} {8} {9} {10} {11} {12} {13}  {}
\hbox to 50pt{\hfill $n$\hskip 20pt}
\hfill}\hfill}
\ligne{\hfill
\hbox to 288pt{%
\demirangeea {} {\bf O} {\bf C} {\bf K} {} {} {} 
\demirangeeb {$f$} {} {} {} {} {\bf H} {\bf B}  {}
\hbox to 50pt{\hfill}
\hfill}\hfill}
\vspace{-4pt}
\demitrait
\vspace{4pt}
\ligne{\hfill\tt
-I- the particle goes to $C$
\hfill}
\demitrait
\vspace{4pt}
\ligne{\hfill
\hbox to 288pt{%
\demirangeea B W W W B W B   \demirangeeb W B B B W B W B
\hbox to 50pt{\hfill 125\hskip 20pt}
\hfill}\hfill}
\ligne{\hfill
\hbox to 288pt{%
\demirangeea B {\small\bf B} W W B W B   \demirangeeb W B B B W B W B
\hbox to 50pt{\hfill 126\hskip 20pt}
\hfill}\hfill}
\ligne{\hfill
\hbox to 288pt{%
\demirangeea B W {\small\bf B} W B W B   \demirangeeb W B B B W B W B
\hbox to 50pt{\hfill 128\hskip 20pt}
\hfill}\hfill}
\demitrait
\vspace{4pt}
\ligne{\hfill\tt
-II- the particle goes to $B$
\hfill}
\demitrait
\vspace{4pt}
\ligne{\hfill
\hbox to 288pt{%
\demirangeea B W W B B W B   \demirangeeb W B B B W W W B
\hbox to 50pt{\hfill 124\hskip 20pt}
\hfill}\hfill}
\ligne{\hfill
\hbox to 288pt{%
\demirangeea B {\small\bf B} W B B W B   \demirangeeb W B B B W W W B
\hbox to 50pt{\hfill 127\hskip 20pt}
\hfill}\hfill}
\ligne{\hfill
\hbox to 288pt{%
\demirangeea B W W B B W B   \demirangeeb W B B B W W {\small\bf B} B
\hbox to 50pt{\hfill 129\hskip 20pt}
\hfill}\hfill}
\demitrait\vspace{15pt}
}

   In particular, we note the presence of a particular neighbour of~$D$,
namely $D.7$, denoted by~$f$ in the table. This cell is usually white and it
remains unchanged during an active crossing of the switch by the particle.
We shall see what happens in the rules for the passive switch.

   For what is an active passage, the rules of the table simply witness the 
passage of the particle. Rules~125 and~124 give the idle configuration which
is different, depending on the states of~$H$ and~$K$ which are always different.
We can notice that these states remain unchanged during the passage of the
particle. Rules~126 and~127 witness the presence of the particle in~$O$
and rules~128 and~129 witness its passage through~$C$ and~$B$ respectively.

   Now, we turn to the rules for~$H$ and~$K$: the cells are the same
as in the flip-flop. As there is no flash of~$D$ at this stage, we have
just the motion rules used in the flip-flop when $H$ or~$K$ can see the particle
and the conservative rules otherwise.

\vspace{-10pt}
\vtop{
\begin{tab}\label{mmB_tabH}
\leurre
Active memory switch. Rules for $H$. 
\end{tab}
\vspace{-10pt}
\grostrait
\ligne{\hfill
\hbox to 288pt{%
\demirangeea 0 1 2 3 4 5 6  \demirangeeb {7} {8} {9} {10} {11} {12} {13}  {}
\hbox to 50pt{\hfill $n$\hskip 20pt}
\hfill}\hfill}
\ligne{\hfill
\hbox to 288pt{%
\demirangeea {} {\bf B} {\bf D} {} {} {} {} 
\demirangeeb {} {} {} {} {} {} {}  {}
\hbox to 50pt{\hfill}
\hfill}\hfill}
\vspace{-4pt}
\demitrait
\vspace{4pt}
\ligne{\hfill\tt
-I- the particle goes to $C$
\hfill}
\demitrait
\vspace{4pt}
\ligne{\hfill
\hbox to 288pt{%
\demirangeea B W B B B B B   \demirangeeb B W W W W W W B
\hbox to 50pt{\hfill 87\hskip 20pt}
\hfill}\hfill}
\demitrait
\vspace{4pt}
\ligne{\hfill\tt
-II- the particle goes to $B$
\hfill}
\demitrait
\vspace{4pt}
\ligne{\hfill
\hbox to 288pt{%
\demirangeea W W B B B B B   \demirangeeb B W W W W W W W
\hbox to 50pt{\hfill 86\hskip 20pt}
\hfill}\hfill}
\ligne{\hfill
\hbox to 288pt{%
\demirangeea W {\small\bf B} B B B B B   \demirangeeb B W W W W W W W
\hbox to 50pt{\hfill 88\hskip 20pt}
\hfill}\hfill}
\demitrait
}

\vtop{
\begin{tab}\label{mmB_tabK}
\leurre
Active memory switch. Rules for $K$. 
\end{tab}
\vspace{-10pt}
\grostrait
\ligne{\hfill
\hbox to 288pt{%
\demirangeea 0 1 2 3 4 5 6  \demirangeeb {7} {8} {9} {10} {11} {12} {13}  {}
\hbox to 50pt{\hfill $n$\hskip 20pt}
\hfill}\hfill}
\ligne{\hfill
\hbox to 288pt{%
\demirangeea {} {\bf C} {} {} {} {} {} 
\demirangeeb {} {} {} {} {} {} {\bf D}  {}
\hbox to 50pt{\hfill}
\hfill}\hfill}
\vspace{-4pt}
\demitrait
\vspace{4pt}
\ligne{\hfill\tt
-I- the particle goes to $C$
\hfill}
\demitrait
\vspace{4pt}
\ligne{\hfill
\hbox to 288pt{%
\demirangeea W W W B B B B   \demirangeeb B W W W W W B W
\hbox to 50pt{\hfill 91\hskip 20pt}
\hfill}\hfill}
\ligne{\hfill
\hbox to 288pt{%
\demirangeea W {\small\bf B} W B B B B   \demirangeeb B W W W W W B W
\hbox to 50pt{\hfill 93\hskip 20pt}
\hfill}\hfill}
\demitrait
\vspace{4pt}
\ligne{\hfill\tt
-II- the particle goes to $B$
\hfill}
\demitrait
\vspace{4pt}
\ligne{\hfill
\hbox to 288pt{%
\demirangeea B W W B B B B   \demirangeeb B W W W W W B B
\hbox to 50pt{\hfill 92\hskip 20pt}
\hfill}\hfill}
\demitrait\vspace{15pt}
}

\vspace{7pt}
\noindent
$\underline{\hbox{\bf The passive memory switch}}$
\vspace{7pt}

   We can now turn to the rules devoted to the passive memory switch. As
known from Fig.\ref{idle_configs}, this switch looks a bit like the fixed switch:
its cell~$T$ has the same neighbouring as the cell~$O$ in the fixed switch.
Now, as here the switch has a very different behaviour, we change the names of
the cells. As just mentioned, the central cell of the switch is the cell~$T$.
The tracks arriving to the switch arrive at the neighbours~$X$ and~$Y$ of~$T$
and in between~$X$ and~$Y$ we exactly have the neighbour~$Z$ of~$T$. The cells~$I$
and~$J$ are placed as a common neighbour for~$B$ with~$Z$ and for~$C$ with~$Z$
respectively. They play the role of~$H$ and~$K$ in the flip-flop.

   We first look at the rules for~$T$, when the particle already arrived at the
switch.

\vtop{
\begin{tab}\label{mmX_tabT}
\leurre
Passive memory switch. Rules for $T$. The particle comes from $X$.
\end{tab}
\vspace{-10pt}
\grostrait
\ligne{\hfill
\hbox to 288pt{%
\demirangeea 0 1 2 3 4 5 6  \demirangeeb {7} {8} {9} {10} {11} {12} {13}  {}
\hbox to 50pt{\hfill $n$\hskip 20pt}
\hfill}\hfill}
\ligne{\hfill
\hbox to 288pt{%
\demirangeea {} {} {} {\bf V} {} {} {} 
\demirangeeb {} {\bf Y} {\bf Z} {\bf X} {} {} {}  {}
\hbox to 50pt{\hfill}
\hfill}\hfill}
\vspace{-4pt}
\demitrait
\vspace{4pt}
\ligne{\hfill
\hbox to 288pt{%
\demirangeea W B B W B W W   \demirangeeb B W B W B W W W
\hbox to 50pt{\hfill 97\hskip 20pt}
\hfill}\hfill}
\ligne{\hfill
\hbox to 288pt{%
\demirangeea W B B W B W W   \demirangeeb B W B {\small\bf B} B W W B
\hbox to 50pt{\hfill 101\hskip 20pt}
\hfill}\hfill}
\demitrait
\vspace{4pt}
\ligne{\hfill\tt
-I- $X$ was not selected
\hfill}
\demitrait
\vspace{4pt}
\ligne{\hfill
\hbox to 288pt{%
\demirangeea {\small\bf B} B B W B W W   \demirangeeb B W {\small\bf W} W B W W W
\hbox to 50pt{\hfill 150\hskip 20pt}
\hfill}\hfill}
\demitrait
\vspace{4pt}
\ligne{\hfill\tt
-II- $X$ was selected
\hfill}
\demitrait
\vspace{4pt}
\ligne{\hfill
\hbox to 288pt{%
\demirangeea {\small\bf B} B B W B W W   \demirangeeb B W B W B W W W
\hbox to 50pt{\hfill 98\hskip 20pt}
\hfill}\hfill}
\demitrait
\vspace{4pt}
\ligne{\hfill
\hbox to 288pt{%
\demirangeea W B B {\small\bf B} B W W   \demirangeeb B W B W B W W W
\hbox to 50pt{\hfill 99\hskip 20pt}
\hfill}\hfill}
\demitrait\vspace{15pt}
}

   In Table~\ref{mmX_tabT}, the particle arrives to~$T$ from~$X$. Now, there
are two cases, depending on whether $X$~was or not on the selected track.
The first row of the table gives the conservative rule for~$T$: rule~97, exactly the
same rule as for~$O$ in the fixed switch, see Table~\ref{fix_tabO}. Then, 
as in the fixed switch, rule~101 attracts the particle into~$T$. But this time,
rule~150 is applied to return the state of~$T$ from black to white because 
$Z$~is flashing at this moment: $X$ was not on the selected track. Next, rule~99
witnesses that the particle left~$T$ and is now in~$V$. 

   If~$X$ was the selected track, there is no flash and so, the rules of the 
fixed switch apply, see Table~\ref{fix_tabO}: rule~101 again draws the particle to~$T$,
then rule~98 witnesses its migration to~$V$, returning~$T$ to the white state. 
Again, rule~99 witnesses that the particle is now in~$V$.

   In both cases, later, the conservative rule~97 applies to~$T$.

   Table~\ref{mmY_tabT} gives the rules for~$T$ when the particle arrives there
coming from~$Y$. There are also two cases, as previously, depending on whether
$Y$~was or was not the selected track. The organization of Table~\ref{mmY_tabT}
is parallel to that of Table~\ref{mmX_tabT}. Note that the conservative rule
is of course the same. The rule which attracts the particle to~$T$ is different:
it is here rule~100{} in both cases. Next, the rule which turns~$T$ back to
white is rule~150 as previously if $Y$~was not selected because $Z$~is flashing.
If $Y$~was selected, rule~98 applies as in the fixed switch. Then, in both
cases, the conservative rule~97 again applies.

\vtop{
\begin{tab}\label{mmY_tabT}
\leurre
Passive memory switch. Rules for $T$. The particle goes from $Y$.
\end{tab}
\vspace{-10pt}
\grostrait
\ligne{\hfill
\hbox to 288pt{%
\demirangeea 0 1 2 3 4 5 6  \demirangeeb {7} {8} {9} {10} {11} {12} {13}  {}
\hbox to 50pt{\hfill $n$\hskip 20pt}
\hfill}\hfill}
\ligne{\hfill
\hbox to 288pt{%
\demirangeea {} {} {} {\bf V} {} {} {} 
\demirangeeb {} {\bf Y} {\bf Z} {\bf X} {} {} {}  {}
\hbox to 50pt{\hfill}
\hfill}\hfill}
\vspace{-4pt}
\demitrait
\vspace{4pt}
\ligne{\hfill
\hbox to 288pt{%
\demirangeea W B B W B W W   \demirangeeb B W B W B W W W
\hbox to 50pt{\hfill 97\hskip 20pt}
\hfill}\hfill}
\ligne{\hfill
\hbox to 288pt{%
\demirangeea W B B W B W W   \demirangeeb B {\small\bf B} B W B W W B
\hbox to 50pt{\hfill 100\hskip 20pt}
\hfill}\hfill}
\demitrait
\vspace{4pt}
\ligne{\hfill\tt
-I- $Y$ was not selected
\hfill}
\demitrait
\vspace{4pt}
\ligne{\hfill
\hbox to 288pt{%
\demirangeea {\small\bf B} B B W B W W   \demirangeeb B W {\small\bf W} W B W W W
\hbox to 50pt{\hfill 150\hskip 20pt}
\hfill}\hfill}
\demitrait
\vspace{4pt}
\ligne{\hfill\tt
-II- $Y$ was selected
\hfill}
\demitrait
\vspace{4pt}
\ligne{\hfill
\hbox to 288pt{%
\demirangeea {\small\bf B} B B W B W W   \demirangeeb B W B W B W W W
\hbox to 50pt{\hfill 98\hskip 20pt}
\hfill}\hfill}
\demitrait
\vspace{4pt}
\ligne{\hfill
\hbox to 288pt{%
\demirangeea W B B {\small\bf B} B W W   \demirangeeb B W B W B W W W
\hbox to 50pt{\hfill 99\hskip 20pt}
\hfill}\hfill}
\demitrait\vspace{15pt}
}

   Let us now turn to the cells~$X$ and~$Y$. We start with the cell~$X$.
Table~\ref{mmY_tabX} gives the rule when the particle comes from~$Y$.

\vtop{
\begin{tab}\label{mmY_tabX}
Passive memory switch. Rules for $X$. The particle comes from $Y$.
\end{tab}
\vspace{-10pt}
\grostrait
\ligne{\hfill
\hbox to 288pt{%
\demirangeea 0 1 2 3 4 5 6  \demirangeeb {7} {8} {9} {10} {11} {12} {13}  {}
\hbox to 50pt{\hfill $n$\hskip 20pt}
\hfill}\hfill}
\ligne{\hfill
\hbox to 288pt{%
\demirangeea {} {\bf T} {\bf Z} {\bf I} {} {} {} 
\demirangeeb {} {} {} {$tr$} {} {} {}  {}
\hbox to 50pt{\hfill}
\hfill}\hfill}
\vspace{-4pt}
\demitrait
\vspace{4pt}
\ligne{\hfill\tt
-I- $Y$ was not selected
\hfill}
\demitrait
\vspace{4pt}
\ligne{\hfill
\hbox to 288pt{%
\demirangeea W W B W W B B   \demirangeeb B B B W B B B W
\hbox to 50pt{\hfill 102\hskip 20pt}
\hfill}\hfill}
\ligne{\hfill
\hbox to 288pt{%
\demirangeea W {\small\bf B} {\small\bf W} W W B B   \demirangeeb B B B W B B B W
\hbox to 50pt{\hfill 103\hskip 20pt}
\hfill}\hfill}
\ligne{\hfill
\hbox to 288pt{%
\demirangeea W W B {\small\bf B} W B B   \demirangeeb B B B W B B B W
\hbox to 50pt{\hfill 106\hskip 20pt}
\hfill}\hfill}
\demitrait
\vspace{4pt}
\ligne{\hfill\tt
-I- $Y$ was selected
\hfill}
\demitrait
\vspace{4pt}
\ligne{\hfill
\hbox to 288pt{%
\demirangeea W W B B W B B   \demirangeeb B B B W B B B W
\hbox to 50pt{\hfill 106\hskip 20pt}
\hfill}\hfill}
\ligne{\hfill
\hbox to 288pt{%
\demirangeea W {\small\bf B} B B W B B   \demirangeeb B B B W B B B W
\hbox to 50pt{\hfill 107\hskip 20pt}
\hfill}\hfill}
\demitrait\vspace{15pt}
}

   The conservative rules are rule~102 when $Y$~was not in the selected tracks and
rule~106 when $Y$ was in the selected tracks. There is a difference as~$I$ is 
not in the same state: $I$~is white when $Y$~is not selected and it is black
when $Y$~is selected. When $Y$~was selected, rule~107 is performed after rule~106,
and rule~107 witnesses that the particle is now in~$T$. Later, rule~106 is again 
in action, as the particle is no more visible. If $Y$~was not selected, when the 
particle is in~$T$, $Z$~is flashing because the particle was just before in~$Y$. 
Rule~103 witnesses both the occurrence of the particle in~$T$ and the flash of~$Z$. 
Later the particle is no more visible but rule~106 is the conservative rule for~$X$ 
when $I$~is black as now the selected track is~$Y$.

\vtop{
\begin{tab}\label{mmX_tabX}
\leurre
Passive memory switch. Rules for $X$. The particle comes from $X$.
\end{tab}
\vspace{-10pt}
\grostrait
\ligne{\hfill
\hbox to 288pt{%
\demirangeea 0 1 2 3 4 5 6  \demirangeeb {7} {8} {9} {10} {11} {12} {13}  {}
\hbox to 50pt{\hfill $n$\hskip 20pt}
\hfill}\hfill}
\ligne{\hfill
\hbox to 288pt{%
\demirangeea {} {\bf T} {\bf Z} {\bf I} {} {} {} 
\demirangeeb {} {} {} {$tr$} {} {} {}  {}
\hbox to 50pt{\hfill}
\hfill}\hfill}
\vspace{-4pt}
\demitrait
\vspace{4pt}
\ligne{\hfill\tt
-I- $X$ was not selected
\hfill}
\demitrait
\vspace{4pt}
\ligne{\hfill
\hbox to 288pt{%
\demirangeea W W B B W B B   \demirangeeb B B B W B B B W
\hbox to 50pt{\hfill 106\hskip 20pt}
\hfill}\hfill}
\ligne{\hfill
\hbox to 288pt{%
\demirangeea W W B B W B B   \demirangeeb B B B {\small\bf B} B B B B
\hbox to 50pt{\hfill 108\hskip 20pt}
\hfill}\hfill}
\ligne{\hfill
\hbox to 288pt{%
\demirangeea {\small\bf B} W B B W B B   \demirangeeb B B B W B B B W
\hbox to 50pt{\hfill 109\hskip 20pt}
\hfill}\hfill}
\ligne{\hfill
\hbox to 288pt{%
\demirangeea W {\small\bf B} {\small\bf W} B W B B   \demirangeeb B B B W B B B W
\hbox to 50pt{\hfill 110\hskip 20pt}
\hfill}\hfill}
\ligne{\hfill
\hbox to 288pt{%
\demirangeea W W B {\small\bf W} W B B   \demirangeeb B B B W B B B W
\hbox to 50pt{\hfill 102\hskip 20pt}
\hfill}\hfill}
\demitrait
\vspace{4pt}
\ligne{\hfill\tt
-II- $X$ was selected
\hfill}
\demitrait
\vspace{4pt}
\ligne{\hfill
\hbox to 288pt{%
\demirangeea W W B W W B B   \demirangeeb B B B W B B B W
\hbox to 50pt{\hfill 102\hskip 20pt}
\hfill}\hfill}
\ligne{\hfill
\hbox to 288pt{%
\demirangeea W W B W W B B   \demirangeeb B B B {\small\bf B} B B B B
\hbox to 50pt{\hfill 105\hskip 20pt}
\hfill}\hfill}
\ligne{\hfill
\hbox to 288pt{%
\demirangeea {\small\bf B} W B W W B B   \demirangeeb B B B W B B B W
\hbox to 50pt{\hfill 112\hskip 20pt}
\hfill}\hfill}
\ligne{\hfill
\hbox to 288pt{%
\demirangeea W {\small\bf B} B W W B B   \demirangeeb B B B W B B B W
\hbox to 50pt{\hfill 104\hskip 20pt}
\hfill}\hfill}
\demitrait\vspace{15pt}
}

   Table~\ref{mmX_tabX} gives the rules for~$X$ when the particle
comes from~$X$. The situation is not completely symmetric with the situation
of Table~\ref{mmY_tabX}. In Table~\ref{mmY_tabX} we can remark that $X$~is
always white, while in Table~\ref{mmX_tabX}, there two times when $X$~is
black, one when $X$~is on the selected tracks and another when when it is not.

    When $X$~is on the selected tracks, the conservative rule is rule~102.
Then we simply have motion rules adapted to~$X$. Indeed, 
Rule~105 detects the occurrence of the particle on the next cell of the tracks 
after~$X$. This makes~$X$ change to black. Then, rule~112 changes~$X$ back to white
while rule~104 witnesses that the particle has reached~$T$.

   When $X$~is not on the selected tracks, the conservative rule is rule~106.
This time, to the motion rules adapted to the configuration of~$X$ in which
$I$~is black, we have the flash of~$Z$ which occurs when the particle is in~$T$.
Indeed, rule~108 detects the particle coming from the tracks and then $X$~turns
to black. Rule~109 turns~$X$ back to white and the particle will go to~$T$.
Rule~110 witnesses that the particle is indeed in~$T$ and it also witnesses the
flash of~$Z$ as already mentioned. From this flash, the cell~$I$ turned to white
which is witnessed by rule~102, the other conservative rule for~$X$, when
$I$~is white.

\vtop{
\begin{tab}\label{mmX_tabY}
\leurre
Passive memory switch. Rules for $Y$. The particle comes from $X$. 
\end{tab}
\vspace{-10pt}
\grostrait
\ligne{\hfill
\hbox to 288pt{%
\demirangeea 0 1 2 3 4 5 6  \demirangeeb {7} {8} {9} {10} {11} {12} {13}  {}
\hbox to 50pt{\hfill $n$\hskip 20pt}
\hfill}\hfill}
\ligne{\hfill
\hbox to 288pt{%
\demirangeea {} {\bf T} {} {} {} {$tr$} {} 
\demirangeeb {} {} {} {} {} {\bf J} {\bf Z}  {}
\hbox to 50pt{\hfill}
\hfill}\hfill}
\vspace{-4pt}
\demitrait
\ligne{\hfill\tt
-I- $X$ was not selected
\hfill}
\demitrait
\vspace{4pt}
\ligne{\hfill
\hbox to 288pt{%
\demirangeea W W B B B W B   \demirangeeb B B W W W W B W
\hbox to 50pt{\hfill 113\hskip 20pt}
\hfill}\hfill}
\ligne{\hfill
\hbox to 288pt{%
\demirangeea W {\small\bf B} B B B W B   \demirangeeb B B W W W W {\small\bf W} W
\hbox to 50pt{\hfill 114\hskip 20pt}
\hfill}\hfill}
\ligne{\hfill
\hbox to 288pt{%
\demirangeea W W B B B W B   \demirangeeb B B W W W {\small\bf B} B W
\hbox to 50pt{\hfill 115\hskip 20pt}
\hfill}\hfill}
\demitrait
\vspace{4pt}
\ligne{\hfill\tt
-II- $X$ was selected
\hfill}
\demitrait
\vspace{4pt}
\ligne{\hfill
\hbox to 288pt{%
\demirangeea W W B B B W B   \demirangeeb B B W W W B B W
\hbox to 50pt{\hfill 115\hskip 20pt}
\hfill}\hfill}
\ligne{\hfill
\hbox to 288pt{%
\demirangeea W {\small\bf B} B B B W B   \demirangeeb B B W W W B B W
\hbox to 50pt{\hfill 118\hskip 20pt}
\hfill}\hfill}
\demitrait\vspace{15pt}
}

\vtop{
\begin{tab}\label{mmY_tabY}
Passive memory switch. Rules for $Y$. The particle comes from $Y$.
\end{tab}
\vspace{-10pt}
\grostrait
\ligne{\hfill
\hbox to 288pt{%
\demirangeea 0 1 2 3 4 5 6  \demirangeeb {7} {8} {9} {10} {11} {12} {13}  {}
\hbox to 50pt{\hfill $n$\hskip 20pt}
\hfill}\hfill}
\ligne{\hfill
\hbox to 288pt{%
\demirangeea {} {\bf T} {} {} {} {$tr$} {} 
\demirangeeb {} {} {} {} {} {\bf J} {\bf Z}  {}
\hbox to 50pt{\hfill}
\hfill}\hfill}
\vspace{-4pt}
\demitrait
\vspace{4pt}
\ligne{\hfill\tt
-I- $Y$ was not selected
\hfill}
\demitrait
\vspace{4pt}
\ligne{\hfill
\hbox to 288pt{%
\demirangeea W W B B B W B   \demirangeeb B B W W W B B W
\hbox to 50pt{\hfill 115\hskip 20pt}
\hfill}\hfill}
\ligne{\hfill
\hbox to 288pt{%
\demirangeea W W B B B {\small\bf B} B   \demirangeeb B B W W W B B B
\hbox to 50pt{\hfill 119\hskip 20pt}
\hfill}\hfill}
\ligne{\hfill
\hbox to 288pt{%
\demirangeea {\small\bf B} W B B B W B   \demirangeeb B B W W W B B W
\hbox to 50pt{\hfill 120\hskip 20pt}
\hfill}\hfill}
\ligne{\hfill
\hbox to 288pt{%
\demirangeea W {\small\bf B} B B B W B   \demirangeeb B B W W W B {\small\bf W} W
\hbox to 50pt{\hfill 121\hskip 20pt}
\hfill}\hfill}
\ligne{\hfill
\hbox to 288pt{%
\demirangeea W W B B B W B   \demirangeeb B B W W W {\small\bf W} B W
\hbox to 50pt{\hfill 113\hskip 20pt}
\hfill}\hfill}
\demitrait
\vspace{4pt}
\ligne{\hfill\tt
-II- $Y$ was selected
\hfill}
\demitrait
\vspace{4pt}
\ligne{\hfill
\hbox to 288pt{%
\demirangeea W W B B B W B   \demirangeeb B B W W W W B W
\hbox to 50pt{\hfill 113\hskip 20pt}
\hfill}\hfill}
\ligne{\hfill
\hbox to 288pt{%
\demirangeea W W B B B {\small\bf B} B   \demirangeeb B B W W W W B B
\hbox to 50pt{\hfill 117\hskip 20pt}
\hfill}\hfill}
\ligne{\hfill
\hbox to 288pt{%
\demirangeea {\small\bf B} W B B B W B   \demirangeeb B B W W W W B W
\hbox to 50pt{\hfill 122\hskip 20pt}
\hfill}\hfill}
\ligne{\hfill
\hbox to 288pt{%
\demirangeea W {\small\bf B} B B B W B   \demirangeeb B B W W W W B W
\hbox to 50pt{\hfill 116\hskip 20pt}
\hfill}\hfill}
\demitrait\vspace{15pt}
}

   Tables~\ref{mmX_tabY} and~\ref{mmY_tabY} repeat for~$Y$ what has been done
for~$X$. In this regard, the rules are very parallel to those of 
Tables~\ref{mmX_tabT} and~\ref{mmX_tabX}. The difference lies in the difference
of the patterns of neighbours around~$X$ and~$Y$ as can be seen from the
conservative rules of the tables. When the particle comes from~$X$, $Y$~always
remains white. If $X$ was on the selected tracks, we have the conservative rule,
rule~115 and a witness that the particle passes through~$T$, rule~120. When
$X$~was not on the selected tracks, outside the conservative rule, rule~113,
rule~114 can see both the particle in~$T$ and the flash of~$Z$. As $J$~changes
its state after the flash, the conservative rule~115 is now in action.

When the particle comes from~$Y$, we have pure motion rules when $Y$ is on the 
selected tracks, second part of Table~\ref{mmY_tabT}. When $Y$~was not on the
selected tracks, we have motion rules too but also rule~121 which witnesses
both the passage of the particle in~$T$ and the flash of~$Z$. The latter
event triggers the change of state of~$J$, rule~113 which is the conservative
rule when $J$~is white, which means that $Y$~is now again on the selected tracks.

   Now, we turn to the rules for~$Z$. Remember that $Z$~controls the arrival of
the particle through the non-selected tracks. In case of this event, it flashes,
which triggers both the change of state of the markers~$I$ and~$J$ and the
sending of a signal to~$D$, the controller of the active memory switch. 

   Tables~\ref{mmX_tabZ} and~\ref{mmY_tabZ} display the rules for~$Z$.

\vtop{
\begin{tab}\label{mmX_tabZ}
\leurre
Passive memory switch. Rules for $Z$. The particle comes from $X$.
\end{tab}
\vspace{-10pt}
\grostrait
\ligne{\hfill
\hbox to 288pt{%
\demirangeea 0 1 2 3 4 5 6  \demirangeeb {7} {8} {9} {10} {11} {12} {13}  {}
\hbox to 50pt{\hfill $n$\hskip 20pt}
\hfill}\hfill}
\ligne{\hfill
\hbox to 288pt{%
\demirangeea {} {\bf T} {\bf Y} {\bf J} {} {} {} 
\demirangeeb {} {$g$} {} {} {} {\bf I} {\bf X}  {}
\hbox to 50pt{\hfill}
\hfill}\hfill}
\vspace{-4pt}
\demitrait
\vspace{4pt}
\ligne{\hfill\tt
-I- $X$ was not selected
\hfill}
\demitrait
\vspace{4pt}
\ligne{\hfill
\hbox to 288pt{%
\demirangeea B W W W W B B   \demirangeeb B W B B W B W B
\hbox to 50pt{\hfill 138\hskip 20pt}
\hfill}\hfill}
\ligne{\hfill
\hbox to 288pt{%
\demirangeea B W W W W B B   \demirangeeb B W B B W B {\small\bf B} W
\hbox to 50pt{\hfill 143\hskip 20pt}
\hfill}\hfill}
\ligne{\hfill
\hbox to 288pt{%
\demirangeea {\small\bf W} {\small\bf B} W W W B B   \demirangeeb B W B B W B W B
\hbox to 50pt{\hfill 139\hskip 20pt}
\hfill}\hfill}
\ligne{\hfill
\hbox to 288pt{%
\demirangeea B W W {\small\bf B} W B B   
\demirangeeb B {\small\bf B} B B W {\small\bf W} W B
\hbox to 50pt{\hfill 146\hskip 20pt}
\hfill}\hfill}
\ligne{\hfill
\hbox to 288pt{%
\demirangeea B W W B W B B   \demirangeeb B W B B W W W B
\hbox to 50pt{\hfill 137\hskip 20pt}
\hfill}\hfill}
\demitrait
\vspace{4pt}
\ligne{\hfill\tt
-II- $X$ was selected
\hfill}
\demitrait
\vspace{4pt}
\ligne{\hfill
\hbox to 288pt{%
\demirangeea B W W B W B B   \demirangeeb B W B B W W W B
\hbox to 50pt{\hfill 137\hskip 20pt}
\hfill}\hfill}
\ligne{\hfill
\hbox to 288pt{%
\demirangeea B W W B W B B   \demirangeeb B W B B W W {\small\bf B} B
\hbox to 50pt{\hfill 142\hskip 20pt}
\hfill}\hfill}
\ligne{\hfill
\hbox to 288pt{%
\demirangeea B {\small\bf B} W B W B B   \demirangeeb B W B B W W W B
\hbox to 50pt{\hfill 149\hskip 20pt}
\hfill}\hfill}
\demitrait\vspace{15pt}
}

   In Table~\ref{mmX_tabZ} we look at the first part, when the particle
come from~$X$ in the case the corresponding tracks was not selected. The conservative
rule is here rule~138. Rule~143 witnesses that the particle is in~$X$ and so, the
rule makes~$Z$ turn to flash: it becomes white. At the next time, rule~139 restores
the black state in~$Z$ and witnesses that the particle is in~$T$. Rule~146 witnesses
the change of states in both~$I$ and~$J$ which was caused by the flash of~$Z$
and the sending of a particle to~$D$ through~$g$ which is $Z.8$.
As the particle is no more visible, a conservative rule is now in action: rule~146
which is attached to the configuration when $I$~is white and $J$~is black.

   It is this configuration that we find in the first row of the second part
of Table~\ref{mmX_tabZ} with rule~137. The other rules witness the motion
of the particle: rule~142 can see it in~$X$, rule~149 can see it in~$T$ and then
rule~137 applies again. As $X$ is on the selected track, the passage of the particle
does not cause any change.

   We have a perfectly similar table for the case when the particle arrives to
the switch through~$Y$: Table~\ref{mmY_tabZ}. 

   The first part of the table deals with the case when $Y$~was not on the 
selected tracks. This time rule~137 is the conservative rule, as $I$~is white 
and~$J$ is black.
Rule~144 witnesses that the particle is in~$Y$, making $Z$ flash. Rule~140 witnesses 
that it passed to~$T$. Rule~147 witnesses that~$I$ and~$J$ changed their states
and that a particle was sent through~$g$ to~$D$. Rule~138 applies as a conservative
rule for the new states of~$I$ and~$J$.

\vtop{
\begin{tab}\label{mmY_tabZ}
Passive memory switch. Rules for $Z$. The particle comes from $Y$.
\end{tab}
\vspace{-10pt}
\grostrait
\ligne{\hfill
\hbox to 288pt{%
\demirangeea 0 1 2 3 4 5 6  \demirangeeb {7} {8} {9} {10} {11} {12} {13}  {}
\hbox to 50pt{\hfill $n$\hskip 20pt}
\hfill}\hfill}
\ligne{\hfill
\hbox to 288pt{%
\demirangeea {} {\bf T} {\bf Y} {\bf J} {} {} {} 
\demirangeeb {} {$g$} {} {} {} {\bf I} {\bf X}  {}
\hbox to 50pt{\hfill}
\hfill}\hfill}
\vspace{-4pt}
\demitrait
\vspace{4pt}
\ligne{\hfill\tt
-I- passage through $Y$, $Y$ was not selected
\hfill}
\demitrait
\vspace{4pt}
\ligne{\hfill
\hbox to 288pt{%
\demirangeea B W W B W B B   \demirangeeb B W B B W W W B
\hbox to 50pt{\hfill 137\hskip 20pt}
\hfill}\hfill}
\ligne{\hfill
\hbox to 288pt{%
\demirangeea B W {\small\bf B} B W B B   \demirangeeb B W B B W W W W
\hbox to 50pt{\hfill 144\hskip 20pt}
\hfill}\hfill}
\ligne{\hfill
\hbox to 288pt{%
\demirangeea {\small\bf W} {\small\bf B} W B W B B   \demirangeeb B W B B W W W B
\hbox to 50pt{\hfill 140\hskip 20pt}
\hfill}\hfill}
\ligne{\hfill
\hbox to 288pt{%
\demirangeea B W W {\small\bf W} W B B   
\demirangeeb B {\small\bf B} B B W {\small\bf B} W B
\hbox to 50pt{\hfill 147\hskip 20pt}
\hfill}\hfill}
\ligne{\hfill
\hbox to 288pt{%
\demirangeea B W W W W B B   \demirangeeb B W B B W B W B
\hbox to 50pt{\hfill 138\hskip 20pt}
\hfill}\hfill}
\demitrait
\vspace{4pt}
\ligne{\hfill\tt
-II- $Y$ was selected
\hfill}
\demitrait
\vspace{4pt}
\ligne{\hfill
\hbox to 288pt{%
\demirangeea B W W W W B B   \demirangeeb B W B B W B W B
\hbox to 50pt{\hfill 138\hskip 20pt}
\hfill}\hfill}
\ligne{\hfill
\hbox to 288pt{%
\demirangeea B W {\small\bf B} W W B B   \demirangeeb B W B B W B W B
\hbox to 50pt{\hfill 141\hskip 20pt}
\hfill}\hfill}
\ligne{\hfill
\hbox to 288pt{%
\demirangeea B {\small\bf B} W W W B B   \demirangeeb B W B B W B W B
\hbox to 50pt{\hfill 148\hskip 20pt}
\hfill}\hfill}
\demitrait\vspace{15pt}
}

   The second part of Table~\ref{mmY_tabZ} starts with the configuration
rule associated to a white~$J$ and a black~$I$ as $Y$~is selected: rule~138 again.
Then rule~141 can see the particle in~$Y$, rule~148 can see it in~$T$ and then
rule~138 applies again.

\vskip 10pt
   Next, Tables~\ref{mm_tabI} and~\ref{mm_tabJ} give the rules for~$I$ and~$J$
respectively. 

\vtop{
\begin{tab}\label{mm_tabI}
\leurre
Passive memory switch. Rules for $I$. 
\end{tab}
\vspace{-10pt}
\grostrait
\ligne{\hfill
\hbox to 288pt{%
\demirangeea 0 1 2 3 4 5 6  \demirangeeb {7} {8} {9} {10} {11} {12} {13}  {}
\hbox to 50pt{\hfill $n$\hskip 20pt}
\hfill}\hfill}
\ligne{\hfill
\hbox to 288pt{%
\demirangeea {} {\bf X} {\bf Z} {} {} {} {} 
\demirangeeb {} {} {} {} {} {} {}  {}
\hbox to 50pt{\hfill}
\hfill}\hfill}
\vspace{-4pt}
\demitrait
\vspace{4pt}
\ligne{\hfill\tt
-I- The particle comes from~$X$, $X$ being not selected
\hfill}
\demitrait
\vspace{4pt}
\ligne{\hfill
\hbox to 288pt{%
\demirangeea B W B W W B B   \demirangeeb B B W B B W W B
\hbox to 50pt{\hfill 153\hskip 20pt}
\hfill}\hfill}
\ligne{\hfill
\hbox to 288pt{%
\demirangeea B {\small\bf B} B W W B B   \demirangeeb B B W B B W W B
\hbox to 50pt{\hfill 155\hskip 20pt}
\hfill}\hfill}
\ligne{\hfill
\hbox to 288pt{%
\demirangeea B W {\small\bf W} W W B B   \demirangeeb B B W B B W W W
\hbox to 50pt{\hfill 157\hskip 20pt}
\hfill}\hfill}
\ligne{\hfill
\hbox to 288pt{%
\demirangeea {\small\bf W} W B W W B B   \demirangeeb B B W B B W W W
\hbox to 50pt{\hfill 152\hskip 20pt}
\hfill}\hfill}
\demitrait\vspace{15pt}
}

\vtop{
\noindent
{\bf Table~\ref{mm_tabI} (continued)}
{\leurre\it
Passive memory switch. Rules for $I$. 
\par}
\vskip 0pt
\ligne{\hfill}
\vspace{-10pt}
\grostrait
\ligne{\hfill
\hbox to 288pt{%
\demirangeea 0 1 2 3 4 5 6  \demirangeeb {7} {8} {9} {10} {11} {12} {13}  {}
\hbox to 50pt{\hfill $n$\hskip 20pt}
\hfill}\hfill}
\ligne{\hfill
\hbox to 288pt{%
\demirangeea {} {\bf X} {\bf Z} {} {} {} {} 
\demirangeeb {} {} {} {} {} {} {}  {}
\hbox to 50pt{\hfill}
\hfill}\hfill}
\vspace{-4pt}
\demitrait
\vspace{4pt}
\ligne{\hfill\tt
-II- The particle comes from~$X$, $X$ being selected
\hfill}
\demitrait
\vspace{4pt}
\ligne{\hfill
\hbox to 288pt{%
\demirangeea W W B W W B B   \demirangeeb B B W B B W W W
\hbox to 50pt{\hfill 152\hskip 20pt}
\hfill}\hfill}
\ligne{\hfill
\hbox to 288pt{%
\demirangeea W {\small\bf B} B W W B B   \demirangeeb B B W B B W W W
\hbox to 50pt{\hfill 76\hskip 20pt}
\hfill}\hfill}
\demitrait
\vspace{4pt}
\ligne{\hfill\tt
-III- The particle comes from~$Y$, $Y$ being not selected
\hfill}
\demitrait
\vspace{4pt}
\ligne{\hfill
\hbox to 288pt{%
\demirangeea W W B W W B B   \demirangeeb B B W B B W W W
\hbox to 50pt{\hfill 152\hskip 20pt}
\hfill}\hfill}
\ligne{\hfill
\hbox to 288pt{%
\demirangeea W W {\small\bf W} W W B B   \demirangeeb B B W B B W W B
\hbox to 50pt{\hfill 156\hskip 20pt}
\hfill}\hfill}
\ligne{\hfill
\hbox to 288pt{%
\demirangeea {\small\bf B} W B W W B B   \demirangeeb B B W B B W W B
\hbox to 50pt{\hfill 153\hskip 20pt}
\hfill}\hfill}
\demitrait
\vspace{4pt}
\ligne{\hfill\tt
-IV- The particle comes from~$Y$, $Y$ being selected
\hfill}
\demitrait
\vspace{4pt}
\ligne{\hfill
\hbox to 288pt{%
\demirangeea B W B W W B B   \demirangeeb B B W B B W W B
\hbox to 50pt{\hfill 153\hskip 20pt}
\hfill}\hfill}
\demitrait\vspace{15pt}
}

   In Table~\ref{mm_tabI} we have four cases first, depending on which side 
the particle arrives at the central cell and then, on whether the tracks
was selected or not.

   Indeed, as we can easily understand in the fourth part of the table, if
the particle passes through~$Y$ and if $Y$~is the selected tracks, $Z$~does not
react and so, $I$ cannot see that a particle passed through the switch. This is
why in this case we have a conservative rule only, here rule~153.
If the $Y$ was on the selected tracks and if the particle passes through~$Y$,
then~$Z$ flashes. This is why we have three instructions in the third part of the
table: the conservative rule~152 corresponding to a black state in~$J$,
rule~156 which, seeing the flash of~$Z$ makes~$I$ turn to black and rule~153,
the conservative rule for the new situation when $I$~is black. 

   Now, the fist two parts of Table~\ref{mm_tabI} deal with the case when
the particle passes through~$X$. In the first part, the rules concern the
case when $X$~was not selected. In this case, after the conservative rule~153
as $I$~is black, corresponding to the non selection of~$X$, rule~155 witnesses
that the particle is in~$X$. Then rule~157 can see the flash of~$Z$ and so,
it turns the state of~$I$ to white: from now on, $X$ is selected.
At the next time, rule~152 applies as it is the conservative rule for the case
when $I$~is white.
\vskip 5pt
   Table~\ref{mm_tabJ} is alike Table~\ref{mm_tabI} with, this time the cell~$J$   
being concerned. The two first parts of the table look at which rules are used
when the particle comes from~$X$. As symmetrically in part~IV of Table~\ref{mm_tabI},
$J$~is not concerned by a passage of the particle though~$X$ when $X$ is on the
selected tracks. This is why we have the single conservative rule~153{} in part~II.
In part~I of Table~\ref{mm_tabJ}, where the conservative rule is rule~152, the 
cell~$J$ is concerned by the passage through~$X$: $Z$ flashes, as $X$ is not on the
selected tracks. This flash is witnessed by rule~156 which makes~$J$ take the
black state. Then rule~153 applies as the conservative rule for this new configuration
of~$J$.

   The last two parts of the table deal with the case when the particle passes
through~$Y$. Then, the cell always witnesses the passage of the particle.
When $Y$~is not on the selected tracks, the conservative rule is rule~153.
The passage of the particle through~$Y$ is witnessed by rule~161 and rule~157 
witnesses the flash of~$Z$, making~$J$ take the white state. We now arrive to
the configuration when~$J$ is white. The configuration is kept unchanged by rule~152.

   In the case when $Y$~is on the selected tracks, $Z$ does not flash and so,
$J$~simply witnesses the passage of the particle through~$Y$: the conservative
rule is~152 as just noticed and rule~160 witnesses the passage of~$X$.

\vtop{
\begin{tab}\label{mm_tabJ}
\leurre
Passive memory switch. Rules for $J$. 
\end{tab}
\vspace{-10pt}
\grostrait
\ligne{\hfill
\hbox to 288pt{%
\demirangeea 0 1 2 3 4 5 6  \demirangeeb {7} {8} {9} {10} {11} {12} {13}  {}
\hbox to 50pt{\hfill $n$\hskip 20pt}
\hfill}\hfill}
\ligne{\hfill
\hbox to 288pt{%
\demirangeea {} {\bf Y} {} {} {} {} {} 
\demirangeeb {} {} {} {} {} {} {\bf Z}  {}
\hbox to 50pt{\hfill}
\hfill}\hfill}
\vspace{-4pt}
\demitrait
\vspace{4pt}
\ligne{\hfill\tt
-I- The particle comes from~$X$, $X$ being not selected
\hfill}
\demitrait
\vspace{4pt}
\ligne{\hfill
\hbox to 288pt{%
\demirangeea W W W B B B B   \demirangeeb W B B W W W B W
\hbox to 50pt{\hfill 152\hskip 20pt}
\hfill}\hfill}
\ligne{\hfill
\hbox to 288pt{%
\demirangeea W W W B B B B   \demirangeeb W B B W W W {\small\bf W} B
\hbox to 50pt{\hfill 156\hskip 20pt}
\hfill}\hfill}
\ligne{\hfill
\hbox to 288pt{%
\demirangeea {\small\bf B} W W B B B B   \demirangeeb W B B W W W B B
\hbox to 50pt{\hfill 153\hskip 20pt}
\hfill}\hfill}
\demitrait
\vspace{4pt}
\ligne{\hfill\tt
-II- The particle comes from~$X$, $X$ being selected
\hfill}
\demitrait
\vspace{4pt}
\ligne{\hfill
\hbox to 288pt{%
\demirangeea B W W B B B B   \demirangeeb W B B W W W B B
\hbox to 50pt{\hfill 153\hskip 20pt}
\hfill}\hfill}
\demitrait
\vspace{4pt}
\ligne{\hfill\tt
-III- The particle comes from~$Y$, $Y$ being not selected
\hfill}
\demitrait
\vspace{4pt}
\ligne{\hfill
\hbox to 288pt{%
\demirangeea B W W B B B B   \demirangeeb W B B W W W B B
\hbox to 50pt{\hfill 153\hskip 20pt}
\hfill}\hfill}
\ligne{\hfill
\hbox to 288pt{%
\demirangeea B {\small\bf B} W B B B B   \demirangeeb W B B W W W B B
\hbox to 50pt{\hfill 161\hskip 20pt}
\hfill}\hfill}
\ligne{\hfill
\hbox to 288pt{%
\demirangeea B W W B B B B   \demirangeeb W B B W W W {\small\bf W} W
\hbox to 50pt{\hfill 157\hskip 20pt}
\hfill}\hfill}
\ligne{\hfill
\hbox to 288pt{%
\demirangeea {\small\bf W} W W B B B B   \demirangeeb W B B W W W B W
\hbox to 50pt{\hfill 152\hskip 20pt}
\hfill}\hfill}
\demitrait
\vspace{4pt}
\ligne{\hfill\tt
-IV- The particle comes from~$Y$, $Y$ being selected
\hfill}
\demitrait
\vspace{4pt}
\ligne{\hfill
\hbox to 288pt{%
\demirangeea W W W B B B B   \demirangeeb W B B W W W B W
\hbox to 50pt{\hfill 152\hskip 20pt}
\hfill}\hfill}
\ligne{\hfill
\hbox to 288pt{%
\demirangeea W {\small\bf B} W B B B B   \demirangeeb W B B W W W B W
\hbox to 50pt{\hfill 160\hskip 20pt}
\hfill}\hfill}
\demitrait\vspace{15pt}
}

   At last, and not the least, we have noted that when the particle arrives
to the passive switch through the non selected tracks, this makes~$Z$
flash. We have studied the consequences of this flash on~$I$ and~$J$ but we
also have to look at the consequence on other cells. Indeed, the flash is
also waited by~$Z.8$, a neighbour of~$Z$ which is white but which becomes black
when $Z$~flashes. 

   When $Z$~flashes, $Z.8$~which is called~$Z1$ in the Table~\ref{mmns_tabZ1}
transmits the signal, a particle, to~$D$ through a sequence of ordinary cells
of the tracks. The last cell is~$D1$, the neighbour $D.7$ of~$D$. When $D$
can see that $D1$ is black, it flashes, making both $H$ and~$K$ change their state.

   First, we look at Table~\ref{mmns_tabZ1}, which gives the rules for~$Z1$
when the particle comes from a non selected track. Note that $Z1$~remains white
as long as $Z$~does not flash and $Z1$ does not see another cell outside its 
neighbour on the track. And so, in this situation, the conservative rule~164 only
is in action, first row of the table.

\vtop{
\begin{tab}\label{mmns_tabZ1}
\leurre
Passive memory switch. Rules for $Z1$. 
\end{tab}
\vspace{-10pt}
\grostrait
\ligne{\hfill
\hbox to 288pt{%
\demirangeea 0 1 2 3 4 5 6  \demirangeeb {7} {8} {9} {10} {11} {12} {13}  {}
\hbox to 50pt{\hfill $n$\hskip 20pt}
\hfill}\hfill}
\ligne{\hfill
\hbox to 288pt{%
\demirangeea {} {\bf Z} {} {} {} {} {} 
\demirangeeb {} {} {} {$tr$} {} {} {}  {}
\hbox to 50pt{\hfill}
\hfill}\hfill}
\vspace{-4pt}
\demitrait
\vspace{4pt}
\ligne{\hfill
\hbox to 288pt{%
\demirangeea W B B W B B B   \demirangeeb B B B W B W B W
\hbox to 50pt{\hfill 164\hskip 20pt}
\hfill}\hfill}
\ligne{\hfill
\hbox to 288pt{%
\demirangeea W {\small\bf W} B W B B B   \demirangeeb B B B W B W B B
\hbox to 50pt{\hfill 165\hskip 20pt}
\hfill}\hfill}
\ligne{\hfill
\hbox to 288pt{%
\demirangeea {\small\bf B} B B W B B B   \demirangeeb B B B W B W B W
\hbox to 50pt{\hfill 166\hskip 20pt}
\hfill}\hfill}
\ligne{\hfill
\hbox to 288pt{%
\demirangeea W B B W B B B   \demirangeeb B B B {\small\bf B} B W B W
\hbox to 50pt{\hfill 167\hskip 20pt}
\hfill}\hfill}
\demitrait\vspace{15pt}
}

   When $Z$~flashes, this is witnessed by rule~165, $Z1$ becomes black and
rule~166 returns its state to white. Next, Rule~167 witnesses that a particle is
now on the tracks, on its route to~$D1$.

   Table~\ref{mmns_tabD1} tells us what happens there. As long as $D1$ is white,
the conservative rule~134 applies. When the particle sent by~$Z$ arrives 
close to~$D1$, it is noticed by rule~135 which attracts the signal into~$D1$.
Rule~136 returns the state of~$D1$ to white and rule~70 witnesses the flash of~$D$.

\vtop{
\begin{tab}\label{mmns_tabD1}
\leurre
Passive memory switch. Rules for $D1$. 
\end{tab}
\vspace{-10pt}
\grostrait
\ligne{\hfill
\hbox to 288pt{%
\demirangeea 0 1 2 3 4 5 6  \demirangeeb {7} {8} {9} {10} {11} {12} {13}  {}
\hbox to 50pt{\hfill $n$\hskip 20pt}
\hfill}\hfill}
\ligne{\hfill
\hbox to 288pt{%
\demirangeea {} {} {} {} {} {} {} 
\demirangeeb {} {} {} {} {$tr$} {} {\bf D}  {}
\hbox to 50pt{\hfill}
\hfill}\hfill}
\vspace{-4pt}
\demitrait
\vspace{4pt}
\ligne{\hfill
\hbox to 288pt{%
\demirangeea W B B W W B B   \demirangeeb W W W B W B B W
\hbox to 50pt{\hfill 134\hskip 20pt}
\hfill}\hfill}
\ligne{\hfill
\hbox to 288pt{%
\demirangeea W B B W W B B   \demirangeeb W W W B {\small\bf B} B B B
\hbox to 50pt{\hfill 135\hskip 20pt}
\hfill}\hfill}
\ligne{\hfill
\hbox to 288pt{%
\demirangeea {\small\bf B} B B W W B B   \demirangeeb W W W B W B B W
\hbox to 50pt{\hfill 136\hskip 20pt}
\hfill}\hfill}
\ligne{\hfill
\hbox to 288pt{%
\demirangeea W B B W W B B   \demirangeeb W W W B W B {\small\bf W} W
\hbox to 50pt{\hfill 70\hskip 20pt}
\hfill}\hfill}
\demitrait\vspace{15pt}
}

\vtop{
\begin{tab}\label{mmXns_tabD}
\leurre
Passive memory switch. Rules for $D$. The particle comes from $X$.
\end{tab}
\vspace{-10pt}
\grostrait
\ligne{\hfill
\hbox to 288pt{%
\demirangeea 0 1 2 3 4 5 6  \demirangeeb {7} {8} {9} {10} {11} {12} {13}  {}
\hbox to 50pt{\hfill $n$\hskip 20pt}
\hfill}\hfill}
\ligne{\hfill
\hbox to 288pt{%
\demirangeea {} {\bf O} {\bf C} {\bf K} {} {} {} 
\demirangeeb {\small\bf D1} {} {} {} {} {\bf H} {\bf B}  {}
\hbox to 50pt{\hfill}
\hfill}\hfill}
\vspace{-4pt}
\demitrait
\vspace{4pt}
\ligne{\hfill
\hbox to 288pt{%
\demirangeea B W W W B W B   \demirangeeb W B B B W B W B
\hbox to 50pt{\hfill 125\hskip 20pt}
\hfill}\hfill}
\ligne{\hfill
\hbox to 288pt{%
\demirangeea B W W W B W B   \demirangeeb {\small\bf B} B B B W B W W
\hbox to 50pt{\hfill 130\hskip 20pt}
\hfill}\hfill}
\ligne{\hfill
\hbox to 288pt{%
\demirangeea {\small\bf W} W W W B W B   \demirangeeb W B B B W B W B
\hbox to 50pt{\hfill 131\hskip 20pt}
\hfill}\hfill}
\ligne{\hfill
\hbox to 288pt{%
\demirangeea B W W {\small\bf B} B W B   \demirangeeb W B B B W {\small\bf W} W B
\hbox to 50pt{\hfill 124\hskip 20pt}
\hfill}\hfill}
\demitrait\vspace{15pt}
}

\vtop{
\begin{tab}\label{mmYns_tabD}
Passive memory switch. Rules for $D$. The particle comes from $Y$.
\end{tab}
\vspace{-10pt}
\grostrait
\ligne{\hfill
\hbox to 288pt{%
\demirangeea 0 1 2 3 4 5 6  \demirangeeb {7} {8} {9} {10} {11} {12} {13}  {}
\hbox to 50pt{\hfill $n$\hskip 20pt}
\hfill}\hfill}
\ligne{\hfill
\hbox to 288pt{%
\demirangeea {} {\bf O} {\bf C} {\bf K} {} {} {} 
\demirangeeb {\small\bf D1} {} {} {} {} {\bf H} {\bf B}  {}
\hbox to 50pt{\hfill}
\hfill}\hfill}
\vspace{-4pt}
\demitrait
\vspace{4pt}
\ligne{\hfill
\hbox to 288pt{%
\demirangeea B W W B B W B   \demirangeeb W B B B W W W B
\hbox to 50pt{\hfill 124\hskip 20pt}
\hfill}\hfill}
\ligne{\hfill
\hbox to 288pt{%
\demirangeea B W W B B W B   \demirangeeb {\small\bf B} B B B W W W W
\hbox to 50pt{\hfill 37\hskip 20pt}
\hfill}\hfill}
\ligne{\hfill
\hbox to 288pt{%
\demirangeea {\small\bf W} W W B B W B   \demirangeeb W B B B W W W B
\hbox to 50pt{\hfill 133\hskip 20pt}
\hfill}\hfill}
\ligne{\hfill
\hbox to 288pt{%
\demirangeea B W W {\small\bf W} B W B   \demirangeeb W B B B W {\small\bf B} W B
\hbox to 50pt{\hfill 125\hskip 20pt}
\hfill}\hfill}
\demitrait\vspace{15pt}
}

    When $D$~flashes, this acts upon its sensible neighbours: $H$, $K$, $B$
and~$C$. So that we have to look at possibly additional rules for these cells.
First, we look at the rules for~$D$ indicated by Tables~\ref{mmXns_tabD} 
and~\ref{mmYns_tabD}.

    A quick look at Table~\ref{ff_tabD} shows us that the rules are different.
Indeed, the cell~$D$ in the active memory switch has not the same neighbourhood
as~$D$ in the flip-flop as they do not behave the same way.

   When the particle sent by~$Z$ is not yet arrived at~$D1$, the conservative 
rule~125 applies if $Y$~is selected and the conservative rule~124 is applied
if~$X$ is selected. Still assuming that $X$~is selected, when the signal arrives 
at~$D1$, it is recognized by rule~37 which makes $D$ flash. Note that this rule 
is also used in the crossing for the cell~$B.12$. Next, rule~133 returns the 
state of~$D$ to black,
and rule~125 witnesses the change of state in both~$H$ and~$K$. 
It is also the conservative rule when $Y$~is selected. If $Y$~is
selected, the signal in~$D1$ is recognized by rule~130 which makes $D$~flash.
Then, rule~131 returns the state of~$D$ to black. Rule~124 witnesses the
change of state in both~$H$ and~$K$.
 
\vtop{
\begin{tab}\label{mmns_tabH}
Passive memory switch. Rules for $H$ when $D$ triggered by~$Z$. 
\end{tab}
\vspace{-10pt}
\grostrait
\ligne{\hfill
\hbox to 288pt{%
\demirangeea 0 1 2 3 4 5 6  \demirangeeb {7} {8} {9} {10} {11} {12} {13}  {}
\hbox to 50pt{\hfill $n$\hskip 20pt}
\hfill}\hfill}
\ligne{\hfill
\hbox to 288pt{%
\demirangeea {} {\bf B} {\bf D} {} {} {} {} 
\demirangeeb {} {} {} {} {} {} {}  {}
\hbox to 50pt{\hfill}
\hfill}\hfill}
\vspace{-4pt}
\demitrait
\vspace{4pt}
\ligne{\hfill\tt
-I- The particle comes from~$Y$
\hfill}
\demitrait
\vspace{4pt}
\ligne{\hfill
\hbox to 288pt{%
\demirangeea B W B B B B B   \demirangeeb B W W W W W W B
\hbox to 50pt{\hfill 87\hskip 20pt}
\hfill}\hfill}
\ligne{\hfill
\hbox to 288pt{%
\demirangeea B W {\small\bf W} B B B B   \demirangeeb B W W W W W W W
\hbox to 50pt{\hfill 90\hskip 20pt}
\hfill}\hfill}
\ligne{\hfill
\hbox to 288pt{%
\demirangeea {\small\bf W} W B B B B B   \demirangeeb B W W W W W W W
\hbox to 50pt{\hfill 86\hskip 20pt}
\hfill}\hfill}
\demitrait
\vspace{4pt}
\ligne{\hfill\tt
-II- The particle comes from~$X$
\hfill}
\demitrait
\vspace{4pt}
\ligne{\hfill
\hbox to 288pt{%
\demirangeea W W B B B B B   \demirangeeb B W W W W W W W
\hbox to 50pt{\hfill 86\hskip 20pt}
\hfill}\hfill}
\ligne{\hfill
\hbox to 288pt{%
\demirangeea W W {\small\bf W} B B B B   \demirangeeb B W W W W W W B
\hbox to 50pt{\hfill 89\hskip 20pt}
\hfill}\hfill}
\ligne{\hfill
\hbox to 288pt{%
\demirangeea {\small\bf B} W B B B B B   \demirangeeb B W W W W W W B
\hbox to 50pt{\hfill 87\hskip 20pt}
\hfill}\hfill}
\demitrait\vspace{15pt}
}

   Tables~\ref{mmns_tabH} and~\ref{mmns_tabK} display the rules for~$H$ and~$K$
when $D$~flashes after receiving the signal sent by~$Z$.

   The rules are close to those of the same cells in the flip-flop,
see Tables~\ref{ff_tabH} and~\ref{ff_tabK} with this
difference that rule~88 for~$H$ and rule~93 for~$K$ are not used here
as the particle does not pass through~$B$ nor through~$C$. And so, when 
the particle went through a non selected~$X$, rules~87, 90 and~86 apply as in
the flip-flop and when the particle went through a non selected~$Y$, rules~86,
89 and~87 apply. Note that rules~87 and~86 are conservative rules, when $H$~is
black and when it is white respectively.

\vtop{
\begin{tab}\label{mmns_tabK}
Passive memory switch. Rules for $K$ when $D$ triggered by~$Z$. 
\end{tab}
\vspace{-10pt}
\grostrait
\ligne{\hfill
\hbox to 288pt{%
\demirangeea 0 1 2 3 4 5 6  \demirangeeb {7} {8} {9} {10} {11} {12} {13}  {}
\hbox to 50pt{\hfill $n$\hskip 20pt}
\hfill}\hfill}
\ligne{\hfill
\hbox to 288pt{%
\demirangeea {} {\bf C} {} {} {} {} {} 
\demirangeeb {} {} {} {} {} {} {\bf D}  {}
\hbox to 50pt{\hfill}
\hfill}\hfill}
\vspace{-4pt}
\demitrait
\vspace{4pt}
\ligne{\hfill\tt
-I- The particle comes from~$Y$
\hfill}
\demitrait
\vspace{4pt}
\ligne{\hfill
\hbox to 288pt{%
\demirangeea B W W B B B B   \demirangeeb B W W W W W B B
\hbox to 50pt{\hfill 92\hskip 20pt}
\hfill}\hfill}
\ligne{\hfill
\hbox to 288pt{%
\demirangeea B W W B B B B   \demirangeeb B W W W W W {\small\bf W} W
\hbox to 50pt{\hfill 90\hskip 20pt}
\hfill}\hfill}
\ligne{\hfill
\hbox to 288pt{%
\demirangeea {\small\bf W} W W B B B B   \demirangeeb B W W W W W B W
\hbox to 50pt{\hfill 91\hskip 20pt}
\hfill}\hfill}
\demitrait
\vspace{4pt}
\ligne{\hfill\tt
-II- The particle comes from~$X$
\hfill}
\demitrait
\vspace{4pt}
\ligne{\hfill
\hbox to 288pt{%
\demirangeea W W W B B B B   \demirangeeb B W W W W W B W
\hbox to 50pt{\hfill 91\hskip 20pt}
\hfill}\hfill}
\ligne{\hfill
\hbox to 288pt{%
\demirangeea W W W B B B B   \demirangeeb B W W W W W {\small\bf W} B
\hbox to 50pt{\hfill 89\hskip 20pt}
\hfill}\hfill}
\ligne{\hfill
\hbox to 288pt{%
\demirangeea {\small\bf B} W W B B B B   \demirangeeb B W W W W W B B
\hbox to 50pt{\hfill 92\hskip 20pt}
\hfill}\hfill}
\demitrait\vspace{15pt}
}

\vtop{
\begin{tab}\label{mmns_tabB}
Passive memory switch. Rules for $B$ when $D$ triggered by~$Z$. 
\end{tab}
\vspace{-10pt}
\grostrait
\ligne{\hfill
\hbox to 288pt{%
\demirangeea 0 1 2 3 4 5 6  \demirangeeb {7} {8} {9} {10} {11} {12} {13}  {}
\hbox to 50pt{\hfill $n$\hskip 20pt}
\hfill}\hfill}
\ligne{\hfill
\hbox to 288pt{%
\demirangeea {} {} {} {} {} {} {\bf D} 
\demirangeeb {\bf H} {} {} {} {} {} {}  {}
\hbox to 50pt{\hfill}
\hfill}\hfill}
\vspace{-4pt}
\demitrait
\vspace{4pt}
\ligne{\hfill\tt
-I- The particle comes from~$Y$
\hfill}
\demitrait
\vspace{4pt}
\ligne{\hfill
\hbox to 288pt{%
\demirangeea W B B W B W B   \demirangeeb W W W B B W W W
\hbox to 50pt{\hfill 61\hskip 20pt}
\hfill}\hfill}
\ligne{\hfill
\hbox to 288pt{%
\demirangeea W B B W B W W   \demirangeeb W W W B B W W W
\hbox to 50pt{\hfill 69\hskip 20pt}
\hfill}\hfill}
\ligne{\hfill
\hbox to 288pt{%
\demirangeea W B B W B W B   \demirangeeb B W W B B W W W
\hbox to 50pt{\hfill 66\hskip 20pt}
\hfill}\hfill}
\demitrait
\vspace{4pt}
\ligne{\hfill\tt
-II- The particle comes from~$X$
\hfill}
\demitrait
\vspace{4pt}
\ligne{\hfill
\hbox to 288pt{%
\demirangeea W B B W B W B   \demirangeeb B W W B B W W W
\hbox to 50pt{\hfill 66\hskip 20pt}
\hfill}\hfill}
\ligne{\hfill
\hbox to 288pt{%
\demirangeea W B B W B W W   \demirangeeb B W W B B W W W
\hbox to 50pt{\hfill 63\hskip 20pt}
\hfill}\hfill}
\ligne{\hfill
\hbox to 288pt{%
\demirangeea W B B W B W B   \demirangeeb W W W B B W W W
\hbox to 50pt{\hfill 61\hskip 20pt}
\hfill}\hfill}
\demitrait\vspace{15pt}
}

   We note that for~$K$ we have also rules contained in those used for the cell~$K$
of the flip-flop, as mentioned previously. The conservative rules are here rule~92
and~91 when $K$~is black and white respectively. Rules~90 and~89 recognize the
flash of~$D$ making~$K$ change to white or black respectively.

   We remain with the rules for~$B$ and~$C$.

   The rules for~$B$ are given in Table~\ref{mmns_tabB}. The conservative rules~61
and~66 are those in action when $D$~does not flash and when no particle goes
through~$B$ nor~$C$. These rules are those used in the flip-flop,
see Tables~\ref{ffB_tabB} and~\ref{ffC_tabB}. When the particle passed through~$X$ 
when it was not on the selected tracks, which means that $H$~is black, the rules 
are a part of the rules of Table~\ref{ffB_tabB}: there is no flash of~$D$ in the
situation we consider presently. Now, when the particle passed through~$Y$ 
when it was not on the selected tracks, we have rule~69 which was not yet used. 
A similar remark holds for the rules for~$C$ given in Table~\ref{mmns_tabC}. The 
conservative rules are~66 and~70 as in the flip-flop, see Tables~\ref{ffB_tabC} 
and~\ref{ffC_tabC}. When $K$~is black, the rules are a part of the rules given
in Table~\ref{ffC_tabC} as no flash of~$D$ occurs. However, when $K$~is white,
we have a new rule, rule~77 which was not yet used.

\vtop{
\begin{tab}\label{mmns_tabC}
Passive memory switch. Rules for $C$ when $D$ triggered by~$Z$. 
\end{tab}
\vspace{-10pt}
\grostrait
\ligne{\hfill
\hbox to 288pt{%
\demirangeea 0 1 2 3 4 5 6  \demirangeeb {7} {8} {9} {10} {11} {12} {13}  {}
\hbox to 50pt{\hfill $n$\hskip 20pt}
\hfill}\hfill}
\ligne{\hfill
\hbox to 288pt{%
\demirangeea {} {} {} {} {} {} {} 
\demirangeeb {} {} {\bf K} {\bf D} {} {} {}  {}
\hbox to 50pt{\hfill}
\hfill}\hfill}
\vspace{-4pt}
\demitrait
\vspace{4pt}
\ligne{\hfill\tt
-I- The particle comes from~$Y$
\hfill}
\demitrait
\vspace{4pt}
\ligne{\hfill
\hbox to 288pt{%
\demirangeea W B B W W B B   \demirangeeb W W B B W B W W
\hbox to 50pt{\hfill 66\hskip 20pt}
\hfill}\hfill}
\ligne{\hfill
\hbox to 288pt{%
\demirangeea W B B W W B B   \demirangeeb W W B W W B W W
\hbox to 50pt{\hfill 72\hskip 20pt}
\hfill}\hfill}
\ligne{\hfill
\hbox to 288pt{%
\demirangeea W B B W W B B   \demirangeeb W W W B W B W W
\hbox to 50pt{\hfill 70\hskip 20pt}
\hfill}\hfill}
\demitrait
\vspace{4pt}
\ligne{\hfill\tt
-II- The particle comes from~$X$
\hfill}
\demitrait
\vspace{4pt}
\ligne{\hfill
\hbox to 288pt{%
\demirangeea W B B W W B B   \demirangeeb W W W B W B W W
\hbox to 50pt{\hfill 70\hskip 20pt}
\hfill}\hfill}
\ligne{\hfill
\hbox to 288pt{%
\demirangeea W B B W W B B   \demirangeeb W W W W W B W W
\hbox to 50pt{\hfill 77\hskip 20pt}
\hfill}\hfill}
\ligne{\hfill
\hbox to 288pt{%
\demirangeea W B B W W B B   \demirangeeb W W B B W B W W
\hbox to 50pt{\hfill 66\hskip 20pt}
\hfill}\hfill}
\demitrait\vspace{15pt}
}

   These new rules~69 and~77 witness the flash of~$D$ in a situation which cannot
occur during the crossing of a flip-flop by the particle. Indeed, when the flash
of~$D$ occurs in the flip-flop when $H$~is white, this means that the particle
goes through~$B$, the particle just left~$B$ and is visible from~$B$ as it is on the
next cell of the tracks, see Table~\ref{ffB_tabB} at a moment where rule~64 applies.
We have a symmetric situation for~$C$ when $K$~is white and rule~73.
Here, the next cell of the track is always empty at this moment, this is why
the new rules~69 and~77 are needed.

\subsection{Traces of execution}
\label{traces}

   In this subsection, we complete the study of the rules with traces of execution 
by the program of the rules in all the configurations we have illustrated in
Section~\ref{implement}.

   Such traces could be obtained from the above tables. It is easier to tune the
computer program in order to execute this task too. This what we do now, following the
same plan as in Section~\ref{rules}. 

\subsubsection{Crossings}

   Table~\ref{tr_cross} gives the trace of execution performed by the computer
program for the crossing.

   For displaying reasons, we could not keep the names $B.11$, $B.12$ and $C.5$
of the sensors controlling the correct execution of the branching at a round-about.
We have used other names: $BF$, $BC$ and~$CE$ respectively. Clearly, $BF$
can see both~$B$ and~$F$, $BC$ can see both~$B$ and~$C$ and $CE$ can see both~$C$
and~$E$.

\vtop{
\begin{tab}\label{tr_cross}   
\leurre
Trace of execution for the crossings.
\end{tab}
\vspace{-12pt}
\grostrait
\vspace{2pt}
\setbox110=\vtop{\hsize=220pt
\obeylines
\obeyspaces\global\let =\ \tt
      A   B   C   D   E   F  F1  BF  BC  CE
}
\setbox112=\vtop{\hsize=220pt
\obeylines
\obeyspaces\global\let =\ \tt
  1   W   W   W   W   W   W   W   B   B   B   
  2   W   W   W   W   W   W   W   B   B   B   
  3   W   W   W   W   B   W   W   B   B   B   
  4   W   W   B   W   W   W   W   B   B   W   
  5   W   W   B   B   W   W   W   B   B   B   
  6   B   W   W   B   W   W   W   B   B   B   
  7   B   B   W   W   W   W   W   B   B   B   
  8   W   B   W   W   W   B   W   B   B   B   
  9   W   W   W   W   W   B   B   W   B   B   
 10   W   W   W   W   W   W   W   B   W   B   
 11   W   W   B   W   W   W   W   W   B   B   
 12   W   W   W   B   W   W   W   B   B   B   
 13   B   W   W   W   W   W   W   B   B   B   
 14   W   B   W   W   W   W   W   B   B   B   
 15   W   W   W   W   W   B   W   B   B   B   
 16   W   W   W   W   W   W   B   B   B   B   
 17   W   W   W   W   W   W   W   B   B   B   
 18   W   W   W   W   W   W   W   B   B   B   
}
\ligne{\hskip 40pt\box110\hfill}
\vspace{-2pt}
\demitrait
\vspace{6pt}
\ligne{\hskip 40pt\box112\hfill}
\vspace{10pt}
\demitrait
\vspace{15pt}
}

   Table~\ref{tr_cross} shows another particularity performed by the computer
program. The table represents all the possible cases at a branching. Instead of
simulating three distinct branchings placed around the round-about, the program
used the same branching executing the following scenario. First, the particle
arrives from outside to~$E$: time~3{} in the table. Then, it goes to~$C$ and
then to~$D$. Now, in the program, the next neighbour of~$D$ on the tracks is~$A$.
As an additional particle was created in~$C$, when the first particle arrives at~$A$
of the same branching, the situation is exactly the same as if it arrived at the 
next one. And so, when both particles are erased at time~10, see the table,
a new one is created in~$C$ at time~11. Now, this particle travels to~$D$ and then
again to~$A$, again at the same branching. But the situation is exactly the same as
if it would arrive at the next one. And so, at the second arrival from~$A$, we can see
that the single particle goes to~$B$, then to~$F$ and so on, without being destroyed.

\subsubsection{Fixed switch}

   Table~\ref{tr_fix} gives two executions for the fixed switch. In one of them,
the particle comes from~$B$, in the other, it comes from~$C$. As can be seen,
in both cases, it leaves the switch through~$A$.

\vtop{
\begin{tab}\label{tr_fix}   
\leurre
Traces of execution for the fixed switch. Left-hand side: the particle comes
from~$B$; right-hand side: it comes from~$C$.
\end{tab}
\vspace{-12pt}
\grostrait
\vspace{2pt}
\setbox110=\vtop{\hsize=160pt
\obeylines
\obeyspaces\global\let =\ \tt
     aA   A   O   B  bB   C  bC
}
\setbox112=\vtop{\hsize=160pt
\obeylines
\obeyspaces\global\let =\ \tt
  1   W   W   W   W   W   W   W   
  2   W   W   W   W   B   W   W   
  3   W   W   W   B   W   W   W   
  4   W   W   B   W   W   W   W   
  5   W   B   W   W   W   W   W   
  6   B   W   W   W   W   W   W   
  7   W   W   W   W   W   W   W   
}
\setbox114=\vtop{\hsize=160pt
\obeylines
\obeyspaces\global\let =\ \tt
  1   W   W   W   W   W   W   W   
  2   W   W   W   W   W   W   B   
  3   W   W   W   W   W   B   W   
  4   W   W   B   W   W   W   W   
  5   W   B   W   W   W   W   W   
  6   B   W   W   W   W   W   W   
  7   W   W   W   W   W   W   W   
}
\ligne{\hskip-20pt\copy110\hskip 20pt\box110\hfill}
\vspace{-2pt}
\demitrait
\vspace{6pt}
\ligne{\hskip-20pt\box112\hskip 20pt\box114\hfill}
\vspace{10pt}
\demitrait
\vspace{15pt}
}

\subsubsection{Flip-flop}

   Table~\ref{tr_basc} gives two executions for the flip-flop. In the first one,
the selected tracks is~$C$, in the second one, it is~$B$. 

\vtop{
\begin{tab}\label{tr_basc}   
\leurre
Traces of execution for the flip-flop. Above: to~$C$; below: to~$B$.
\end{tab}
\vspace{-12pt}
\grostrait
\vspace{2pt}
\setbox110=\vtop{\hsize=220pt
\obeylines
\obeyspaces\global\let =\ \tt
     bA   A   O   B  aB   C  aC   D   H   K
}
\setbox112=\vtop{\hsize=220pt
\obeylines
\obeyspaces\global\let =\ \tt
  1   W   W   W   W   W   W   W   B   B   W   
  2   W   W   W   W   W   W   W   B   B   W   
  3   B   W   W   W   W   W   W   B   B   W   
  4   W   B   W   W   W   W   W   B   B   W   
  5   W   W   B   W   W   W   W   B   B   W   
  6   W   W   W   W   W   B   W   B   B   W   
  7   W   W   W   W   W   W   B   W   B   W   
  8   W   W   W   W   W   W   W   B   W   B   
}
\setbox114=\vtop{\hsize=220pt
\obeylines
\obeyspaces\global\let =\ \tt
  1   W   W   W   W   W   W   W   B   W   B   
  2   W   W   W   W   W   W   W   B   W   B   
  3   B   W   W   W   W   W   W   B   W   B   
  4   W   B   W   W   W   W   W   B   W   B   
  5   W   W   B   W   W   W   W   B   W   B   
  6   W   W   W   B   W   W   W   B   W   B   
  7   W   W   W   W   B   W   W   W   W   B   
  8   W   W   W   W   W   W   W   B   B   W   
}
\ligne{\hskip 40pt\box110\hfill}
\vspace{-2pt}
\demitrait
\vspace{6pt}
\ligne{\hskip 40pt\box112\hfill}
\vspace{10pt}
\demitrait
\vspace{6pt}
\ligne{\hskip 40pt\box114\hfill}
\vspace{10pt}
\demitrait
\vspace{15pt}
}

   In both cases, we can notice that $H$ and~$K$ both change their states, at
time~8 exactly.

\subsubsection{Active memory switch}

   Table~\ref{tr_mma} gives the traces of execution for the active memory
switch. As mentioned in Section~\ref{implement}, the switch looks like
a {\bf passive programmable} flip-flop. At this stage, the flip-flop
mechanism exists but it is frozen as $D$~is not the same cell as in the flip-flop.

\vtop{
\begin{tab}\label{tr_mma}   
\leurre
Traces of execution for the active memory switch. Above: to~$C$; below: to~$B$.
\end{tab}
\vspace{-12pt}
\grostrait
\vspace{2pt}
\setbox110=\vtop{\hsize=220pt
\obeylines
\obeyspaces\global\let =\ \tt
     bA   A   O   B  aB   C  aC   D   H   K
}
\setbox112=\vtop{\hsize=220pt
\obeylines
\obeyspaces\global\let =\ \tt
  1   W   W   W   W   W   W   W   B   B   W   
  2   W   W   W   W   W   W   W   B   B   W   
  3   B   W   W   W   W   W   W   B   B   W   
  4   W   B   W   W   W   W   W   B   B   W   
  5   W   W   B   W   W   W   W   B   B   W   
  6   W   W   W   W   W   B   W   B   B   W   
  7   W   W   W   W   W   W   B   B   B   W   
  8   W   W   W   W   W   W   W   B   B   W   
}
\setbox114=\vtop{\hsize=220pt
\obeylines
\obeyspaces\global\let =\ \tt
  1   W   W   W   W   W   W   W   B   W   B   
  2   W   W   W   W   W   W   W   B   W   B   
  3   B   W   W   W   W   W   W   B   W   B   
  4   W   B   W   W   W   W   W   B   W   B   
  5   W   W   B   W   W   W   W   B   W   B   
  6   W   W   W   B   W   W   W   B   W   B   
  7   W   W   W   W   B   W   W   B   W   B   
  8   W   W   W   W   W   W   W   B   W   B   
}
\ligne{\hskip 40pt\box110\hfill}
\vspace{-2pt}
\demitrait
\vspace{6pt}
\ligne{\hskip 40pt\box112\hfill}
\vspace{10pt}
\demitrait
\vspace{6pt}
\ligne{\hskip 40pt\box114\hfill}
\vspace{10pt}
\demitrait
\vspace{15pt}
}

   We notice that, in this execution, $H$ and~$K$ are not affected by the
passage of the particle, which is not the case in the flip-flop.

   In the next subsubsection, devoted to the passive memory switch we shall see the
action on the flip-flop mechanism of the active memory switch.

\subsubsection{Passive memory switch}

   In the computer program, we simplified the execution of the situation when the
particle arrives from the non selected tracks. We know that the flash of~$Z$,
the controller of the passive switch, triggers a signal to~$D$, the controller
of the active switch. This signals is performed as a particle which follows a
path which looks like the usual tracks exactly. We call here locomotive the
particle which runs over the circuit to simulate a computation. As we can see in 
Fig.~\ref{idle_configs}, the path of the messenger particle crosses two other 
paths followed by the locomotive particle. And so there are two crossings. As we have 
previously checked that the program correctly performs the crossings, it is
useless to check it again. And so, we connected $Z$ and~$D$ directly by a 
small path of five cells including~$Z1$ and~$D1$. One of these cells, $M$, is 
represented in Tables~\ref{tr_mmpX} and~\ref{tr_mmpY}.

\vtop{
\begin{tab}\label{tr_mmpX}   
\leurre
Traces of execution for the passive memory switch when the particle comes 
from~$X$. Above: $X$~was not selected; below: $X$~is selected. 
\end{tab}
\vspace{-12pt}
\grostrait
\vspace{2pt}
\setbox110=\vtop{\hsize=300pt
\obeylines
\obeyspaces\global\let =\ \tt
     aV  V  O  X bX  Y bY  Z  I  J  D  H  K Z1  M D1 
}
\setbox112=\vtop{\hsize=300pt
\obeylines
\obeyspaces\global\let =\ \tt
  1   W  W  W  W  W  W  W  B  B  W  B  B  W  W  W  W  
  2   W  W  W  W  W  W  W  B  B  W  B  B  W  W  W  W  
  3   W  W  W  W  B  W  W  B  B  W  B  B  W  W  W  W  
  4   W  W  W  B  W  W  W  B  B  W  B  B  W  W  W  W  
  5   W  W  B  W  W  W  W  W  B  W  B  B  W  W  W  W  
  6   W  B  W  W  W  W  W  B  W  B  B  B  W  B  W  W  
  7   B  W  W  W  W  W  W  B  W  B  B  B  W  W  W  W  
  8   W  W  W  W  W  W  W  B  W  B  B  B  W  W  B  W  
  9   W  W  W  W  W  W  W  B  W  B  B  B  W  W  W  W  
 10   W  W  W  W  W  W  W  B  W  B  B  B  W  W  W  B  
 11   W  W  W  W  W  W  W  B  W  B  W  B  W  W  W  W  
 12   W  W  W  W  W  W  W  B  W  B  B  W  B  W  W  W  
 13   W  W  W  W  W  W  W  B  W  B  B  W  B  W  W  W  
 14   W  W  W  W  W  W  W  B  W  B  B  W  B  W  W  W  
 15   W  W  W  W  W  W  W  B  W  B  B  W  B  W  W  W  
}
\setbox114=\vtop{\hsize=300pt
\obeylines
\obeyspaces\global\let =\ \tt
  1   W  W  W  W  W  W  W  B  W  B  B  W  B  W  W  W  
  2   W  W  W  W  W  W  W  B  W  B  B  W  B  W  W  W  
  3   W  W  W  W  B  W  W  B  W  B  B  W  B  W  W  W  
  4   W  W  W  B  W  W  W  B  W  B  B  W  B  W  W  W  
  5   W  W  B  W  W  W  W  B  W  B  B  W  B  W  W  W  
  6   W  B  W  W  W  W  W  B  W  B  B  W  B  W  W  W  
  7   B  W  W  W  W  W  W  B  W  B  B  W  B  W  W  W  
  8   W  W  W  W  W  W  W  B  W  B  B  W  B  W  W  W  
}
\ligne{\hskip 20pt\box110\hfill}
\vspace{-2pt}
\demitrait
\vspace{6pt}
\ligne{\hskip 20pt\box112\hfill}
\vspace{10pt}
\demitrait
\vspace{6pt}
\ligne{\hskip 20pt\box114\hfill}
\vspace{10pt}
\demitrait
\vspace{15pt}
}

   In both tables, we can notice that when the particle comes from the non-selected
tracks, the message of~$Z$ to~$D$ is sent at time~6 as $Z$~flashes at time~5,
the locomotive particle being in~$X$ or~$Y$ at time~4. The messenger particle 
arrives in~$D1$ at time~10. This corresponds to the running over the path of five 
cells between~$Z$ and~$D$. Accordingly, $D$~is able to flash at time~11. And so, 
at time~12, $H$ and~$K$ have both changed their states. Accordingly, the passive 
and the active memory switch now indicate the same selected tracks.

   The real length of the path between~$Z$ and~$D$ is much longer. Looking at 
Fig.~\ref{idle_configs}, consider that
the central tile of the tiling is placed at the central cell of the round-about~$R_0$
which is the closest to both the active and the passive switch. We can imagine
at least three cells between the neighbour of~$X$ on the tracks and the cell~$E$ of 
the branching to~$R_0$. Consider $R_p$, the round-about closer to the passive 
memory switch. We can imagine that its central cell is at three cells from the
central cell of~$R_0$. The path from~$Z1$ can go around the immediate neighbours 
of~$Z$ towards~$X$. Then, it goes around the immediate neighbours of~$X$ and, at a 
few cells from the tracks which leaves~$X$. We may consider that the cells of the
path are on the third level of the tree rooted at the central cell of~$R_0$. 
Now, the number of cells on the third level of the tree rooted at the central cell 
of~$R_0$ is 711. Now, the path between $R_p$ and $R_a$, the round-about closer
to the active switch is still longer. This means that the simulation could not 
consider the real path and that the artifact we presented is a good solution
for the simulation.

\vtop{
\begin{tab}\label{tr_mmpY}   
\leurre
Traces of execution for the passive memory switch when the particle comes 
from~$Y$. Above: $Y$~was not selected; below: $Y$~is selected. 
\end{tab}
\vspace{-12pt}
\grostrait
\vspace{2pt}
\setbox110=\vtop{\hsize=300pt
\obeylines
\obeyspaces\global\let =\ \tt
     aV  V  O  X bX  Y bY  Z  I  J  D  H  K Z1  M D1 
}
\setbox112=\vtop{\hsize=300pt
\obeylines
\obeyspaces\global\let =\ \tt
  1   W  W  W  W  W  W  W  B  W  B  B  W  B  W  W  W  
  2   W  W  W  W  W  W  B  B  W  B  B  W  B  W  W  W  
  3   W  W  W  W  W  B  W  B  W  B  B  W  B  W  W  W  
  4   W  W  B  W  W  W  W  W  W  B  B  W  B  W  W  W  
  5   W  B  W  W  W  W  W  B  B  W  B  W  B  B  W  W  
  6   B  W  W  W  W  W  W  B  B  W  B  W  B  W  W  W  
  7   W  W  W  W  W  W  W  B  B  W  B  W  B  W  B  W  
  8   W  W  W  W  W  W  W  B  B  W  B  W  B  W  W  W  
  9   W  W  W  W  W  W  W  B  B  W  B  W  B  W  W  B  
 10   W  W  W  W  W  W  W  B  B  W  W  W  B  W  W  W  
 11   W  W  W  W  W  W  W  B  B  W  B  B  W  W  W  W  
 12   W  W  W  W  W  W  W  B  B  W  B  B  W  W  W  W  
 13   W  W  W  W  W  W  W  B  B  W  B  B  W  W  W  W  
 14   W  W  W  W  W  W  W  B  B  W  B  B  W  W  W  W  
 15   W  W  W  W  W  W  W  B  B  W  B  B  W  W  W  W  
}
\setbox114=\vtop{\hsize=300pt
\obeylines
\obeyspaces\global\let =\ \tt
  1   W  W  W  W  W  W  W  B  B  W  B  B  W  W  W  W  
  2   W  W  W  W  W  W  B  B  B  W  B  B  W  W  W  W  
  3   W  W  W  W  W  B  W  B  B  W  B  B  W  W  W  W  
  4   W  W  B  W  W  W  W  B  B  W  B  B  W  W  W  W  
  5   W  B  W  W  W  W  W  B  B  W  B  B  W  W  W  W  
  6   B  W  W  W  W  W  W  B  B  W  B  B  W  W  W  W  
  7   W  W  W  W  W  W  W  B  B  W  B  B  W  W  W  W  
  8   W  W  W  W  W  W  W  B  B  W  B  B  W  W  W  W  
}
\ligne{\hskip 20pt\box110\hfill}
\vspace{-2pt}
\demitrait
\vspace{6pt}
\ligne{\hskip 20pt\box112\hfill}
\vspace{10pt}
\demitrait
\vspace{6pt}
\ligne{\hskip 20pt\box114\hfill}
\vspace{10pt}
\demitrait
\vspace{15pt}
}

\section{The rules for $\{p,3\}$, $p\geq 17$}
\label{rulesgene}

   As we completed the study of the case $=13$, we can now turn to the study
of the general case of~ $\{p,3\}$, when $p\geq 17$.

   We shall follow the scenario described in Subsection~\ref{scenario}. There will
be some changes as certain cells which were black in the case of $\{13,3\}$
are now white and as for them, flashing means turning from white to black and 
then return to white at the next step.

   We shall follow the same plan as in the case of $\{13,3\}$. We start by 
defining the patterns and the rules for the tracks. From that we can define
the patterns and rules for the crossings and then for the switches: fixed, 
flip-flop and memory switch in this order.

   Before turning to the detailed study of each case, we briefly indicate 
the general idea. Define the {\bf context} of a cell as the pattern defined by the 
state of its neighbours as a word on $\{\tt B,W\}$, up to circular permutations:
considering the various rotated forms, we shall always take the smallest one
in the order where \hbox{\tt B $<$ W}. We shall 
devise the patterns in the following conditions: in the patterns of the cells
of the tracks, the context contains at most one block of two or three contiguous black
states and no block with more contiguous black states. For the central cells of 
the switches, cells~$O$ or~$T$ in Section~\ref{implement}, there are exactly two 
blocks of three contiguous black cells and they do not contain blocks with more
contiguous black states. There may be an additional black of two cells. The other 
cells have at least one block of four contiguous black states. This block of four 
black cells will be called the {\bf anchor}. With the orientation which we may assume
to be known by the cell, this gives a direction: there will be
a first cell of the block while turning around the cell. This will allow us to use
the cells which are after the four one of the block to encode a number which
defines which kind of cell we have. As we have twenty kinds of cells outside
those of the tracks and the central cells of the switches, we need five cells
for encoding their number: we shall speak of the $i^{\rm th}$ cell of the encoding. 
As not all possible codes from $\{O,1\}^5$ are needed, we can avoid any other block 
of four contiguous black states. It will be possible to have one block of five 
contiguous cells: the four ones of the anchor and the fifth one being the first cell 
of the encoding. The first two cells of the encoding are~{\tt W} for crossings,
they are {\tt BW} in this order for the memory switch and {\tt WB} in this
order for the flip-flop. This block of cells defined by this just described
encoding is the {\bf type} of the cell. As outside~$O$ all the cells of a fixed 
switch are cells of the track, the encoding is {\tt W$^5$} for the fixed switch. i
This requires
that the last three cells of the encoding contain at least one black cell in the
case of crossings. This constraint can be observed.  

   In what follows, the format of the rules will follow the general form given
in Section~\ref{hypCA} which we remind here for the convenience of the reader:

\vskip 3pt
\ligne{\hfill 
$\eta_0$, $\eta_1$, $\ldots$, $\eta_{p} \rightarrow \eta_0^1$,\hfill}
\vskip 2pt
\noindent 
where $\eta_0$ is the state of the cell,
$\eta_i$ the state of its neighbour~$i$ and~$\eta_0^1$ is its new state. 
We assume that the rule is given in such a way that
\hbox{$\eta_1$, $\ldots$, $\eta_{p}$} is the minimal form among the rotated
forms of the rule obtained by circular permutation on $[1..p]$.

   We shall systematically use the following format given as a word 
in $\{\tt B,W\}^*$:

\vskip 3pt
\ligne{\hfill 
$\underline{\eta_0}$$\eta_1$$\ldots$$\eta_{p}\underline{\eta_0^1}$.\hfill}
\vskip 2pt
\noindent
Clearly, the context of the cell is \hbox{$\eta_1$$\ldots$$\eta_{p}$}.

\subsection{Patterns and rules for the tracks}
\label{trackpattrules}

   We can now define the configurations and the rules for the tracks.
The general constraints which we have defined force us to change the contexts of
the cells of the tracks.

   The context of an ordinary cell of the track will be {\tt BWWBWWWBW$^k$}
or {\tt BWWBW$^k$BWWW} with $k=p$$-$8. Note that these patterns cannot be deduced
from each other by any circular permutation. The patterns are illustrated by the first
row of Fig.~\ref{trackgene} for the case when $p=17$. The yellow cells are those
through which the particle may enter the cell. The orange ones are those through
which the particle exits from the cell. This possible choice gives more flexibility
for the construction of the tracks near the switches.

\vtop{
\ligne{\hskip 20pt
\includegraphics[scale=0.70]{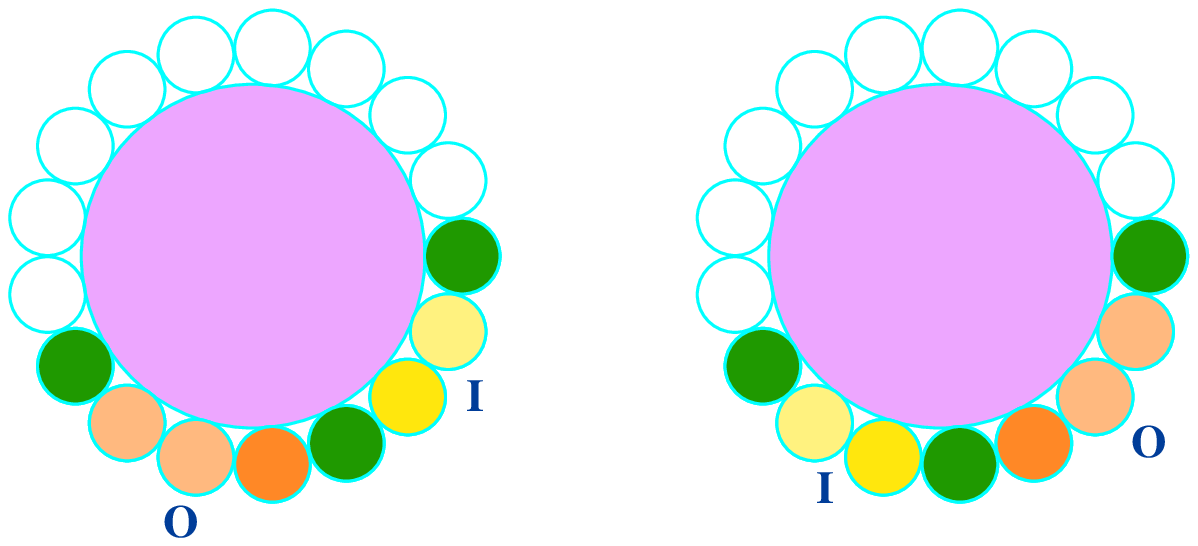}
\hfill}
\vspace{-15pt}
\begin{fig}\label{trackgene}
\leurre
The patterns for the cells of the tracks. Here, the illustration
when \hbox{$p=17$}. The patterns for ordinary cells
are \hbox{\tt BWWBW$^9$BWWW}, to left, and \hbox{\tt BWWBWWWBW$^9$},
to right. In more intense colour: the favorite entrance and exit. 
\end{fig}
}

\vtop{
\ligne{\hskip 20pt
\includegraphics[scale=0.70]{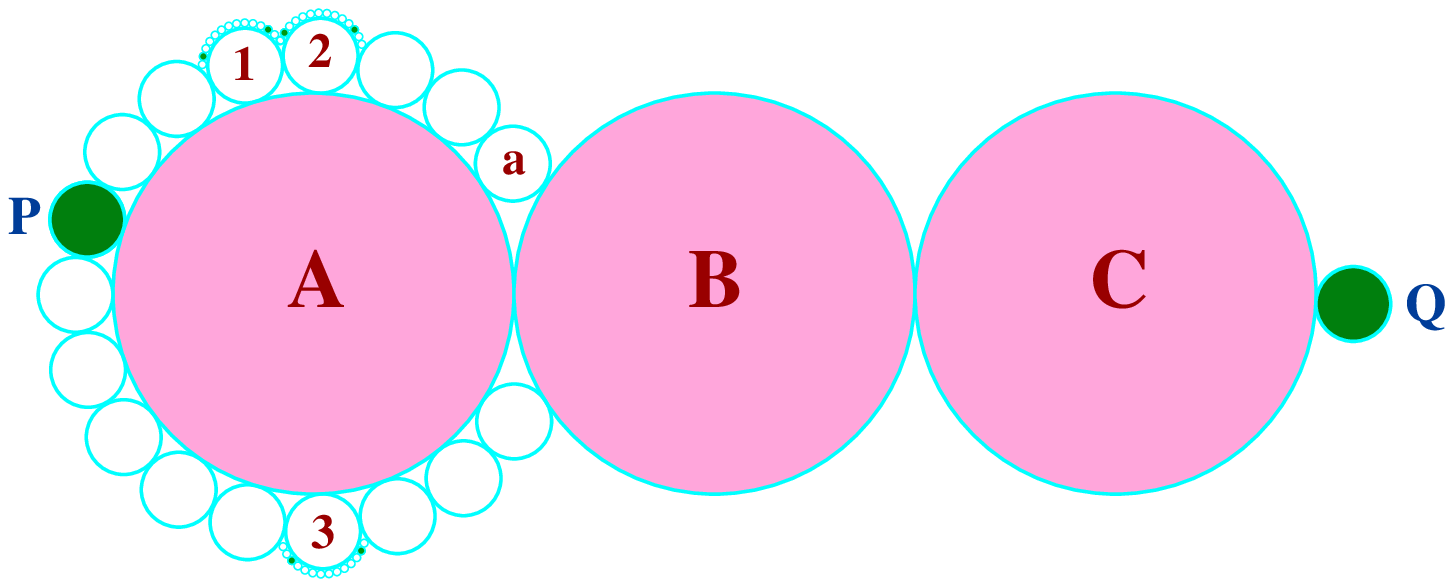}
\hfill}
\vspace{-15pt}
\begin{fig}\label{voiegene_1}
\leurre
Sketch of the paths from~$P$ to~$Q$. The shortest path is given by the cells~$A$,
$B$ and~$C$. We have given a part of the path around cell~$A$. We give the context
only for cells~$1$ and~$2$ in one direction, for cell~$3$ in the other one.
Note the cell~$a$ which has two consecutive black neighbours.
\end{fig}
}

   As in Section~\ref{tracks}, we can go from one cell~$P$ to any other one~$Q$ by 
first fixing a shortest path~$\pi$ from~$P$ to~$Q$ and then by going along the upper
neighbours of~$\pi$ in the direction form~$P$ to~$Q$ and by going along the lower
neighbours of~$\pi$ in the direction from~$Q$ to~$P$. Fig.~\ref{voiegene_1} gives
an attempt of illustration for such a path. We can see that even when $p=17$
drawings become almost useless as details are very difficult to see: the neighbours
of the cell~$A$ can be seen but many neighbours of cell~$1$ or~$2$ are hardly
visible, especially to distinguish clearly those which are black.

   The figure allows us to notice that here two, we have the phenomenon already
noticed in Section~\ref{tracks}: a cell which belongs to the tracks constructed
along a shortest path may have two contiguous black neighbours. We can see that 
the cell~$a$ in Fig.~\ref{voiegene_1} belongs to this case. We say that
such a cell is a {\bf corner}.
For this situation, we need to adapt the previous neighbourhoods. This new contexts
are illustrated by Fig.~\ref{trackcorners} and the corresponding contexts
are: \hbox{\tt BBWWWBW$^k$BWW} in one direction and \hbox{\tt BBWWBW$^k$BWWW} 
in the other. In both cases, \hbox{$k=p$$-$9}.

\vtop{
\ligne{\hskip 20pt
\includegraphics[scale=0.70]{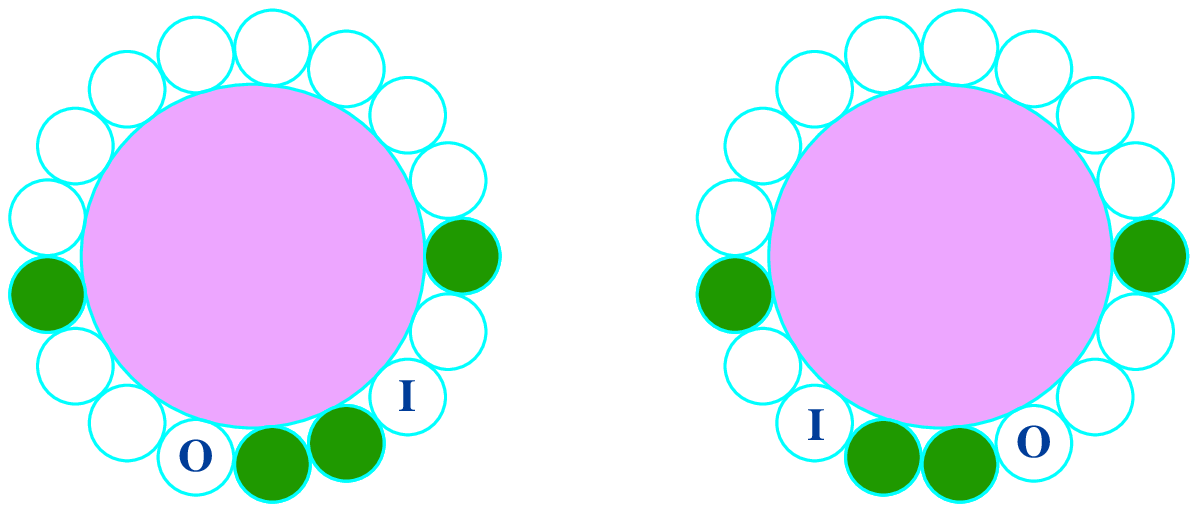}
\hfill}
\vspace{-15pt}
\begin{fig}\label{trackcorners}
\leurre
The patterns for the cells of the tracks having two contiguous black cells
as 'basic' neighbours. Here, the illustration
when \hbox{$p=17$}. The patterns for these cells are 
\hbox{\tt BBWWBW$^8$BWWW}, to left,
and \hbox{\tt BBWWWBW$^8$BWW}, to right.
\end{fig}
}

   Fig.~\ref{voiegene_1} also shows us that in consecutive ordinary cells, the
position of the entrance and the exit of the particle during the motion
are around the black cell which is in between the two others, here the black
cell around which the path is constructed. This position of the entrance and exit
is the more used, it is called {\it favorite} in the caption of Fig.~\ref{trackgene}.
Other positions for the entrance and exit are rare and used only to tune the
approach or leaving of the particle around a switch or near a round-about.

   The rules for the motion of the particle are easy. For ordinary cells
they are:

\vskip 7pt
\ligne{\hfill
$\vcenter{\vtop{\hsize=325pt
\ligne{\hfill
\hbox to 63pt{\tt $\underline{\tt W}$BWWBWWWBW$^k$$\underline{\tt W}$\hfill},
\hfill
\hbox to 63pt{\tt $\underline{\tt W}$BWBBWWWBW$^k$$\underline{\tt B}$\hfill},
\hfill
\hbox to 63pt{\tt $\underline{\tt B}$BWWBWWWBW$^k$$\underline{\tt W}$\hfill},
\hfill
\hbox to 63pt{\tt $\underline{\tt W}$BWWBBWWBW$^k$$\underline{\tt W}$\hfill},
\hfill}
\ligne{\hfill
\hbox to 63pt{\tt $\underline{\tt W}$BBWBWWWBW$^k$$\underline{\tt B}$\hfill},
\hskip 10pt
\hbox to 63pt{\tt $\underline{\tt W}$BBWWBWWBW$^k$$\underline{\tt W}$\hfill},
\hskip 10pt
\hbox to 63pt{\tt $\underline{\tt W}$BWBWBWWBW$^k$$\underline{\tt W}$\hfill},
\hfill}
}}$
\hfill$(a)$}
\vskip 5pt
\noindent
and, for the opposite direction:

\vskip 7pt
\ligne{\hfill
$\vcenter{\vtop{\hsize=325pt
\ligne{\hfill
\hbox to 63pt{\tt $\underline{\tt W}$BWWBW$^k$BWWW$\underline{\tt W}$\hfill},
\hfill
\hbox to 63pt{\tt $\underline{\tt W}$BBWBW$^k$BWWW$\underline{\tt B}$\hfill},
\hfill
\hbox to 63pt{\tt $\underline{\tt B}$BWWBW$^k$BWWW$\underline{\tt W}$\hfill},
\hfill
\hbox to 63pt{\tt $\underline{\tt W}$BWWBW$^k$BWWB$\underline{\tt W}$\hfill.}
\hfill}
\ligne{\hfill
\hbox to 63pt{\tt $\underline{\tt W}$BWBBW$^k$BWWW$\underline{\tt B}$\hfill},
\hskip 10pt
\hbox to 63pt{\tt $\underline{\tt W}$BWWBW$^k$BBWW$\underline{\tt W}$\hfill.}
\hskip 10pt
\hbox to 63pt{\tt $\underline{\tt W}$BWWBW$^k$BWBW$\underline{\tt W}$\hfill.}
\hfill}
}}$
\hfill$(b)$}
\vskip 5pt
In both cases, we have that \hbox{$k=p$-$8$}.
Note that here, the rules are not exactly conforming to the requirement of minimality
of context in the writing of the rules. The rule 
\hbox{\tt$\underline{\tt W}$BWBBWWWBW$^k$$\underline{\tt B}$} is not written
in its minimal form. It should be written:
\hbox{\tt$\underline{\tt W}$BBWWWBW$^k$BW$\underline{\tt B}$}. Accordingly,
the conforming presentation of the rules is: 

\vskip 7pt
\ligne{\hfill
$\vcenter{\vtop{\hsize=325pt
\ligne{\hfill
\hbox to 63pt{\tt $\underline{\tt W}$BWWBWWWBW$^k$$\underline{\tt W}$\hfill},
\hfill
\hbox to 63pt{\tt $\underline{\tt W}$BBWWWBW$^k$BW$\underline{\tt B}$\hfill},
\hfill
\hbox to 63pt{\tt $\underline{\tt B}$BWWBWWWBW$^k$$\underline{\tt W}$\hfill},
\hfill
\hbox to 63pt{\tt $\underline{\tt W}$BBWWBW$^k$BWW$\underline{\tt W}$\hfill},
\hfill}
\ligne{\hfill
\hbox to 63pt{\tt $\underline{\tt W}$BBWBWWWBW$^k$$\underline{\tt B}$\hfill},
\hskip 10pt
\hbox to 63pt{\tt $\underline{\tt W}$BBW$^k$BWWBWW$\underline{\tt W}$\hfill},
\hskip 10pt
\hbox to 63pt{\tt $\underline{\tt W}$BWBWBW$^k$BWW$\underline{\tt W}$.\hfill}
\hfill}
}}$
\hfill$(c)$}
\vskip 5pt
\ligne{\hfill
$\vcenter{\vtop{\hsize=325pt
\ligne{\hfill
\hbox to 63pt{\tt $\underline{\tt W}$BWWBW$^k$BWWW$\underline{\tt W}$\hfill},
\hfill
\hbox to 63pt{\tt $\underline{\tt W}$BBWBW$^k$BWWW$\underline{\tt B}$\hfill},
\hfill
\hbox to 63pt{\tt $\underline{\tt B}$BWWBW$^k$BWWW$\underline{\tt W}$\hfill},
\hfill
\hbox to 63pt{\tt $\underline{\tt W}$BBWWBW$^k$BWW$\underline{\tt W}$\hfill},
\hfill}
\ligne{\hfill
\hbox to 63pt{\tt $\underline{\tt W}$BBW$^k$BWWWBW$\underline{\tt B}$\hfill},
\hskip 10pt
\hbox to 63pt{\tt $\underline{\tt W}$BBWWBWWBW$^k$$\underline{\tt W}$\hfill},
\hskip 10pt
\hbox to 63pt{\tt $\underline{\tt W}$BWBWBWWBW$^k$$\underline{\tt W}$.\hfill}
\hfill}
}}$
\hfill$(d)$}
\vskip 5pt

   We can note that in lines~$(a)$ and~$(b)$, the fourth rules have the same minimal
form as the fourth rules in lines~$(c)$ and~$(d)$. The reason is that the 
corresponding contexts are symmetric.

   Similarly, for the corners we get the following rules, this time in minimal form:

\vskip 7pt
\ligne{\hfill
\hbox to 63pt{\tt $\underline{\tt W}$BBWWWBW$^k$BWW$\underline{\tt W}$\hfill},
\hfill
\hbox to 63pt{\tt $\underline{\tt W}$BBBWWWBW$^k$BW$\underline{\tt B}$\hfill},
\hfill
\hbox to 63pt{\tt $\underline{\tt B}$BBWWWBW$^k$BWW$\underline{\tt W}$\hfill},
\hfill
\hbox to 63pt{\tt $\underline{\tt W}$BBBWWBW$^k$BWW$\underline{\tt W}$\hfill},
\hfill$(e)$}
\vskip 7pt
\ligne{\hfill
\hbox to 63pt{\tt $\underline{\tt W}$BBWWBW$^k$BWWW$\underline{\tt W}$\hfill},
\hfill
\hbox to 63pt{\tt $\underline{\tt W}$BBBWBW$^k$BWWW$\underline{\tt B}$\hfill},
\hfill
\hbox to 63pt{\tt $\underline{\tt B}$BBWWBW$^k$BWWW$\underline{\tt W}$\hfill},
\hfill
\hbox to 63pt{\tt $\underline{\tt W}$BBBWWBW$^k$BWW$\underline{\tt W}$\hfill.}
\hfill$(f)$}
\vskip 5pt

\noindent
and, here, we have \hbox{$k=p$-$9$}. We have a similar remark as previously: the
last two rules in lines~$(e)$ and~$(f)$ have the same minimal form due to a symmetric
neighbourhood.

   We conclude this subsection with an important remark. We have seen that, in the
crossings, two consecutive particles may travel on the round-about. This requires
new rules,namely the following ones:

For ordinary cells:

\vskip 7pt
\ligne{\hfill
\hbox to 63pt{\tt $\underline{\tt B}$BBWWWBW$^k$BW$\underline{\tt B}$\hfill},
\hfill 
\hbox to 63pt{\tt $\underline{\tt B}$BBWWBW$^k$BWW$\underline{\tt W}$\hfill},
\hfill 
\hbox to 63pt{\tt $\underline{\tt B}$BBWBW$^k$BWWW$\underline{\tt B}$\hfill},
\hfill 
\hbox to 63pt{\tt $\underline{\tt B}$BBWWBW$^k$BWW$\underline{\tt W}$\hfill},
\hfill$(g)$
}
\vskip 5pt

\noindent
and for corners:

\vskip 7pt
\ligne{\hfill
\hbox to 63pt{\tt $\underline{\tt B}$BBBWWWBW$^k$BW$\underline{\tt B}$\hfill},
\hfill
\hbox to 63pt{\tt $\underline{\tt B}$BBBWWBW$^k$BWW$\underline{\tt W}$\hfill},
\hfill
\hbox to 63pt{\tt $\underline{\tt B}$BBBWBW$^k$BWWW$\underline{\tt B}$\hfill},
\hfill
\hbox to 63pt{\tt $\underline{\tt B}$BBBWWBW$^k$BWW$\underline{\tt W}$.\hfill}
\hfill$(h)$
}
\vskip 5pt
  We remark that these new rules have the same contexts than several rules of
lines~$(c)$, $(d)$, $(e)$ and~$(f)$. However, in these cases, the new state is 
always the same. So that the new rules are compatible with the others.

   We also remark that in these new rules we assume the favorite entrance and exit
in the cells: this will be the case in the round-about as we shall soon see.

\subsection{Patterns and rules for the crossings}

   The configuration of the round about is basically similar to that of
Fig.~\ref{crossconfig} when $p=13$. Fig.~\ref{cross_config_17} illustrates the 
new configuration when \hbox{$p=17$}.  

\vtop{
\ligne{\hfill
\includegraphics[scale=0.80]{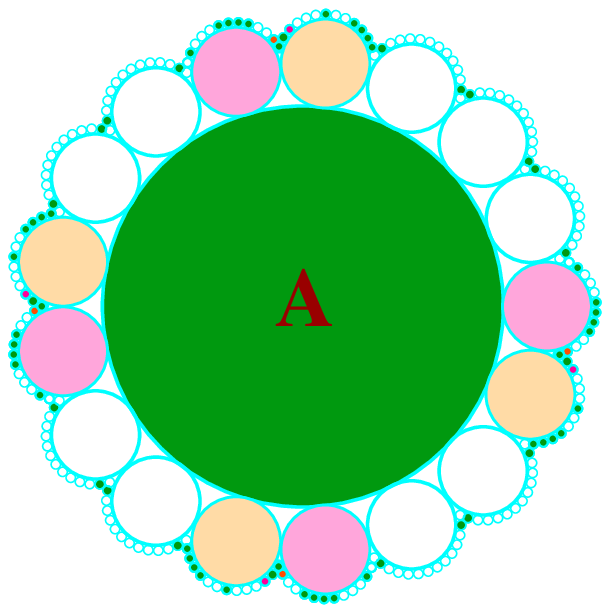}
\hfill}
\vspace{-15pt}
\begin{fig}\label{cross_config_17}
\leurre
The configuration of a round-about. The configurations are adapted to the 
new setting of Fig.~{\rm\ref{trackgene}}. The configurations of the cells~$B$ (orange)
and~$C$ (pink) are very different from those of Fig.~{\rm\ref{crossconfig}} for the
cells with the same names.
\end{fig}
}

In order to better explain the situation,
we give the same names for the cells $A$, $B$, $C$, $D$, $E$ and~$F$ as in
the case when $p=13$. However, here, $B.11$, $B.12$ have a different meaning than
in the case when $p=13$. We call $BF$ the neighbour of~$B$ which can also see~$F$
and~$BC$ that which can also see~$C$. Although $C.5$ has the same meaning here,
we call it $CE$.

   We fix the patterns for each cell which are not tracks of the cell with the 
patterns defined in Subsection~\ref{trackpattrules}.

   First, the black cell which is the core of the round-about, we have white neighbours
and at most two consecutive black cells. We shall consider that rules for which
the context contains at most two black cells are always conservative.

   Remember that at the beginning of Section~\ref{rulesgene}, we said that
in the block of cells which encodes its type, we have two cells devoted to
the identification of the situation, which switch or crossing, while the three 
other cells are a numbering of the types in the considered situation. 

In a round-about, the type of the cell is: {\tt BBBBWWXYZ} where {\tt XYZ}
is given  in Table~\ref{crossthetypes} as well as the corresponding types.

\vtop{
\vspace{7pt}
\begin{tab}\label{crossthetypes}
\leurre
The types of the cells of the round point which have patterns different from
those of Subsection~{\rm\ref{trackpattrules}}.
\end{tab}
\ligne{\hfill
\vtop{\hsize=300pt
\vspace{-12pt}
\grostrait
\ligne{\hfill\hbox to 55pt{\hfill $B$\hfill}
\hbox to 55pt{\hfill $C$\hfill}
\hbox to 55pt{\hfill $BC$\hfill}
\hbox to 55pt{\hfill $BF$\hfill}
\hbox to 55pt{\hfill $CE$\hfill}
\hfill}
\vspace{-4pt}
\demitrait
\vspace{-11.5pt}
\ligne{\hrulefill}
\ligne{\hfill\hbox to 55pt{\hfill\tt BWW\hfill}
\hbox to 55pt{\hfill\tt WBW\hfill}
\hbox to 55pt{\hfill\tt BBW\hfill}
\hbox to 55pt{\hfill\tt WWB\hfill}
\hbox to 55pt{\hfill\tt BWB\hfill}
\hfill}
\ligne{\hfill\hbox to 55pt{\hfill\tt BBBBWWBWW\hfill}
\hbox to 55pt{\hfill\tt BBBBWWWBW\hfill}
\hbox to 55pt{\hfill\tt BBBBWWBBW\hfill}
\hbox to 55pt{\hfill\tt BBBBWWWWB\hfill}
\hbox to 55pt{\hfill\tt BBBBWWBWB\hfill}
\hfill}
\vspace{-4pt}
\demitrait
}
\hfill}
\vspace{7pt}
}

   Note that Table~\ref{crossthetypes} contains five cells only. This means that 
the patterns of the others are among those described in 
Subsection~\ref{trackpattrules}.

\vskip 7pt
\noindent
$\underline{\hbox{\bf  The rules for~$B$}}$
\vskip 7pt

   From Fig.~\ref{cross_config_17}, we know that its context is
\hbox{\tt BWW$^a$BBBBWWBWWW$^b$WBBW}, with $a,b\geq1$. The type 
is here \hbox{\tt BBBBWWBWW} as indicated by Table~\ref{crossthetypes}.

The rules are defined by each type of situation where~$B$ is concerned.
Its the always case: if a particle arrives from outside, it goes through~$E$
and then~$C$ so~$B$ can see the particle passing through~$C$. Also, it is concerned
by the arrival of one or two particles from~$A$, see Fig.~\ref{crossing1} and
the following ones. In the rules, when it is present, we indicate by a bold letter
which black cell is the particle.

  Consider the case when a particle is present at~$A$. We have two rules:

\vskip 5pt
\ligne{\hfill\tt
\hbox{$\underline{\tt W}$BBBBWWBWWW$^b$WBBWBWW$^a$$\underline{\tt W}$}
\hskip 15pt
\hbox{$\underline{\tt W}$BBBBWWBWWW$^b$WBBWB{\small\bf B}W$^a$$\underline{\tt B}$}
\hfill$(cr.B.a)$}
\vskip 5pt

The left-hand side one is the conservative rule of the cell. The right-hand side one
makes the particle enter~$B$. The bold~$B$ indicates the particle. If we have a 
single particle, we have then:

\vskip 5pt
\ligne{\hfill\tt
\hbox{$\underline{\small\bf B}$BBBBWWBWWW$^b$WBBWBWW$^a$$\underline{\tt W}$} 
\hskip 15pt
\hbox{$\underline{\tt W}$BBBBWWBWWW$^b${\small\bf B}BBWBWW$^a$$\underline{\tt W}$}
\hfill$(cr.B.1b)$}
\vskip 5pt
              
The left-hand one makes the particle leave the cell. The second one witnesses
that it entered~$F$. 

   Now, at this point, as $BF$ checks the number of particles passing through~$B$,
it can see a particle in~$F$ and none in~$B$, so~$BF$ remains the same. At the next 
time, as the particle left~$F$, the conservative rule again applies.

   Consider the case when two particles arrive at~$B$. The first one is dealt
with the rules of $(cr.B.a)$. But the rules of $(cr.B.1b)$ cannot apply as
a particle is still present in~$A$. We have the rules:

\vskip 5pt
\ligne{\hfill\tt
\hbox{$\underline{\small\bf B}$BBBBWWBWWW$^b$WBBWB{\small\bf B}W$^a$%
$\underline{\tt B}$}
\hskip 15pt
\hbox{%
$\underline{\small\bf B}$BBBBWWBWWW$^b${\small\bf B}BBWBWW$^a$$\underline{\tt W}$}
\hfill$(cr.B.2b)$}
\vskip 5pt

   The left-hand side rule can see that there is still a particle~in $A$
and this second particle is admitted to~$B$. The right-hand side rule
deals with this second particle which will leave~$B$ and it witnesses that the
first particle is in~$F$.

   And so, at this time, one particle is in~$B$ and the other is in~$F$. The
cell~$BF$ can check this situation and so it flashes and then $BC$ flashes
and then a particle appears in~$C$. All these events are witnessed by~$B$
thanks to the following rules in this order:

\vskip 5pt
\ligne{\hfill
$\vcenter{
\vtop{\hsize=303pt
\ligne{\hfill\tt
\hbox{$\underline{\small\bf W}$BBBBWWBWWW$^b${\small\bf B}WB{\small\bf W}BWW$^a$%
$\underline{\tt W}$}
\hskip 5pt
\hbox{$\underline{\small\bf W}$BBBBWWBWWW$^b$WB{\small\bf W}WBWW$^a$%
$\underline{\tt W}$}
\hfill}
\ligne{\hfill\tt
\hbox{$\underline{\small\bf W}$BB{\small\bf B}BWW$^a$BBBBWWBWWW$^b$%
$\underline{\tt W}$}
\hfill}
}}$
\hfill$(cr.B.2c)$}

\vskip 7pt
\noindent
$\underline{\hbox{\bf  The rules for~$C$}}$
\vskip 7pt

   From Fig.~\ref{cross_config_17}, we know that its context is
\hbox{\tt WBWWW$^a$BBBBWWWBWWW$^b$W}, with \hbox{$a,b\geq1$}. The type 
is here \hbox{\tt BBBBWWWBW} as indicated by Table~\ref{crossthetypes}.

The rules are defined by each type of situation where~$C$ is concerned.
Now, $C$ is concerned in two cases only. When a particle arrives from outside, 
it goes through~$E$ and then~$C$ can see the particle passing through~$C$. 
Also, when two particles arrive at~$A$, eventually one particle is created
in~$C$ by the flash of~$BC$. However, when a single particle arrives at~$A$,
$C$~just witnesses that it passes through~$B$ and is no more concerned. But
this witnessing was already required when two particles arrive.
   
   First, consider the case when a particle arrives at~$E$. We have first
the conservative rule and then the rule which makes the particle pass from~$E$ to~$C$:
\vskip 5pt
\ligne{\hfill\tt
\hbox{$\underline{\tt W}$BBBBWWWBWWW$^b$WWBWWW$^a$$\underline{\tt W}$}
\hskip 5pt
\hbox{$\underline{\tt W}$BBBBWWWBWWW$^b$WWB{\small\bf B}WW$^a$$\underline{\tt B}$}
\hfill$(cr.C.a)$}
\vskip 5pt

   The next rule shows that $CE$ is flashing so that as one~$B$ is leaving~$C$
by the motion rules which apply to the ordinary cell~$D$, a second~$B$ is created
in~$C$. The second rule in~$(cr.C.b)$ witnesses that the second~$B$ leaves
the cell while the first one can be seen in~$D$. The last rule of~$(cr.C.b)$
witnesses that the second~$B$ is in~$D$.
\vskip 5pt
\ligne{\hfill
$\vcenter{
\vtop{\hsize=303pt
\ligne{\hfill\tt
\hbox{$\underline{\tt B}$BBBBWWWBWWW$^b$WWBW{\small\bf B}W$^a$$\underline{\tt B}$}
\hskip 5pt
\hbox{$\underline{\tt B}$BBBBWWWBWWW$^b${\small\bf B}WBWWW$^a$$\underline{\tt W}$}
\hfill}
\ligne{\hfill\tt
\hbox{$\underline{\tt W}$BBBBWWWBWWW$^b${\small\bf B}WBWWW$^a$$\underline{\tt W}$}
\hfill}
}}$
\hfill$(cr.C.b)$}
\vskip 5pt

   Consider now the case when two particles arrive at~$A$. The presence of the
first particle and then the second one in~$B$ is witnessed by the first rule
of~$(cr.C.c)$ which is applied twice. The second rule of~$(cr.C.c)$ is in action
when $BC$~is flashing: this means that a particle must be created in~$C$ which is
the result of the application of the rule. The third rule of~$(cr.C.c)$ contributes
to the move of the particle from~$C$ to~$D$ and witnesses the return of~$BC$
to its normal black state. The rule which witnesses that
the particle is now in~$D$ is the last one of~$(cr.C.b)$.

\vskip 5pt
\ligne{\hfill
$\vcenter{
\vtop{\hsize=303pt
\ligne{\hfill\tt
\hbox{$\underline{\tt W}$BBBBWWWBWWW$^b$W{\small\bf B}BWWW$^a$$\underline{\tt W}$}
\hskip 5pt
\hbox{$\underline{\tt W}$BBBBWWWBWWW$^b$WWWWWW$^a$$\underline{\tt B}$}
\hfill}
\ligne{\hfill\tt
\hbox{$\underline{\tt B}$BBBBWWWBWWW$^b$WWBWWW$^a$$\underline{\tt W}$}
\hfill}
}}$
\hfill$(cr.C.c)$}
\vskip 5pt

   We remember that $E$ and~$F$ are on the tracks and that they are ordinary cells.
We simply note that when some sensors are flashing, there are only two black
cells in the neighbourhood of the cell of the tracks. In such a case, the 
state of the cell is unchanged. Now, it may happen that the flash cancels a black
cell and that a new one occurs as the particle going to enter or to leave the cell.
Two rules are in use in this case. The first one is in the neighbour~$F1$ of~$F$ which
is on the tracks. When $BF$~flashes, the two particles are still present, one in~$F$
and one already in~$F1$. The particles are destroyed: as one milestone is
missing for~$F$ the presence of the particle in~$F1$ restores the configuration of
a cell of the tracks so that the particle present at~$F$ is erased by a motion
rule of the tracks. Now, in~$F1$, the flash also cancels a milestone. But the
presence of the particle in~$F$ creates a black neighbour at an unusual place for 
a cell of the tracks. This also requires a new rule: this is the left-hand side rule
in~$(cr.tr)$. Another situation appears in~$E$ where the creation of the particle
in~$C$ and the flash of~$CE$ creates a neighbourhood with five black cells which
has no counter part in the motion rules of the track for ordinary cells. The
appropriate rule is the right-hand side one in~$(cr.tr)$.

\vskip 5pt
\ligne{\hfill\tt
\hbox{$\underline{\tt B}$BWBWWWBWW$^k${\small\bf W}W$\underline{\tt W}$} 
\hskip 5pt
\hbox{$\underline{\tt W}$BBBWWBW$^k$BW$\underline{\tt W}$} 
\hfill$(cr.tr)$}
\vskip 5pt

   Now, we turn to the auxiliary cells which are not cells of the tracks.

\vskip 7pt
\noindent
$\underline{\hbox{\bf  The rules for~$BF$}}$
\vskip 7pt

   This cell is a milestone of~$F$ and also of~$F1$, as we have already noticed.
The role of this cell is to check whether one or two particles arrive at~$B$.
From Fig.~\ref{cross_config_17}, we know that its context is
\hbox{\tt WWWW$^a$BBBBWWWWBWW$^b$WB}, with \hbox{$a,b\geq1$}. The type 
is here \hbox{\tt BBBBWWWWB} as indicated by Table~\ref{crossthetypes}.

    The conservative rule is the first one in~$(cr.BF.A)$. If a particle
arrive in~$B$, it is noticed by the second rule of~$(cr.BF.A)$ and nothing happens
as $F$~is still white. And then, the particle goes to~$F$, which is witnessed
by the third rule of~$(cr.BF.a)$. If the particle is alone, then the particle
left~$F$ and is now in~$F1$ as witnessed by the fourth rule of~$(cr.BF.a)$.

\vskip 5pt
\ligne{\hfill
$\vcenter{
\vtop{\hsize=303pt
\ligne{\hfill\tt
\hbox{$\underline{\tt B}$BBBBWWWWBWW$^b$WBWWWW$^a$$\underline{\tt B}$} 
\hskip 5pt 
\hbox{$\underline{\tt B}$BBBBWWWWBWW$^b$WB{\small\bf B}WWW$^a$$\underline{\tt B}$} 
\hfill}
\ligne{\hfill\tt
\hbox{$\underline{\tt B}$BBBBWWWWBWW$^b$WBW{\small\bf B}WW$^a$$\underline{\tt B}$} 
\hskip 5pt 
\hbox{$\underline{\tt B}$BBBBWWWWBWW$^b$WBWW{\small\bf B}W$^a$$\underline{\tt B}$} 
\hfill}
}}$
\hfill$(cr.BF.a)$}
\vskip 5pt

   If two particles are present, then after the application of the first two
rules of~$(cr.BF.a)$, there is a particle in~$B$ and one in~$F$. At this moment,
$BF$ must flash, which is obtained by the first rule of~$(cr.BF.b)$. But, still
at this time, the motion rules apply to other cells, $F$ and~$F1$ in particular
so now, both particles are now in~$F$ and in~$F1$.  This situation is witnessed
by the second rule of~$(cr.BF.b)$ which returns $BF$ to black as $BF$ flashes
only one step. But, as $F$, $F1$ and~$BC$ can now see the flash of~$F$,
they all change their state to white, which is witnessed by the third rule
of~$(cr.BF.b)$.

\vskip 5pt
\ligne{\hfill
$\vcenter{
\vtop{\hsize=303pt
\ligne{\hfill\tt
\hbox{$\underline{\tt B}$BBBBWWWWBWW$^b$WB{\small\bf B}{\small\bf B}WW$^a$%
$\underline{\tt W}$} 
\hskip 5pt 
\hbox{$\underline{\tt W}$BBBBWWWWBWW$^b$WBW{\small\bf B}{\small\bf B}W$^a$%
$\underline{\tt B}$}
\hfill}
\ligne{\hfill\tt
\hbox{$\underline{\tt B}$BBBBWWWWBWW$^b$WWWWWW$^a$$\underline{\tt B}$}
\hfill}
}}$
\hfill$(cr.BF.b)$}
\vskip 5pt

\vskip 7pt
\noindent
$\underline{\hbox{\bf  The rules for~$BC$}}$
\vskip 7pt

   We know that the flash of~$BF$ is transmitted to~$C$ by~$BC$.
From Fig.~\ref{cross_config_17}, we know that the context of~$BC$ is
\hbox{\tt WBW$^a$BBBBWWBBWWW$^b$WWW}, with \hbox{$a,b\geq1$}. The type 
is here \hbox{\tt BBBBWWBBW} as indicated by Table~\ref{crossthetypes}.

   The first rule of~$(cr.BC)$ is conservative. The second rule witnesses that 
a particle is in~$B$. It is applied twice of two particles arrive at~$B$.
The third rule detects the flash o~$BF$ and so, it changes the state of~$BC$
to white. Now, the fourth rule restores the black state of~$BC$ as the flash is
for one step only. The fifth rule witnesses the effect of the flash of~$BC$
and~$C$: a black state appeared both in~$C$ and~$BC$ after the flash of~$BC$.
After that, the conservative rule again applies.

\vskip 5pt
\ligne{\hfill
$\vcenter{
\vtop{\hsize=303pt
\ligne{\hfill\tt
\hbox{$\underline{\tt B}$BBBBWWBBWW$^b$WWWWWBW$^a$$\underline{\tt B}$} 
\hskip 5pt 
\hbox{$\underline{\tt B}$BBBBWWBBWW$^b$WWWW{\small\bf B}BW$^a$$\underline{\tt B}$}
\hfill}
\ligne{\hfill\tt
\hbox{$\underline{\tt B}$BBBBWWBBWW$^b$WWWWW{\small\bf W}W$^a$$\underline{\tt W}$}
\hskip 5pt 
\hbox{$\underline{\tt W}$BBBBWWBBWW$^b$WWWWWBW$^a$$\underline{\tt B}$}
\hfill}
\ligne{\hfill\tt
\hbox{$\underline{\tt B}$BBBBWWBBWW$^b$WWW{\small\bf B}WBW$^a$$\underline{\tt B}$}
\hfill}
}}$
\hfill$(cr.BC)$}
\vskip 5pt

\vskip 7pt
\noindent
$\underline{\hbox{\bf  The rules for~$CE$}}$
\vskip 7pt

   When  a particle arrives from outside to the round-about, it enters the
round-about through~$C$ but a second particle has to accompany the initial one
until the next branching. The role of~$CE$ is to control and to perform this
task.
From Fig.~\ref{cross_config_17}, we know that its context is
\hbox{\tt WWWW$^a$BBBBWWWWBWW$^b$WB}, with \hbox{$a,b\geq1$}. The type 
is here \hbox{\tt BBBBWWWWB} as indicated by Table~\ref{crossthetypes}.
 
   The first rule of~$(cr.CE)$ is the conservative rule. The second rule is 
triggered by the occurrence of the particle in~$E$. This is why the rule makes 
$CE$~to flash by turning to black. The third instruction returns~$CE$ to white and
it witnesses the presence of the particle in~$C$. The effect of the flash
is a second particle in~$C$ which is witnessed by the fourth rule of~$(cr.CE)$.

\vskip 5pt
\ligne{\hfill
$\vcenter{
\vtop{\hsize=303pt
\ligne{\hfill\tt
\hbox{$\underline{\tt W}$BBBBWWBWBWW$^b$WWWWWW$^a$$\underline{\tt W}$} 
\hskip 5pt 
\hbox{$\underline{\tt W}$BBBBWWBWBWW$^b$WWWW{\small\bf B}W$^a$$\underline{\tt B}$}
\hfill}
\ligne{\hfill\tt
\hbox{$\underline{\tt B}$BBBBWWBWBWW$^b$WWW{\small\bf B}WW$^a$$\underline{\tt W}$}
\hskip 5pt 
\hbox{$\underline{\tt W}$BBBBWWBWBWW$^b$WWW{\small\bf B}WW$^a$$\underline{\tt W}$}
\hfill}
}}$
\hfill$(cr.CE)$}
\vskip 5pt

\subsection{Patterns and rules for the fixed switch}

   In the fixed switch, all the cells around~$O$ which are on the tracks are
ordinary cells. This can be checked on Fig.~\ref{trackfixed} which illustrates
the idle configuration of the central cell of the switch when \hbox{$p=17$}.

\vtop{
\vspace{-20pt}
\ligne{\hfill
\includegraphics[scale=1.0]{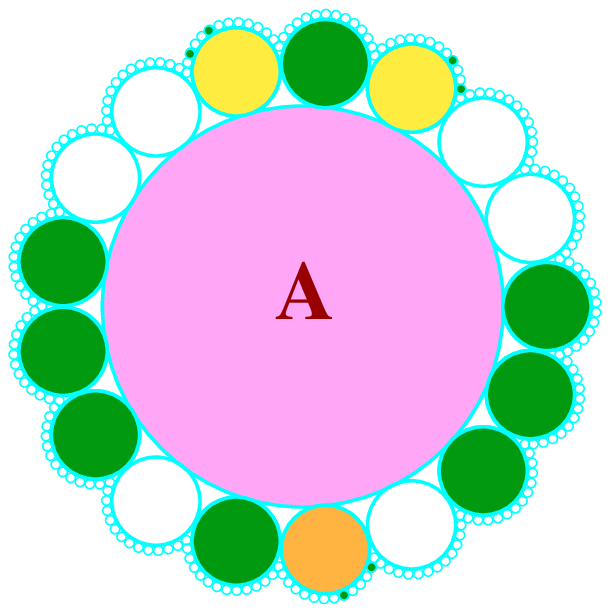}
\hfill}
\vspace{-15pt}
\begin{fig}\label{trackfixed}
\leurre
The idle configuration of a fixed switch. The configuration is adapted from the 
new setting of Fig.~{\rm\ref{trackgene}}. The cells~$B$ and~$C$ are in yellow.
The cell~$A$ is in orange. We can check that they are cells of the tracks.
\end{fig}
}

   In these cells of the tracks, as indicated in Subsection~\ref{trackpattrules},
we take advantage on the flexibility given by the possible choice of the exact 
neighbour through which the particle enters or leaves the cell.

   Accordingly, we only have to look at the rules for the cell~$O$. From
Fig.~\ref{trackfixed}, we can see that the context of the rules is
\hbox{\tt BWWBBBWW$^a$WBWW$^b$WBBBW}, with \hbox{$a,b\geq1$}. As mentioned
at the beginning of Section~\ref{rulesgene}, this configuration is characterized
by the occurrence of two blocks of consecutive three black cells. Here, the
shortest distance between the blocks is 4~cells.

   The rules for~$O$ are simply motion rules as no sensor is present around the
cell. They are given by formulas~$(fx.O)$.

   The rule of the first line of~$(fx.O)$ is the conservative rule of~$O$ when the
particle is far from the switch.
\vskip 5pt
\ligne{\hfill
$\vcenter{
\vtop{\hsize=303pt
\ligne{\hfill\tt
\hbox{$\underline{\tt W}$BBBWBWWBBBWW$^a$WBWW$^b$W$\underline{\tt W}$} 
\hfill}
\ligne{\hfill\tt
\hbox{$\underline{\tt W}$BBBWBWWBBBWW$^a${\small\bf B}BWW$^b$W$\underline{\tt B}$} 
\hskip 5pt 
\hbox{$\underline{\tt W}$BBBWBWWBBBWW$^a$WB{\small\bf B}W$^b$W$\underline{\tt B}$} 
\hfill}
\ligne{\hfill\tt
\hbox{$\underline{\tt\small\bf B}$BBBWBWWBBBWW$^a$WBWW$^b$W$\underline{\tt W}$} 
\hskip 5pt 
\hbox{$\underline{\tt W}$BBBWB{\small\bf B}WBBBWW$^a$WBWW$^b$W$\underline{\tt W}$} 
\hfill}
}}$
\hfill$(fx.O)$}
\vskip 5pt
   The second line gives the rules which allow the particle to enter the cell:
in the left-hand side rule, the particle is in~$C$, in the right-hand side one,
it is in~$B$. In the third line,the first rule changes the state of~$O$ from black
to white: the particle is ejected from the cell. The second rule witnesses that
the particle went to~$A$.

\subsection{Patterns and rules for the flip-flop}

   The idle configuration of the flip-flop is illustrated by Fig.~\ref{ffconfig}.
We can notice that here too, we have two blocks of consecutive three black cells.
However, here, the distance between the blocks is 5~cells.

\vtop{
\vspace{-20pt}
\ligne{\hfill
\includegraphics[scale=1.0]{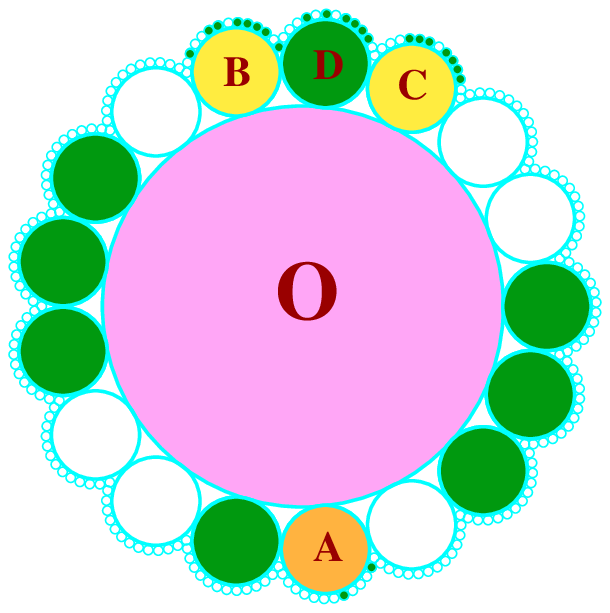}
\hfill}
\vspace{-15pt}
\begin{fig}\label{ffconfig}
\leurre
The idle configuration of a flip-flop. The configuration is adapted from the 
new setting of Fig.~{\rm\ref{trackgene}}. The cells~$B$ and~$C$ are in yellow.
The cell~$A$ is in orange. Only $A$ is an ordinary cell of the tracks.
\end{fig}
}

   Indeed, the context is now \hbox{\tt BWWBBBWW$^a$WBWWBBBW$^b$W}. But the big 
difference is that outside~$A$ which is still an ordinary cell of the tracks,
the cells~$B$ and~$C$ are no more ordinary cells of the tracks: they have sensors
and a specific surrounding. Also, the cell~$D$, which is in between~$B$ and~$C$
is the controller of the switch and it has too a specific surrounding. Now, in 
between~$B$ and~$D$ as well as in between~$C$ and~$D$ there is a sensor, $H$ and~$K$
respectively. These latter cells also have a complex neighbourhood. We know
that the context of all these cells contain the pattern \hbox{\tt BBBBWB}
where {\tt WB} characterizes the flip-flop. We now turn
to the study of the rules of all these particular cells.

\vskip 7pt
\noindent
$\underline{\hbox{\bf  The rules for~$B$}}$
\vskip 7pt

    The context of~$B$ is defined by \hbox{\tt WBBW$^a$BBBBWBBWWBWW$^b$W} 
where \hbox{$a,b\geq 1$}. This is illustrated by Fig.~\ref{ffconfig}
when \hbox{$p=17$}.

    The rules for~$B$ are given by~$(ff.Ba)$. The first rule is the conservative
rule. Then, in the second line, the first rule witnesses the passage of the particle 
in the other tracks: this means that it is the case when $B$~is not selected, which
means that $H$~is black. Still when $B$~is not selected, the third rule of~$(ff.Ba)$ 
witnesses the flash of~$D$ 
caused by the particle after its passage through~$C$, two steps later.

\vskip 5pt
\ligne{\hfill
$\vcenter{
\vtop{\hsize=303pt
\ligne{\hfill\tt
\hbox{$\underline{\tt W}$BBBBWBBWWBWW$^b$WWBBW$^a$$\underline{\tt W}$} 
\hfill}
\ligne{\hfill\tt
\hbox{$\underline{\tt W}$BBBBWBBWWBWW$^b$W{\small\bf B}BBW$^a$$\underline{\tt W}$} 
\hskip 5pt 
\hbox{$\underline{\tt W}$BBBBWBBWWBWW$^b$WW{\small\bf W}BW$^a$$\underline{\tt W}$} 
\hfill}
}}$
\hfill$(ff.Ba)$}
\vskip 5pt

   Next, the rules deals with the case when $B$~is selected by the switch: this means
that~$H$ is white. The first rule of~$(ff.Bb)$ is the conservative rule of this
situation. The second rule can see the particle in~$O$. As $B$~is selected, it
attracts the particle: the rule changes the state of~$B$ to black. The third rule
returns the state of~$B$ to white as the particle goes further on the tracks.
This is witnessed by the fourth rule which can see the particle in the next cell
of the tracks. The same rule also witnesses the flash of~$D$ which occurs at this very
moment.
   
\vskip 5pt
\ligne{\hfill
$\vcenter{
\vtop{\hsize=303pt
\ligne{\hfill\tt
\hbox{$\underline{\tt W}$BBBBWBBWWBWW$^b$WWBWW$^a$$\underline{\tt W}$} 
\hskip 5pt 
\hbox{$\underline{\tt W}$BBBBWBBWWBWW$^b$W{\small\bf B}BWW$^a$$\underline{\tt B}$} 
\hfill}
\ligne{\hfill\tt
\hbox{$\underline{\tt \small\bf B}$BBBBWBBWWBWW$^b$WWBWW$^a$$\underline{\tt W}$} 
\hskip 5pt 
\hbox{$\underline{\tt W}$BBBBWBBWWB{\small\bf B}W$^b$WW{\small\bf W}WW$^a$%
$\underline{\tt W}$} 
\hfill}
}}$
\hfill$(ff.Bb)$}

\vskip 7pt
\noindent
$\underline{\hbox{\bf  The rules for~$D$}}$
\vskip 7pt

    The context of~$D$ is defined by \hbox{\tt WBBW$^a$BBBBWBBWWBWW$^b$W} 
where \hbox{$a,b\geq 1$}. This is illustrated by Fig.~\ref{ffconfig}
when \hbox{$p=17$}.

    The context of~$D$ is \hbox{\tt WWWW$^a$BBBBWBWBWW$^b$WBW} when $B$~is not selected
and it is  \hbox{\tt WWBW$^a$BBBBWBWBWW$^b$WWW} when $B$~is selected.

   The rules of~$(ff.Da)$ are the conservative rules of~$D$. There are two such rules
depending on which tracks between~$B$ and~$C$ is the selected one.
\vskip 5pt
\ligne{\hfill
$\vcenter{
\vtop{\hsize=303pt
\ligne{\hfill\tt
\hbox{$\underline{\tt B}$BBBBWBWBWW$^b$WBWWWWW$^a$$\underline{\tt B}$} 
\hskip 5pt 
\hbox{$\underline{\tt B}$BBBBWBWBWW$^b$WWWWWBW$^a$$\underline{\tt B}$} 
\hfill}
}}$
\hfill$(ff.Da)$}
\vskip 5pt
   The rules of the first line of~$(ff.Db)$ gives the rules which make~$D$
flash. In both cases, the particle occurs in the neighbour belonging to the
selected tracks.
\vskip 5pt
\ligne{\hfill
$\vcenter{
\vtop{\hsize=303pt
\ligne{\hfill\tt
\hbox{$\underline{\tt B}$BBBBWBWBWW$^b$WBWW{\small\bf B}WW$^a$$\underline{\tt W}$} 
\hskip 5pt 
\hbox{$\underline{\tt B}$BBBBWBWBWW$^b$WW{\small\bf B}WWBW$^a$$\underline{\tt W}$} 
\hfill}
\ligne{\hfill\tt
\hbox{$\underline{\tt W}$BBBBWBWBWW$^b$WBWWWWW$^a$$\underline{\tt B}$} 
\hskip 5pt 
\hbox{$\underline{\tt W}$BBBBWBWBWW$^b$WWWWWBW$^a$$\underline{\tt B}$} 
\hfill}
}}$
\hfill$(ff.Db)$}
\vskip 5pt
The rules of the second row of~$(ff.Db)$ bring back~$D$ to the black state as the
flash is intended for one step only.

\vskip 7pt
\noindent
$\underline{\hbox{\bf  The rules for~$C$}}$
\vskip 7pt

   We turn to the rules for~$C$ which have a similar role as those for~$B$.
The number of the cell is {\tt BBW}, so that it context contains the
pattern \hbox{\tt BBBBWBBBW}. The context of the cell is
\hbox{\tt WWWW$^a$BBBBWBBBWW$^b$WBB}
when $C$~is not selected and 
\hbox{\tt WWWW$^a$BBBBWBBBWW$^b$WWB}, with \hbox{$a,b\geq1$}. 
The rules for~$C$ are displayed in
$(ff.Ca)$ and $(ff.Cb)$.

   The first rule of $(ff.Ca)$ is the conservative rule for~$C$ when the cell
is not selected. Then, if a particle passes through~$O$, it does not move to~$C$,
which is attested by the second rule: the state of~$C$ is not change. The third rule
witnesses a flash of~$D$ which occurs when a particle crosses the switch as it will 
pass through~$B$ and this will trigger the flash of~$D$.

\vskip 5pt
\ligne{\hfill
$\vcenter{
\vtop{\hsize=303pt
\ligne{\hfill\tt
\hbox{$\underline{\tt W}$BBBBWBBBWW$^b$WBBWWWW$^a$$\underline{\tt W}$} 
\hfill}
\ligne{\hfill\tt
\hbox{$\underline{\tt W}$BBBBWBBBWW$^b$WBB{\small\bf B}WWW$^a$$\underline{\tt W}$} 
\hskip 5pt 
\hbox{$\underline{\tt W}$BBBBWBBBWW$^b$WB{\small\bf W}WWWW$^a$$\underline{\tt W}$} 
\hfill}
}}$
\hfill$(ff.Ca)$}
\vskip 5pt

   The first rule in $(ff.Cb)$ is the conservative rule when $C$ is selected.
Now, the second rule shows that if the particle is present in~$O$ it then passes
to~$C$. The third rule witnesses the flash of~$D$ and it also turns back the
state of~$C$ to white. The fourth rule witnesses that the particle left the cell
and is now in the neighbouring cell of the tracks.
\vskip 5pt
\ligne{\hfill
$\vcenter{
\vtop{\hsize=303pt
\ligne{\hfill\tt
\hbox{$\underline{\tt W}$BBBBWBBBWWW$^b$WBWWWW$^a$$\underline{\tt W}$} 
\hskip 5pt
\hbox{$\underline{\tt W}$BBBBWBBBWWW$^b$WB{\small\bf B}WWW$^a$$\underline{\tt B}$} 
\hfill}
\ligne{\hfill\tt
\hbox{$\underline{\tt \small\bf B}$BBBBWBBBWWW$^b$WBWWWW$^a$$\underline{\tt W}$} 
\hskip 5pt
\hbox{$\underline{\tt W}$BBBBWBBBWWW$^b$W{\small\bf W}WW{\small\bf B}W$^a$%
$\underline{\tt W}$} 
\hfill}
}}$
\hfill$(ff.Cb)$}
\vskip 5pt

\vskip 7pt
\noindent
$\underline{\hbox{\bf  The rules for~$H$}}$
\vskip 7pt

   Now, we arrive to the sensors of~$D$: $H$ and~$K$. These cells show the
non-selected track and they change their state to the other one when $D$~flashes.

   The cell~$H$ has the number {\tt WWB}, so that its context contains the
pattern \hbox{\tt BBBBWBWWB}. The context of~$H$ is
\hbox{\tt WBW$^a$BBBBWBWWBWWW$^b$WW}, whatever the sate of~$H$.

   The rules for~$H$ are displayed in $(ff.H)$.

   The first two rules are conservative: one for the white state, the other for the
black one. The third rule witnesses the passage of the particle in~$B$ when $H$~is
white. The last two rules change the state of~$H$ to its opposite one as $D$~is
flashing in both cases.
\vskip 5pt
\ligne{\hfill
$\vcenter{
\vtop{\hsize=303pt
\ligne{\hfill\tt
\hbox{$\underline{\tt W}$BBBBWBWWBWWWWWWBW$\underline{\tt W}$} 
\hskip 5pt 
\hbox{$\underline{\tt B}$BBBBWBWWBWWWWWWBW$\underline{\tt B}$} 
\hfill}
\ligne{\hfill\tt
\hbox{$\underline{\tt W}$BBBBWBWWBWWWWW{\small\bf B}BW$\underline{\tt W}$} 
\hfill}
\ligne{\hfill\tt
\hbox{$\underline{\tt W}$BBBBWBWWBWWWWWW{\small\bf W}W$\underline{\tt B}$} 
\hskip 5pt 
\hbox{$\underline{\tt B}$BBBBWBWWBWWWWWW{\small\bf W}W$\underline{\tt W}$} 
\hfill}
}}$
\hfill$(ff.H)$}

\vskip 7pt
\noindent
$\underline{\hbox{\bf  The rules for~$K$}}$
\vskip 7pt

   The number of the cell is {\tt BWB} and so its context contains the pattern
\hbox{\tt BBBBWBBWB}. The context of~$K$ is
\hbox{\tt WWW$^a$BBBBWBBWBWWW$^b$WB}, with \hbox{$a,b\geq1$}, whatever the
state of~$K$.

    The structure of the rules is very similar to that of the rules for~$H$.
The first two rules are the conservative rules for~$K$: one when it is white, the other
when it is black.

   Here too, the first two rules are conservative for~$K$: one for the white state, the
other for the black one. The third rule witnesses the passage of the particle
in~$C$. The last two rules trigger the change of state: from white to black and from
black to white. In both cases, the rule witnesses the flashing of~$D$.
\vskip 5pt
\ligne{\hfill
$\vcenter{
\vtop{\hsize=303pt
\ligne{\hfill\tt
\hbox{$\underline{\tt W}$BBBBWBBWBWWW$^b$WBWWW$^a$$\underline{\tt W}$} 
\hskip 5pt 
\hbox{$\underline{\tt B}$BBBBWBBWBWWW$^b$WBWWW$^a$$\underline{\tt B}$} 
\hfill}
\ligne{\hfill\tt
\hbox{$\underline{\tt W}$BBBBWBBWBWWW$^b$WB{\small\bf B}WW$^a$$\underline{\tt W}$} 
\hfill}
\ligne{\hfill\tt
\hbox{$\underline{\tt W}$BBBBWBBWBWWW$^b$W{\small\bf W}WWW$^a$$\underline{\tt B}$} 
\hskip 5pt 
\hbox{$\underline{\tt B}$BBBBWBBWBWWW$^b$W{\small\bf W}WWW$^a$$\underline{\tt W}$} 
\hfill}
}}$
\hfill$(ff.K)$}
\vskip 5pt
\vskip 5pt

   We can now turn to the memory switch.

\subsection{Patterns and rules for the memory switch}

The idle configurations of both parts of the memory switch are represented by
Fig.~\ref{mmaconfig} and~\ref{mmpconfig} for the active switch and the
passive one respectively. The figures illustrate the case when \hbox{$p=17$}. 
As mentioned in the captions, we detail 
the neighbourhood of the cells which 
contribute to the working of the switch. However, $H$, $K$, $I$ and~$J$ cannot 
be named as their representation is too small. They can be deduced from what 
we explain in the rules.

   We first look at the rules for the active switch in Subsubsection~\ref{activmemo}.
Then, in Subsubsection~\ref{passivmemo}, we look at the rules for the passive
memory switch. Remember that the names of the cells are very different in both cases 
so that there will be no confusion between the cells of one one-way switch and those
of the other.

\vtop{
\vspace{-30pt}
\ligne{\hfill
\includegraphics[scale=1.0]{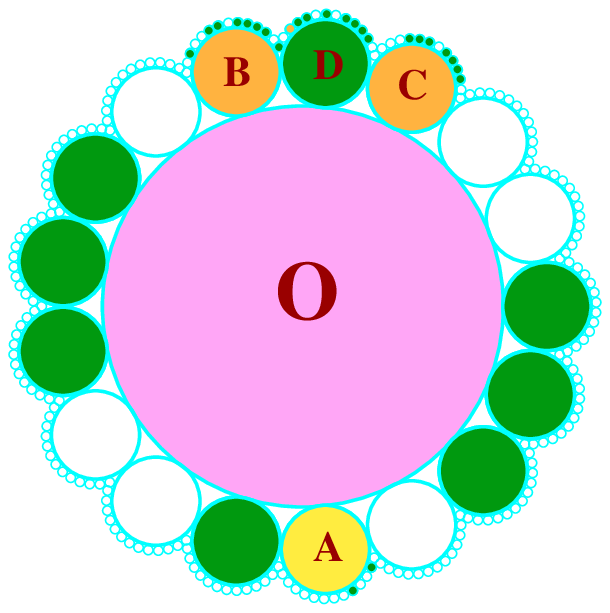}
\hfill}
\vspace{-22pt}
\begin{fig}\label{mmaconfig}
\leurre
The idle configuration of the active memory switch. The configuration is adapted 
from the new setting of Fig.~{\rm\ref{trackgene}}. 
The cell~$A$, the entrance, is in yellow.  The cells~$B$ and~$C$, the exits, are 
in orange. The cell~$A$ only is an ordinary cell of the tracks. Note $D1$, the
neighbour in orange of~$D$, waiting the signal of~$Z$.
\end{fig}
}

\vtop{
\vspace{-20pt}
\ligne{\hfill
\includegraphics[scale=1.0]{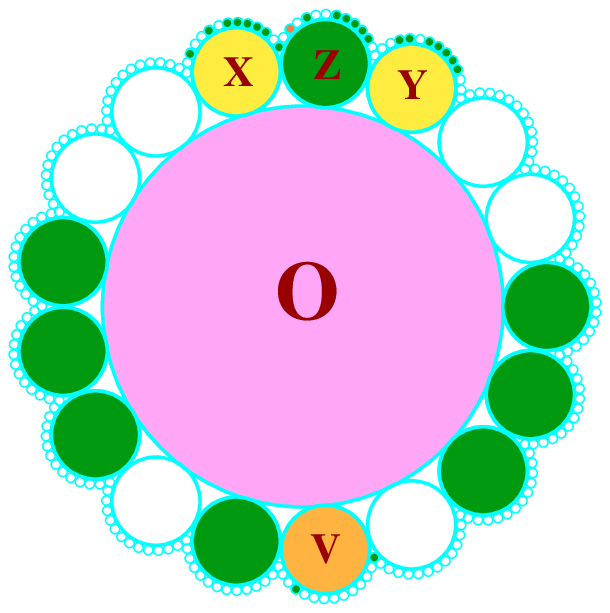}
\hfill}
\vspace{-22pt}
\begin{fig}\label{mmpconfig}
\leurre
The idle configuration of the passive memory switch. The configuration is adapted 
from the new setting of Fig.~{\rm\ref{trackgene}}. The cells~$X$ and~$Y$, the
entries, are in yellow. The cell~$V$, the exit, is in orange. The cell~$V$ only
is an ordinary cell of the tracks.
Among the neighbours of~$Z$, note the one in orange: it is~$Z1$ which leads to
the path to~$D1$.
\end{fig}
}

\subsubsection{The active memory switch}
\label{activmemo}

 As mentioned in the
case when \hbox{$p=13$}, we can reproduce the configurations of the cells implied
in the flip-flop for the active memory switch with one exception: the cell~$D$
which has a different behaviour in the active memory switch. Accordingly,
the rules which we have established for~$B$, $C$, $O$, $H$ and~$K$ are also in
action here. We have just to give the rules for~$D$. As this switch has some
similarity with the flip-flop and as a single cell requires specific rules,
the cell shares the pattern \hbox{\tt BBBBWB} with the cells of the flip-flop.
However, it receives the number~{\tt WBB} so that it characteristic pattern
is \hbox{\tt BBBBWBWBB}.

   From Fig.~\ref{mmaconfig}, we can see that the context of~$D$ is given by
the word \hbox{\tt WWWW$^a$BBBBWBWBBWW$^b$BW} when $C$~is the selected track and
\hbox{\tt WWBW$^a$BBBBWBWBBWW$^b$WW} when~$B$ is the selected track. The
big difference with $D$~in a flip-flop is that here, $D$~does not react to the
passage of the particle through~$B$ or~$C$. The concerned rules do not change
the state of~$D$ which remains black. The flash of~$D$ is here triggered by~$D1$.

The rules of~$(ma.Da)$ deal with the case when $C$~is selected. The first rule is
the conservative rule. The second rule witnesses the particle is in~$O$ and the
third rule witnesses that the particle is in~$C$. Later, the particle is gone on 
the tracks and so, the conservative rule again applies.
\vskip 5pt
\ligne{\hfill
$\vcenter{
\vtop{\hsize=303pt
\ligne{\hfill\tt
\hbox{$\underline{\tt B}$BBBBWBWBBWW$^b$BWWWWW$^a$$\underline{\tt B}$} 
\hfill}
\ligne{\hfill\tt
\hbox{$\underline{\tt B}$BBBBWBWBBWW$^b$BW{\small\bf B}WWW$^a$$\underline{\tt B}$} 
\hskip 5pt 
\hbox{$\underline{\tt B}$BBBBWBWBBWW$^b$BWW{\small\bf B}WW$^a$$\underline{\tt B}$}
\hfill}
}}$
\hfill$(ma.Da)$}
\vskip 5pt
 The rules of $(ma.Db)$ are exactly parallel to those of $(mq.Da)$: they deal with
the case when $B$~is selected. First, the conservative rule, then the witnessing
of the particle at~$O$ and then of its presence in~$B$.

\vskip 5pt
\ligne{\hfill
$\vcenter{
\vtop{\hsize=303pt
\ligne{\hfill\tt
\hbox{$\underline{\tt B}$BBBBWBWBBWW$^b$WWWWBW$^a$$\underline{\tt B}$} 
\hfill}
\ligne{\hfill\tt
\hbox{$\underline{\tt B}$BBBBWBWBBWW$^b$WWBW{\small\bf B}W$^a$$\underline{\tt B}$}
\hskip 5pt 
\hbox{$\underline{\tt B}$BBBBWBWBBWW$^b$W{\small\bf B}WWBW$^a$$\underline{\tt B}$}
\hfill}
}}$
\hfill$(ma.Db)$}
\vskip 5pt

   The rules of $(ma.Dc)$ deal with the flash of~$D1$. It may occur whatever the
selected track is. In the first line, $B$~is selected. The first rule witnesses
the flash of~$D1$ so that the rule makes~$D$ flash too. The second rule
returns~$D$ to the black state. In the second line, $C$~is selected. Again
the first rule makes~$D$ flash when the flash of~$D1$ is detected. The second rule
also returns~$D$ to the black state. In both cases, the change of states in~$H$
and~$K$ is witnessed by the appropriate conservative rule, see~$(ma.Da)$
and~$(ma.Db)$.
   
\vskip 5pt
\ligne{\hfill
$\vcenter{
\vtop{\hsize=303pt
\ligne{\hfill\tt
\hbox{$\underline{\tt B}$BBBBWBWBB{\small\bf B}W$^b$WWWWBW$^a$$\underline{\tt W}$} 
\hskip 5pt 
\hbox{$\underline{\tt W}$BBBBWBWBB{\small\bf W}W$^b$WWWWBW$^a$$\underline{\tt B}$}
\hfill}
\ligne{\hfill\tt
\hbox{$\underline{\tt B}$BBBBWBWBB{\small\bf B}W$^b$BWWWWW$^a$$\underline{\tt W}$}
\hskip 5pt 
\hbox{$\underline{\tt W}$BBBBWBWBB{\small\bf W}W$^b$BWWWWW$^a$$\underline{\tt B}$}
\hfill}
}}$
\hfill$(ma.Dc)$}
\vskip 5pt

\subsubsection{The passive memory switch}
\label{passivmemo}

   In the memory switch, many cells have a specific surrounding. However, the central
cell~$O$ of the switch is the same as that of the fixed switch as the latter switch
is also a one-way passive switch. Also, the cell~$V$, the exit of~$O$, is an
ordinary cell of the tracks. So that We need not give its rules.
But for the other cells, $X$, $Y$, $Z$, $I$, $J$, $Z1$
and~$D1$, we have to look at the
rules specifically. We not that here, the rules contain the pattern
\hbox{\tt BBBBBW} where {\tt BW} characterizes the memory switch.

\vskip 7pt
\noindent
$\underline{\hbox{\bf The rules for $X$}}$
\vskip 7pt

   This cell receives the number~{\tt BWW} so that its characteristic
pattern is the word \hbox{\tt BBBBBWBWW}. From Fig.~\ref{mmpconfig}, we know that 
\hbox{\tt WBBW$^a$BBBBBWBWWBWW$^b$W} 
is the context of~$X$ when it is not selected and 
\hbox{\tt WBWW$^a$BBBBBWBWWBWW$^b$W} when it is, where, in both cases,
\hbox{$a,b\geq 1$}. 

   First, consider the case when $X$~is not selected. The first 
rule of $(mp.Xa)$ is the conservative rule corresponding to this configuration.
The second rule detects the arrival of the particle from the tracks and
it makes it enter~$X$. The third rule returns~$X$ to white.
The fourth rule witnesses that the particle is now in~$O$ and it also witnesses   
the flash of~$Z$ as $Z$ has detected that a particle passed through non selected
tracks.
\vskip 5pt
\ligne{\hfill
$\vcenter{
\vtop{\hsize=303pt
\ligne{\hfill\tt
\hbox{$\underline{\tt W}$BBBBBWBWWBWW$^b$WWBBW$^a$$\underline{\tt W}$} 
\hskip 5pt 
\hbox{$\underline{\tt W}$BBBBBWBWWB{\small\bf B}W$^b$WWBBW$^a$$\underline{\tt B}$} 
\hfill}
\ligne{\hfill\tt
\hbox{$\underline{\tt \small\bf B}$BBBBBWBWWBWW$^b$WWBBW$^a$$\underline{\tt W}$} 
\hskip 5pt 
\hbox{$\underline{\tt W}$BBBBBWBWWBWW$^b$W{\small\bf B}WBW$^a$$\underline{\tt W}$} 
\hfill}
}}$
\hfill$(mp.Xa)$}
\vskip 5pt
    Next, we look at the case when $X$~is selected. The first rule of $(mp.Xb)$
is the conservative rule of this situation. The next three rules are ordinary
motion rules adapted to the context of~$X$: the first rule attracts the particle
into~$X$, the second one returns the state of~$X$ to white and the third one
witnesses that the particle is in the neighbour of~$X$ which is the next cell
on the tracks. Note that here, as $X$~is selected, there is no flash of~$Z$ which
is always black.
\vskip 5pt
\ligne{\hfill
$\vcenter{
\vtop{\hsize=303pt
\ligne{\hfill\tt
\hbox{$\underline{\tt W}$BBBBBWBWWBWW$^b$WWBWW$^a$$\underline{\tt W}$} 
\hskip 5pt 
\hbox{$\underline{\tt W}$BBBBBWBWWB{\small\bf B}W$^b$WWBWW$^a$$\underline{\tt B}$} 
\hfill}
\ligne{\hfill\tt
\hbox{$\underline{\tt \small\bf B}$BBBBBWBWWBBW$^b$WWBWW$^a$$\underline{\tt W}$} 
\hskip 5pt 
\hbox{$\underline{\tt W}$BBBBBWBWWBBW$^b$WBBWW$^a$$\underline{\tt W}$} 
\hfill}
}}$
\hfill$(mp.Xb)$}
\vskip 5pt
   A last rule is needed when $X$ is selected and when the particle passes 
through~$Y$. Nothing happens in~$X$, except when $Z$ flashes as the particle
crosses~$Y$ which was not selected. This flash is also noticed by~$X$ as
the particle is already in~$O$, whence the rule $(mp.Xc)$.
\vskip 5pt
\ligne{\hfill
$\vcenter{
\vtop{\hsize=303pt
\ligne{\hfill\tt
\hbox{$\underline{\tt W}$BBBBBWBWWBBW$^b$WB{\small\bf W}WW$^a$$\underline{\tt W}$} 
\hfill}
}}$
\hfill$(mp.Xc)$}
\vskip 5pt

\vskip 7pt
\noindent
$\underline{\hbox{\bf The rules for $Y$}}$
\vskip 7pt

   This cell receives the number~{\tt BBW} so that its characteristic
pattern is the word \hbox{\tt BBBBBWBBW}. From Fig.~\ref{mmpconfig}, we know that 
\hbox{\tt WWW$^a$WWBBBBBWBBWW$^b$BB} 
is the context of~$Y$ when it is not selected and 
\hbox{\tt WWW$^a$WWBBBBBWBBWW$^b$WB} when it is, where, in both cases,
\hbox{$a,b\geq 1$}. 

   First, we look at the case when $Y$~is not selected. The concerned rules
are given in $(mp.Ya)$. The first rule is the conservative rule of this case.
The neighbour~$J$ is black. The second rule attracts the particle into~$Y$.
The third rule returns the state of~$Y$ to white. The fourth state witnesses
that the particle is in~$O$. It also witnesses that $Z$ flashes as $Z$ has
seen that the particle went through the non selected tracks at the previous step.
\vskip 5pt
\ligne{\hfill
$\vcenter{
\vtop{\hsize=303pt
\ligne{\hfill\tt
\hbox{$\underline{\tt W}$BBBBBWBBWW$^b$BBWWW$^a$WW$\underline{\tt W}$} 
\hskip 5pt 
\hbox{$\underline{\tt W}$BBBBBWBBWW$^b$BBWWW$^a${\small\bf B}W$\underline{\tt B}$} 
\hfill}
\ligne{\hfill\tt
\hbox{$\underline{\tt \small\bf B}$BBBBBWBBWW$^b$BBWWW$^a$WW$\underline{\tt W}$} 
\hskip 5pt 
\hbox{$\underline{\tt W}$BBBBBWBBWW$^b$B{\small\bf W}{\small\bf B}WW$^a$WW%
$\underline{\tt W}$} 
\hfill}
}}$
\hfill$(mp.Ya)$}
\vskip 5pt

   Next, we look at the rules when $Y$~is selected which are displayed by
$(mp.Yb)$. Again, the first rule is conservative, adapted to this configuration,
where $J$~is white. The other rules are ordinary motion rules adapted to the
context of~$Y$. The second rule of $(mp.Yb)$ attracts the particle into~$Y$.
The third rule witnesses that the particle goes out from~$Y$ as the state of~$Y$
returns to white. The fourth rule witnesses that the particle is now in~$O$.
\vskip 5pt
\ligne{\hfill
$\vcenter{
\vtop{\hsize=303pt
\ligne{\hfill\tt
\hbox{$\underline{\tt W}$BBBBBWBBWW$^b$WBWWW$^a$WW$\underline{\tt W}$} 
\hskip 5pt 
\hbox{$\underline{\tt W}$BBBBBWBBWW$^b$WBWWW$^a${\small\bf B}W$\underline{\tt B}$} 
\hfill}
\ligne{\hfill\tt
\hbox{$\underline{\tt \small\bf B}$BBBBBWBBWW$^b$WBWWW$^a$WW$\underline{\tt W}$} 
\hskip 5pt 
\hbox{$\underline{\tt W}$BBBBBWBBWW$^b$WB{\small\bf B}WW$^a$WW$\underline{\tt W}$} 
\hfill}
}}$
\hfill$(mp.Yb)$}
\vskip 5pt
   This case still requires another rule: when the particle passes through~$X$,
as $X$ is not selected, this triggers the flash of~$Z$ which is seen by~$Y$
at the time when the particle is in~$O$. Whence the rule of $(mp.Yc)$.

\vskip 5pt
\ligne{\hfill
$\vcenter{
\vtop{\hsize=303pt
\ligne{\hfill\tt
\hbox{$\underline{\tt W}$BBBBBWBBWW$^b$W{\small\bf W}{\small\bf B}WW$^a$WW%
$\underline{\tt W}$} 
\hfill}
}}$
\hfill$(mp.Yc)$}

\vskip 7pt
\noindent
$\underline{\hbox{\bf The rules for $Z$}}$
\vskip 7pt

   This cell receives the number~{\tt WBW} so that its characteristic
pattern is the word \hbox{\tt BBBBBWWBW}. From Fig.~\ref{mmpconfig}, we know that 
\hbox{\tt WWWW$^a$BBBBBWWBWWW$^b$BW} is the context of~$Z$ when~$Y$ is selected and 
\hbox{\tt WWBW$^a$BBBBBWWBWWW$^b$WW} is the context of~$Z$ when~$X$
is selected and, in both cases, \hbox{$a,b\geq 1$}. 

   First, we look at the case when the particle goes through the non-selected tracks.
The rules are displayed in $(mp.ZXn)$ when the particle goes through~$X$
and in $(mp.ZYn)$ when the particle goes through~$Y$.

   In both cases, the first rule is the conservative rule of the idle
configuration for~$Z$. The second rule witnesses that the particle goes through
the non selected tracks: $X$~in $(mp.ZXn)$, $Y$~in $(mp.ZXn)$. accordingly, the rule
makes~$Z$ flash. The third rule restores the black state in~$Z$ and it witnesses
that the particle is now in~$O$.
\vskip 5pt
\ligne{\hfill
$\vcenter{
\vtop{\hsize=303pt
\ligne{\hfill\tt
\hbox{$\underline{\tt B}$BBBBBWWBWWW$^b$BWWWWW$^a$$\underline{\tt B}$} 
\hfill}
\ligne{\hfill\tt
\hbox{$\underline{\tt B}$BBBBBWWBWWW$^b$B{\small\bf B}WWWW$^a$$\underline{\tt W}$} 
\hskip 5pt 
\hbox{$\underline{\tt \small\bf W}$BBBBBWWBWBW$^b$BW{\small\bf B}WWW$^a$%
$\underline{\tt B}$} 
\hfill}
}}$
\hfill$(mp.ZXn)$}
\vskip 5pt
\ligne{\hfill
$\vcenter{
\vtop{\hsize=303pt
\ligne{\hfill\tt
\hbox{$\underline{\tt B}$BBBBBWWBWWW$^b$WWWWBW$^a$$\underline{\tt B}$} 
\hfill}
\ligne{\hfill\tt
\hbox{$\underline{\tt B}$BBBBBWWBWWW$^b$WWW{\small\bf B}BW$^a$$\underline{\tt W}$} 
\hskip 5pt
\hbox{$\underline{\tt \small\bf W}$BBBBBWWBWBWW$^b$W{\small\bf B}WBW$^a$%
$\underline{\tt B}$} 
\hfill}
}}$
\hfill$(mp.ZYn)$}
\vskip 5pt
   It is important to notice that the next situation, when $I$ and~$J$ both change
their state, is witnessed by the fact that another conservative rule applies.

   Presently, we look at the situation when the particle goes through the selected
tracks. The corresponding rules are displayed in $(mp.ZXs)$ and $(mp.ZYs)$.

When the selected track is~$Y$, the first rule witnesses the presence of the particle
in~$Y$ and the next rule witnesses that it is now in~$O$.
\vskip 5pt
\ligne{\hfill
$\vcenter{
\vtop{\hsize=303pt
\ligne{\hfill\tt
\hbox{$\underline{\tt B}$BBBBBWWBWWWBWWBWW$\underline{\tt B}$} 
\hskip 5pt
\hbox{$\underline{\tt B}$BBBBBWWBWWWBWBWWW$\underline{\tt B}$} 
\hfill}
}}$
\hfill$(mp.ZYs)$}
\vskip 5pt
The same is performed when the selected track is~$X$ by the rules of $(mp.ZXs)$. 
The first rule can see that the particle is in~$X$ and the next rule can see that 
it is in~$O$.
\vskip 5pt
\ligne{\hfill
$\vcenter{
\vtop{\hsize=303pt
\ligne{\hfill\tt
\hbox{$\underline{\tt B}$BBBBBWWBWWWWBWWBW$\underline{\tt B}$} 
\hskip 5pt
\hbox{$\underline{\tt B}$BBBBBWWBWWWWWBWBW$\underline{\tt B}$} 
\hfill}
}}$
\hfill$(mp.ZXs)$}

\vskip 7pt
\noindent
$\underline{\hbox{\bf The rules for $I$}}$
\vskip 7pt

    With~$I$, we arrive to the sensors which contribute, with~$Z$, to the
management of the switch. This sensor is in between~$X$ and~$Z$ and it
allows~$Z$ to detect the passage of the particle on the non-selected tracks
when~$Y$ is selected.

   As a cell of the passive memory switch, $I$~receives the number
\hbox{\tt WWB}, so that its context contains the pattern \hbox{\tt BBBBBWWWB}.
whether the state of~$I$ is white or black, the context of~$I$ is always 
\hbox{\tt WBW$^a$BBBBBWWWBWW$^b$WWW}. 
In $(mp.Ib)$, we can see the two corresponding conservative
rules.

\vskip 5pt
\ligne{\hfill
$\vcenter{
\vtop{\hsize=303pt
\ligne{\hfill\tt
\hbox{$\underline{\tt W}$BBBBBWWWBWW$^b$WWWWBW$^a$$\underline{\tt W}$} 
\hskip 5pt
\hbox{$\underline{\tt B}$BBBBBWWWBWW$^b$WWWWBW$^a$$\underline{\tt B}$} 
\hfill}
}}$
\hfill$(mp.Ia)$}
\vskip 5pt
In $(mp.Ib)$ the rules witness the presence of the particle in~$X$, whatever the
state of~$I$ which is unchanged.
\vskip 5pt
\ligne{\hfill
$\vcenter{
\vtop{\hsize=303pt
\ligne{\hfill\tt
\hbox{$\underline{\tt W}$BBBBBWWWBWW$^b$WWW{\small\bf B}BW$^a$$\underline{\tt W}$} 
\hskip 5pt
\hbox{$\underline{\tt B}$BBBBBWWWBWW$^b$WWW{\small\bf B}BW$^a$$\underline{\tt B}$} 
\hfill}
}}$
\hfill$(mp.Ib)$}
\vskip 5pt
At last, we know that $I$ changes its colour, depending on the flash of~$Z$.
The rules are given in $(mp.Ic)$: when $I$~is white it becomes black and
when it is black, it becomes white.
\vskip 5pt
\ligne{\hfill
$\vcenter{
\vtop{\hsize=303pt
\ligne{\hfill\tt
\hbox{$\underline{\tt B}$BBBBBWWWBWW$^b$WWWW{\small\bf W}W$^a$$\underline{\tt W}$} 
\hskip 5pt
\hbox{$\underline{\tt W}$BBBBBWWWBWW$^b$WWWW{\small\bf W}W$^a$$\underline{\tt B}$} 
\hfill}
}}$
\hfill$(mp.Ic)$}

\vskip 7pt
\noindent
$\underline{\hbox{\bf The rules for $J$}}$
\vskip 7pt

   The cell~$J$ plays with~$Y$ and~$Z$ the same role as $I$~plays with~$X$ and~$Z$.
Its number is {\tt BWB}, so that its context contains the pattern
\hbox{\tt BBBBBWBWB}. Accordingly, the context of the cell is now
\hbox{\tt WWW$^a$BBBBBWBWBWW$^b$WWB} whether its state is white or black.

   In $(mp.Ja)$, we have the conservative rules, one for the white state, the
other for the black one. 
\vskip 5pt
\ligne{\hfill
$\vcenter{
\vtop{\hsize=303pt
\ligne{\hfill\tt
\hbox{$\underline{\tt W}$BBBBBWBWBWW$^b$WWBWWW$^a$$\underline{\tt W}$} 
\hskip 5pt
\hbox{$\underline{\tt B}$BBBBBWBWBWW$^b$WWBWWW$^a$$\underline{\tt B}$} 
\hfill}
}}$
\hfill$(mp.Ja)$}
\vskip 5pt
   In $(mp.Jb)$, we have the rules which detects the presence of the particle
in~$Y$, whatever the state of~$J$.
\vskip 5pt
\ligne{\hfill
$\vcenter{
\vtop{\hsize=303pt
\ligne{\hfill\tt
\hbox{$\underline{\tt W}$BBBBBWBWBWW$^b$WWB{\small\bf B}WW$^a$$\underline{\tt W}$} 
\hskip 5pt
\hbox{$\underline{\tt B}$BBBBBWBWBWW$^b$WWB{\small\bf B}WW$^a$$\underline{\tt B}$} 
\hfill}
}}$
\hfill$(mp.Jb)$}
\vskip 5pt
   In $(mp.Jc)$, we have the rules which manage the effect of the flash of~$Z$.
When $Z$~flashes, the particle is in~$O$ but cannot be seen by~$J$. One rule
changes the state from black to white, the other from white to black.
\vskip 5pt
\ligne{\hfill
$\vcenter{
\vtop{\hsize=303pt
\ligne{\hfill\tt
\hbox{$\underline{\tt B}$BBBBBWBWBWW$^b$WW{\small\bf W}WWW$^a$$\underline{\tt W}$} 
\hskip 5pt
\hbox{$\underline{\tt W}$BBBBBWBWBWW$^b$WW{\small\bf W}WWW$^a$$\underline{\tt B}$} 
\hfill}
}}$
\hfill$(mp.Jc)$}

\vskip 7pt
\noindent
$\underline{\hbox{\bf The rules for $Z1$}}$
\vskip 7pt

    When $Z$ flashes, it also sends a signal to the active memory switch in order
to change the states of~$H$ and~$K$, both at the same time. This is performed
by sending a particle along a path from~$Z$ to~$D$. This path crosses twice other
paths thanks to round-abouts as we have seen in Subsections~\ref{scenario}
and~\ref{implement}. Outside its
first and its last cells and outside the round-abouts, the path consists of
ordinary cells of the tracks and of corners. Its first cell is~$Z1$, a neighbour
of~$Z$. Its last cell is~$D1$, a neighbour of~$D$. We presently see the rules
of~$Z1$.

   As a cell of the passive memory switch, its number is {\tt WBB}, so that its
context contains the pattern \hbox{\tt BBBBBWWBB}. The context itself is
\hbox{\tt BWW$^a$BBBBBWWBBWWWW$^b$W}.
   
\vskip 5pt
\ligne{\hfill
$\vcenter{
\vtop{\hsize=303pt
\ligne{\hfill\tt
\hbox{$\underline{\tt W}$BBBBBWWBBWWWW$^b$WBWW$^a$$\underline{\tt W}$} 
\hskip 5pt
\hbox{$\underline{\tt W}$BBBBBWWBBWWWW$^b$W{\small\bf W}WW$^a$$\underline{\tt B}$} 
\hfill}
\ligne{\hfill\tt
\hbox{$\underline{\tt \small\bf B}$BBBBBWWBBWWWW$^b$WBWW$^a$$\underline{\tt W}$} 
\hskip 5pt
\hbox{$\underline{\tt W}$BBBBBWWBBW{\small\bf B}WW$^b$WBWW$^a$$\underline{\tt W}$} 
\hfill}
}}$
\hfill$(mp.Z1)$}
\vskip 5pt

   The rules for~$Z1$ are displayed by $(mp.Z1)$. The first rule is conservative
and the second one reacts to the flash of~$Z$ by changing the state of~$Z1$ from
white to black. The third rule turns back the sate of~$Z1$ to white. Now, the motion
rules of the neighbour of~$Z1$ on the tracks has taken the particle emitted
by~$Z$: this is witnessed by the fourth rule which can see the particle in this
neighbour.

\vskip 7pt
\noindent
$\underline{\hbox{\bf The rules for $D1$}}$
\vskip 7pt

   Just before reaching~$D$~of the active memory switch, the particle sent by~$Z$
passes through~$D1$, the neighbour of~$D$ which is on the path from~$Z$ to~$D$.
And so, when the particle is in~$D$, the next time $D$~flashes, see the
rules in $(ma.Dc)$.

   The cell~$D1$ has the number {\tt BBB} so that its contexts contains the
pattern \hbox{\tt BBBBBWBBB}. The context is:
\hbox{\tt BBW$^a$BBBBBWBBBWWWW$^b$W}, with \hbox{$a,b\geq 1$}.
The rules are displayed in $(mp.D1)$.

   We can see that the first rule is a conservative rule for~$D1$. The second rule
applies when $D1$~can see a particle in its neighbouring cell of the tracks: it
changes the state of~$D1$ to black. The third rule restores the white state of~$D1$.
The fourth rule can witness the fact that $D$ is flashing as it has seen $D1$ black
the time before.
\vskip 5pt
\ligne{\hfill
$\vcenter{
\vtop{\hsize=303pt
\ligne{\hfill\tt
\hbox{$\underline{\tt W}$BBBBBWBBBWWWW$^b$WBBW$^a$$\underline{\tt W}$} 
\hskip 5pt
\hbox{$\underline{\tt W}$BBBBBWBBBW{\small\bf B}WW$^b$WBBW$^a$$\underline{\tt B}$} 
\hfill}
\ligne{\hfill\tt
\hbox{$\underline{\tt \small\bf B}$BBBBBWBBBWBWW$^b$WBBW$^a$$\underline{\tt W}$} 
\hskip 5pt
\hbox{$\underline{\tt W}$BBBBBWBBBWBWW$^b$W{\small\bf W}BW$^a$$\underline{\tt W}$} 
\hfill}
}}$
\hfill$(mp.D1)$}
\vskip 5pt

   The effect of the flash of~$D$ on~$H$ and~$K$ has been studied on the
rules for these cells, namely in $(ff.H)$ and in $(ff.K)$. Now, this
flash of~$D$ requires a rule which was not present in those of~$B$ and~$C$
in the flip-flop. Indeed, when $B$ or~$C$ can see the flash of~$D$ in the flip-flop,
the particle can also be seen by the cell, see the last rule of $(ff.Bb)$
and the last one of $(ff.Cb)$. Here, in the memory switch, when the flash of~$Z$ 
eventually reaches~$D$, there is no particle in the switch so that when the flash
of~$D$ occurs, there is no particle seen by $B$ nor~$C$. And so, this requires
the rules of $(ma.BC)$. The left-hand side rule applies to~$B$ and the
right-hand side one applies to~$C$.

\vskip 5pt
\ligne{\hfill
$\vcenter{
\vtop{\hsize=303pt
\ligne{\hfill\tt
\hbox{$\underline{\tt W}$BBBBWBBWWBWW$^b$WWWWW$^a$$\underline{\tt W}$} 
\hskip 5pt
\hbox{$\underline{\tt W}$BBBBWBBBWW$^b$WWWWWWW$^a$$\underline{\tt W}$} 
\hfill}
}}$
\hfill$(ma.BC)$}
\vskip 5pt

\ifnum 1=0 {

} \fi

\vskip 10pt
   With these last rules, we have examined all possible cases. Accordingly,
   this
 completes the proof of Theorem~\ref{universal}. \cqfd 

\section{Conclusion}

   We have now reached the minimal number of states in order to get a universal
cellular automaton on the heptagrid with a true planar cellular automaton. 
However, we have not exactly all possible tessellations of the hyperbolic plane.
A few of them are missing in the $\{p,3\}$ family: the values of~$p$
from~7 up to~12, which means 6~cases. Moreover, the question arises of what can 
be said for tilings $\{p,4\}$ which are tightly connected to the tilings
$\{p,3\}$: we now that $\{p,4\}$ and $\{p$+$2,3\}$ have the same spanning tree,
see~\cite{mmbook1} where other references can also be found.


   Accordingly, we remain with some hard work ahead.



\end{document}